\documentclass[aps,prd,twocolumn,showpacs,superscriptaddress,groupedaddress]{revtex4}
\usepackage{ads}
\usepackage{jjzhang}
\usepackage{xcolor}
\usepackage{graphicx,color,bm}

\usepackage{amsmath,amssymb}
\usepackage[colorlinks,linkcolor=blue,citecolor=blue]{hyperref}
\usepackage{ulem}
\usepackage{array}
\usepackage{verbatim}
\usepackage[utf8]{inputenc}
\usepackage{rotating}
\newcolumntype{K}[1]{>{\centering\arraybackslash}m{#1}}

\usepackage{soul}

\begin{document}

\title{A study on Cubic Galileon Gravity Using N-body Simulations}

\author{Jiajun Zhang}
\email{liambx@ibs.re.kr}
\affiliation{Center for Theoretical Physics of the Universe, Institute for Basic Science (IBS), Daejeon 34126, Korea}
\author{Bikash R. Dinda}
\email{bikashd18@gmail.com} 
\affiliation{Department of Theoretical Physics, Tata Institute of Fundamental Research,\\Dr Homi Bhabha Road, Navy Nagar, Colaba, Mumbai-400005, India.}
\author{Md. Wali Hossain}
\email{wali.h@knu.ac.kr}
\affiliation{Center for High Energy Physics, Kyungpook National University, Daegu, Korea}
\author{Anjan A. Sen}
\email{aasen@jmi.ac.in}
\affiliation{Centre For Theoretical Physics, Jamia Millia Islamia, New Delhi, 110025, India}
\author{Wentao Luo}
\email{wentao.luo@ipmu.jp} 
\affiliation{Kavli Institute for the Physics and Mathematics of the Universe (Kavli IPMU, WPI), University of Tokyo, Chiba 277-8582, Japan}

\begin{abstract}
We use N-body simulation to study the structure formation in the Cubic Galileon Gravity model where along with the usual kinetic and potential term we also have a higher derivative self-interaction term. We find that the large scale structure provides a unique constraining power for this model. The matter power spectrum, halo mass function, galaxy-galaxy weak lensing signal, marked density power spectrum as well as count in cell are measured. The simulations show that there are less massive halos in the Cubic Galileon Gravity model than corresponding $\Lambda$CDM model and the marked density power spectrum in these two models are different by more than $10\%$. Furthermore, the Cubic Galileon model shows significant differences in voids compared to $\Lambda$CDM. The number of low density cells is far higher in the Cubic Galileon model than that in the $\Lambda$CDM model. Therefore, it would be interesting to put constraints on this model using future large scale structure observations, especially in void regions. 
\end{abstract}
\date{\today}

\maketitle
\section{Introduction}\label{sec:introduction}
Since the first observational evidence for late time acceleration in our Universe was confirmed in 1998 \cite{Riess:1998cb,Perlmutter:1998np,Tonry:2003zg}, we are still in search for a correct theoretical model that can explain this accelerated expansion as well as is also consistent with hosts of different cosmological observations. Although the simplest concordance $\Lambda$CDM model \cite{Sahni:1999gb} has been successful in both these counts, but the latest tension (which is currently at more than $4\sigma$ \cite{Riess:2020sih}) in measurements of Hubble constant $H_{0}$ from local observations \cite{Riess:2019cxk,Wong:2019kwg,Pesce:2020xfe} and from CMB by Planck \cite{Aghanim:2018eyx}, lands $\Lambda$CDM model in serious trouble. In simple words, the constrained value of $H_{0}$ parameter (Hubble constant at $z=0$) for  $\Lambda$CDM model by Planck observation for CMB \cite{Aghanim:2018eyx} is more than $4\sigma$ away from the model independent local measurements by Riess et al \cite{Riess:2019cxk}. Recently, this has resulted renewed interests in models beyond $\Lambda$CDM.

To construct models beyond $\Lambda$CDM that can explain the late time acceleration in the Universe, one can approach in two different ways. The first approach is to modify the energy content in the Universe to include an unknown component with negative pressure called "dark energy". Scalar fields that are ubiquitous in standard model for particle physics, are the most suitable candidates for dark energy \cite{Wetterich:1987fk,Wetterich:1987fm,Copeland:2006wr}. With sufficiently flat potentials, they can mimic the negative pressure that can result the repulsive gravity to start late time acceleration in the Universe. Although this approach works at the phenomenological level to explain late time acceleration, we are still in search for scalar fields with suitable potentials that can arise in standard models for particle physics or its various extensions. 
Also ensuring that these scalar fields do not give rise to fifth force effects that spoil the local gravity constraints, is equally challenging.

The second approach is to modify the gravity at large cosmological scale in such a way so that it becomes repulsive at large scales resulting accelerated cosmological expansion \cite{Clifton:2011jh,deRham:2014zqa,deRham:2012az,DeFelice:2010aj}. One of such attempt was made by Dvali, Gabadadze and Porrati (DGP) where a 4D Minkowsky brane is located on
an infinitely large extra dimension and gravity is localized in the 4D Minkowsky brane \cite{Dvali:2000hr}. Even though this scenario gives rise to late time acceleration its self-accelerating branch has a ghost \cite{Luty:2003vm,Nicolis:2004qq}. But the decoupling limit of the DGP model gives rise to a Lagrangian of the form $(\nabla \phi)^2 \Box \phi$ \cite{Luty:2003vm}. Despite of having higher order term this Lagrangian gives second order equation of motion and hence free from ghost \cite{Luty:2003vm,Nicolis:2004qq,Nicolis:2008in}. This Lagrangian, in the Minkowski background, possesses the Galilean shift symmetry $\phi\to\phi+b_\mu x^\mu+c$, whre $b_\mu$ and $c$ are the constants, and hence dubbed as the "Galileon" \cite{Nicolis:2008in}. In the Minkowski background there exists five such terms including the usual canonical kinetic term and a linear term in $\phi$ which can possess the above mentioned shift symmetry and give second order equation of motion \cite{Nicolis:2008in}. In curved background we need to include some nonminimal terms in the Galileon Lagrangian to keep the equation of motion second order\cite{Deffayet:2009wt}. Galileon models can be realized as the sub-classes of the more general scalar-tensor theory known as the Horndeski theory \cite{Horndeski:1974wa} and can give rise to late time cosmic acceleration \cite{Chow:2009fm,Silva:2009km,Kobayashi:2010wa,Kobayashi:2009wr,Gannouji:2010au,DeFelice:2010gb,DeFelice:2010pv,Ali:2010gr,Mota:2010bs,Deffayet:2010qz,deRham:2010tw,deRham:2011by,Hossain:2012qm,Ali:2012cv} while being consistent with the local astrophysical bounds by implementing the Vainshtein mechanism \cite{Vainshtein:1972sx} which suppresses the fifth force locally.

The detection of the event of binary neutron star merger GW170817, using both gravitational waves (GW) \cite{TheLIGOScientific:2017qsa} as well as its electromagnetic counterpart \cite{Monitor:2017mdv,GBM:2017lvd} rules out a large class of Horndeski theories that predicts the speed of GW propagation different from that of speed of light \cite{Ezquiaga:2017ekz,Zumalacarregui:2020cjh}. In Galileon models, the only higher derivative term that survives is $(\nabla \phi)^2 \Box \phi$, the cubic term in the Galileon Lagrangian which does not modify the speed of GW. This cubic term along with the usual kinetic term and the term linear in $\phi$ (linear potential) forms the Cubic Galileon model. Replacing the linear potential with a general potential breaks the shift symmetry but still the eqaution of motion is second order. This kind of models are known as the Light Mass Galileon models \cite{Hossain:2012qm,Ali:2012cv}. The Cubic Galileon model without potential can not give rise to a stable late time acceleration \cite{Gannouji:2010au}. The Cubic Galileon model has been studied extensively in the context of late time acceleration \cite{Chow:2009fm,Silva:2009km,Hossain:2012qm,Ali:2012cv,Brahma:2019kch} in the Universe as well as in the context of growth of matter fluctuations in both sub-horizon and super-horizon scales \cite{Bartolo:2013ws,Bellini:2013hea,Barreira:2013eea,Hossain:2017ica,Dinda:2017lpz}. The current constraints and models of modified gravity is well summarized in \citet{ishakreview}.

Although the background expansion and growth of linear fluctuations of the matter density field have been extensively studied in Cubic Galileon model, a detail analysis of structure formation in nonlinear regime using N-body simulations is necessary to study evolution of voids and clusters in this model and to compare them with the prediction from $\Lambda$CDM model. It has been proved that N-body simulation is essential to investigate the structure formation and put constraints on modified gravity models like $f(R)$ gravity model \citep{He:2018nature} or Interacting Dark Energy models \citep{zhang2019apjl,an2019mnras}. The deeply nonlinear structure formation process disclosed by the N-body simulation provides the accurate prediction of large scale structures, which can be used to compare with observations like SDSS \citep{sdss7,luo2017apj}.

The nonlinear structure formation of Cubic Galileon model using N-body simulation has been studied without potential\citep{Barreira:2013eea,barreira2014jcap}. However, a further study into the Cubic Galileon model with a potential is still lack of nonlinear investigation.
Using ME-Gadget code\citep{megadget}\footnote{the public version of ME-Gadget is available at https://github.com/liambx/ME-Gadget-public}, we investigate the Cubic Galileon model using N-body simulation and study the large scale structure in this model. A comparison between the simulation results of Cubic Galileon model and $\Lambda$CDM model will allow us to locate our future focusing point when trying to get constraints from observations. As we are expecting a large class accurate data from different future surveys like, LSST \citep{lsst}, Euclid \citep{euclid1,euclid2}, DESI \citep{desilegacy}, JPAS \citep{jpas1,jpas2} and others, such study is particularly relevant for any viable modified gravity models.

The background expansion calculation is introduced in Sec.\ref{sec:background}. The perturbation calculation is introduced in Sec.\ref{sec:perturbation}, including the linear perturbation equations for each components and the linear matter power spectrum results. In Sec.\ref{sec:nbody}, we explained the simulations we have set for comparison in the analysis. We show the results of the simulations in Sec.\ref{sec:result}, including the density field, matter power spectrum, marked density, halo mass function, count in cell and galaxy-galaxy lensing. Finally, we give the conclusion in Sec.\ref{sec:conclusion}. In summary, we have found that voids is more important than we expected and it might be the focus for our future work.
\section{Background Cosmology}\label{sec:background}

To study the background and perturbation history of the Universe, we consider Cubic Galileon model. The evolutionary dynamics of the Cubic Galileon field, $\phi$ is described by the action given by \citep{Hossain:2012qm,Ali:2012cv}

\begin{eqnarray}
S &=& \int d^4x\sqrt{-g}\Bigl [\frac{M^2_{\rm{pl}}}{2} R -  \frac{1}{2}(\nabla \phi)^2\Bigl(1 + \beta \Box \phi\Bigr) - V(\phi) \Bigr] \nonumber\\
&& + \mathcal{S}_m \, ,
\label{eq:action}
\end{eqnarray}

\noindent
where $M_{\rm pl}$ is the reduced Planck mass. $g$ is the determinant of the metric describing the Universe. $R$ is the corresponding Ricci scalar. $\mathcal{S}_m$ is the action for the total matter counterpart. The action \eqref{eq:action} is a subclass of a more general action namely the Horndeski action \cite{Horndeski:1974wa}. $V(\phi)$ is the potential of the Galileon field. Here, we consider only linear potential which is the case for the original Galileon model. $\beta$ is a cubic Galileon parameter (for more details see Appendix~\ref{sec-actioncubgal}). For $\beta = 0$ the action \eqref{eq:action} reduces to the standard quintessence action with linear potential \citep{Wetterich:1987fk,Wetterich:1987fm,Caldwell:2005tm,Linder:2006sv,Tsujikawa:2010sc,Scherrer:2007pu,Dinda:2016ibo}.

For the background cosmology, we consider flat FRW metric given by $ds^{2} = - dt^{2} + a^{2} (t) d\vec{r}.d\vec{r}$, where $t$ is the cosmic time, $\vec{r}$ is the comoving coordinate vector and $a$ is the cosmic scale factor. Varying the action \eqref{eq:action} with respect to the metric, the background Einstein equations become
\begin{eqnarray}
3M_{\rm pl}^2H^2 &=& \bar{\rho}_{\rm m}+\frac{\dot{\phi}^2}{2}\Bigl(1-6\beta H\dot{\phi}\Bigr)+V{(\phi)},
\label{eq:first_Friedmanonumber} \\
M_{\rm pl}^2(2\dot H + 3H^2) &=& -\frac{\dot{\phi}^2}{2}\Bigl(1+2\beta \ddot{\phi}\Bigr)+V(\phi),
\label{eq:second_Friedmanonumber}
\end{eqnarray}
 
\noindent
where overdot is the derivative with respect to the cosmic time $t$. $H$ is the Hubble parameter. $ \bar{\rho}_{\rm m} $ is the background matter energy density. The background Euler-Lagrangian equation for the Galileon field, $ \phi $ is given by
\begin{equation}
\ddot{\phi} + 3H\dot{\phi}-3\beta \dot{\phi}\Bigl(3H^2\dot{\phi}+\dot{H}\dot{\phi}+2H\ddot{\phi}\Bigr)+ V_{\phi}=0,
\label{eq:E-L_eq}
\end{equation}

\noindent
where subscript $\phi$ is the derivative with respect to the field $\phi$. Note that for the simplicity of the notation, we have considered same $\phi$ as the background field.
\\
All the above-mentioned equations can be rewritten in a system of differential equations with respect to some dimensionless quantities given by \citep{Dinda:2017lpz,Hossain:2012qm,Hossain:2017ica,Ali:2012cv,Dinda:2018eyt}

\begin{eqnarray}
x &=& \frac{\Big{(} \dfrac{d \phi}{d N} \Big{)}}{\sqrt{6} M_{\rm pl}}, \hspace{0.3 cm}  y = \frac{\sqrt{V}}{\sqrt{3} H M_{\rm pl}}, \hspace{0.3 cm} \epsilon = -6 \beta H^{2} \Big{(} \dfrac{d \phi}{d N} \Big{)}, \nonumber\\
\lambda &=& - M_{\rm pl} \frac{V_{\phi}}{V}, \hspace{0.3 cm} \text{with} \hspace{0.3 cm}  \Gamma = V \frac{V_{\phi \phi}}{V_{\phi}^{2}} = 0\hspace{0.2 cm} (\text{Here}),
\label{eq:dimless_var_bkg}
\end{eqnarray}

\noindent
where $N=\ln \hspace{0.02 cm} a$ is the number of e-foldings. The expressions for the system of differential equations can be found in Appendix~\ref{sec-autosys} (see first to fourth lines in Eq.~\eqref{eq:dynsys}). To solve all the differential equation, we consider initial conditions at an initial redshift, $z=z_{i}=49$. The subscipt, $i$ represents the initial value (at $z_{i}=49$) corresponding to a quantity. Among all the quantities in Eq.~\eqref{eq:dimless_var_bkg}, the $\epsilon$ (or $\epsilon_{i}$ i.e. the initial value of it) quantifies the difference between cubic galileon and quintessence. So, in all our subsequent sections, we vary only $\epsilon_{i}$ parameter keeping all the other parameters fixed accordingly. For the details of the initial conditions, see Appendix~\ref{sec-initial} (see point no. 1 to 4).

The expressions for some relevant background quantities are given by

\begin{eqnarray}
w_{\phi} &=& \frac{3 x^2 (\epsilon  (\epsilon +8)+4)-2 \sqrt{6} \lambda  x y^2 \epsilon -12 y^2 (\epsilon +1)}{3 \left(\epsilon  \left(x^2 \epsilon +4\right)+4\right) \left(x^2 (\epsilon +1)+y^2\right)}, \nonumber\\
\Omega_{\phi} &=& x^2 (\epsilon +1)+y^2, \nonumber\\
\Omega_{m} &=& 1 - \Omega_{\phi}, \nonumber\\
H^{2} &=& H_{0}^{2} \frac{\Omega_{m}^{(0)} (1+ z)^{3}}{\Omega_{m}},
\label{eq:imp_bkg_qnt}
\end{eqnarray}

\noindent
where $w_{\phi}$ is the equation of state of the Galileon field. $\Omega_{m}$ is the energy density parameter of the total matter and $\Omega_{m}^{(0)}$ is its present value. $\Omega_{\phi}$ is the energy density parameter of the Galileon field.

\begin{figure}[tbp]
\centering
\includegraphics[width=.45\textwidth]{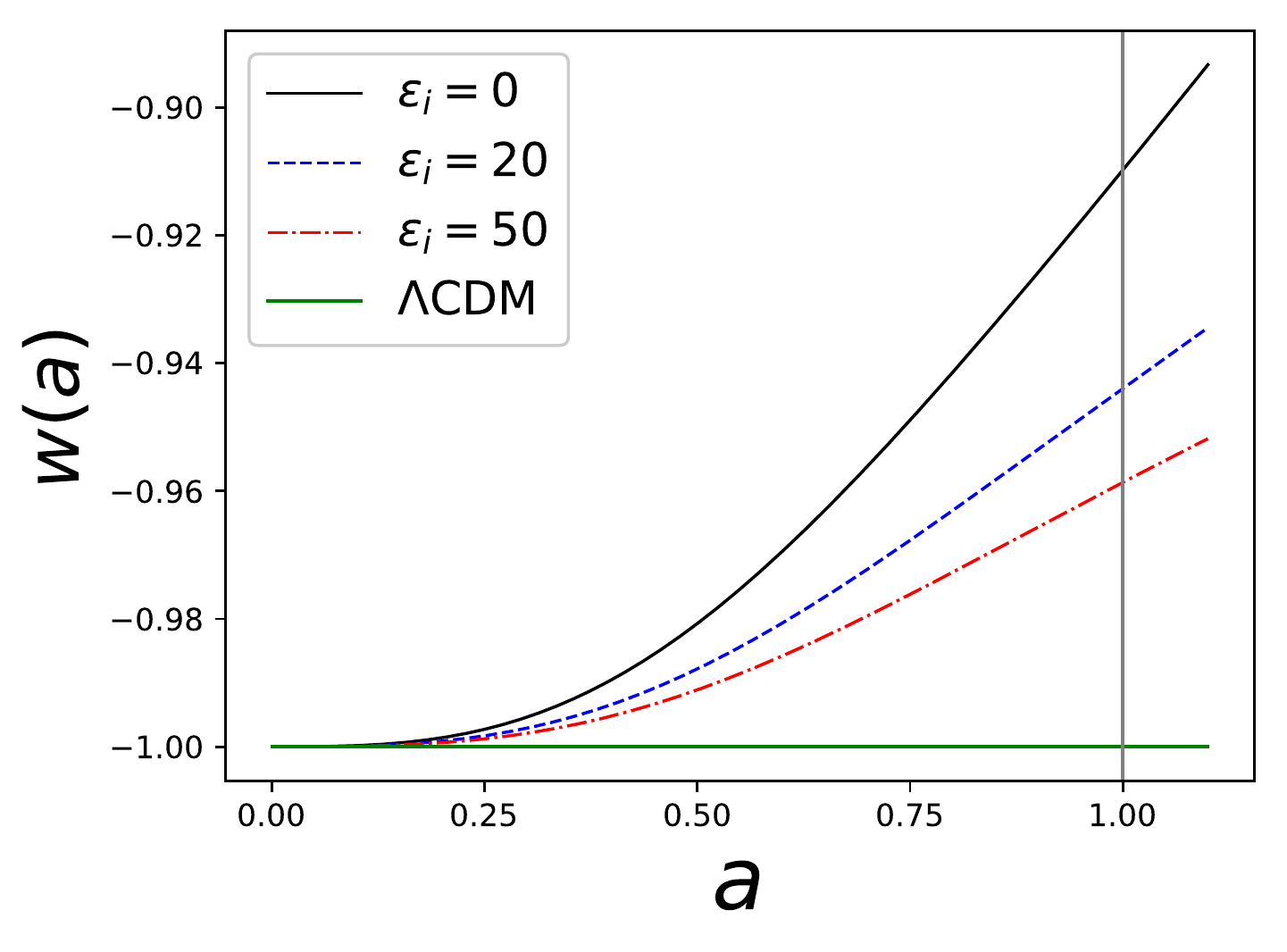}
\caption{\label{fig:eosw} Behaviour of the Equation of state ($w_{\phi}$) of the Cubic Galileon field as a function of the scale factor ($a$) for different $\epsilon_{i}$. We can see that, irrespective of values of $\epsilon_{i}$, $w_{\phi}\approx -1$ at early times ($a\ll1$). At late times ($a\approx1$), the equation of state becomes non-phantom ($w_{\phi}>-1$). The value of $w_{\phi}$ is the largest for the quintessence model ($\epsilon_{i}=0$). The value of $w_{\phi}$ decrease with increasing $\epsilon_{i}$ and finally approach towards cosmological constant behaviour ($w_{\phi}=-1$) for very high value of $\epsilon_{i}$.}
\end{figure}

\begin{figure}[tbp]
\centering
\includegraphics[width=.45\textwidth]{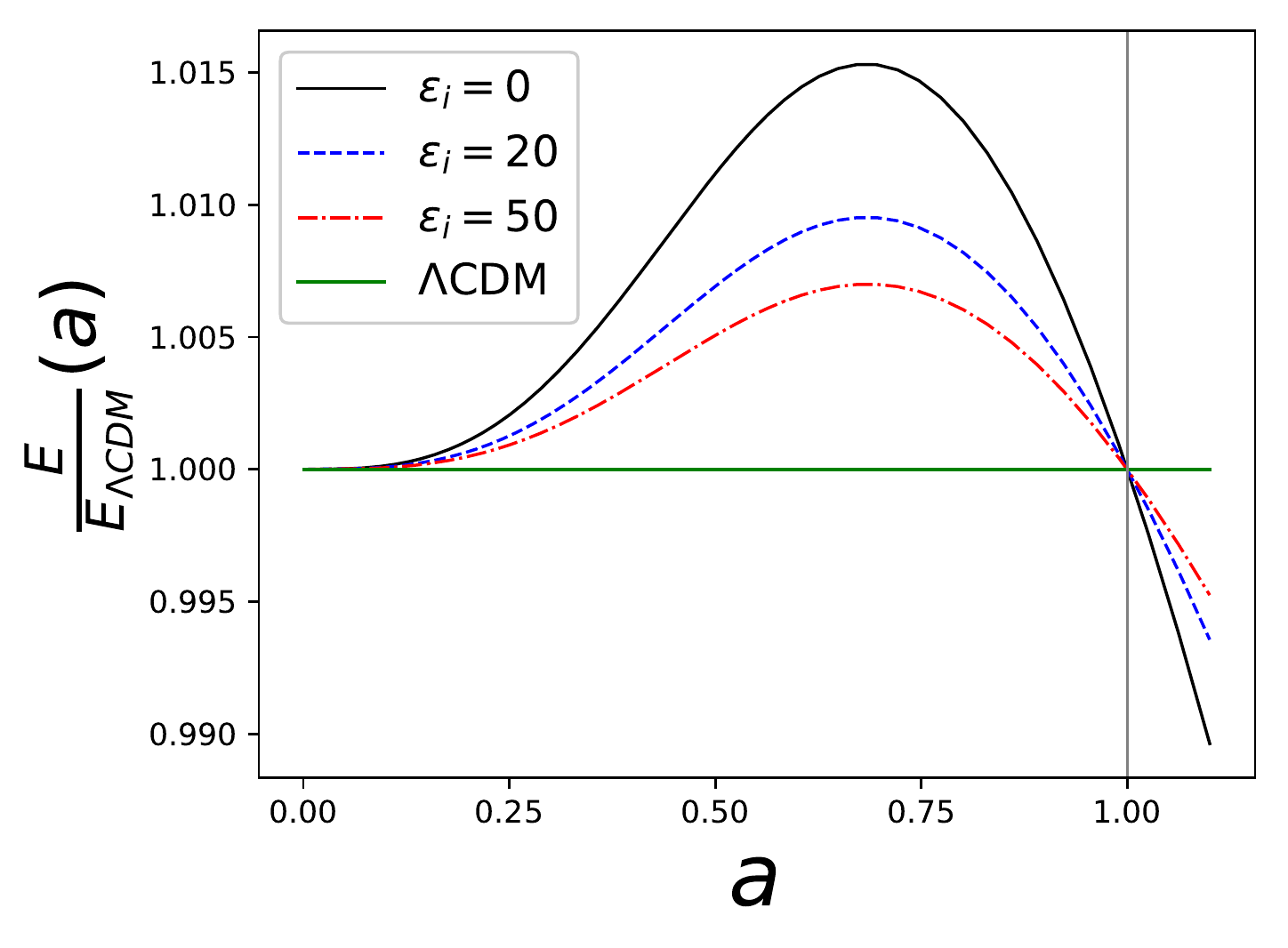}
\caption{\label{fig:Eofa} Behaviour of the normalized Hubble parameter ($E$) as a function of the scale factor (a) for different $\epsilon_{i}$. Similar to the Fig.~\ref{fig:eosw}, the deviation in $E$ from the $\Lambda$CDM model is the highest for $\epsilon_{i}=0$. The deviations decrease with increasing $\epsilon_{i}$.}
\end{figure}

\noindent
In Fig.~\ref{fig:eosw}, we have plotted the Equation of state ($w_{\phi}$) of the Cubic Galileon field as a function of the scale factor ($a$) for different $\epsilon_{i}$. Black (solid), blue (dashed) and red (dashed-dotted) lines are for $\epsilon_{i}$ values $0$, $20$ and $50$ respectively. The horizontal green (solid) line is for the corresponding value in $\Lambda$CDM model. We can see that, irrespective of values of $\epsilon_{i}$, $w_{\phi}\approx -1$ at early times ($a\ll1$). At late times ($a\approx1$), the equation of state becomes non-phantom ($w_{\phi}>-1$). This should be the case as we have chosen the thawing class of initial conditions (discussed in the Subsection~\ref{sec-initial}). The value of $w_{\phi}$ is the largest for the quintessence model ($\epsilon_{i}=0$). The value of $w_{\phi}$ decrease with increasing $\epsilon_{i}$ and finally approach towards cosmological constant behaviour ($w_{\phi}=-1$) for very high value of $\epsilon_{i}$.

\noindent
In Fig.~\ref{fig:Eofa}, we have plotted the normalized Hubble parameter ($E=H/H_{0}$ with $H_{0}$ being the present day ($z=0$ or $a=1$) Hubble constant.) as a function of the scale factor (a) for different $\epsilon_{i}$. Colour codes are same as in Fig.~\ref{fig:eosw}. Similar to the Fig.~\ref{fig:eosw}, the deviation in $E$ from the $\Lambda$CDM model is the highest for $\epsilon_{i}=0$. The deviations decrease with increasing $\epsilon_{i}$.

\begin{figure*}
\includegraphics[width=0.45\textwidth]{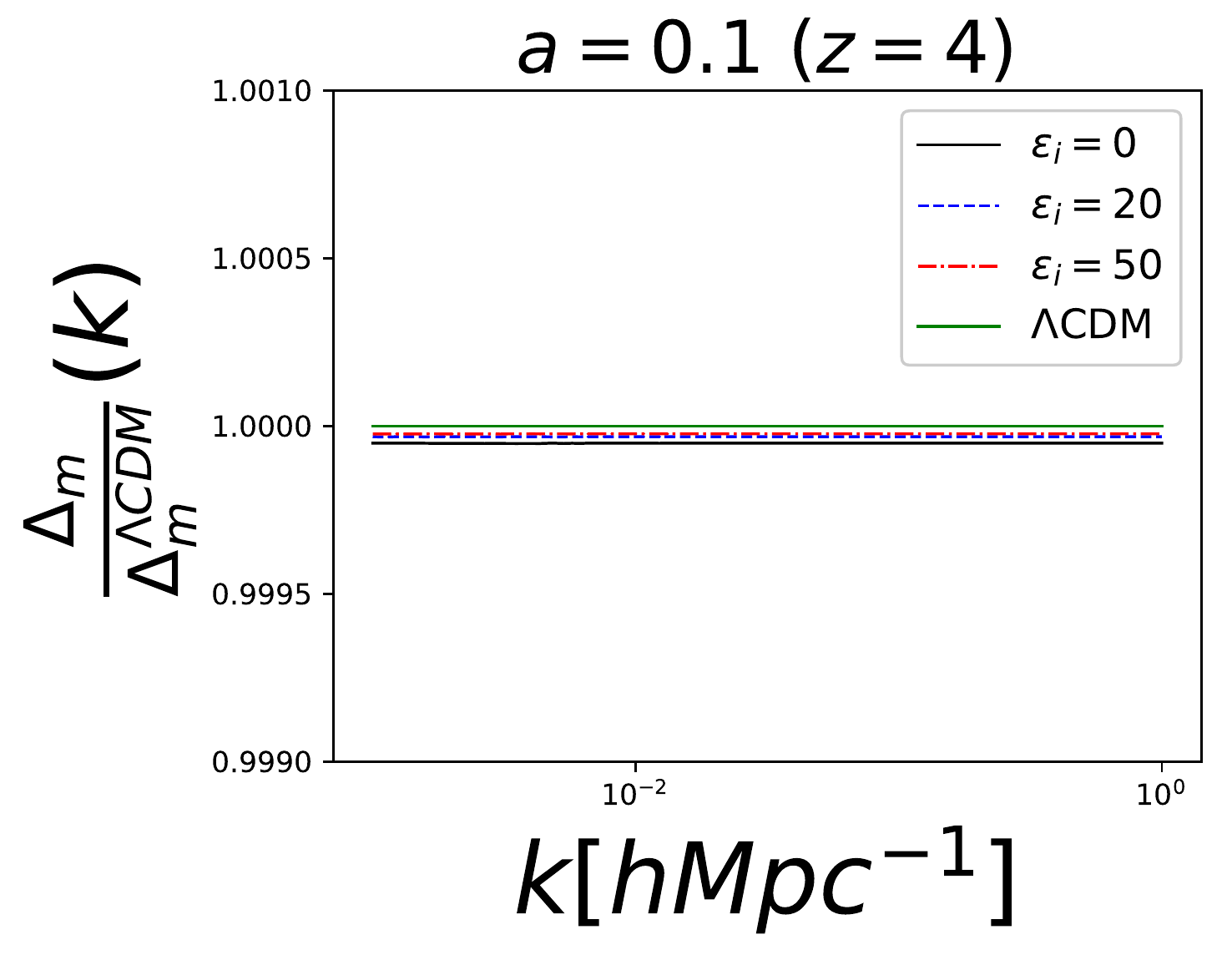}
\includegraphics[width=0.45\textwidth]{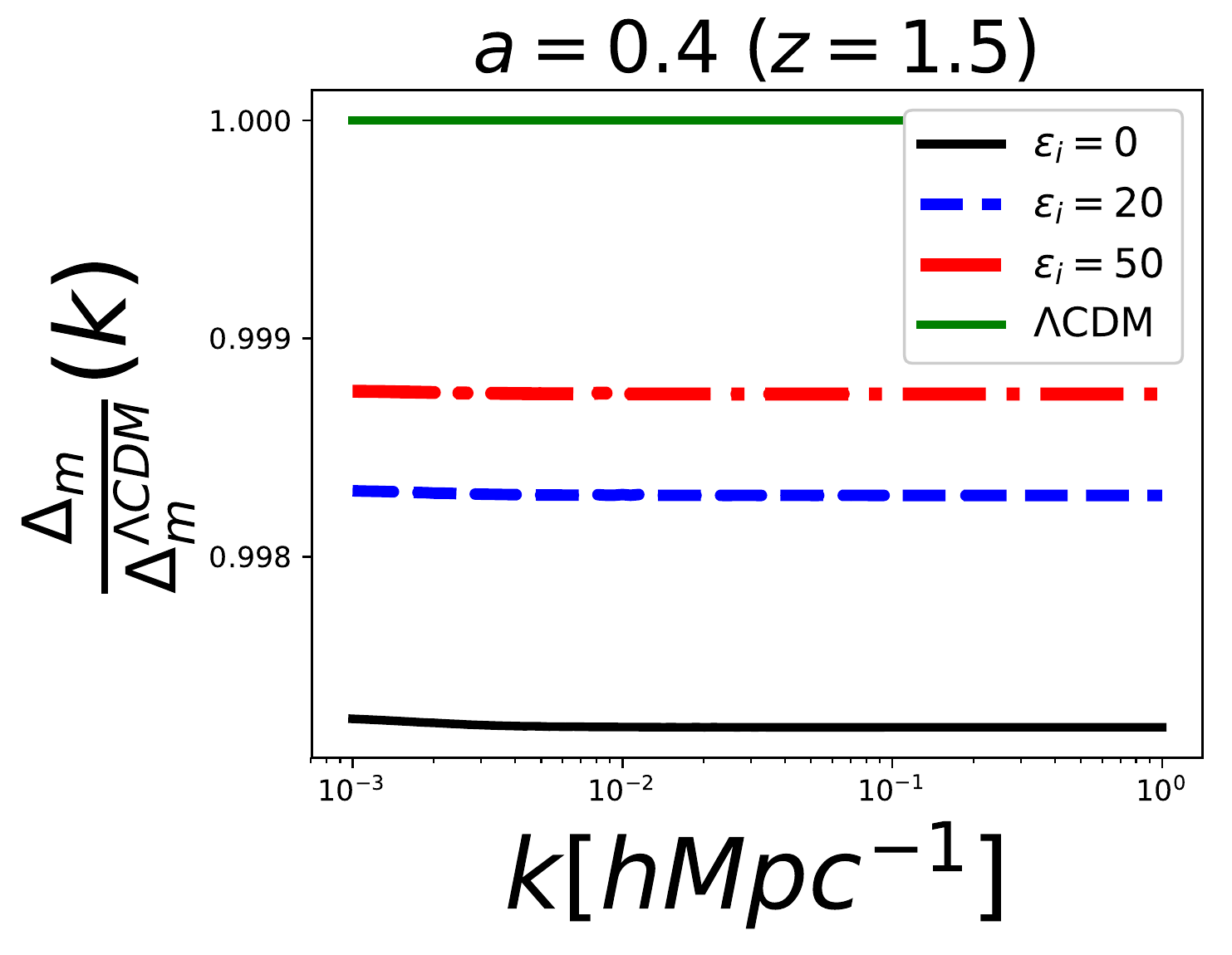}\\
\includegraphics[width=0.45\textwidth]{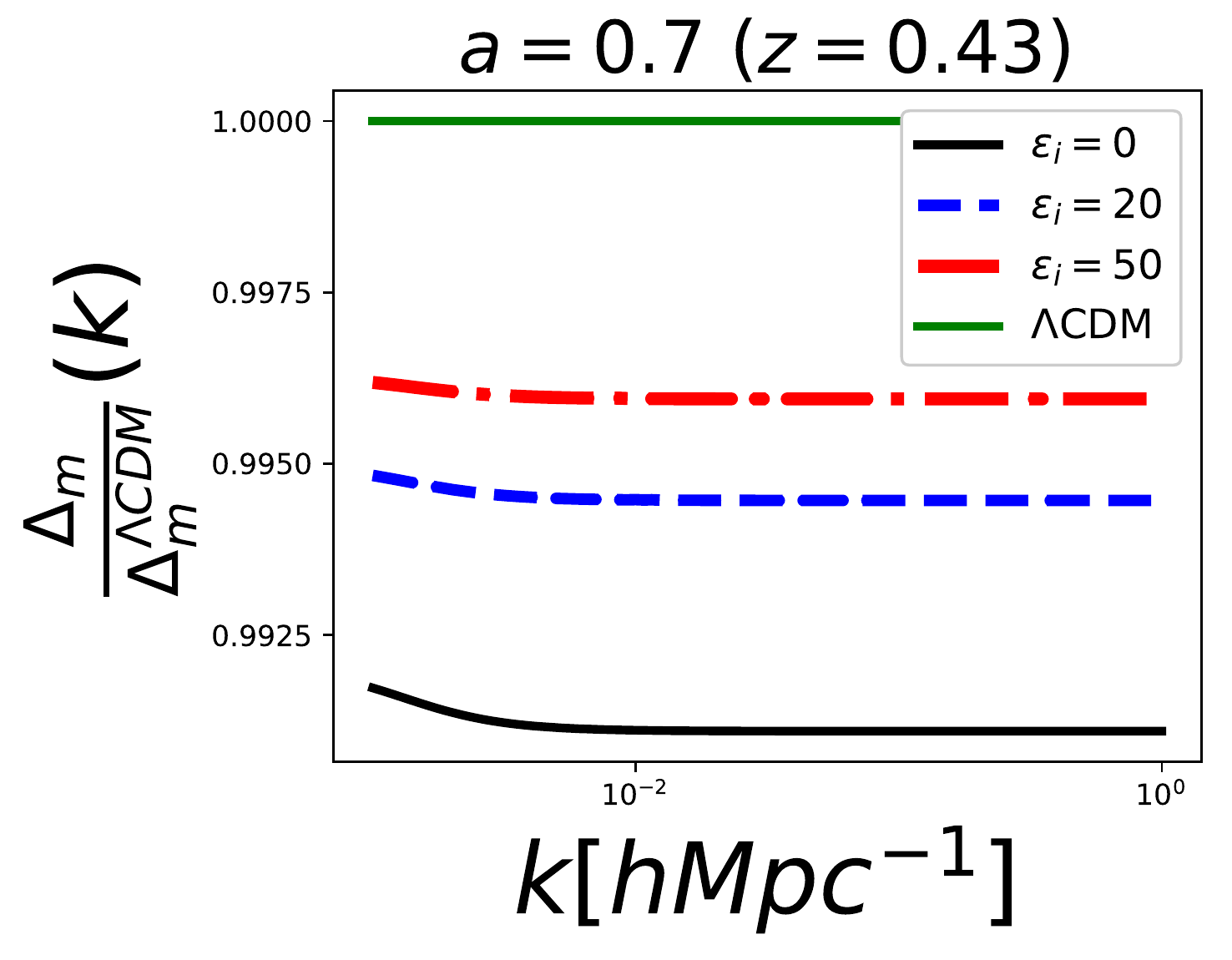}
\includegraphics[width=0.45\textwidth]{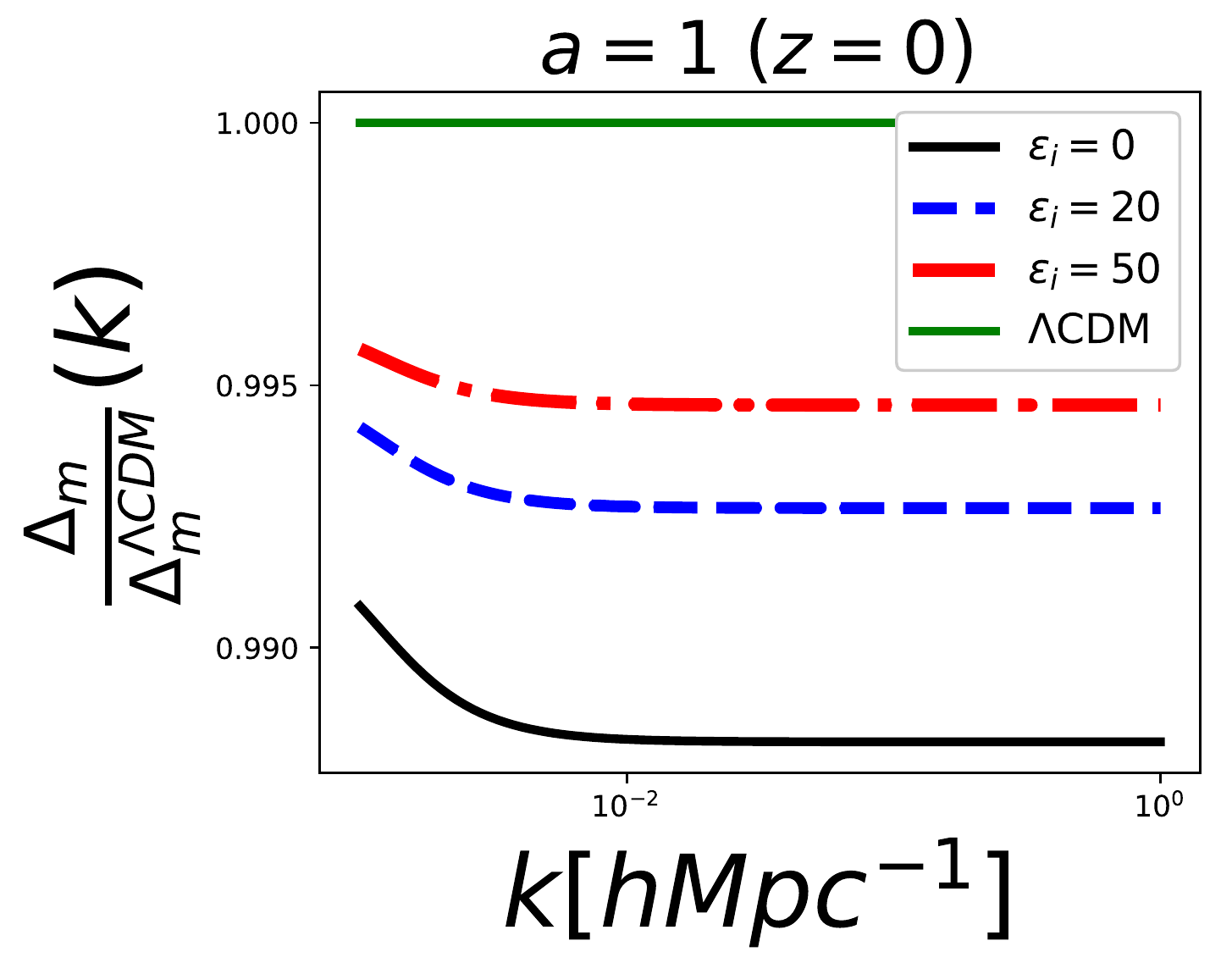}
\caption{Behaviour of the comoving matter energy density contrast ($\Delta_{m}$) as function of wave number ($k$) at different redshifts ($z$) for different $ \epsilon_{i} $. The deviations in $\Delta_{m}$ from $\Lambda$CDM model is the highest at present ($z=0$) for a particular $\epsilon_{i}$ value. The deviations decrease with increasing redshifts. At a particular redshift, the deviation is the highest for $\epsilon_{i}=0$ and decreases with increasing $\epsilon_{i}$.}
\label{fig:Delmk}
\end{figure*}

\begin{figure}[tbp]
\centering
\includegraphics[width=.45\textwidth]{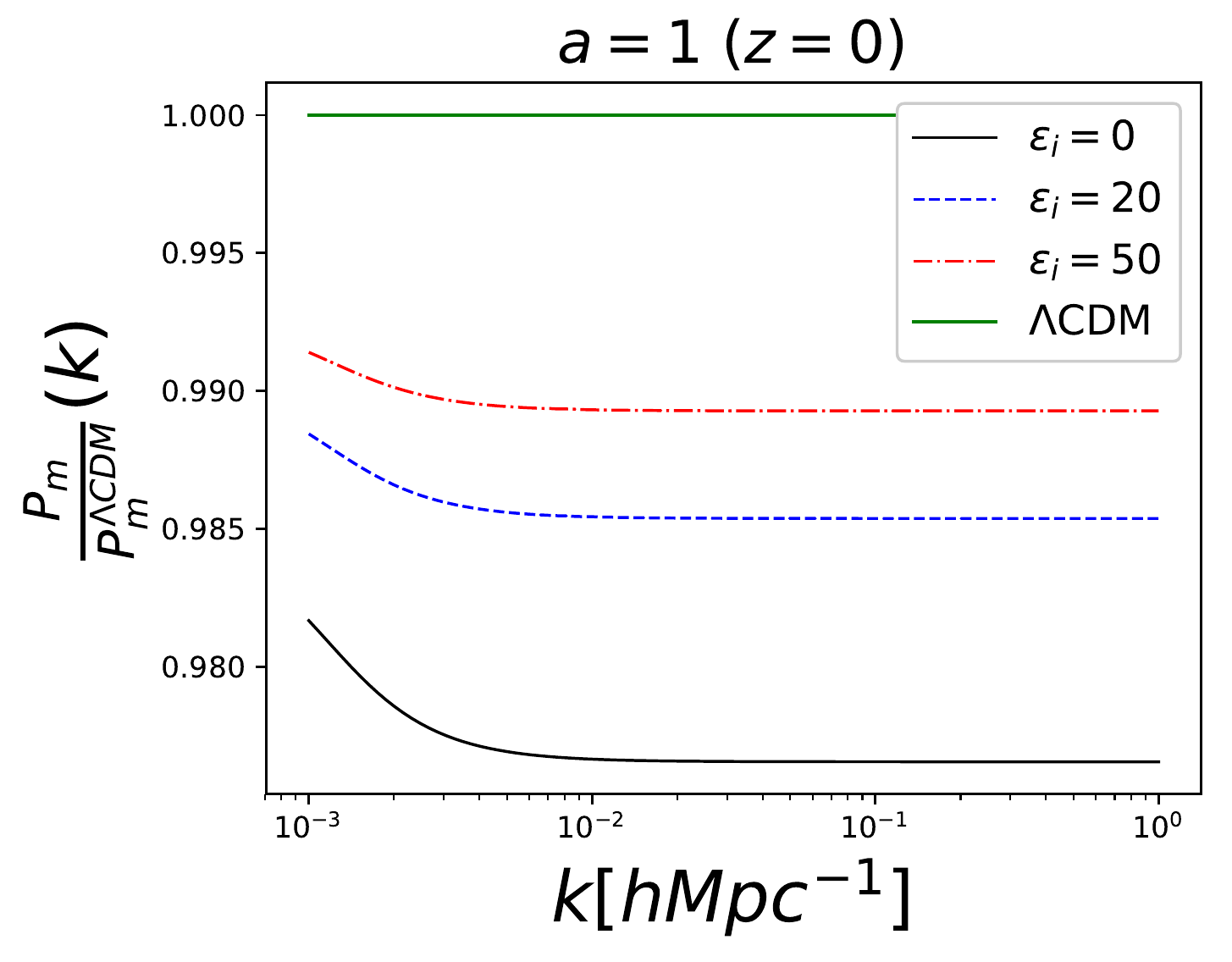}
\caption{\label{fig:Pmz0} Deviations in the linear matter power spectrum for Cubic Galileon models from $\Lambda$CDM model as a function of wave number ($k$) at $z=0$ for different $\epsilon_{i}$. The deviation is the highest for $\epsilon_{i}=0$. The deviations decrease with increasing $\epsilon_{i}$. This behaviour is consistent with the bottom-right panel of the Fig.~\ref{fig:Delmk}.}
\end{figure}

\section{Perturbation Calculation}\label{sec:perturbation}

In the linear perturbation theory, the scalar perturbations can be studied independently with two scalar degrees of freedom. We consider conformal Newtonian gauge, in which the perturbed space-time is given by

\begin{equation}
ds^{2} = (1+2\Psi) dt^{2} - a(t)^{2}(1-2\Phi) d\vec{r}.d\vec{r},
\label{eq:sptm}
\end{equation}

\noindent
where $\Phi$ is the gravitational potential. $\Psi$ is an another scalar potential. For Cubic Galileon, there is no gravitational slip {\it i.e.} $ \Psi = \Phi $ in the Fourier space \cite{Hossain:2017ica}. So, we are left with one scalar degree of freedom which is $ \Phi $. All the relevant perturbation equations are mentioned in Appendix~\ref{sec-detailed_perturbation}.

Similar to the background case, the perturbation equations can also be written in a system of dynamical differential equations (See Appendix~\ref{sec-autosys} for details), where we have introduced two extra dimensionless variables given by \citep{Dinda:2017lpz}

\begin{equation}
q=(\delta \phi)/\Big{(} \dfrac{d \phi}{d N} \Big{)}, \hspace{0.3 cm} \text{and} \hspace{0.3 cm} \tilde{\mathcal{H}} = \frac{\mathcal{H}}{\mathcal{H}_{0}},
\label{eq:dimless_var_pert}
\end{equation}

\noindent
where $\mathcal{H}_{0}=H_{0}$.
\\
\noindent
For the details of the initial conditions, see Appendix~\ref{sec-initial}.
\\
\noindent
The matter density contrast is given by

\begin{eqnarray}
\delta_{m} &&= -\frac{1}{\Omega_{m}} \bigg[(2-x^2\epsilon) \Phi_{1} + 2 \Big{(} 1+L-x^2(1+2\epsilon) \Big{)} \Phi \nonumber\\
&& + x^2(2+3\epsilon) q_{1} + x^2 \Big{(} (2+3 \epsilon)A - 2 J + L \epsilon \Big{)} q \bigg] \, ,
\label{eq:delm}
\end{eqnarray}

\noindent
where $A$ is given in Eq.~\eqref{eq:dlm_A} in Appendix~\ref{sec-deltam}.
\\
\noindent
The pecular velocity for the matter is given by

\begin{eqnarray}
y_{m} = 3 \mathcal{H} v_{m} &=& \frac{1}{\Omega_{m}} \bigg[2 \Phi_{1} + (2-x^2\epsilon) \Phi + x^2\epsilon q_{1} \nonumber\\
&& -x^2 \Big{(} 6+\epsilon (3-A) \Big{)} q \bigg].
\label{eq:ym}
\end{eqnarray}

\noindent
The comoving matter energy density contrast (from Eqs.~\eqref{eq:delm} and \eqref{eq:ym} with the definition in Eq.~\eqref{eq:gnrlcmvngD} for matter) is given by

\begin{equation}
\Delta_{\rm m} = \delta_{\rm m} + y_{\rm m}.
\label{eq:Deltam}
\end{equation}

\noindent
In Fig.~\ref{fig:Delmk}, we have plotted the comoving matter energy density contrast ($\Delta_{m}$) as function of wave number ($k$) at different redshifts ($z$) for different $ \epsilon_{i} $. The deviations in $\Delta_{m}$ from $\Lambda$CDM model is the highest at present ($z=0$) for a particular $\epsilon_{i}$ value. This behaviour is consistent with Fig.~\ref{fig:eosw}. At early matter dominated era, all the models have similar behaviour like $\Lambda$CDM model. At late times, they deviate sufficiently from $\Lambda$CDM behaviour. The deviations decrease with increasing redshifts. At a particular redshift, the deviation is the highest for $\epsilon_{i}=0$ and decreases with increasing $\epsilon_{i}$. This behaviour is also consistent with Figs.~\ref{fig:eosw} and~\ref{fig:Eofa}.

The linear matter power spectrum ($P_{m}$) is proportional to square of the Comoving matter energy density contrast i.e. $P_{m} \propto \Delta_{m}^{2}$ \citep{Dinda:2017lpz,Dinda:2016ibo}. So, if we fix initial power spectrum to be $P_{m}^{i}$, we can rewrite

\begin{equation}
P_{\rm m}(k,z) = \left[ \frac{\Delta_{\rm m}^{2}(k,z)}{\Delta_{\rm m}^{2}(k,z_{i})} \right] P_{\rm m}^{i}(k,z_{i}).
\label{eq:linrPm}
\end{equation}

\noindent
Eq.~\eqref{eq:linrPm} is valid on all scales. On small scales, $\Delta_{m}(k,z)$ can be approximated by $\delta_{m}(k,z)$ in above equation.

In Fig.~\ref{fig:Pmz0}, we have plotted the deviations in the linear matter power spectrum for Cubic Galileon models from $\Lambda$CDM model as a function of wave number ($k$) at $z=0$ for different $\epsilon_{i}$. To plot these deviations, we have considered the same initial matter power spectrum ($P_{\rm m}^{i}(k,z_{i}=49)$) for all the models. The initial linear matter power spectrum ($P_{\rm m}^{i}(k,z_{i}=49)$) is computed by the CAMB code \footnote[1]{https://camb.info/} with $\Lambda$CDM model with $\Omega_{\rm m}^{(0)}=0.3156$, $\Omega_{\phi}^{(0)}=0.6844$, $\Omega_{\rm b}^{0}=0.0491$ (baryon energy density parameter at present), $h=0.6727$, $\sigma_{8}=0.831$ (at $z=0$) and $n_{s}=0.96$. These values are consistent with Planck15, BAO, SNIa and H0 data \citep{Costa:2016tpb}. The deviation is the highest for $\epsilon_{i}=0$. The deviations decrease with increasing $\epsilon_{i}$. This behaviour is consistent with the bottom-right panel of the Fig.~\ref{fig:Delmk}.

\section{N-body Simulation} \label{sec:nbody}
N-body simulation has long been used to study the structure formation of the Universe. With N-body simulation, we may be able to study the structure formation in deeply nonlinear regime. The generic simulation pipeline was introduced in \citet{megadget}. In this pipeline, the modification of structure formation can be classified into three kinds in the Cubic Galileon Gravity, which is
\begin{itemize}
    \item[1] Modification of the initial condition for the simulation,
    \item[2] Modification of the hubble parameter, which affect the expansion history,
    \item[3] Modification of the effective gravitational constant, which is both time and scale dependant in Cubic Galileon.
\end{itemize}

\begin{figure}
    \includegraphics[width=0.5\textwidth]{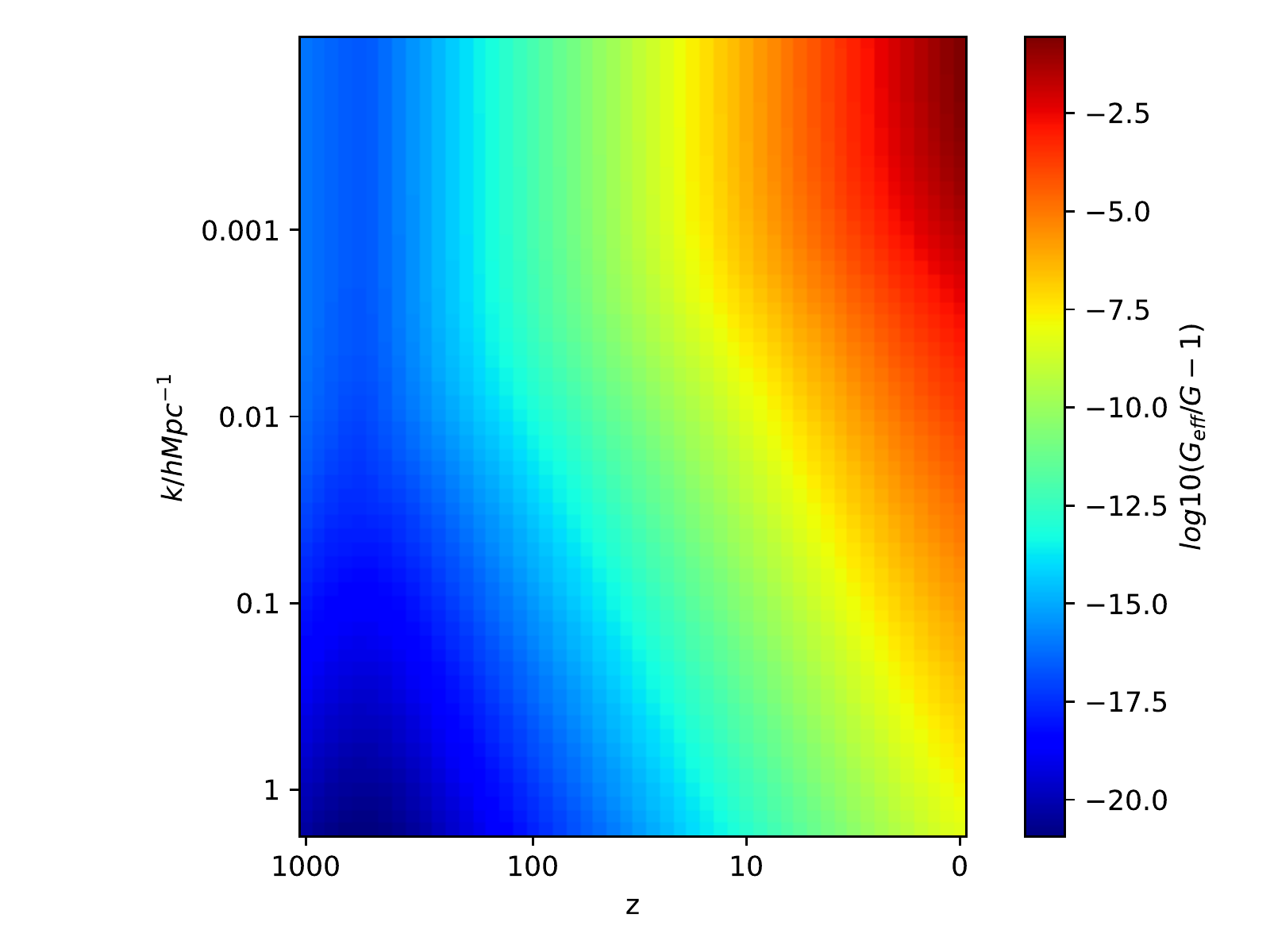}
    \caption{$G_{\rm eff}$ as a function of redshift $z$ and wave number $k$ is shown. At larger scale (smaller $k$) and lower redshift, the deviation of effective gravity from GR is larger.}
    \label{fig:geff}
\end{figure}

We have run two sets of simulations to see the effect of Cubic Galileon Gravity. First, with the same initial condition files generated for $\Lambda$CDM model, using Planck15 cosmology, with $\epsilon_{i}=0,20,50$. Second, the effect of Cubic Galileon, in the case of $\epsilon_{i}=0$, was separated into changing the initial condition for simulation, changing the expansion history and changing the effective gravity. The simulations are:
\begin{itemize}
    \item CGIC, only the initial condition of the simulation is changed. The $\sigma_8$ calculated by linear perturbation theory is controlled to be the same as $\Lambda$CDM at $z=0$. Therefore the matter power spectrum at $z=49$, when we started the simulation, is different.
    \item CGHz, only the expansion history is changed. The change of expansion is represented in the hubble parameter, illustrated in Fig.\ref{fig:Eofa}.
    \item CGGeff, only the Poisson equation is changed. The change of Poisson equation is expressed in Eqs.\ref{eq:Pssn}. If we rewrite the equation as $\vec{\nabla}^{2} \Phi = 4 \pi G_{\rm eff}(k,z) a^{2} \bar{\rho_{\rm m}} \Delta_{\rm m}$, $G_{\rm eff}(k,z)$ for $\epsilon_{i}=0$ is illustrated in Fig.\ref{fig:geff}. 
\end{itemize}
We would like to see how much difference will this difference of choice contribute to the final results. We have used the ME-Gadget simulation code\citep{megadget} for all the simulations. The boxsize is $400Mpc/h$ and the number of particles is $512^3$, the softening length is $25kpc/h$. The initial condition is generated using 2LPTic\citep{2lptic} at $z=49$, and the pre-initial condition file is generated using CCVT\citep{liao2018ccvt}. 

\section{Result}\label{sec:result}
\subsection{Marked Density}
\begin{figure*}
\includegraphics[width=0.45\textwidth]{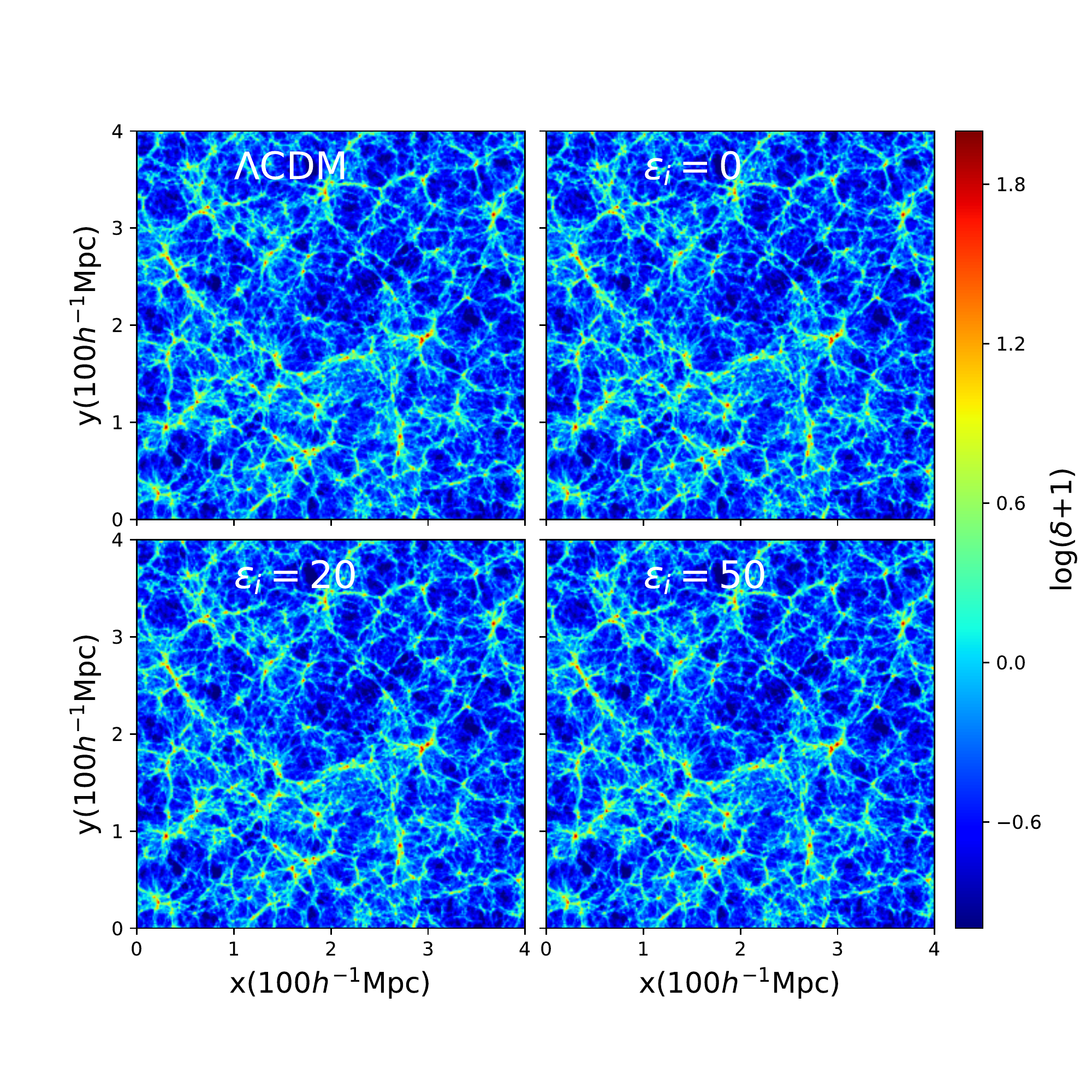}
\includegraphics[width=0.45\textwidth]{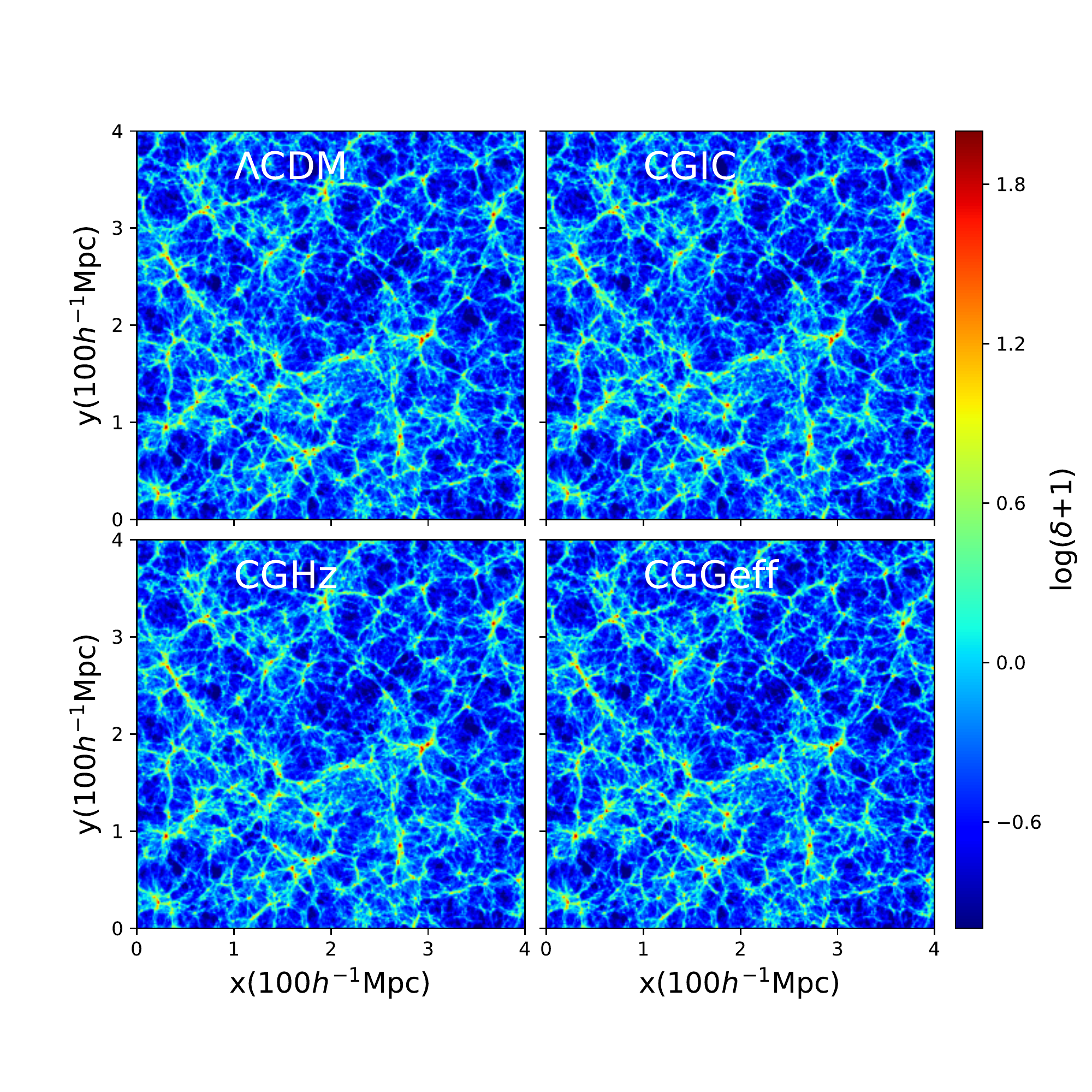}
\caption{The dark matter density distribution in a 2D slice of the simulation box at $z=0$, shows the comparison among $\Lambda$CDM and $\epsilon_{i}=0,20,50$ CG models on the left panel, and the comparison among $\Lambda$CDM, CGIC, CGHz and CGGeff simulations on the right panel. Since we use the same initial condition random seed, the distribution looks very similar. The difference introduced by CG is also quite small so that it is not distinguishable.}
\label{fig:density}
\end{figure*}

We have shown the density field slice in Fig.~\ref{fig:density}. The colorbar shows the dark matter over density, where $\delta=\rho/\bar{\rho}-1$. We have chosen the same initial condition random seed for the simulations, so the overall large scale structure looks quite similar between different simulations. We also notice that the difference between different simulations are really tiny and not distinguishable by eye. This means the overall difference between different simulations are quite small.
\begin{figure*}
\includegraphics[width=0.45\textwidth]{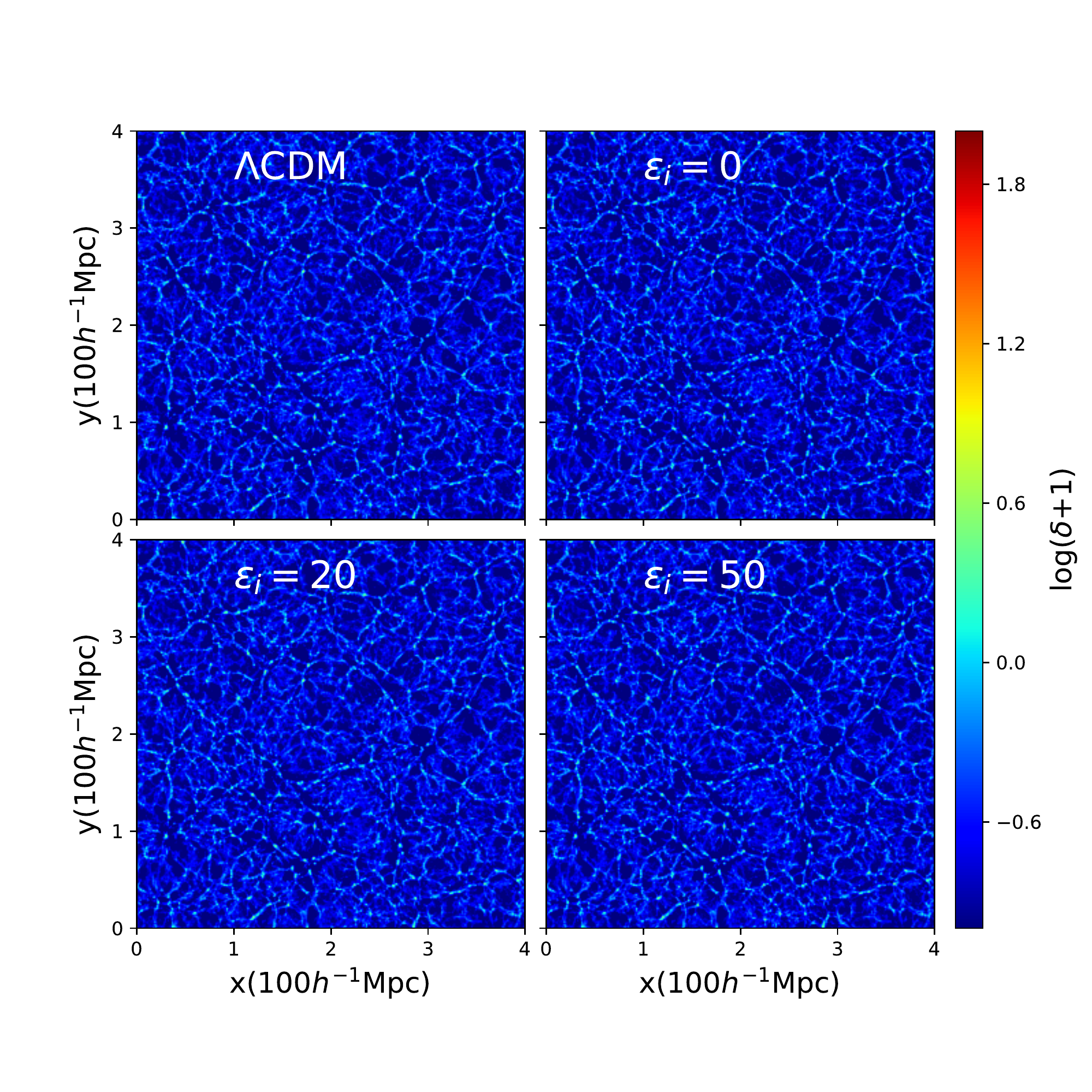}
\includegraphics[width=0.45\textwidth]{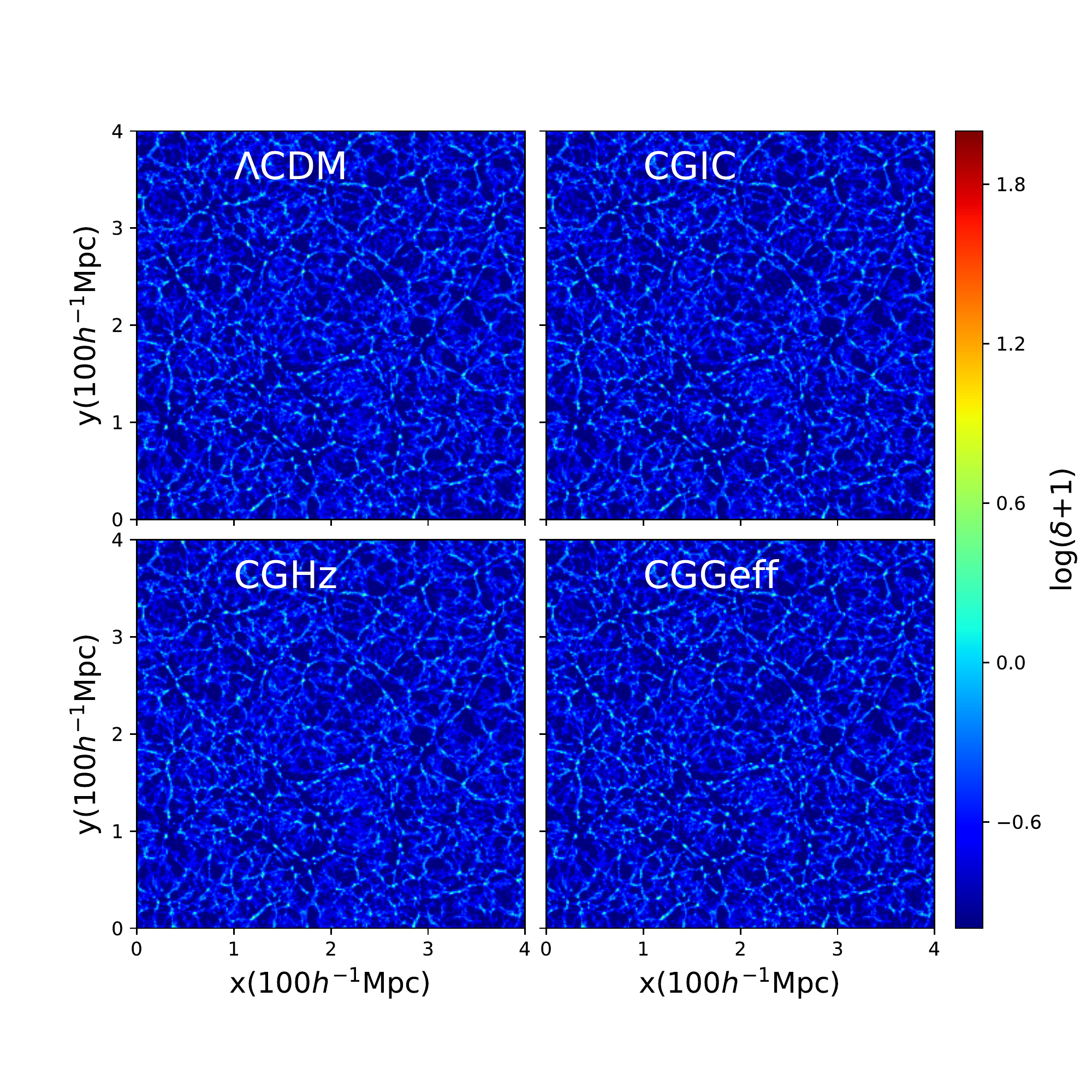}
\caption{The marked dark matter density distribution shows the comparison among the simulations at $z=0$. Comparing to Fig.\ref{fig:density}, we can see that the high density region is clearly suppressed and the fluctuations in the voids are much more clear. The overall difference between the simulations are still not very clear.}
\label{fig:mark}
\end{figure*}
Marked density field and power spectrum were used recently\citep{massara2020} to highlight the signature of massive neutrinos. The marking of density field depends on its "environment". We define the mark
\begin{equation}
    m(\vec{x};R,p,\delta_s)=\Bigg(\dfrac{1+\delta_s}{1+\delta_s+\delta_R(\vec{x})}\Bigg)^p,
\end{equation}
and the marked over density is $m(\vec{x};R,p,\delta_s)\delta$, where we have chosen $R=10Mpc/h,p=2,\delta_s=0.25$. $\delta_R(\vec{x})$ is the over density at position $\vec{x}$ smoothed by a Top Hat filter with radius $R$. Under this choice, the density field in low density environment, like voids, receives higher weight and the density field in high density environment, like clusters, receives lower weight. The overall density field will become more Gaussian\citep{massara2020}. We have shown the marked density field in Fig.~\ref{fig:mark}. We can see that, compared to Fig.~\ref{fig:density}, the color looks more uniform and blue, which means the fluctuation is much smaller than density field, the difference between high density regions and low density regions is less significant. However, the comparison between different simulations is still not very clear by eye. We need to calculate the power spectrum to see the difference more clearly.

\subsection{Matter Power Spectrum}
\begin{figure*}
\includegraphics[width=0.45\textwidth]{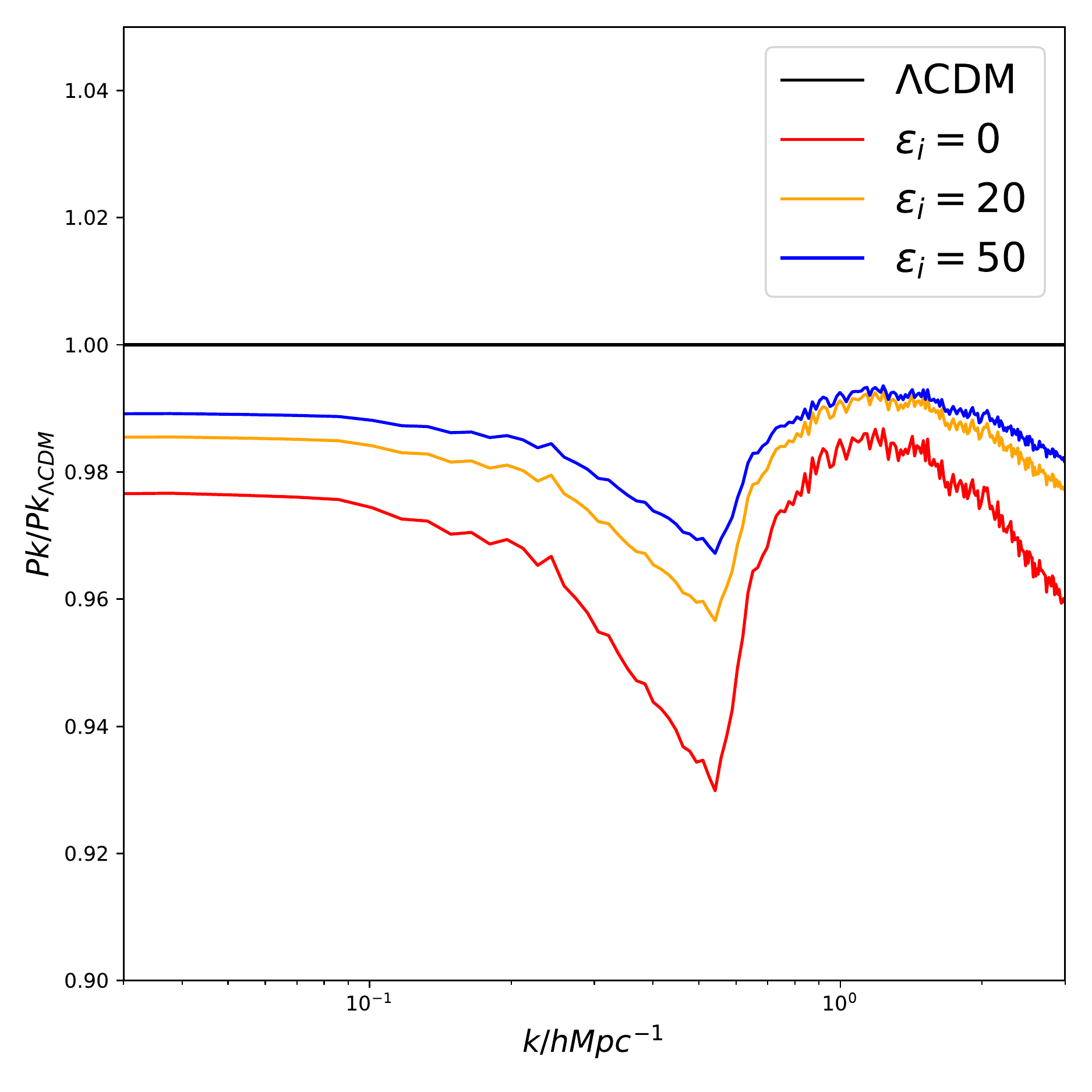}
\includegraphics[width=0.45\textwidth]{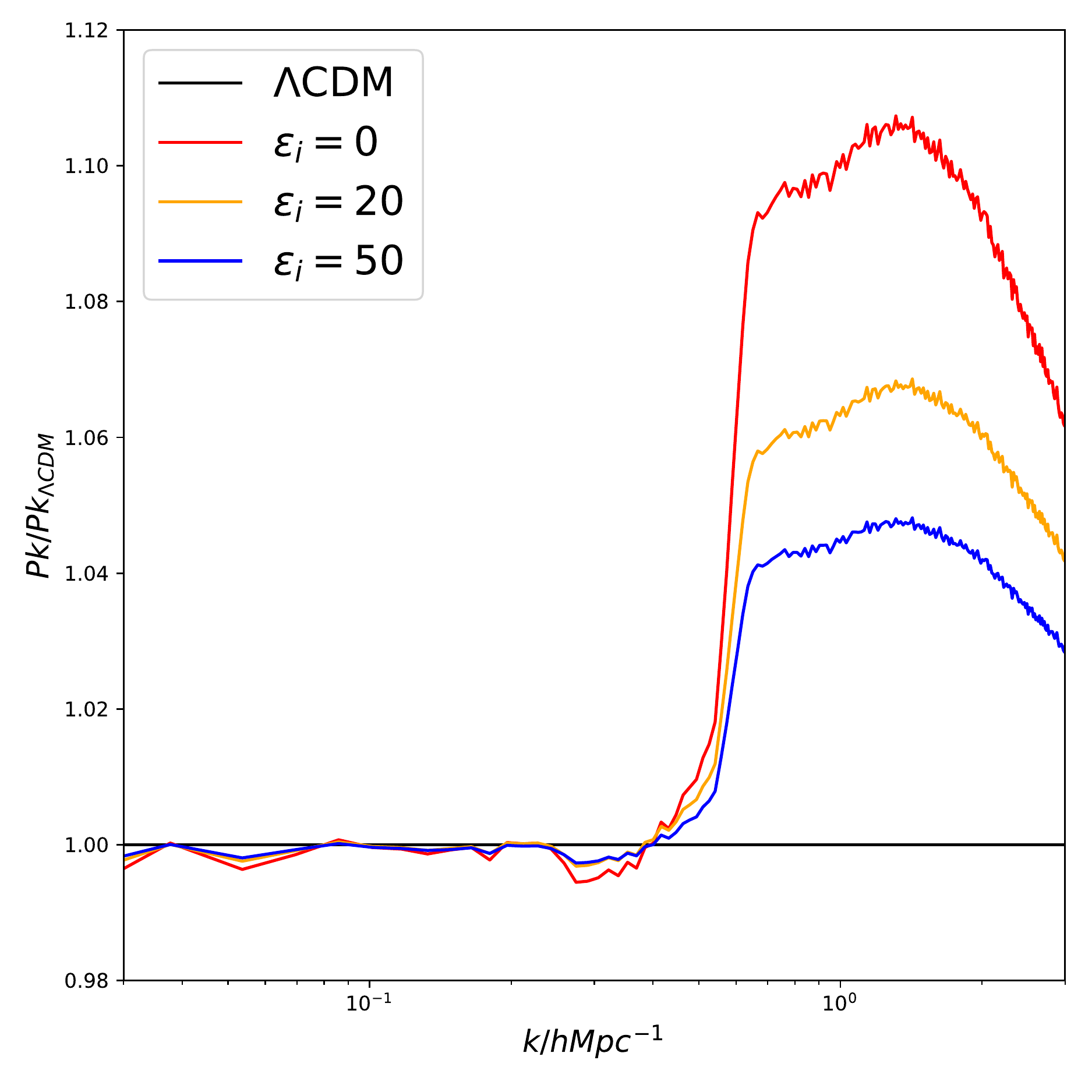}
\caption{On the left panel, we show the matter power spectrum ratio of $\epsilon_{i}=0,20,50$ CG models and $\Lambda$CDM model. The solid lines are results from simulations, while the dashed lines show the results calculated by modified HMcode with halo model\citep{Dinda:2018eyt}. The difference is largest at about $k=0.5h/Mpc$, with no more than $-7\%$. On the right panel, we show the marked matter power spectrum ratio of $\epsilon_{i}=0,20,50$ CG models and $\Lambda$CDM model. The difference is largest at about $k=1h/Mpc$, with at most $11\%$. The mark process enlarge the difference by about a factor of two.}
\label{fig:pk}
\end{figure*}
\begin{figure*}
\includegraphics[width=0.45\textwidth]{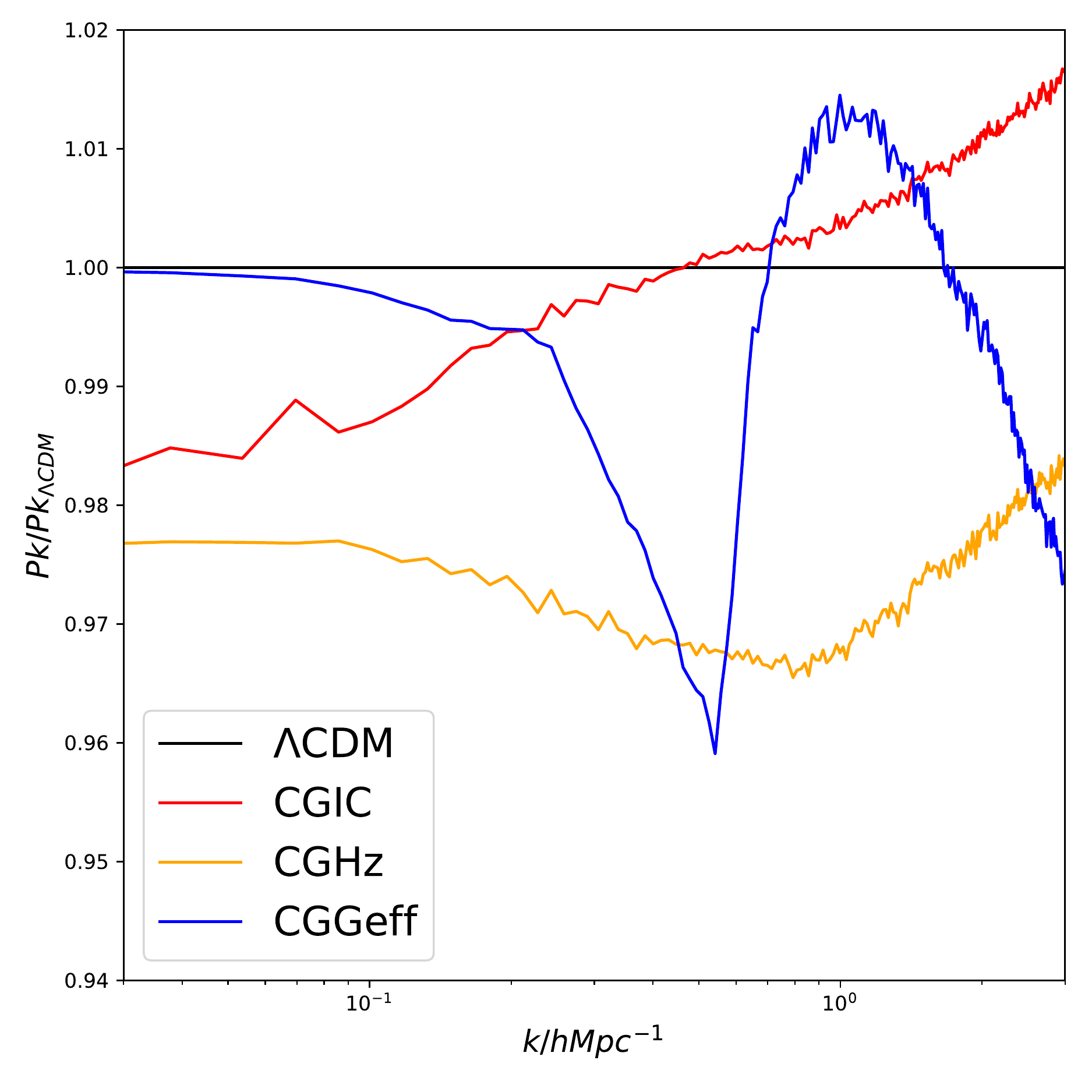}
\includegraphics[width=0.45\textwidth]{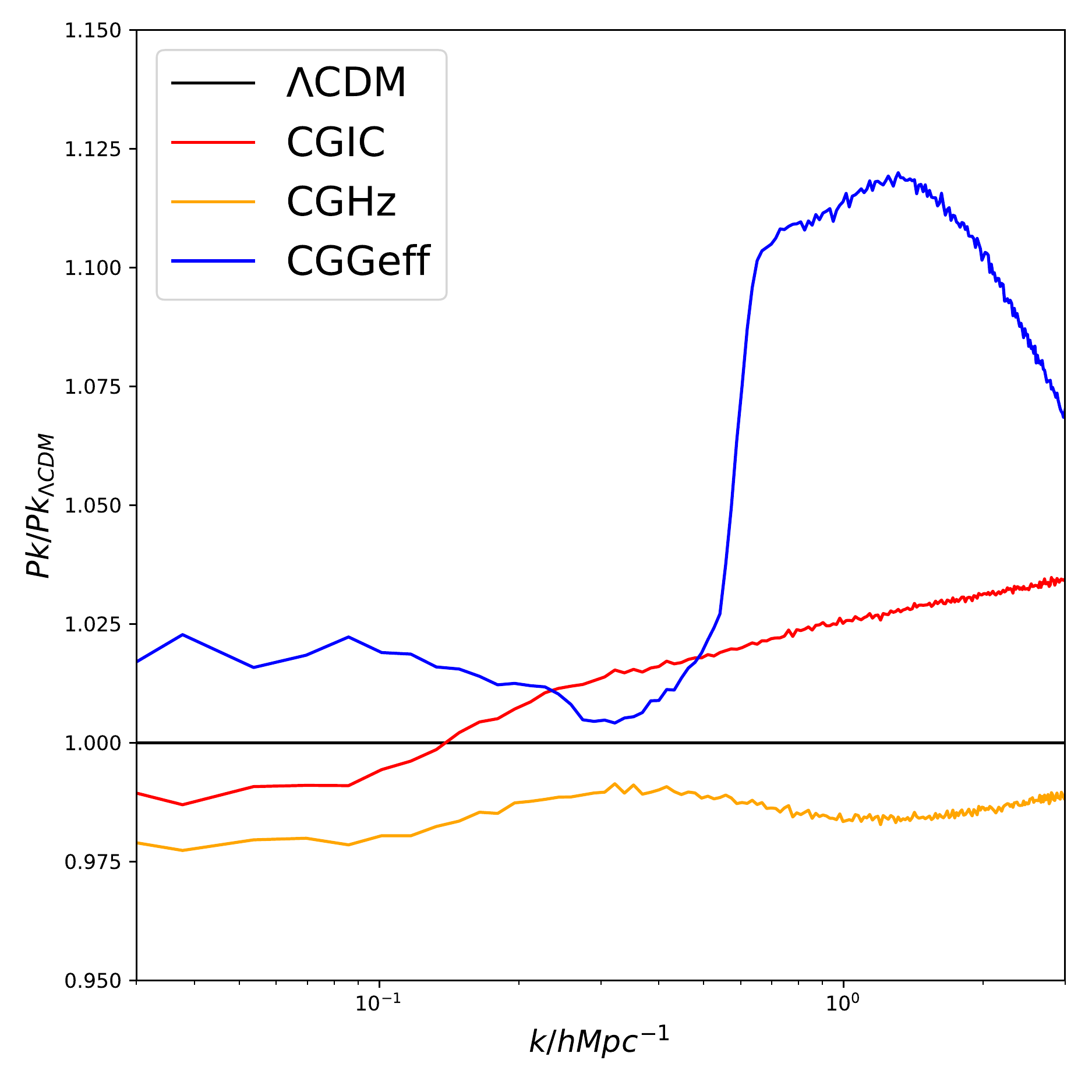}
\caption{On the left panel, we show the matter power spectrum ratio of CGIC, CGHz, CGGeff and $\Lambda$CDM. The difference is no larger than $4\%$, and the change of expansion history provides the largest difference at all scales. On the right panel, we show the marked matter power sepctrum ratio of CGIC, CGHz, CGGeff and $\Lambda$CDM. The difference is largest in the CGGeff simulation, which is about $11\%$. This means mark according to the large scale environment is very useful in distinguishing the modified gravitation constant.}
\label{fig:part_pk}
\end{figure*}
Power spectra is the measurement of the correlation of a given density field in k space. We have used Pylians python library\citep{pylians}\footnote{https://github.com/franciscovillaescusa/Pylians} to measure the power spectrum. The comparison between $\Lambda$CDM and $\epsilon_{i}=0,20,50$ is provided in Fig.~\ref{fig:pk}. We can see that, compared to $\Lambda$CDM, CG models is lower in power spectrum. At large scale, the suppression is $1-2\%$, $\epsilon_{i}=0$ is the lowest. This trend and amount of suppression is very well predicted by the linear perturbation theory in Fig. \ref{fig:Pmz0}. This means the predictions from simulation and linear calculation are consistent. We have also shown the comparison between the simulation results in solid lines and halo model calculated results by the modified HMcode\citep{Dinda:2018eyt} in dashed lines. At large scale, the solid lines and dashed lines are very consistent as expected. At smaller scale, there is an additional suppression of power spectrum in CG models. In simulations, a sharp drop at around $k=0.5h/Mpc$ can be noticed. While in the halo model calculations, we can see similar drop, but at smaller scale around $k=0.9h/Mpc$. This is the scale of large clusters. We suspected that this is a unique feature for the Cubic Galileon Gravity near high density clusters. The additional suppression of power spectrum in CG models is due to the suppression of very massive halo formation, which is shown in Fig.\ref{fig:hmf}. Therefore, the additional suppression is physical, can be identified both in simulations and in halo model calculations. 
\begin{figure}
    \centering
    \includegraphics[width=0.45\textwidth]{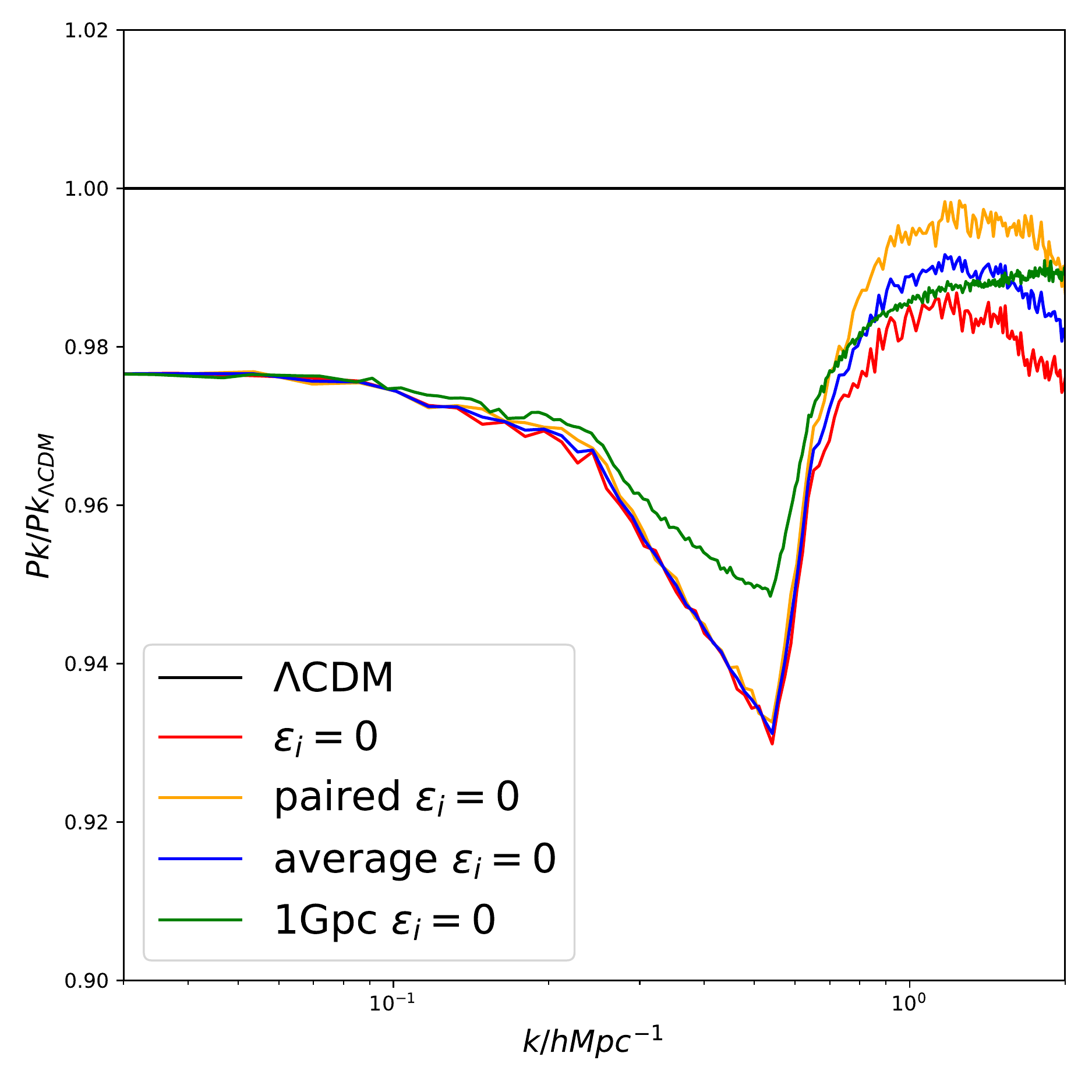}
    \caption{The real space power spectrum ratio of $\epsilon_i=0$ and $\Lambda$CDM are shown, compared with paired simulation, average value and larger boxsize. The clear sharp bottom at around $k=0.5h/Mpc$ exist in all different cases, which means we cannot rule out the possibility that such sharp kink is numerical and not physical.}
    \label{fig:pair_pk}
\end{figure}
We also notice that the power spectrum ratio measured from simulations are very sharp at the bottom, which is likely due to limited number of realizations. In order to answer whether such sharp kink is physical or not, we have done the following discussion. If the kink is due to cosmic variance, then a simulations with paired initial condition and the average value between a simulation with its paired one should remove the kink. Paired-and-fixed simulation is a technique to get the mean value of observable like power spectrum from only two simulations, without a lot of realizations\citep{Angulo2016MNRAS,Klypin2020MNRAS}. The paired initial condition has the anti-phase of the desired initial condition, which means where there is a void in the simulation, there is a cluster in the paired simulation. Therefore, the average of the simulation and its paired part can provide a good estimate of the mean value of any observable. The paired simulated power spectrum ratio of CG model and $\Lambda$CDM is shown in yellow line in Fig.\ref{fig:pair_pk}, the average value is shown in blue line. They all show the clear kink feature. It is also possible that the kink may come from numerical issues such as PM solver in the code. If so, the position of the kink will be different or disappear if we have a different boxsize. We show the power spectrum ratio between CG model and $\Lambda$CDM model with the boxsize of $1Gpc/h$ in the green line. Though the shape of the curve is different due to the lower resolution, the position of the kink remains the same. Therefore, it is also not likely to be numerical reason in the code. However, such kink is still hard to believe as physical and it is not at where 2-halo and 1-halo transition happens. So whether the kink is physical or not remains mysterious to us. It remains as an open question to be answered in the future study. 

On the other hand, the difference is at most $7\%$ for the $\epsilon_{i}=0$ case. The errorbar of shear correlation in DES Y1 METACALLIBRATION catalog is no smaller than $10\%$, so that the constraints on matter power spectrum is also no better than $10\%$. Therefore, such $7\%$ difference is not easy to be identified in observations\citep{DES1,DES2,Sheldon2017ApJ}. With the marked matter density, we can see about twice significant difference power spectrum. For the $\epsilon_{i}=0$ case, the difference can be as large as $11\%$ at around $k=1h/Mpc$. Even for the $\epsilon_{i}=50$ case, the difference is smallest, is also about $5\%$. By down-weighting the high density regions and highlight the low density regions, the difference between $\Lambda$CDM and CG models is also increased. This indicates that the density fluctuation in the voids might be crucial to tell $\Lambda$CDM and CG apart.

In order to investigate in detail about the reason of such difference, we compared the power spectrum and marked power spectrum among $\Lambda$CDM, CGIC, CGHz and CGGeff simulations in Fig.\ref{fig:part_pk}. We chose $\epsilon_{i}=0$ for the test of CGIC, CGHz and CGGeff simulations. Because we have found that the difference between $\epsilon_{i}=0$ and $\Lambda$CDM is the most significant, it is easier for us to measure the difference. The effect of changing the initial condition of the simulation is not very significant, both for the power spectrum and the marked power spectrum. It is between $-2\%$at large scale to $+2\%$ at small scale. So changing the initial condition is not the major cause of the noticeable difference. Changing the expansion rate will suppress the power spectrum and marked power spectrum at all scales by about $2\%$. This is well expected because over all, in the $\epsilon_{i}=0$ case, the universe expands faster than $\Lambda$CDM as shown in Fig.\ref{fig:Eofa}, therefore the growth should be less in all scales. The change of the effective gravitational constant, can clearly produce the sharp drop of power spectrum at $k=0.5h/Mpc$ and increase the marked power spectrum at around $k=1h/Mpc$ by about $12\%$. Therefore, we clearly know that the major contribution of the difference we see in Fig.\ref{fig:pk} is caused by the modification of Poisson equation. Because of the special behavior of gravity at different scale, we can see the difference of power spectrum and more clearly, the difference of marked power spectrum. 

\subsection{Count in Cell}
\begin{figure*}
\includegraphics[width=0.45\textwidth]{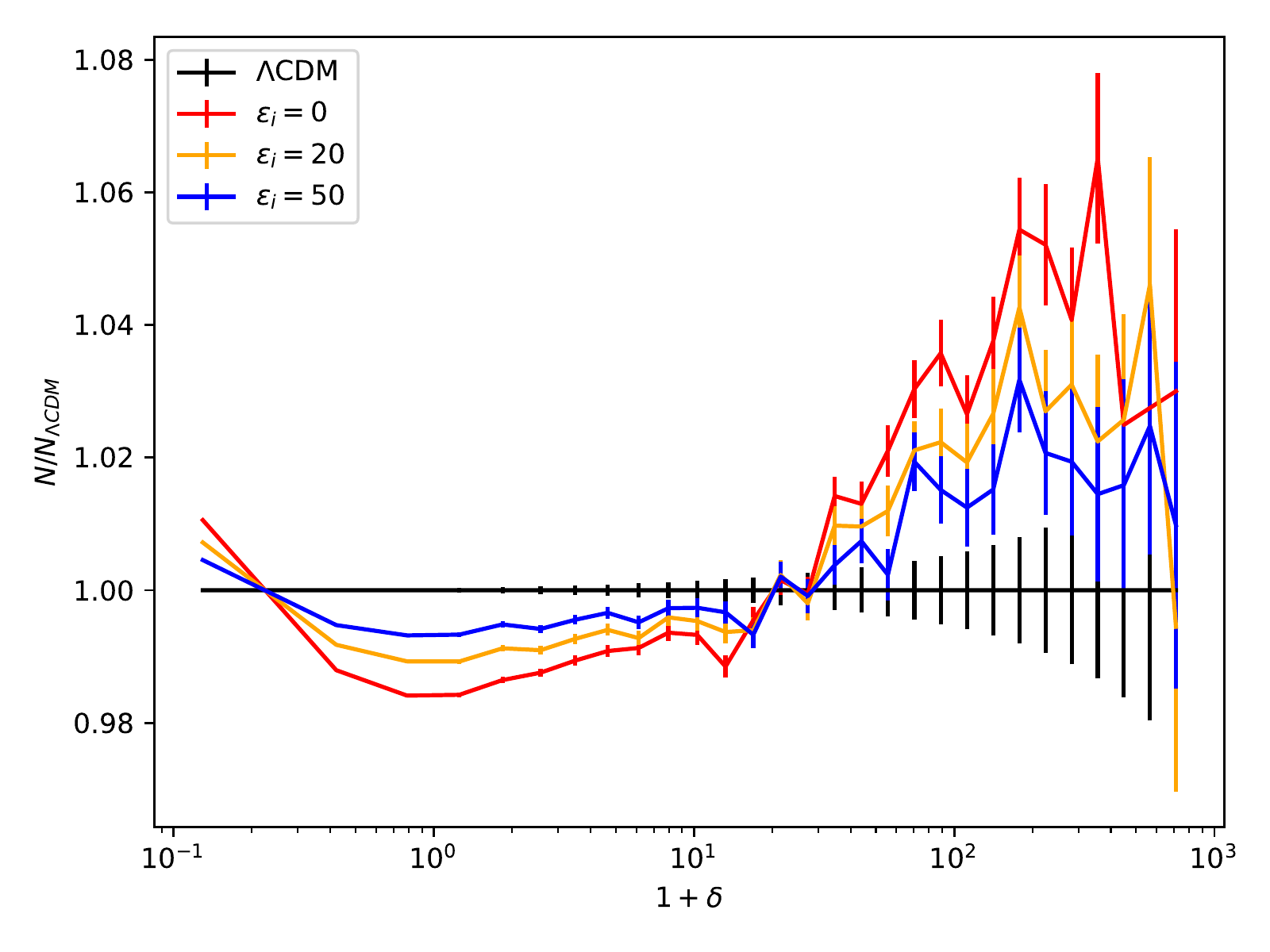}
\includegraphics[width=0.45\textwidth]{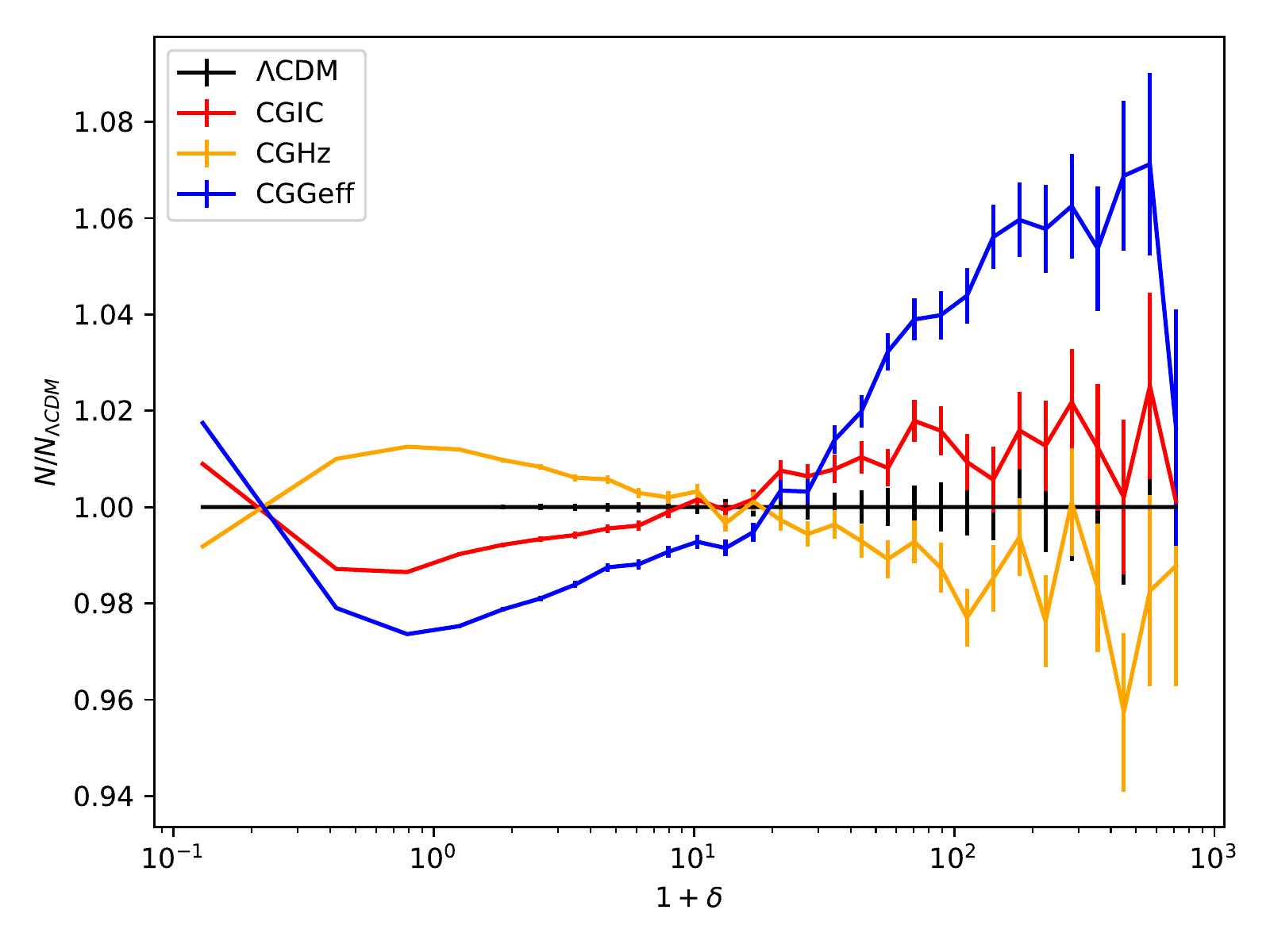}
\caption{We show the ratio of number count of cells among CG models and $\Lambda$CDM. On the left panel, we show the comparison between $\epsilon_{i}=0,20,50$ CG models and $\Lambda$CDM model. The CG models have at most $1\%$ more very low density cells ($\delta<-0.7$) than $\Lambda$CDM, and $1\%$ less medium density cells ($-0.7<\delta<10$), which is very significant comparing to the small error bar. In the right panel, we show the effect of CGIC, CGHz and CGGeff on the number counting. CGHz has less extreme density cells, while CGGeff has more extreme cells, the effect of CGIC is less important.}
\label{fig:cic}
\end{figure*}

The dark matter density was calculated in cells. We can also compare the number of cells with difference over density among difference models. We calculated the density in $512^3$ cells from each simulation box. The ratio was taken between the CG model simulations and the $\Lambda$CDM model. The result is shown in Fig.\ref{fig:cic}. The error bar was estimated by $1/\sqrt{N}$ by assuming the Poisson error. There are much more low density cells than high density cells, so the error bar for low density cell number is too small to be shown. We can see that the overall trend of number of cells is similar for CG models. There are more void ($\delta<-0.7$) cells, less average cells ($-0.7<\delta<10$) and more cluster ($\delta>10$) cells, in the CG model than $\Lambda$CDM model. The error bar for the number of cluster cells is large, so the significance of the difference is not too surprising, and the difference of number of void cells is more significant. Both the high density cells and low density cells can leave clear weak lensing effects. We may be able to tell the difference by this count in cell measurement, focusing on voids, from weak lensing observations\citep{baker2018prd,dong2019apj}. We can again see that the effect of modifying Poisson equation is the most significant, changing the expansion history is less significant, with opposite trend. 
\subsection{Halo Mass Function}
\begin{figure*}
\includegraphics[width=0.45\textwidth]{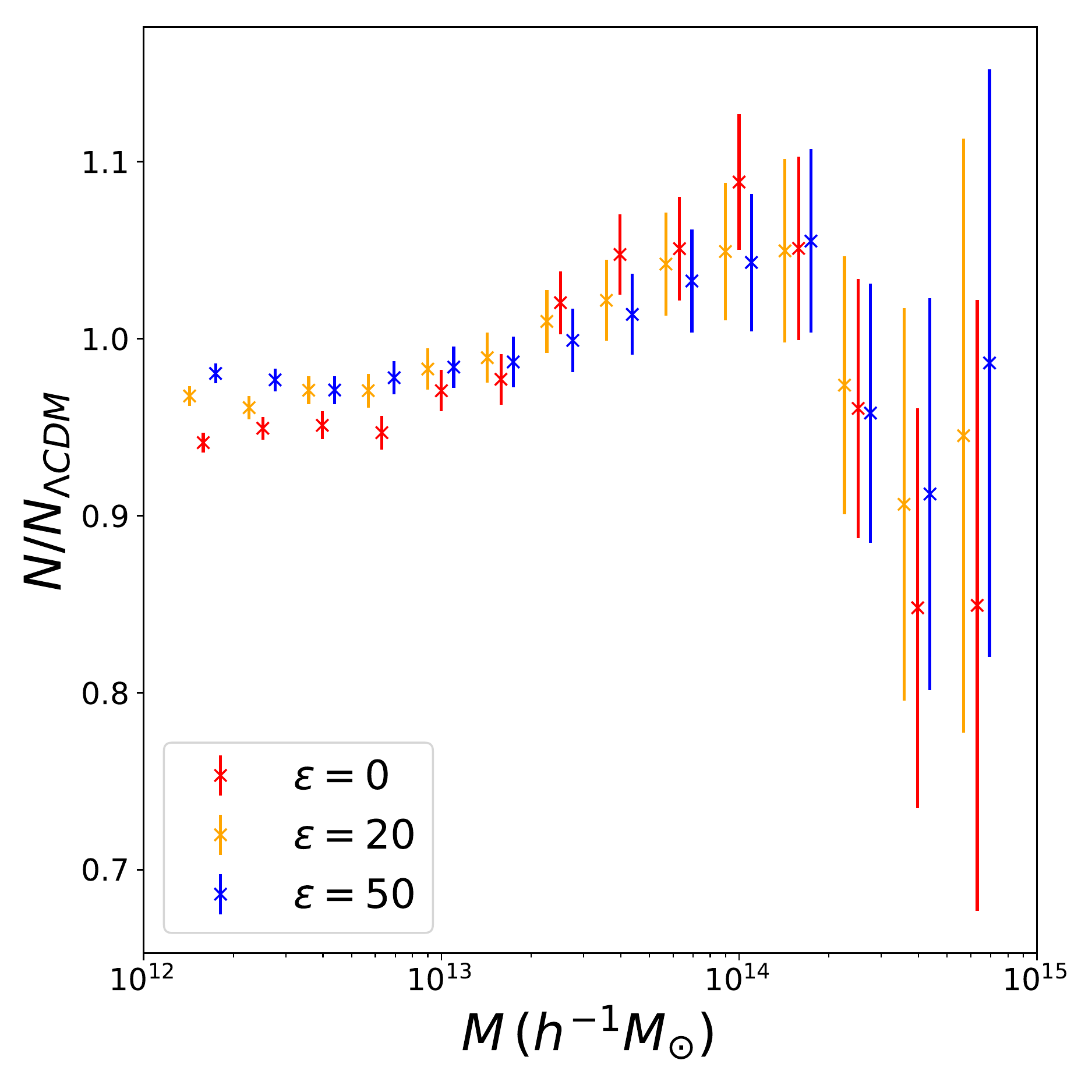}
\includegraphics[width=0.45\textwidth]{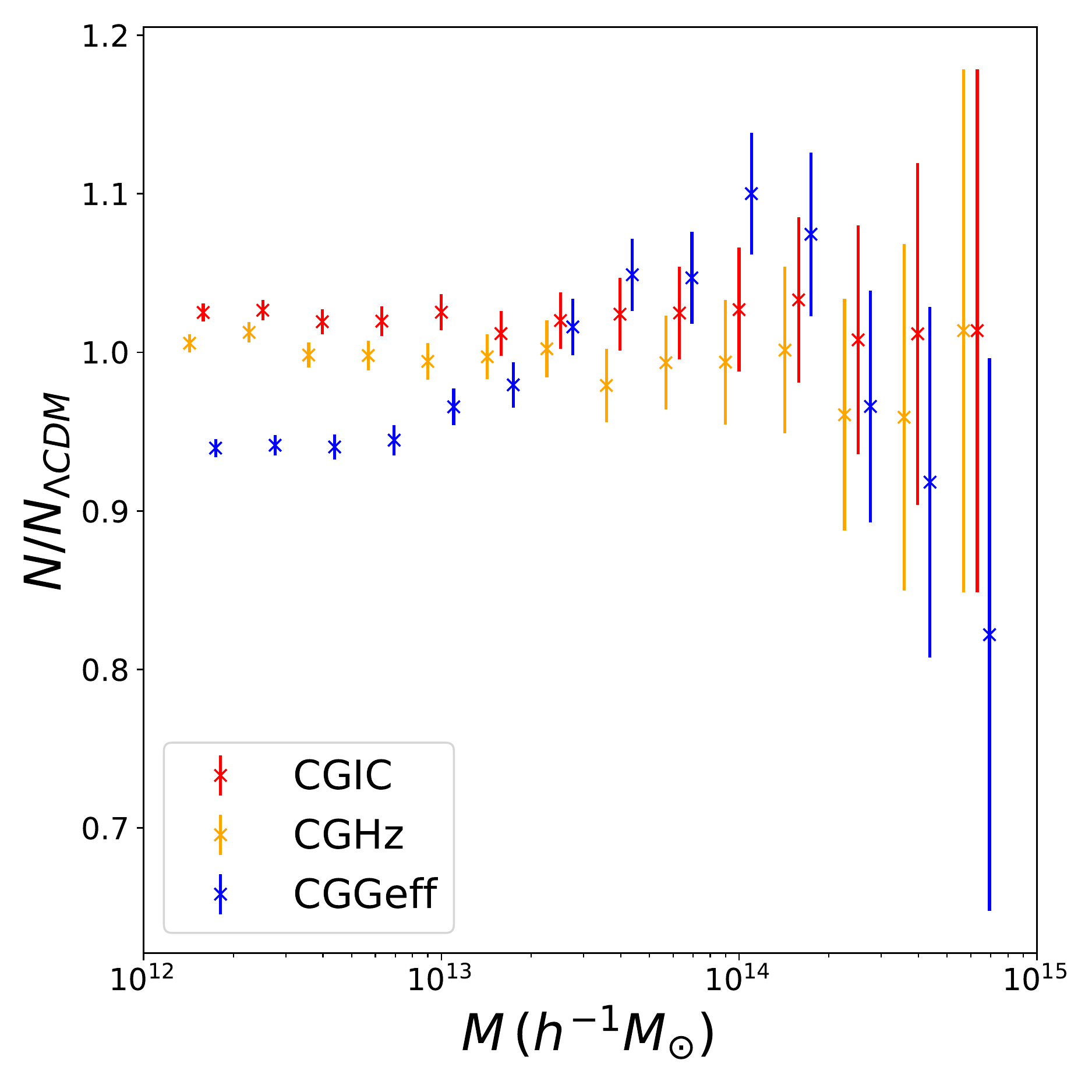}
\caption{We show the ratio of halo mass function between CG models and $\Lambda$CDM model. The errorbar is given by poisson noise. The difference is mainly at high mass end. There are less very massive halos ($>10^{14}h^{-1}M_{\odot}$) in the CG model, which is mainly caused by the change of gravitational constant, shown in the lower panel by CGGeff.}
\label{fig:hmf}
\end{figure*}
Halo mass function was used to show the abundance of dark matter halos with different mass. It is a good measure of the structure formation. It is also believed that galaxies lies in dark matter halos, therefore taking the statistics of the halos is a good way to link simulations with observations. We use AHF halo finder\citep{ahf} to identify the dark matter halos in the simulation. 

The halo mass function comparison is shown in Fig.\ref{fig:hmf}. We show the ratio of halo mass function between CG models and $\Lambda$CDM model. In CG models, there are less halos with $10^{12}h^{-1}M_{\odot}<M<3\times 10^{13}h^{-1},M_{\odot},M>10^{14}h^{-1}M_{\odot}$ and more halos with $3\times 10^{13}h^{-1}M_{\odot}<M<2\times 10^{14}h^{-1}$.  The difference is more clear with a smaller $\epsilon_{i}$ value. By studying the halo mass function in CGIC, CGHz and CGGeff simulations, we can see that the major effect is coming from changing the Poisson equation. The change of initial condition and the expansion has very limited effect in halo mass function. It is also understandable why high mass halo is more sensitive to the change, because around the high mass halos, the gravitational field is strong. Therefore, the effect of changing the gravity is more clear. There are less high mass halos in CG models, in other words, the halos in CG models are less massive than those in the $\Lambda$CDM model. We should expect that galaxy-galaxy lensing may be able to distinguish such difference.

\subsection{Galaxy-Galaxy Lensing}
Gravitational lensing is the phenomenon where the light rays from distant galaxies are distorted by  intervening gravitational potentials traced by galaxies and dark matter halos. Assuming an isotropic distribution of both the galaxy shape and orientation, the non-zero average tangential shear residual,  $\gamma_T$, can be related to the foreground potential. In galaxy-galaxy lensing, this signal is interpreted as the combination of $\gamma_T$ and the geometry of the lensing system,  $\Sigma_{crit}(z_l,z_s)=\frac{c^2}{4\pi G}\frac{D_s}{D_{ls}D_l}$, where $z_l, z_s$ denote the redshifts of the lens and the source.  $D_l$, 
$D_s$ and $D_{ls}$ are the angular diameter distances of the lens, source galaxy and the difference between them respectively. The galaxy-galaxy lensing signal is reflecting the differential change of 2D surface density, Excess Surface Density (ESD), \begin{equation}
\Delta\Sigma(R)=\Sigma(\leq R)-\Sigma(R)=\gamma_t \Sigma_{crit}(z_l,z_s),
\end{equation}
here $\Sigma(\leq R)$ is the average surface density inside the projected distance $R$ and $\Sigma(R)$ is the surface density at the projected distance $R$. Therefore, the ESD provide the link between simulations and observations. By comparing the ESD signal measured from simulations and that measured from observations, we may be able to tell different models apart. And galaxy-galaxy lensing signals have already
been applied to constrain various modified gravity models, e.g. \citet{brouwer2017MNR},  \citet{luo2020} as well as test General Relativity at galactic scale \citet{chen2019arx}.
\begin{figure*}
\includegraphics[width=0.45\textwidth]{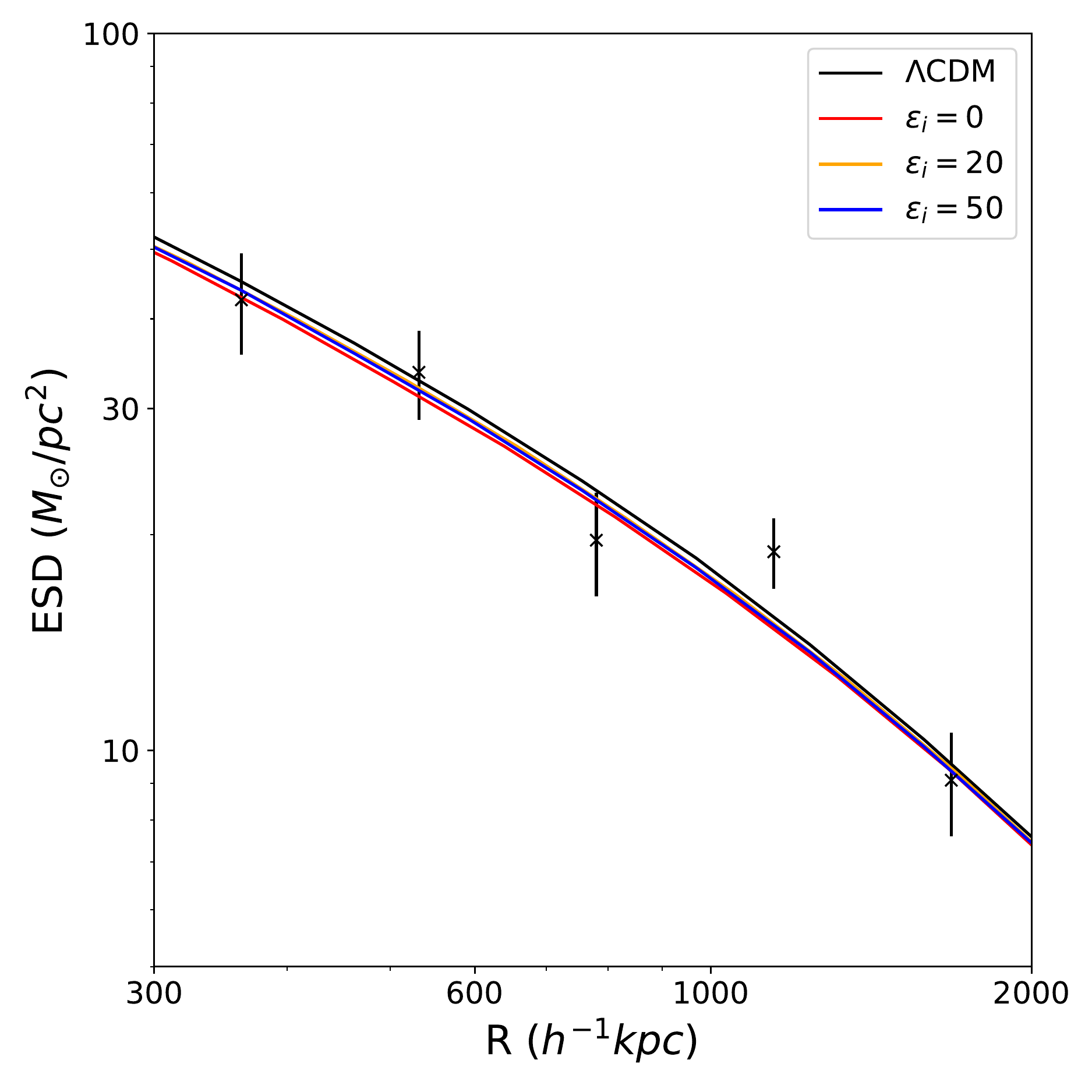}
\includegraphics[width=0.45\textwidth]{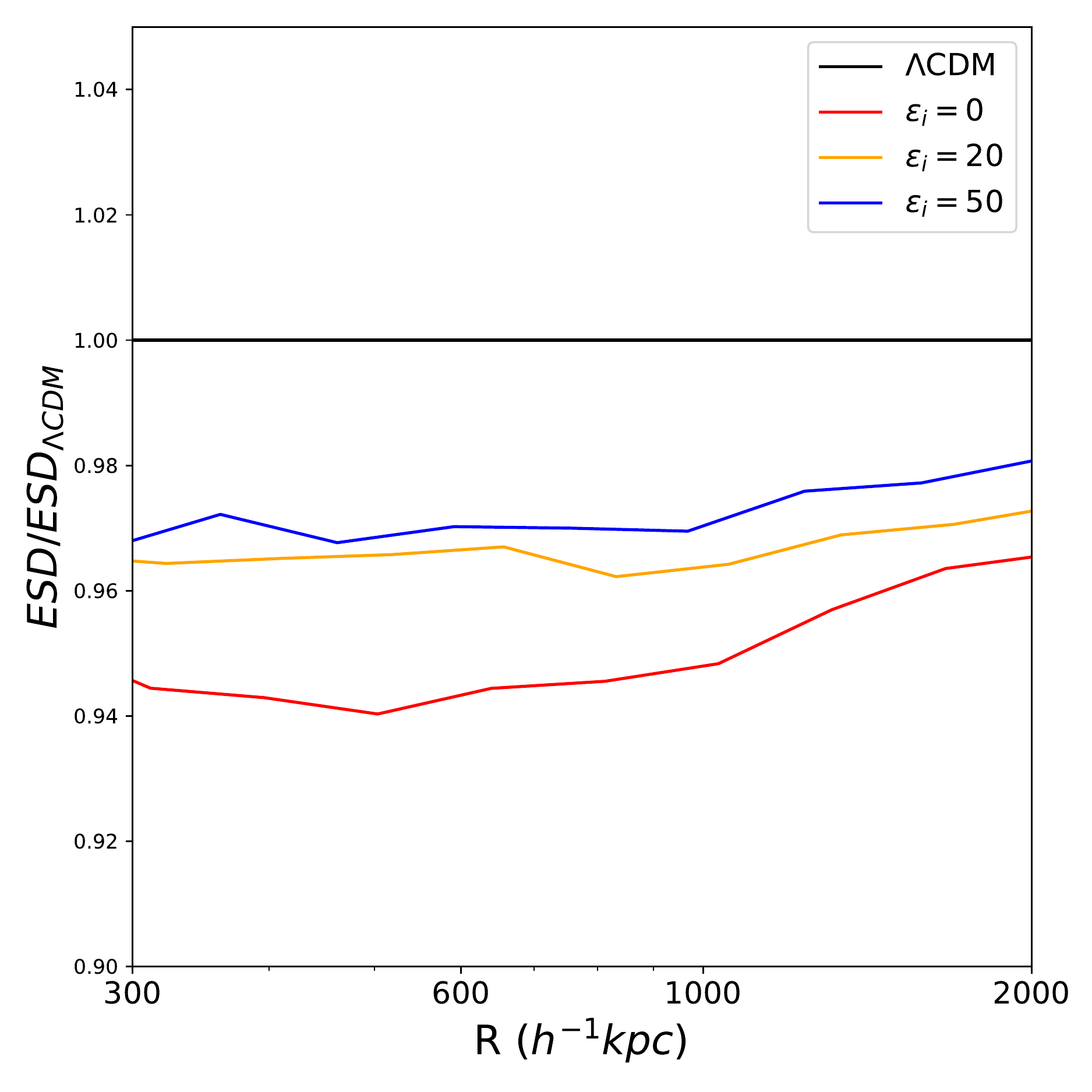}
\caption{We show the excess surface density (ESD) of $\epsilon_{i}=0,20,50$ CG models and $\Lambda$CDM models on the left panel, comparing to the observational data points in black cross. The ESD in CG models are lower than that in $\Lambda$CDM model, but the difference is still within the errorbar of the current observational data. On the right panel, we show the ratio of ESD signal from CG models and $\Lambda$CDM model. The difference is at most about $-6\%$ in the $\epsilon_{i}=0$ case. This difference might be distinguishable in the future studies.}
\label{fig:esd}
\end{figure*}
We follow the calculation introduced in \citet{zhang2019apjl} for both the simulation and observation. In the simulations, We cut off a cylinder near the selected halos (10 Mpc/h), compress the cylinder in the line-of-sight direction, stack them and calculate the ESD signal. By stacking the most massive 1771 halos in each simulation at z=0.1, we can measure the ESD signal for different models. We have also measured the ESD signal at $z=0.2$ to take the redshift evolution into account. The evolution of halos is an important uncertainty in the ESD signal prediction from simulations. However, since the errorbar from observations are much larger than the difference between models, we only show the curves measured at $z=0.1$ for better illustration. In the observation, we use the shear catalog from \citet{luo2017apj}, which is based on the SDSS DR7 image data\citep{sdss7}. For foreground galaxies, we employ the catalog from \citet{yang2007apj} to identify the lens systems. Following the galaxy-galaxy lensing measurement procedure in \citet{luo2018apj},  we select the most luminous 3660 galaxy groups in the group catalog from redshift 0.01-0.2 as the lens. 

The weak lensing measurements and simulation predictions of each cosmological model are shown in Fig.~\ref{fig:esd}. The mean value measured from stacked halos is given in solid lines and the measured data points from observations are given in black cross, together with the error bar. We can see that the ESD signals in CG models are lower than that of $\Lambda$CDM model. This is in agreement with what we have found in the halo mass function, that the halos in CG models are less massive than that in $\Lambda$CDM model with the same initial condition. However, such difference is so small that they are all within the error bar range of the observational data. The ESD difference between each model is only about $3-6\%$, which is hard to obtain in the current data sets. In the future, we may have enough observational data to constrain CG models. 

\section{Conclusion}\label{sec:conclusion}
We have studied the effect of Cubic Galileon Gravity on the large scale structure using N-body simulation. Though the overall difference between Cubic Galileon Gravity and $\Lambda$CDM is not very big, we still see that following difference which might be useful for constraints in the future.
\begin{itemize}
    \item[1] The major difference introduced in the Cubic Galileon Gravity is the expansion history and the modification of Poisson equation.
    \item[2] The difference in matter power spectrum is at most $7\%$ at $k=0.5h/Mpc$, while the marked matter power spectrum can be at most $11\%$ different at $k=1.0h/Mpc$. 
    \item[3] The number of low density or void cells is significantly different.
    \item[4] The Cubic Galileon Gravity tends to produce less massive halos than $\Lambda$CDM model.
    \item[5] The galaxy-galaxy lensing signal difference is at most $6\%$, which is not enough to be distinguished by the current observational data.
    \item[6] The difference is mainly caused by the time and scale dependant effective gravitational constant.
\end{itemize}
From the simulation results, we can see that the modified Poisson equation in the Cubic Galileon Gravity model clearly introduced different structure formation from $\Lambda$CDM model. However, the effect in high density region, represented by galaxy-galaxy lensing signal and halo mass function, shows that it is not distinguishable in the uncertainty range. The void region, instead, shows more promising future. We can tell from the result of marked density and marked matter power spectrum that, when we suppress the weight of high density region and raise the weight in low density region, the difference is enhanced by about a factor of 2. We can also see that the number counting of void cells clearly shows the difference due to their large number and small error bar. It has also been reported that the void is crucial for telling the difference of modified gravity and $\Lambda$CDM model\citep{cai2014mnras,lam2015mnras,Voivodic2017PhRvD,baker2018prd,Sahl2018PhRvD}. In future, the void should be taken more seriously for constraining the Cubic Galileon Gravity models and maybe also other modified gravity models.

In future, We plan to study further the possibility of using voids to test modified gravity models. By combining the void lensing\citep{falck2018mnras,baker2018prd,dong2019apj,davies2019mnras} from observation and N-body simulation, we may be able to better constrain Cubic Galileon Gravity. 

\section*{Acknowledgement}
\noindent
J. Z was supported by IBS under the project code, IBS-R018-D1. AAS acknowledges funding from DST-SERB, Govt of India, under the project NO. MTR/20l9/000599. BRD would like to acknowledge DAE, Govt. of India for financial support through Visiting Fellow through TIFR. MWH was supported by the National Research Foundation of Korea (NRF No-2018R1A6A1A06024970) funded by the Ministry of Education, Korea. MWH also thanks Anjan A. Sen and acknowledges the warm hospitality of the Centre for Theoretical Physics, JMI, New Delhi, where part of the work was done. We would also like to thank Eoin \'O Colg\'ain fo organising the APCTP lecture series on (evidence for) physics beyond Lambda-CDM at APCTP, Pohang, Korea where the discussion and collaboration initiated.

\appendix

\section{Action of the cubic Galileon model}
\label{sec-actioncubgal}

The evolutionary dynamics of the Cubic Galileon field, $\phi$ is described by the action given by \citep{Nicolis:2008in,Deffayet:2009wt}

\begin{equation}
S=\int d^4x\sqrt{-g}\Bigl [\frac{M^2_{\rm{pl}}}{2} R +  \frac{1}{2} \sum_{i=1}^{3} c_{i} \mathcal{L}_{i} \Bigr] + \mathcal{S}_m \, ,
\label{eq:actionmain}
\end{equation}

\noindent
with $ \mathcal{L}_{1} = M^{3} \phi $, $ \mathcal{L}_{2} = (\nabla \phi)^2 $ and $ \mathcal{L}_{3} = \frac{(\nabla \phi)^2}{M^{3}} \Box \phi $. Where M is a mass dimensional constant $c_{i}^{,}s$ are dimensionless constants. For simplicity, we take $ c_{2} = - 1 $ since this does not change the essence of the Cubic Galileon model. Also, we define $\frac{c_{3}}{M^{3}} = - \beta$. We consider the linear term in the action \eqref{eq:actionmain} in a way that it looks like a potential given by $V(\phi) = - \frac{1}{2} c_{1} M^{3} \phi$. So, the action \eqref{eq:actionmain} looks like $S = \int d^4x\sqrt{-g}\Bigl [\frac{M^2_{\rm{pl}}}{2} R -  \frac{1}{2}(\nabla \phi)^2\Bigl(1 + \beta \Box \phi\Bigr) - V(\phi) \Bigr] + \mathcal{S}_m \,$ \citep{Hossain:2012qm,Ali:2012cv} which is exactly the Eq.~\eqref{eq:action}. The purpose to write down the action in this form is that for $\beta = 0$ the action reduces to the standard quintessence action with linear potential \citep{Wetterich:1987fk,Wetterich:1987fm,Caldwell:2005tm,Linder:2006sv,Tsujikawa:2010sc,Scherrer:2007pu,Dinda:2016ibo}.

\section{Detailed perturbation calculation}
\label{sec-detailed_perturbation}

We present the detailed perturbation calculations here (mainly which have not discussed in the main text). The first order Einstein equations (with the metric \eqref{eq:sptm}) are given by \citep{Unnikrishnan:2008qe}:

\begin{eqnarray}
\vec{\nabla}^{2} \Phi - 3 a^{2} H (\dot{\Phi} + H \Phi) &=& 4 \pi G a^{2} \sum_{i} \delta \rho_{i} \, , 
\label{eq:ein_per_1}\\
\dot{\Phi} + H \Phi &=& 4 \pi G a \sum_{i} (\bar{\rho_{i}} + \bar{P_{i}}) v_{i} \, , 
\label{eq:ein_per_2}\\
\ddot{\Phi} + 4 H \dot{\Phi} + (2 \dot{H} + 3 H^{2}) \Phi &=& 4 \pi G \sum_{i} \delta P_{i} \, ,
\label{eq:ein_per_3}
\end{eqnarray}

\noindent
where the summation is over matter and Galileon field. Any quantity with $bar$ corresponds to the background counterpart. $\delta \rho_{i} $, $ \delta P_{i} $ and $ v_{i} $ are the perturbations of the individual component's ($i=m$ for matter and $i=\phi$ for Galileon) energy density, pressure and velocity field respectively. Combining Eqs.~\eqref{eq:ein_per_1} and \eqref{eq:ein_per_2}, we have the relativistic Poisson equation given by

\begin{equation}
\vec{\nabla}^{2} \Phi = 4 \pi G a^{2} \sum_{i} \bar{\rho_{i}} \Delta_{i},
\label{eq:Pssn}
\end{equation}

\noindent
where $ \Delta_{i} $ is given by

\begin{equation}
\Delta_{i} = \delta_{i} + 3 \mathcal{H} (1 + w_{i}) v_{i},
\label{eq:gnrlcmvngD}
\end{equation}

\noindent
where $\delta_{i}$ is the individual component's energy density contrast, defined through $ \delta \rho_{i} = \bar{\rho}_{i} \delta_{i} $. $\Delta_{i}$ is a gauge invariant quantity and it is called the comoving energy density contrast for a particular component (i.e. either for matter or for Galileon). Here $\mathcal{H}$ is the conformal Hubble parameter ($ \mathcal{H} = a H $).
\\
\noindent
With the space-time~\eqref{eq:sptm} and from the action \eqref{eq:action}, the first order perturbed energy density, pressure and velocity for the Galileon field $\phi$ become \citep{Dinda:2017lpz,Hossain:2017ica}

\begin{eqnarray}
\delta \rho_{\phi} &=& (1-9 \beta H \dot{\phi}) \dot{\phi} \dot{\delta \phi} + \beta \dot{\phi}^{2} \frac{\vec{\nabla}^{2} \delta \phi}{a^{2}}  + V_{\phi} \delta \phi \nonumber\\
&& - (1-12 \beta H \dot{\phi}) \dot{\phi}^{2} \Phi + 3 \beta \dot{\phi}^{3} \dot{\Phi},
\label{eq:del_rho_phi} \\
\delta P_{\phi} &=& \beta \dot{\phi}^{2} \ddot{\delta \phi} + (1+2 \beta \ddot{\phi}) \dot{\phi} \dot{\delta \phi} \nonumber\\
&& - (1+4 \beta \ddot{\phi}) \dot{\phi}^{2} \Phi - \beta \dot{\phi}^{3} \dot{\Phi} - V_{\phi} \delta \phi, 
\label{eq:del_p_phi}\\
a (\bar{\rho_{\phi}} + \bar{P_{\phi}}) v_{\phi} &=& \beta \dot{\phi}^{2} \dot{\delta \phi} + (1-3 \beta H \dot{\phi}) \dot{\phi} \delta \phi \nonumber\\
&& - \beta \dot{\phi}^{3} \Phi,
\label{eq:v_phi}
\end{eqnarray}

\noindent
where $ \delta \phi $ is the first order perturbation to the background field, $\phi$.
\\
\noindent
Now putting Eq.~\eqref{eq:del_p_phi} into Eq.~\eqref{eq:ein_per_3}, we get evolution equation for the gravitational potential $ \Phi $. And by varying the action \eqref{eq:action}, we calculate the Euler-Lagrangian equation order by order and in the first order perturbation we get evolution equation for the $ \delta \phi $. We are not explicitly writing down these two equations separately because of their large expressions. These are mentioned in the last four lines of Eq.~\eqref{eq:dynsys} in Appendix~\ref{sec-autosys}.

\subsection{Autonomous system of equations}
\label{sec-autosys}

Using both background and perturbed dimensionless quantities (mentioned in Eqs.~\eqref{eq:dimless_var_bkg} and.~\eqref{eq:dimless_var_pert}), we form the following autonomous system of equations (including background and perturbation quantities together) \citep{Dinda:2017lpz}:

\begin{eqnarray}
\dfrac{d x}{d N} &=& f_{1}(x,y,\epsilon,\lambda), \nonumber\\
\dfrac{d y}{d N} &=& f_{2}(x,y,\epsilon,\lambda), \nonumber\\
\dfrac{d \epsilon}{d N} &=& f_{3}(x,y,\epsilon,\lambda), \nonumber\\
\dfrac{d \lambda}{d N} &=& \sqrt{6}x\lambda^2(1-\Gamma),\nonumber\\
\dfrac{d \tilde{\mathcal{H}}}{d N} &=& f_{4}(x,y,\epsilon,\lambda) \tilde{\mathcal{H}}, \nonumber\\
\dfrac{d \Phi}{d N} &=& \Phi_{1}, \nonumber\\
\dfrac{d q}{d N} &=& q_{1}, \nonumber\\
\dfrac{d \Phi_{1}}{d N} &=& f_{5}(x,y,\epsilon,\lambda,\Phi,q,\Phi_{1},q_{1}), \nonumber\\
\dfrac{d q_{1}}{d N } &=& f_{6}(x,y,\epsilon,\lambda,\Phi,q,\Phi_{1},q_{1}).
\label{eq:dynsys}
\end{eqnarray}

\noindent
Note that for simplicity of the notations, in the above set of equations, we have kept the same notations for $\Phi$ and $q$ in the Fourier space corresponding to the same quantities in the real space. $f_{1}$ to $f_{6}$ are given in the Appendix~\ref{sec-f1tof6}.

\subsection{Initial conditions}
\label{sec-initial}

We choose initial conditions at sufficiently large redshift, $z$ in early matter-dominated era. For this purpose $z=49$ is large enough to be considered. At this large redshift, the dark energy density contribution is negligible to the total energy density.

\begin{itemize}
\item
(1) Here, we consider thawing class of initial conditions \citep{Caldwell:2005tm,Linder:2006sv,Tsujikawa:2010sc,Scherrer:2007pu,Dinda:2016ibo}. In thawing class of scalar field models, due to the large Hubble friction in the early matter-dominated era, the scalar field is initially frozen to a value $ w_{\phi} \approx -1 $. At late times, the scalar field thaws away from its initial frozen state. The equation of state of the scalar field becomes larger towards non-phantom values ($ w_{\phi} > -1 $). For Cubic Galileon field, this thawing behaviour is possible if $ x\ll1 $ (this can be seen through first line of Eq. \eqref{eq:imp_bkg_qnt}: at $ x\ll1 $, $ w_{\phi} \approx \frac{-12 y^{2}(\epsilon+1)}{3(4 \epsilon+4)y^{2}} = -1 $). So, we restrict ourselves to $x_{i}=10^{-8}$. The subscript 'i' refers to the corresponding initial value of any quantity at initial redshift ($z_{i}=49$). Note that the evolution of the background quantities has no significant dependence on $x_{i}$ as long as $x_{i}\ll1$.
\item
(2) The initial condition in $y$ is chosen in such a way that the present value of $\Omega_{\phi}$ becomes a relevant specific value (this can be seen through second line of Eq. \eqref{eq:imp_bkg_qnt}). So, we compute $y_{i}$ by solving back $\Omega_{\phi}^{(0)} = 0.6844$. This value is consistent with Planck15, BAO, SNIa and H0 data \citep{Costa:2016tpb}.
\item
(3) The initial slope of the potential is controlled by the initial value of $\lambda$. For $\lambda_{i}\ll1$, the equation of state of the Galileon field does not deviate much from its initial value $-1$ (i.e. initially, it always stays very close to the cosmological constant behavior). For higher values of $\lambda_{i}$, the Galileon field sufficiently thaws away from the cosmological constant behavior accordingly. So, in our analysis, we consider $\lambda_{i}=0.7$ throughout.
\item
(4) We keep $\epsilon_{i}$ to be a free parameter.
\item
(5) The initial value of $\tilde{\mathcal{H}}$ is chosen such that it becomes $1$ at present.
\item
(6) Initially, at redshift $z_{i}=49$, there is hardly any contribution from the Galileon field to the evolution. So, we set $q_{i}=0$.
\item
(7) For the same reason (same to the previous point), we put $ q_{1}|_{i} = \dfrac{d q}{dN} \Big{|}_{i} = 0 $.
\item
(8) One can check that, during the matter dominated era, $\Phi$ is constant i.e. $ \Phi_{1}|_{i} = \dfrac{d \Phi}{dN} \Big{|}_{i} = 0 $.
\item
(9) Also, during the matter dominated era, we have $\Delta_{m} \sim a$ (can be seen through Eq.~\eqref{eq:gnrlcmvngD} or Eq.~\eqref{eq:Deltam}). Considering this and using the Poisson equation, Eq.~\eqref{eq:Pssn}, we get the initial condition in $\Phi$ given by

\begin{equation}
\Phi_{i} = - \frac{3}{2} \frac{\mathcal{H}^{2}_{i}}{k^2} a_{i} = - \frac{3}{2} \left[ \frac{h^{2}}{3000} \frac{\tilde{\mathcal{H}}_{i}}{ \left(k \hspace{0.1 cm} \text{in} \hspace{0.1 cm} h \hspace{0.1 cm} \text{Mpc}^{-1}\right) } \right]^{2} a_{i},
\label{eq:Phi_ini}
\end{equation}
\end{itemize}

\noindent
where $h$ is related to the present value of Hubble parameter given by $H_{0}=100 \hspace{0.1 cm} h \hspace{0.1 cm} \text{km} \hspace{0.1 cm} \text{s}^{-1} \text{Mpc}^{-1}$.

\subsection{matter energy density contrast}
\label{sec-deltam}

By putting Eq.~\eqref{eq:del_rho_phi} into Eq.~\eqref{eq:ein_per_1} and going to the Fourier space, we get the matter density contrast given in Eq.~\eqref{eq:delm}. The expression of quantity $A$ in Eq.~\eqref{eq:delm} is given by

\begin{eqnarray}
A &=& \frac{\big{(} \frac{d^2 \phi}{d N^2} \big{)}}{\big{(} \frac{d \phi}{d N} \big{)}} = \frac{-3 B \epsilon -2 B+2 J+6 \epsilon }{2 (\epsilon +1)},
\label{eq:dlm_A} \\
& \text{with} & \nonumber\\
B &=& 3+\frac{1}{H} \left(\frac{dH}{dN}\right) = 2+\frac{1}{\mathcal{H}} \left(\frac{d\mathcal{H}}{dN}\right) \nonumber\\
&=& \frac{3}{2}(1-\omega_{\phi} \Omega_{\phi}),
\label{eq:dlm_B}
\end{eqnarray}

\noindent
where $L = \frac{k^{2}}{3 \mathcal{H}^{2}} = \frac{1}{3} \left[ \frac{3000}{h^{2}} \frac{1}{\tilde{\mathcal{H}}} \left( k \hspace{0.1 cm} \text{in} \hspace{0.1 cm} h \hspace{0.1 cm} \text{Mpc}^{-1} \right) \right]^{2}$ and $J=\sqrt{\frac{3}{2}} \lambda \frac{y^{2}}{x}$.
\\
\noindent
Similarly, by putting Eq.~\eqref{eq:v_phi} into Eq.~\eqref{eq:ein_per_2} and going to the Fourier space, we get the pecular velocity for the matter given in Eq.~\eqref{eq:ym}.

\subsection{Dark energy density contrast}
\label{sec-deltaDE}

The dark energy density contrast can be computed as

\begin{equation}
\delta_{de} = \frac{x T_A}{T_B},
\label{eq:delde}
\end{equation}

where $T_{A}$ and $T_{B}$ are given by

\begin{eqnarray}
T_{A} &=& T_{A}^{(1)}+T_{A}^{(2)}, \nonumber\\
T_{B} &=& 3 \left(x^2 (\epsilon +1)+y^2\right), \nonumber\\
&& \text{with} \nonumber\\
T_{A}^{(1)} &=& -3 \sqrt{6} \lambda  q y^2-6 \Phi  x (2 \epsilon +1)-3 \Phi _1 x \epsilon + 3 L q x \epsilon, \nonumber\\
T_{A}^{(2)} &=& \frac{3 (3 \epsilon +2) (T_{A}^{21}+q T_{A}^{22})}{x^2 \epsilon ^2+4 \epsilon +4}, \nonumber\\
&& \text{with} \nonumber\\
T_{A}^{21} &=& q_1 x \left(x^2 \epsilon ^2+4 \epsilon +4\right), \nonumber\\
T_{A}^{22} &=& 3 x \left(x^2 \left(\epsilon ^2+5 \epsilon +2\right)+\epsilon -2\right) \nonumber\\
&& -x y^2 \left(\epsilon  \left(\sqrt{6} \lambda  x+9\right)+6\right)+2 \sqrt{6} \lambda  y^2.
\label{eq:deldeTs}
\end{eqnarray}

\section{$f_{1}$ to $f_{6}$ in Eq.~\eqref{eq:dynsys}}
\label{sec-f1tof6}

$f_{1}$ to $f_{3}$ in Eq.~\eqref{eq:dynsys} are given by

\begin{eqnarray}
f_{1} &=& \frac{1}{f_{d}} \Big{[} 3 x^3 \left(2+5 \epsilon +\epsilon^2\right)-3 x \left(2-\epsilon +y^2 (2+3 \epsilon )\right) \nonumber\\
&& +2 \sqrt{6} y^2 \lambda -\sqrt{6} x^2 y^2 \epsilon  \lambda \Big{]},
\end{eqnarray}

\begin{eqnarray}
f_{2} &=& - \frac{y}{2 f_{d}} \Big{[} 12 \left(-1+y^2\right) (1+\epsilon )-6 x^2 \left(2+4 \epsilon +\epsilon^2\right) \nonumber\\
&& +\sqrt{6} x^3 \epsilon^2 \lambda +2 \sqrt{6} x \left(2+\left(2+y^2\right) \epsilon \right) \lambda \Big{]},
\end{eqnarray}

\begin{eqnarray}
f_{3} &=& - \frac{\epsilon}{x f_{d}} \Big{[} -3 x \left(-3+y^2\right) (2+\epsilon )+3 x^3 \left(2+3 \epsilon +\epsilon^2\right) \nonumber\\
&& -2 \sqrt{6} y^2 \lambda -\sqrt{6} x^2 y^2 \epsilon  \lambda \Big{]},
\end{eqnarray}

\noindent
with

\begin{equation}
f_{d} = 4+4 \epsilon +x^2 \epsilon^2.
\end{equation}

\noindent
$f_{4}$ is given by

\begin{eqnarray}
f_{4} &=& - \frac{1}{2} \left( 1 + \frac{f_{n}}{ f_{d} } \right) = -\frac{1}{2} (1 + 3 w_{\phi} \Omega_{\phi}) \nonumber\\
&& \text{with} \nonumber\\
&& f_{n} = 3 x^2 (\epsilon  (\epsilon +8)+4)-2 \sqrt{6} \lambda  x y^2 \epsilon \nonumber\\
&& -12 y^2 (\epsilon +1).
\end{eqnarray}

\noindent
$f_{5}$ and $f_{6}$ in Eq.~\eqref{eq:dynsys} are given by

\begin{eqnarray}
f_{5} &=& A_{2}^{-1} [ x^2 (\epsilon  (4 \epsilon ^2 (-2 (J-3) x^2+L-3) \nonumber\\
&& +4 \epsilon  (-4 J+L+6 x^2-6)+L x^2 \epsilon ^3-48) \nonumber\\
&& -12 Q^2 (\epsilon  (\epsilon  (x^2 (2 \epsilon +3)+4)+8)+4)) ] \Phi \nonumber\\
&& - A_{1}^{-1} [ 2 (\epsilon +1) \left(A_4 x^2 \epsilon -2 A_3\right) ] q \nonumber\\
&& - A_{2}^{-1} [ 2 x^4 \epsilon ^2 \left(\epsilon  (J+2 \epsilon )+3 Q^2-3\right) \nonumber\\
&& +2 x^2 (\epsilon ^2 (8 J+10 \epsilon -11)+4 (J-6) \epsilon \nonumber\\
&& +12 Q^2 (\epsilon +1)^2-12)+40 (\epsilon +1)^2 ] \Phi_{1} \nonumber\\
&& + A_{2}^{-1} [ 2 x^2 (\epsilon  (2 J \left(\epsilon  \left(x^2 \epsilon -2\right)-4\right) \nonumber\\
&& +3 \epsilon  (x^2 \left(Q^2 (3 \epsilon +4)-2 (\epsilon +1)\right) \nonumber\\
&& +3 \epsilon +20)+84)+24) ] q_{1},
\label{eq:f1tof5}
\end{eqnarray}

\begin{eqnarray}
f_{6} &=& A_{2}^{-1} [ 8 J \left(\epsilon  \left(3 x^2 \epsilon +8\right)+4\right) \nonumber\\
&& -2 x^2 \epsilon ^3 \left(L+\left(6 Q^2-3\right) x^2+3\right) \nonumber\\
&& -8 \epsilon ^2 \left(L+3 \left(Q^2+2\right) x^2\right)-8 \epsilon  \left(L+3 x^2+9\right) ] \Phi \nonumber\\
&& + A_{1}^{-1} [ 2 A_3 \epsilon +4 A_4 (\epsilon +1) ] q \nonumber\\
&& + A_{2}^{-1} [ \epsilon  (16 J+\epsilon  (2 x^2 \left(-6 Q^2+7 \epsilon +16\right) \nonumber\\
&& +x^4 \epsilon ^2+28)+56)+64 ] \Phi_{1} \nonumber\\
&& + A_{2}^{-1} [ 2 J \left(\epsilon  \left(x^2 \left(\epsilon  \left(x^2 \epsilon -8\right)+4\right)-24\right)-16\right) \nonumber\\
&& -3 x^4 \epsilon ^2 \left(-2 Q^2 (3 \epsilon +1)+\epsilon  (\epsilon +6)+2\right) \nonumber\\
&& + 6 x^2 \left(Q^2 \left(6 \epsilon ^2+8 \epsilon +4\right)+\epsilon  \left(-\epsilon ^2+\epsilon -8\right)-4\right) \nonumber\\
&& -12 ((\epsilon -4) \epsilon -2) ] q_{1},
\label{eq:f1tof5}
\end{eqnarray}

\noindent
with

\begin{eqnarray}
Q &=& \frac{y}{x} \nonumber\\
J &=& \sqrt{\frac{3}{2}} \lambda \frac{y^{2}}{x} \nonumber\\
A_{1} &=& f_{d} \nonumber\\
A_{2} &=& f_{d}^{2}.
\label{eq:A1toA2}
\end{eqnarray}

\noindent
Finally, $A_{3}$ and $A_{4}$ are given by

\begin{eqnarray}
A_{3} &=& - Q^{-2} A_{1}^{-3} x^{2} [Q^2 ( 4 J^2 \epsilon  (\epsilon  (x^6 \epsilon ^3+4 x^4 \epsilon  (\epsilon +1) \nonumber\\
&& -4 x^2 (7 \epsilon +6)+8 )+16 ) \nonumber\\
&& + 6 J (\epsilon  (-x^6 \epsilon ^3 (5 \epsilon +4)+x^4 \epsilon  (\epsilon  ((\epsilon -24) \epsilon -40)-16) \nonumber\\
&& +16 x^2 (\epsilon +1) (2 \epsilon  (\epsilon +6)+5)-8 (\epsilon  (\epsilon +16)+26) )-64 ) \nonumber\\
&& + 9 (x^6 \epsilon ^3 ( 3 \epsilon  (\epsilon +2)^2+4 )+x^4 \epsilon  (\epsilon  (\epsilon  (\epsilon  (23 \epsilon +112)+156) \nonumber\\
&& +80)+16)-x^2 (\epsilon  (\epsilon  (\epsilon  (\epsilon \nonumber\\
&& (9 \epsilon +94)+380)+480)+208)+32) \nonumber\\
&& - 2 \epsilon ^3 (3 \epsilon +26)+96 \epsilon +32 ) )+2 \Gamma  J^2 \epsilon \nonumber\\
&& (x^2 \epsilon -2 ) (\epsilon  (x^2 \epsilon +4 )+4 )^2 \nonumber\\
&& +3 Q^4 x^2 (\epsilon  (8 J (\epsilon  (x^2 (\epsilon +1)  \nonumber\\
&& (\epsilon  (x^2 \epsilon +8 )+4 )-2 (7 \epsilon +12) )-8 ) \nonumber\\
&& - 3 x^4 \epsilon ^2 (3 \epsilon  (\epsilon +2) (\epsilon +3)+8)-6 x^2 (\epsilon  (\epsilon  (\epsilon  (15 \epsilon +88) \nonumber\\
&& +132)+72)+16)+12 (\epsilon  (\epsilon  (26-3 \epsilon ) \nonumber\\
&& +60)+36) )+96 ) \nonumber\\
&& + 9 Q^6 x^4 \epsilon  (\epsilon  (\epsilon  (x^2 (3 \epsilon  (\epsilon +2)+4) \nonumber\\
&& +42 \epsilon +92 )+64 )+16 )],
\label{eq:apndxA3}
\end{eqnarray}

\begin{eqnarray}
A_{4} &=& Q^{-2} (1+\epsilon)^{-1} A_{1}^{-3} [-2 J^2 x^2 \epsilon  (2 Q^2 (\epsilon  \nonumber\\
&& (x^2 (\epsilon  (\epsilon +2) (x^2 \epsilon +8 )+8 )-44 \epsilon -80 )-32 ) \nonumber\\
&& + \Gamma  (3 \epsilon +2) (\epsilon  (x^2 \epsilon +4)+4)^2) \nonumber\\
&& -4 J Q^2 (x^4 \epsilon  (\epsilon  (\epsilon  (\epsilon  ((L+21) \epsilon \nonumber\\
&& -45)-192)-168) \nonumber\\
&& +6 Q^2 (\epsilon  (\epsilon  (13 \epsilon +34)+28)+8)-48 ) \nonumber\\
&& + 2 x^2 (\epsilon  (\epsilon  (\epsilon  (4 L (\epsilon +1)+75 \epsilon +390)+612)+360) \nonumber\\
&& -24 Q^2 (2 \epsilon +1) (\epsilon +1)^2+72 )+16 (\epsilon +1)^2 ((L+6) \epsilon +3) \nonumber\\
&& + 3 x^6 \epsilon ^3 (Q^2 (\epsilon +1) (\epsilon +4)+\epsilon  ((\epsilon -1) \epsilon -7)-4 ) ) \nonumber\\
&& +Q^2 (9 (x^2 (16 Q^2 (3 \epsilon ^2+\epsilon +2 ) (\epsilon +1)^2 \nonumber\\
&& + \epsilon  (\epsilon  (\epsilon  (3 \epsilon  (\epsilon +16)+284)+456)+240)+32 ) \nonumber\\
&& +x^6 \epsilon ^2 (Q^4 (3 \epsilon ^3 \nonumber\\
&& -12 \epsilon -8 )+Q^2 (3 \epsilon +2) (\epsilon  (\epsilon  (\epsilon +3)+8)+8) \nonumber\\
&& - 2 (\epsilon +1) (\epsilon  (\epsilon  (2 \epsilon +7)+10)+4) ) \nonumber\\
&& -x^4 (16 Q^4 (\epsilon +1)^2 (\epsilon  (3 \epsilon +4)+2) \nonumber\\
&& -2 Q^2 (\epsilon  (\epsilon  (\epsilon  (\epsilon  (27 \epsilon +184)+384)+352)+160)+32) \nonumber\\
&& + \epsilon  (\epsilon  (\epsilon  (\epsilon  (\epsilon +7) (3 \epsilon +50) \nonumber\\
&& +624)+496)+192)+32 )+48 \epsilon  (\epsilon +1)^2 ) \nonumber\\
&& - L (\epsilon  (x^2 \epsilon +4 )+4 )^2 (\epsilon  (\epsilon  \nonumber\\
&& (x^2 (-3 Q^2+2 \epsilon +6 )+5 )+8 )+12 ) )].
\label{eq:apndxA4}
\end{eqnarray}

\bibliographystyle{apsrev4-1}
\bibliography{cgnbody}

\begin{thebibliography}{93}%
\makeatletter
\providecommand \@ifxundefined [1]{%
 \@ifx{#1\undefined}
}%
\providecommand \@ifnum [1]{%
 \ifnum #1\expandafter \@firstoftwo
 \else \expandafter \@secondoftwo
 \fi
}%
\providecommand \@ifx [1]{%
 \ifx #1\expandafter \@firstoftwo
 \else \expandafter \@secondoftwo
 \fi
}%
\providecommand \natexlab [1]{#1}%
\providecommand \enquote  [1]{``#1''}%
\providecommand \bibnamefont  [1]{#1}%
\providecommand \bibfnamefont [1]{#1}%
\providecommand \citenamefont [1]{#1}%
\providecommand \href@noop [0]{\@secondoftwo}%
\providecommand \href [0]{\begingroup \@sanitize@url \@href}%
\providecommand \@href[1]{\@@startlink{#1}\@@href}%
\providecommand \@@href[1]{\endgroup#1\@@endlink}%
\providecommand \@sanitize@url [0]{\catcode `\\12\catcode `\$12\catcode
  `\&12\catcode `\#12\catcode `\^12\catcode `\_12\catcode `\%12\relax}%
\providecommand \@@startlink[1]{}%
\providecommand \@@endlink[0]{}%
\providecommand \url  [0]{\begingroup\@sanitize@url \@url }%
\providecommand \@url [1]{\endgroup\@href {#1}{\urlprefix }}%
\providecommand \urlprefix  [0]{URL }%
\providecommand \Eprint [0]{\href }%
\providecommand \doibase [0]{http://dx.doi.org/}%
\providecommand \selectlanguage [0]{\@gobble}%
\providecommand \bibinfo  [0]{\@secondoftwo}%
\providecommand \bibfield  [0]{\@secondoftwo}%
\providecommand \translation [1]{[#1]}%
\providecommand \BibitemOpen [0]{}%
\providecommand \bibitemStop [0]{}%
\providecommand \bibitemNoStop [0]{.\EOS\space}%
\providecommand \EOS [0]{\spacefactor3000\relax}%
\providecommand \BibitemShut  [1]{\csname bibitem#1\endcsname}%
\let\auto@bib@innerbib\@empty
\bibitem [{\citenamefont {Riess}\ \emph {et~al.}(1998)\citenamefont {Riess}
  \emph {et~al.}}]{Riess:1998cb}%
  \BibitemOpen
  \bibfield  {author} {\bibinfo {author} {\bibfnamefont {A.~G.}\ \bibnamefont
  {Riess}} \emph {et~al.} (\bibinfo {collaboration} {Supernova Search Team}),\
  }\href {\doibase 10.1086/300499} {\bibfield  {journal} {\bibinfo  {journal}
  {Astron. J.}\ }\textbf {\bibinfo {volume} {116}},\ \bibinfo {pages} {1009}
  (\bibinfo {year} {1998})},\ \Eprint {http://arxiv.org/abs/astro-ph/9805201}
  {arXiv:astro-ph/9805201} \BibitemShut {NoStop}%
\bibitem [{\citenamefont {Perlmutter}\ \emph {et~al.}(1999)\citenamefont
  {Perlmutter} \emph {et~al.}}]{Perlmutter:1998np}%
  \BibitemOpen
  \bibfield  {author} {\bibinfo {author} {\bibfnamefont {S.}~\bibnamefont
  {Perlmutter}} \emph {et~al.} (\bibinfo {collaboration} {Supernova Cosmology
  Project}),\ }\href {\doibase 10.1086/307221} {\bibfield  {journal} {\bibinfo
  {journal} {Astrophys. J.}\ }\textbf {\bibinfo {volume} {517}},\ \bibinfo
  {pages} {565} (\bibinfo {year} {1999})},\ \Eprint
  {http://arxiv.org/abs/astro-ph/9812133} {arXiv:astro-ph/9812133} \BibitemShut
  {NoStop}%
\bibitem [{\citenamefont {Tonry}\ \emph {et~al.}(2003)\citenamefont {Tonry}
  \emph {et~al.}}]{Tonry:2003zg}%
  \BibitemOpen
  \bibfield  {author} {\bibinfo {author} {\bibfnamefont {J.~L.}\ \bibnamefont
  {Tonry}} \emph {et~al.} (\bibinfo {collaboration} {Supernova Search Team}),\
  }\href {\doibase 10.1086/376865} {\bibfield  {journal} {\bibinfo  {journal}
  {Astrophys. J.}\ }\textbf {\bibinfo {volume} {594}},\ \bibinfo {pages} {1}
  (\bibinfo {year} {2003})},\ \Eprint {http://arxiv.org/abs/astro-ph/0305008}
  {arXiv:astro-ph/0305008} \BibitemShut {NoStop}%
\bibitem [{\citenamefont {Sahni}\ and\ \citenamefont
  {Starobinsky}(2000)}]{Sahni:1999gb}%
  \BibitemOpen
  \bibfield  {author} {\bibinfo {author} {\bibfnamefont {V.}~\bibnamefont
  {Sahni}}\ and\ \bibinfo {author} {\bibfnamefont {A.~A.}\ \bibnamefont
  {Starobinsky}},\ }\href {\doibase 10.1142/S0218271800000542} {\bibfield
  {journal} {\bibinfo  {journal} {Int. J. Mod. Phys. D}\ }\textbf {\bibinfo
  {volume} {9}},\ \bibinfo {pages} {373} (\bibinfo {year} {2000})},\ \Eprint
  {http://arxiv.org/abs/astro-ph/9904398} {arXiv:astro-ph/9904398} \BibitemShut
  {NoStop}%
\bibitem [{\citenamefont {Riess}(2019)}]{Riess:2020sih}%
  \BibitemOpen
  \bibfield  {author} {\bibinfo {author} {\bibfnamefont {A.~G.}\ \bibnamefont
  {Riess}},\ }\href {\doibase 10.1038/s42254-019-0137-0} {\bibfield  {journal}
  {\bibinfo  {journal} {Nature Rev. Phys.}\ }\textbf {\bibinfo {volume} {2}},\
  \bibinfo {pages} {10} (\bibinfo {year} {2019})},\ \Eprint
  {http://arxiv.org/abs/2001.03624} {arXiv:2001.03624 [astro-ph.CO]}
  \BibitemShut {NoStop}%
\bibitem [{\citenamefont {Riess}\ \emph {et~al.}(2019)\citenamefont {Riess},
  \citenamefont {Casertano}, \citenamefont {Yuan}, \citenamefont {Macri},\ and\
  \citenamefont {Scolnic}}]{Riess:2019cxk}%
  \BibitemOpen
  \bibfield  {author} {\bibinfo {author} {\bibfnamefont {A.~G.}\ \bibnamefont
  {Riess}}, \bibinfo {author} {\bibfnamefont {S.}~\bibnamefont {Casertano}},
  \bibinfo {author} {\bibfnamefont {W.}~\bibnamefont {Yuan}}, \bibinfo {author}
  {\bibfnamefont {L.~M.}\ \bibnamefont {Macri}}, \ and\ \bibinfo {author}
  {\bibfnamefont {D.}~\bibnamefont {Scolnic}},\ }\href {\doibase
  10.3847/1538-4357/ab1422} {\bibfield  {journal} {\bibinfo  {journal}
  {Astrophys. J.}\ }\textbf {\bibinfo {volume} {876}},\ \bibinfo {pages} {85}
  (\bibinfo {year} {2019})},\ \Eprint {http://arxiv.org/abs/1903.07603}
  {arXiv:1903.07603 [astro-ph.CO]} \BibitemShut {NoStop}%
\bibitem [{\citenamefont {Wong}\ \emph {et~al.}(2019)\citenamefont {Wong} \emph
  {et~al.}}]{Wong:2019kwg}%
  \BibitemOpen
  \bibfield  {author} {\bibinfo {author} {\bibfnamefont {K.~C.}\ \bibnamefont
  {Wong}} \emph {et~al.},\ }\href@noop {} {\  (\bibinfo {year} {2019})},\
  \Eprint {http://arxiv.org/abs/1907.04869} {arXiv:1907.04869 [astro-ph.CO]}
  \BibitemShut {NoStop}%
\bibitem [{\citenamefont {Pesce}\ \emph {et~al.}(2020)\citenamefont {Pesce}
  \emph {et~al.}}]{Pesce:2020xfe}%
  \BibitemOpen
  \bibfield  {author} {\bibinfo {author} {\bibfnamefont {D.}~\bibnamefont
  {Pesce}} \emph {et~al.},\ }\href {\doibase 10.3847/2041-8213/ab75f0}
  {\bibfield  {journal} {\bibinfo  {journal} {Astrophys. J.}\ }\textbf
  {\bibinfo {volume} {891}},\ \bibinfo {pages} {L1} (\bibinfo {year} {2020})},\
  \Eprint {http://arxiv.org/abs/2001.09213} {arXiv:2001.09213 [astro-ph.CO]}
  \BibitemShut {NoStop}%
\bibitem [{\citenamefont {Aghanim}\ \emph {et~al.}(2018)\citenamefont {Aghanim}
  \emph {et~al.}}]{Aghanim:2018eyx}%
  \BibitemOpen
  \bibfield  {author} {\bibinfo {author} {\bibfnamefont {N.}~\bibnamefont
  {Aghanim}} \emph {et~al.} (\bibinfo {collaboration} {Planck}),\ }\href@noop
  {} {\  (\bibinfo {year} {2018})},\ \Eprint {http://arxiv.org/abs/1807.06209}
  {arXiv:1807.06209 [astro-ph.CO]} \BibitemShut {NoStop}%
\bibitem [{\citenamefont {Wetterich}(1988{\natexlab{a}})}]{Wetterich:1987fk}%
  \BibitemOpen
  \bibfield  {author} {\bibinfo {author} {\bibfnamefont {C.}~\bibnamefont
  {Wetterich}},\ }\href {\doibase 10.1016/0550-3213(88)90192-7} {\bibfield
  {journal} {\bibinfo  {journal} {Nucl. Phys. B}\ }\textbf {\bibinfo {volume}
  {302}},\ \bibinfo {pages} {645} (\bibinfo {year}
  {1988}{\natexlab{a}})}\BibitemShut {NoStop}%
\bibitem [{\citenamefont {Wetterich}(1988{\natexlab{b}})}]{Wetterich:1987fm}%
  \BibitemOpen
  \bibfield  {author} {\bibinfo {author} {\bibfnamefont {C.}~\bibnamefont
  {Wetterich}},\ }\href {\doibase 10.1016/0550-3213(88)90193-9} {\bibfield
  {journal} {\bibinfo  {journal} {Nucl. Phys. B}\ }\textbf {\bibinfo {volume}
  {302}},\ \bibinfo {pages} {668} (\bibinfo {year} {1988}{\natexlab{b}})},\
  \Eprint {http://arxiv.org/abs/1711.03844} {arXiv:1711.03844 [hep-th]}
  \BibitemShut {NoStop}%
\bibitem [{\citenamefont {Copeland}\ \emph {et~al.}(2006)\citenamefont
  {Copeland}, \citenamefont {Sami},\ and\ \citenamefont
  {Tsujikawa}}]{Copeland:2006wr}%
  \BibitemOpen
  \bibfield  {author} {\bibinfo {author} {\bibfnamefont {E.~J.}\ \bibnamefont
  {Copeland}}, \bibinfo {author} {\bibfnamefont {M.}~\bibnamefont {Sami}}, \
  and\ \bibinfo {author} {\bibfnamefont {S.}~\bibnamefont {Tsujikawa}},\ }\href
  {\doibase 10.1142/S021827180600942X} {\bibfield  {journal} {\bibinfo
  {journal} {Int. J. Mod. Phys. D}\ }\textbf {\bibinfo {volume} {15}},\
  \bibinfo {pages} {1753} (\bibinfo {year} {2006})},\ \Eprint
  {http://arxiv.org/abs/hep-th/0603057} {arXiv:hep-th/0603057} \BibitemShut
  {NoStop}%
\bibitem [{\citenamefont {Clifton}\ \emph {et~al.}(2012)\citenamefont
  {Clifton}, \citenamefont {Ferreira}, \citenamefont {Padilla},\ and\
  \citenamefont {Skordis}}]{Clifton:2011jh}%
  \BibitemOpen
  \bibfield  {author} {\bibinfo {author} {\bibfnamefont {T.}~\bibnamefont
  {Clifton}}, \bibinfo {author} {\bibfnamefont {P.~G.}\ \bibnamefont
  {Ferreira}}, \bibinfo {author} {\bibfnamefont {A.}~\bibnamefont {Padilla}}, \
  and\ \bibinfo {author} {\bibfnamefont {C.}~\bibnamefont {Skordis}},\ }\href
  {\doibase 10.1016/j.physrep.2012.01.001} {\bibfield  {journal} {\bibinfo
  {journal} {Phys. Rept.}\ }\textbf {\bibinfo {volume} {513}},\ \bibinfo
  {pages} {1} (\bibinfo {year} {2012})},\ \Eprint
  {http://arxiv.org/abs/1106.2476} {arXiv:1106.2476 [astro-ph.CO]} \BibitemShut
  {NoStop}%
\bibitem [{\citenamefont {de~Rham}(2014)}]{deRham:2014zqa}%
  \BibitemOpen
  \bibfield  {author} {\bibinfo {author} {\bibfnamefont {C.}~\bibnamefont
  {de~Rham}},\ }\href {\doibase 10.12942/lrr-2014-7} {\bibfield  {journal}
  {\bibinfo  {journal} {Living Rev. Rel.}\ }\textbf {\bibinfo {volume} {17}},\
  \bibinfo {pages} {7} (\bibinfo {year} {2014})},\ \Eprint
  {http://arxiv.org/abs/1401.4173} {arXiv:1401.4173 [hep-th]} \BibitemShut
  {NoStop}%
\bibitem [{\citenamefont {de~Rham}(2012)}]{deRham:2012az}%
  \BibitemOpen
  \bibfield  {author} {\bibinfo {author} {\bibfnamefont {C.}~\bibnamefont
  {de~Rham}},\ }\href {\doibase 10.1016/j.crhy.2012.04.006} {\bibfield
  {journal} {\bibinfo  {journal} {Comptes Rendus Physique}\ }\textbf {\bibinfo
  {volume} {13}},\ \bibinfo {pages} {666} (\bibinfo {year} {2012})},\ \Eprint
  {http://arxiv.org/abs/1204.5492} {arXiv:1204.5492 [astro-ph.CO]} \BibitemShut
  {NoStop}%
\bibitem [{\citenamefont {De~Felice}\ and\ \citenamefont
  {Tsujikawa}(2010{\natexlab{a}})}]{DeFelice:2010aj}%
  \BibitemOpen
  \bibfield  {author} {\bibinfo {author} {\bibfnamefont {A.}~\bibnamefont
  {De~Felice}}\ and\ \bibinfo {author} {\bibfnamefont {S.}~\bibnamefont
  {Tsujikawa}},\ }\href {\doibase 10.12942/lrr-2010-3} {\bibfield  {journal}
  {\bibinfo  {journal} {Living Rev. Rel.}\ }\textbf {\bibinfo {volume} {13}},\
  \bibinfo {pages} {3} (\bibinfo {year} {2010}{\natexlab{a}})},\ \Eprint
  {http://arxiv.org/abs/1002.4928} {arXiv:1002.4928 [gr-qc]} \BibitemShut
  {NoStop}%
\bibitem [{\citenamefont {Dvali}\ \emph {et~al.}(2000)\citenamefont {Dvali},
  \citenamefont {Gabadadze},\ and\ \citenamefont {Porrati}}]{Dvali:2000hr}%
  \BibitemOpen
  \bibfield  {author} {\bibinfo {author} {\bibfnamefont {G.}~\bibnamefont
  {Dvali}}, \bibinfo {author} {\bibfnamefont {G.}~\bibnamefont {Gabadadze}}, \
  and\ \bibinfo {author} {\bibfnamefont {M.}~\bibnamefont {Porrati}},\ }\href
  {\doibase 10.1016/S0370-2693(00)00669-9} {\bibfield  {journal} {\bibinfo
  {journal} {Phys. Lett. B}\ }\textbf {\bibinfo {volume} {485}},\ \bibinfo
  {pages} {208} (\bibinfo {year} {2000})},\ \Eprint
  {http://arxiv.org/abs/hep-th/0005016} {arXiv:hep-th/0005016} \BibitemShut
  {NoStop}%
\bibitem [{\citenamefont {Luty}\ \emph {et~al.}(2003)\citenamefont {Luty},
  \citenamefont {Porrati},\ and\ \citenamefont {Rattazzi}}]{Luty:2003vm}%
  \BibitemOpen
  \bibfield  {author} {\bibinfo {author} {\bibfnamefont {M.~A.}\ \bibnamefont
  {Luty}}, \bibinfo {author} {\bibfnamefont {M.}~\bibnamefont {Porrati}}, \
  and\ \bibinfo {author} {\bibfnamefont {R.}~\bibnamefont {Rattazzi}},\ }\href
  {\doibase 10.1088/1126-6708/2003/09/029} {\bibfield  {journal} {\bibinfo
  {journal} {JHEP}\ }\textbf {\bibinfo {volume} {09}},\ \bibinfo {pages} {029}
  (\bibinfo {year} {2003})},\ \Eprint {http://arxiv.org/abs/hep-th/0303116}
  {arXiv:hep-th/0303116} \BibitemShut {NoStop}%
\bibitem [{\citenamefont {Nicolis}\ and\ \citenamefont
  {Rattazzi}(2004)}]{Nicolis:2004qq}%
  \BibitemOpen
  \bibfield  {author} {\bibinfo {author} {\bibfnamefont {A.}~\bibnamefont
  {Nicolis}}\ and\ \bibinfo {author} {\bibfnamefont {R.}~\bibnamefont
  {Rattazzi}},\ }\href {\doibase 10.1088/1126-6708/2004/06/059} {\bibfield
  {journal} {\bibinfo  {journal} {JHEP}\ }\textbf {\bibinfo {volume} {06}},\
  \bibinfo {pages} {059} (\bibinfo {year} {2004})},\ \Eprint
  {http://arxiv.org/abs/hep-th/0404159} {arXiv:hep-th/0404159} \BibitemShut
  {NoStop}%
\bibitem [{\citenamefont {Nicolis}\ \emph {et~al.}(2009)\citenamefont
  {Nicolis}, \citenamefont {Rattazzi},\ and\ \citenamefont
  {Trincherini}}]{Nicolis:2008in}%
  \BibitemOpen
  \bibfield  {author} {\bibinfo {author} {\bibfnamefont {A.}~\bibnamefont
  {Nicolis}}, \bibinfo {author} {\bibfnamefont {R.}~\bibnamefont {Rattazzi}}, \
  and\ \bibinfo {author} {\bibfnamefont {E.}~\bibnamefont {Trincherini}},\
  }\href {\doibase 10.1103/PhysRevD.79.064036} {\bibfield  {journal} {\bibinfo
  {journal} {Phys. Rev. D}\ }\textbf {\bibinfo {volume} {79}},\ \bibinfo
  {pages} {064036} (\bibinfo {year} {2009})},\ \Eprint
  {http://arxiv.org/abs/0811.2197} {arXiv:0811.2197 [hep-th]} \BibitemShut
  {NoStop}%
\bibitem [{\citenamefont {Deffayet}\ \emph {et~al.}(2009)\citenamefont
  {Deffayet}, \citenamefont {Esposito-Farese},\ and\ \citenamefont
  {Vikman}}]{Deffayet:2009wt}%
  \BibitemOpen
  \bibfield  {author} {\bibinfo {author} {\bibfnamefont {C.}~\bibnamefont
  {Deffayet}}, \bibinfo {author} {\bibfnamefont {G.}~\bibnamefont
  {Esposito-Farese}}, \ and\ \bibinfo {author} {\bibfnamefont {A.}~\bibnamefont
  {Vikman}},\ }\href {\doibase 10.1103/PhysRevD.79.084003} {\bibfield
  {journal} {\bibinfo  {journal} {Phys. Rev. D}\ }\textbf {\bibinfo {volume}
  {79}},\ \bibinfo {pages} {084003} (\bibinfo {year} {2009})},\ \Eprint
  {http://arxiv.org/abs/0901.1314} {arXiv:0901.1314 [hep-th]} \BibitemShut
  {NoStop}%
\bibitem [{\citenamefont {Horndeski}(1974)}]{Horndeski:1974wa}%
  \BibitemOpen
  \bibfield  {author} {\bibinfo {author} {\bibfnamefont {G.~W.}\ \bibnamefont
  {Horndeski}},\ }\href {\doibase 10.1007/BF01807638} {\bibfield  {journal}
  {\bibinfo  {journal} {Int. J. Theor. Phys.}\ }\textbf {\bibinfo {volume}
  {10}},\ \bibinfo {pages} {363} (\bibinfo {year} {1974})}\BibitemShut
  {NoStop}%
\bibitem [{\citenamefont {Chow}\ and\ \citenamefont
  {Khoury}(2009)}]{Chow:2009fm}%
  \BibitemOpen
  \bibfield  {author} {\bibinfo {author} {\bibfnamefont {N.}~\bibnamefont
  {Chow}}\ and\ \bibinfo {author} {\bibfnamefont {J.}~\bibnamefont {Khoury}},\
  }\href {\doibase 10.1103/PhysRevD.80.024037} {\bibfield  {journal} {\bibinfo
  {journal} {Phys. Rev. D}\ }\textbf {\bibinfo {volume} {80}},\ \bibinfo
  {pages} {024037} (\bibinfo {year} {2009})},\ \Eprint
  {http://arxiv.org/abs/0905.1325} {arXiv:0905.1325 [hep-th]} \BibitemShut
  {NoStop}%
\bibitem [{\citenamefont {Silva}\ and\ \citenamefont
  {Koyama}(2009)}]{Silva:2009km}%
  \BibitemOpen
  \bibfield  {author} {\bibinfo {author} {\bibfnamefont {F.~P.}\ \bibnamefont
  {Silva}}\ and\ \bibinfo {author} {\bibfnamefont {K.}~\bibnamefont {Koyama}},\
  }\href {\doibase 10.1103/PhysRevD.80.121301} {\bibfield  {journal} {\bibinfo
  {journal} {Phys. Rev. D}\ }\textbf {\bibinfo {volume} {80}},\ \bibinfo
  {pages} {121301} (\bibinfo {year} {2009})},\ \Eprint
  {http://arxiv.org/abs/0909.4538} {arXiv:0909.4538 [astro-ph.CO]} \BibitemShut
  {NoStop}%
\bibitem [{\citenamefont {Kobayashi}(2010)}]{Kobayashi:2010wa}%
  \BibitemOpen
  \bibfield  {author} {\bibinfo {author} {\bibfnamefont {T.}~\bibnamefont
  {Kobayashi}},\ }\href {\doibase 10.1103/PhysRevD.81.103533} {\bibfield
  {journal} {\bibinfo  {journal} {Phys. Rev. D}\ }\textbf {\bibinfo {volume}
  {81}},\ \bibinfo {pages} {103533} (\bibinfo {year} {2010})},\ \Eprint
  {http://arxiv.org/abs/1003.3281} {arXiv:1003.3281 [astro-ph.CO]} \BibitemShut
  {NoStop}%
\bibitem [{\citenamefont {Kobayashi}\ \emph {et~al.}(2010)\citenamefont
  {Kobayashi}, \citenamefont {Tashiro},\ and\ \citenamefont
  {Suzuki}}]{Kobayashi:2009wr}%
  \BibitemOpen
  \bibfield  {author} {\bibinfo {author} {\bibfnamefont {T.}~\bibnamefont
  {Kobayashi}}, \bibinfo {author} {\bibfnamefont {H.}~\bibnamefont {Tashiro}},
  \ and\ \bibinfo {author} {\bibfnamefont {D.}~\bibnamefont {Suzuki}},\ }\href
  {\doibase 10.1103/PhysRevD.81.063513} {\bibfield  {journal} {\bibinfo
  {journal} {Phys. Rev. D}\ }\textbf {\bibinfo {volume} {81}},\ \bibinfo
  {pages} {063513} (\bibinfo {year} {2010})},\ \Eprint
  {http://arxiv.org/abs/0912.4641} {arXiv:0912.4641 [astro-ph.CO]} \BibitemShut
  {NoStop}%
\bibitem [{\citenamefont {Gannouji}\ and\ \citenamefont
  {Sami}(2010)}]{Gannouji:2010au}%
  \BibitemOpen
  \bibfield  {author} {\bibinfo {author} {\bibfnamefont {R.}~\bibnamefont
  {Gannouji}}\ and\ \bibinfo {author} {\bibfnamefont {M.}~\bibnamefont
  {Sami}},\ }\href {\doibase 10.1103/PhysRevD.82.024011} {\bibfield  {journal}
  {\bibinfo  {journal} {Phys. Rev. D}\ }\textbf {\bibinfo {volume} {82}},\
  \bibinfo {pages} {024011} (\bibinfo {year} {2010})},\ \Eprint
  {http://arxiv.org/abs/1004.2808} {arXiv:1004.2808 [gr-qc]} \BibitemShut
  {NoStop}%
\bibitem [{\citenamefont {De~Felice}\ \emph {et~al.}(2010)\citenamefont
  {De~Felice}, \citenamefont {Mukohyama},\ and\ \citenamefont
  {Tsujikawa}}]{DeFelice:2010gb}%
  \BibitemOpen
  \bibfield  {author} {\bibinfo {author} {\bibfnamefont {A.}~\bibnamefont
  {De~Felice}}, \bibinfo {author} {\bibfnamefont {S.}~\bibnamefont
  {Mukohyama}}, \ and\ \bibinfo {author} {\bibfnamefont {S.}~\bibnamefont
  {Tsujikawa}},\ }\href {\doibase 10.1103/PhysRevD.82.023524} {\bibfield
  {journal} {\bibinfo  {journal} {Phys. Rev. D}\ }\textbf {\bibinfo {volume}
  {82}},\ \bibinfo {pages} {023524} (\bibinfo {year} {2010})},\ \Eprint
  {http://arxiv.org/abs/1006.0281} {arXiv:1006.0281 [astro-ph.CO]} \BibitemShut
  {NoStop}%
\bibitem [{\citenamefont {De~Felice}\ and\ \citenamefont
  {Tsujikawa}(2010{\natexlab{b}})}]{DeFelice:2010pv}%
  \BibitemOpen
  \bibfield  {author} {\bibinfo {author} {\bibfnamefont {A.}~\bibnamefont
  {De~Felice}}\ and\ \bibinfo {author} {\bibfnamefont {S.}~\bibnamefont
  {Tsujikawa}},\ }\href {\doibase 10.1103/PhysRevLett.105.111301} {\bibfield
  {journal} {\bibinfo  {journal} {Phys. Rev. Lett.}\ }\textbf {\bibinfo
  {volume} {105}},\ \bibinfo {pages} {111301} (\bibinfo {year}
  {2010}{\natexlab{b}})},\ \Eprint {http://arxiv.org/abs/1007.2700}
  {arXiv:1007.2700 [astro-ph.CO]} \BibitemShut {NoStop}%
\bibitem [{\citenamefont {Ali}\ \emph {et~al.}(2010)\citenamefont {Ali},
  \citenamefont {Gannouji},\ and\ \citenamefont {Sami}}]{Ali:2010gr}%
  \BibitemOpen
  \bibfield  {author} {\bibinfo {author} {\bibfnamefont {A.}~\bibnamefont
  {Ali}}, \bibinfo {author} {\bibfnamefont {R.}~\bibnamefont {Gannouji}}, \
  and\ \bibinfo {author} {\bibfnamefont {M.}~\bibnamefont {Sami}},\ }\href
  {\doibase 10.1103/PhysRevD.82.103015} {\bibfield  {journal} {\bibinfo
  {journal} {Phys. Rev. D}\ }\textbf {\bibinfo {volume} {82}},\ \bibinfo
  {pages} {103015} (\bibinfo {year} {2010})},\ \Eprint
  {http://arxiv.org/abs/1008.1588} {arXiv:1008.1588 [astro-ph.CO]} \BibitemShut
  {NoStop}%
\bibitem [{\citenamefont {Mota}\ \emph {et~al.}(2010)\citenamefont {Mota},
  \citenamefont {Sandstad},\ and\ \citenamefont {Zlosnik}}]{Mota:2010bs}%
  \BibitemOpen
  \bibfield  {author} {\bibinfo {author} {\bibfnamefont {D.~F.}\ \bibnamefont
  {Mota}}, \bibinfo {author} {\bibfnamefont {M.}~\bibnamefont {Sandstad}}, \
  and\ \bibinfo {author} {\bibfnamefont {T.}~\bibnamefont {Zlosnik}},\ }\href
  {\doibase 10.1007/JHEP12(2010)051} {\bibfield  {journal} {\bibinfo  {journal}
  {JHEP}\ }\textbf {\bibinfo {volume} {12}},\ \bibinfo {pages} {051} (\bibinfo
  {year} {2010})},\ \Eprint {http://arxiv.org/abs/1009.6151} {arXiv:1009.6151
  [astro-ph.CO]} \BibitemShut {NoStop}%
\bibitem [{\citenamefont {Deffayet}\ \emph {et~al.}(2010)\citenamefont
  {Deffayet}, \citenamefont {Pujolas}, \citenamefont {Sawicki},\ and\
  \citenamefont {Vikman}}]{Deffayet:2010qz}%
  \BibitemOpen
  \bibfield  {author} {\bibinfo {author} {\bibfnamefont {C.}~\bibnamefont
  {Deffayet}}, \bibinfo {author} {\bibfnamefont {O.}~\bibnamefont {Pujolas}},
  \bibinfo {author} {\bibfnamefont {I.}~\bibnamefont {Sawicki}}, \ and\
  \bibinfo {author} {\bibfnamefont {A.}~\bibnamefont {Vikman}},\ }\href
  {\doibase 10.1088/1475-7516/2010/10/026} {\bibfield  {journal} {\bibinfo
  {journal} {JCAP}\ }\textbf {\bibinfo {volume} {10}},\ \bibinfo {pages} {026}
  (\bibinfo {year} {2010})},\ \Eprint {http://arxiv.org/abs/1008.0048}
  {arXiv:1008.0048 [hep-th]} \BibitemShut {NoStop}%
\bibitem [{\citenamefont {de~Rham}\ \emph {et~al.}(2011)\citenamefont
  {de~Rham}, \citenamefont {Gabadadze}, \citenamefont {Heisenberg},\ and\
  \citenamefont {Pirtskhalava}}]{deRham:2010tw}%
  \BibitemOpen
  \bibfield  {author} {\bibinfo {author} {\bibfnamefont {C.}~\bibnamefont
  {de~Rham}}, \bibinfo {author} {\bibfnamefont {G.}~\bibnamefont {Gabadadze}},
  \bibinfo {author} {\bibfnamefont {L.}~\bibnamefont {Heisenberg}}, \ and\
  \bibinfo {author} {\bibfnamefont {D.}~\bibnamefont {Pirtskhalava}},\ }\href
  {\doibase 10.1103/PhysRevD.83.103516} {\bibfield  {journal} {\bibinfo
  {journal} {Phys. Rev. D}\ }\textbf {\bibinfo {volume} {83}},\ \bibinfo
  {pages} {103516} (\bibinfo {year} {2011})},\ \Eprint
  {http://arxiv.org/abs/1010.1780} {arXiv:1010.1780 [hep-th]} \BibitemShut
  {NoStop}%
\bibitem [{\citenamefont {de~Rham}\ and\ \citenamefont
  {Heisenberg}(2011)}]{deRham:2011by}%
  \BibitemOpen
  \bibfield  {author} {\bibinfo {author} {\bibfnamefont {C.}~\bibnamefont
  {de~Rham}}\ and\ \bibinfo {author} {\bibfnamefont {L.}~\bibnamefont
  {Heisenberg}},\ }\href {\doibase 10.1103/PhysRevD.84.043503} {\bibfield
  {journal} {\bibinfo  {journal} {Phys. Rev. D}\ }\textbf {\bibinfo {volume}
  {84}},\ \bibinfo {pages} {043503} (\bibinfo {year} {2011})},\ \Eprint
  {http://arxiv.org/abs/1106.3312} {arXiv:1106.3312 [hep-th]} \BibitemShut
  {NoStop}%
\bibitem [{\citenamefont {Hossain}\ and\ \citenamefont
  {Sen}(2012)}]{Hossain:2012qm}%
  \BibitemOpen
  \bibfield  {author} {\bibinfo {author} {\bibfnamefont {M.~W.}\ \bibnamefont
  {Hossain}}\ and\ \bibinfo {author} {\bibfnamefont {A.~A.}\ \bibnamefont
  {Sen}},\ }\href {\doibase 10.1016/j.physletb.2012.06.016} {\bibfield
  {journal} {\bibinfo  {journal} {Phys.\ Lett.\ B}\ }\textbf {\bibinfo {volume}
  {713}},\ \bibinfo {pages} {140} (\bibinfo {year} {2012})},\ \Eprint
  {http://arxiv.org/abs/1201.6192} {arXiv:1201.6192 [astro-ph.CO]} \BibitemShut
  {NoStop}%
\bibitem [{\citenamefont {Ali}\ \emph {et~al.}(2012)\citenamefont {Ali},
  \citenamefont {Gannouji}, \citenamefont {Hossain},\ and\ \citenamefont
  {Sami}}]{Ali:2012cv}%
  \BibitemOpen
  \bibfield  {author} {\bibinfo {author} {\bibfnamefont {A.}~\bibnamefont
  {Ali}}, \bibinfo {author} {\bibfnamefont {R.}~\bibnamefont {Gannouji}},
  \bibinfo {author} {\bibfnamefont {M.~W.}\ \bibnamefont {Hossain}}, \ and\
  \bibinfo {author} {\bibfnamefont {M.}~\bibnamefont {Sami}},\ }\href {\doibase
  10.1016/j.physletb.2012.10.009} {\bibfield  {journal} {\bibinfo  {journal}
  {Phys.\ Lett.\ B}\ }\textbf {\bibinfo {volume} {718}},\ \bibinfo {pages} {5}
  (\bibinfo {year} {2012})},\ \Eprint {http://arxiv.org/abs/1207.3959}
  {arXiv:1207.3959 [gr-qc]} \BibitemShut {NoStop}%
\bibitem [{\citenamefont {Vainshtein}(1972)}]{Vainshtein:1972sx}%
  \BibitemOpen
  \bibfield  {author} {\bibinfo {author} {\bibfnamefont {A.}~\bibnamefont
  {Vainshtein}},\ }\href {\doibase 10.1016/0370-2693(72)90147-5} {\bibfield
  {journal} {\bibinfo  {journal} {Phys. Lett. B}\ }\textbf {\bibinfo {volume}
  {39}},\ \bibinfo {pages} {393} (\bibinfo {year} {1972})}\BibitemShut
  {NoStop}%
\bibitem [{\citenamefont {Abbott}\ \emph
  {et~al.}(2017{\natexlab{a}})\citenamefont {Abbott} \emph
  {et~al.}}]{TheLIGOScientific:2017qsa}%
  \BibitemOpen
  \bibfield  {author} {\bibinfo {author} {\bibfnamefont {B.}~\bibnamefont
  {Abbott}} \emph {et~al.} (\bibinfo {collaboration} {LIGO Scientific,
  Virgo}),\ }\href {\doibase 10.1103/PhysRevLett.119.161101} {\bibfield
  {journal} {\bibinfo  {journal} {Phys. Rev. Lett.}\ }\textbf {\bibinfo
  {volume} {119}},\ \bibinfo {pages} {161101} (\bibinfo {year}
  {2017}{\natexlab{a}})},\ \Eprint {http://arxiv.org/abs/1710.05832}
  {arXiv:1710.05832 [gr-qc]} \BibitemShut {NoStop}%
\bibitem [{\citenamefont {Abbott}\ \emph
  {et~al.}(2017{\natexlab{b}})\citenamefont {Abbott} \emph
  {et~al.}}]{Monitor:2017mdv}%
  \BibitemOpen
  \bibfield  {author} {\bibinfo {author} {\bibfnamefont {B.}~\bibnamefont
  {Abbott}} \emph {et~al.} (\bibinfo {collaboration} {LIGO Scientific, Virgo,
  Fermi-GBM, INTEGRAL}),\ }\href {\doibase 10.3847/2041-8213/aa920c} {\bibfield
   {journal} {\bibinfo  {journal} {Astrophys. J.}\ }\textbf {\bibinfo {volume}
  {848}},\ \bibinfo {pages} {L13} (\bibinfo {year} {2017}{\natexlab{b}})},\
  \Eprint {http://arxiv.org/abs/1710.05834} {arXiv:1710.05834 [astro-ph.HE]}
  \BibitemShut {NoStop}%
\bibitem [{\citenamefont {Abbott}\ \emph
  {et~al.}(2017{\natexlab{c}})\citenamefont {Abbott} \emph
  {et~al.}}]{GBM:2017lvd}%
  \BibitemOpen
  \bibfield  {author} {\bibinfo {author} {\bibfnamefont {B.}~\bibnamefont
  {Abbott}} \emph {et~al.} (\bibinfo {collaboration} {LIGO Scientific, Virgo,
  Fermi GBM, INTEGRAL, IceCube, AstroSat Cadmium Zinc Telluride Imager Team,
  IPN, Insight-Hxmt, ANTARES, Swift, AGILE Team, 1M2H Team, Dark Energy Camera
  GW-EM, DES, DLT40, GRAWITA, Fermi-LAT, ATCA, ASKAP, Las Cumbres Observatory
  Group, OzGrav, DWF (Deeper Wider Faster Program), AST3, CAASTRO, VINROUGE,
  MASTER, J-GEM, GROWTH, JAGWAR, CaltechNRAO, TTU-NRAO, NuSTAR, Pan-STARRS,
  MAXI Team, TZAC Consortium, KU, Nordic Optical Telescope, ePESSTO, GROND,
  Texas Tech University, SALT Group, TOROS, BOOTES, MWA, CALET, IKI-GW
  Follow-up, H.E.S.S., LOFAR, LWA, HAWC, Pierre Auger, ALMA, Euro VLBI Team, Pi
  of Sky, Chandra Team at McGill University, DFN, ATLAS Telescopes, High Time
  Resolution Universe Survey, RIMAS, RATIR, SKA South Africa/MeerKAT}),\ }\href
  {\doibase 10.3847/2041-8213/aa91c9} {\bibfield  {journal} {\bibinfo
  {journal} {Astrophys. J.}\ }\textbf {\bibinfo {volume} {848}},\ \bibinfo
  {pages} {L12} (\bibinfo {year} {2017}{\natexlab{c}})},\ \Eprint
  {http://arxiv.org/abs/1710.05833} {arXiv:1710.05833 [astro-ph.HE]}
  \BibitemShut {NoStop}%
\bibitem [{\citenamefont {Ezquiaga}\ and\ \citenamefont
  {Zumalacárregui}(2017)}]{Ezquiaga:2017ekz}%
  \BibitemOpen
  \bibfield  {author} {\bibinfo {author} {\bibfnamefont {J.~M.}\ \bibnamefont
  {Ezquiaga}}\ and\ \bibinfo {author} {\bibfnamefont {M.}~\bibnamefont
  {Zumalacárregui}},\ }\href {\doibase 10.1103/PhysRevLett.119.251304}
  {\bibfield  {journal} {\bibinfo  {journal} {Phys. Rev. Lett.}\ }\textbf
  {\bibinfo {volume} {119}},\ \bibinfo {pages} {251304} (\bibinfo {year}
  {2017})},\ \Eprint {http://arxiv.org/abs/1710.05901} {arXiv:1710.05901
  [astro-ph.CO]} \BibitemShut {NoStop}%
\bibitem [{\citenamefont {Zumalacarregui}(2020)}]{Zumalacarregui:2020cjh}%
  \BibitemOpen
  \bibfield  {author} {\bibinfo {author} {\bibfnamefont {M.}~\bibnamefont
  {Zumalacarregui}},\ }\href@noop {} {\  (\bibinfo {year} {2020})},\ \Eprint
  {http://arxiv.org/abs/2003.06396} {arXiv:2003.06396 [astro-ph.CO]}
  \BibitemShut {NoStop}%
\bibitem [{\citenamefont {Brahma}\ and\ \citenamefont
  {Hossain}(2019)}]{Brahma:2019kch}%
  \BibitemOpen
  \bibfield  {author} {\bibinfo {author} {\bibfnamefont {S.}~\bibnamefont
  {Brahma}}\ and\ \bibinfo {author} {\bibfnamefont {M.~W.}\ \bibnamefont
  {Hossain}},\ }\href {\doibase 10.1007/JHEP06(2019)070} {\bibfield  {journal}
  {\bibinfo  {journal} {JHEP}\ }\textbf {\bibinfo {volume} {06}},\ \bibinfo
  {pages} {070} (\bibinfo {year} {2019})},\ \Eprint
  {http://arxiv.org/abs/1902.11014} {arXiv:1902.11014 [hep-th]} \BibitemShut
  {NoStop}%
\bibitem [{\citenamefont {Bartolo}\ \emph {et~al.}(2013)\citenamefont
  {Bartolo}, \citenamefont {Bellini}, \citenamefont {Bertacca},\ and\
  \citenamefont {Matarrese}}]{Bartolo:2013ws}%
  \BibitemOpen
  \bibfield  {author} {\bibinfo {author} {\bibfnamefont {N.}~\bibnamefont
  {Bartolo}}, \bibinfo {author} {\bibfnamefont {E.}~\bibnamefont {Bellini}},
  \bibinfo {author} {\bibfnamefont {D.}~\bibnamefont {Bertacca}}, \ and\
  \bibinfo {author} {\bibfnamefont {S.}~\bibnamefont {Matarrese}},\ }\href
  {\doibase 10.1088/1475-7516/2013/03/034} {\bibfield  {journal} {\bibinfo
  {journal} {JCAP}\ }\textbf {\bibinfo {volume} {1303}},\ \bibinfo {pages}
  {034} (\bibinfo {year} {2013})},\ \Eprint {http://arxiv.org/abs/1301.4831}
  {arXiv:1301.4831 [astro-ph.CO]} \BibitemShut {NoStop}%
\bibitem [{\citenamefont {Bellini}\ and\ \citenamefont
  {Jimenez}(2013)}]{Bellini:2013hea}%
  \BibitemOpen
  \bibfield  {author} {\bibinfo {author} {\bibfnamefont {E.}~\bibnamefont
  {Bellini}}\ and\ \bibinfo {author} {\bibfnamefont {R.}~\bibnamefont
  {Jimenez}},\ }\href {\doibase 10.1016/j.dark.2013.11.001} {\bibfield
  {journal} {\bibinfo  {journal} {Phys.\ Dark Univ.}\ }\textbf {\bibinfo
  {volume} {2}},\ \bibinfo {pages} {179} (\bibinfo {year} {2013})},\ \Eprint
  {http://arxiv.org/abs/1306.1262} {arXiv:1306.1262 [astro-ph.CO]} \BibitemShut
  {NoStop}%
\bibitem [{\citenamefont {Barreira}\ \emph {et~al.}(2013)\citenamefont
  {Barreira}, \citenamefont {Li}, \citenamefont {Hellwing}, \citenamefont
  {Baugh},\ and\ \citenamefont {Pascoli}}]{Barreira:2013eea}%
  \BibitemOpen
  \bibfield  {author} {\bibinfo {author} {\bibfnamefont {A.}~\bibnamefont
  {Barreira}}, \bibinfo {author} {\bibfnamefont {B.}~\bibnamefont {Li}},
  \bibinfo {author} {\bibfnamefont {W.~A.}\ \bibnamefont {Hellwing}}, \bibinfo
  {author} {\bibfnamefont {C.~M.}\ \bibnamefont {Baugh}}, \ and\ \bibinfo
  {author} {\bibfnamefont {S.}~\bibnamefont {Pascoli}},\ }\href {\doibase
  10.1088/1475-7516/2013/10/027} {\bibfield  {journal} {\bibinfo  {journal}
  {JCAP}\ }\textbf {\bibinfo {volume} {1310}},\ \bibinfo {pages} {027}
  (\bibinfo {year} {2013})},\ \Eprint {http://arxiv.org/abs/1306.3219}
  {arXiv:1306.3219 [astro-ph.CO]} \BibitemShut {NoStop}%
\bibitem [{\citenamefont {Hossain}(2017)}]{Hossain:2017ica}%
  \BibitemOpen
  \bibfield  {author} {\bibinfo {author} {\bibfnamefont {M.~W.}\ \bibnamefont
  {Hossain}},\ }\href {\doibase 10.1103/PhysRevD.96.023506} {\bibfield
  {journal} {\bibinfo  {journal} {Phys.\ Rev.\ D}\ }\textbf {\bibinfo {volume}
  {96}},\ \bibinfo {pages} {023506} (\bibinfo {year} {2017})},\ \Eprint
  {http://arxiv.org/abs/1704.07956} {arXiv:1704.07956 [gr-qc]} \BibitemShut
  {NoStop}%
\bibitem [{\citenamefont {Dinda}\ \emph {et~al.}(2018)\citenamefont {Dinda},
  \citenamefont {Hossain},\ and\ \citenamefont {Sen}}]{Dinda:2017lpz}%
  \BibitemOpen
  \bibfield  {author} {\bibinfo {author} {\bibfnamefont {B.~R.}\ \bibnamefont
  {Dinda}}, \bibinfo {author} {\bibfnamefont {M.~W.}\ \bibnamefont {Hossain}},
  \ and\ \bibinfo {author} {\bibfnamefont {A.~A.}\ \bibnamefont {Sen}},\ }\href
  {\doibase 10.1088/1475-7516/2018/01/045} {\bibfield  {journal} {\bibinfo
  {journal} {JCAP}\ }\textbf {\bibinfo {volume} {01}},\ \bibinfo {pages} {045}
  (\bibinfo {year} {2018})},\ \Eprint {http://arxiv.org/abs/1706.00567}
  {arXiv:1706.00567 [astro-ph.CO]} \BibitemShut {NoStop}%
\bibitem [{\citenamefont {{Ishak}}(2019)}]{ishakreview}%
  \BibitemOpen
  \bibfield  {author} {\bibinfo {author} {\bibfnamefont {M.}~\bibnamefont
  {{Ishak}}},\ }\href {\doibase 10.1007/s41114-018-0017-4} {\bibfield
  {journal} {\bibinfo  {journal} {Living Reviews in Relativity}\ }\textbf
  {\bibinfo {volume} {22}},\ \bibinfo {eid} {1} (\bibinfo {year} {2019})},\
  \Eprint {http://arxiv.org/abs/1806.10122} {arXiv:1806.10122 [astro-ph.CO]}
  \BibitemShut {NoStop}%
\bibitem [{\citenamefont {{He}}\ \emph {et~al.}(2018)\citenamefont {{He}},
  \citenamefont {{Guzzo}}, \citenamefont {{Li}},\ and\ \citenamefont
  {{Baugh}}}]{He:2018nature}%
  \BibitemOpen
  \bibfield  {author} {\bibinfo {author} {\bibfnamefont {J.-h.}\ \bibnamefont
  {{He}}}, \bibinfo {author} {\bibfnamefont {L.}~\bibnamefont {{Guzzo}}},
  \bibinfo {author} {\bibfnamefont {B.}~\bibnamefont {{Li}}}, \ and\ \bibinfo
  {author} {\bibfnamefont {C.~M.}\ \bibnamefont {{Baugh}}},\ }\href {\doibase
  10.1038/s41550-018-0573-2} {\bibfield  {journal} {\bibinfo  {journal} {Nature
  Astronomy}\ }\textbf {\bibinfo {volume} {2}},\ \bibinfo {pages} {967}
  (\bibinfo {year} {2018})},\ \Eprint {http://arxiv.org/abs/1809.09019}
  {arXiv:1809.09019 [astro-ph.CO]} \BibitemShut {NoStop}%
\bibitem [{\citenamefont {{Zhang}}\ \emph {et~al.}(2019)\citenamefont
  {{Zhang}}, \citenamefont {{An}}, \citenamefont {{Luo}}, \citenamefont {{Li}},
  \citenamefont {{Liao}},\ and\ \citenamefont {{Wang}}}]{zhang2019apjl}%
  \BibitemOpen
  \bibfield  {author} {\bibinfo {author} {\bibfnamefont {J.}~\bibnamefont
  {{Zhang}}}, \bibinfo {author} {\bibfnamefont {R.}~\bibnamefont {{An}}},
  \bibinfo {author} {\bibfnamefont {W.}~\bibnamefont {{Luo}}}, \bibinfo
  {author} {\bibfnamefont {Z.}~\bibnamefont {{Li}}}, \bibinfo {author}
  {\bibfnamefont {S.}~\bibnamefont {{Liao}}}, \ and\ \bibinfo {author}
  {\bibfnamefont {B.}~\bibnamefont {{Wang}}},\ }\href {\doibase
  10.3847/2041-8213/ab133f} {\bibfield  {journal} {\bibinfo  {journal} {\apjl}\
  }\textbf {\bibinfo {volume} {875}},\ \bibinfo {eid} {L11} (\bibinfo {year}
  {2019})},\ \Eprint {http://arxiv.org/abs/1807.05522} {arXiv:1807.05522
  [astro-ph.CO]} \BibitemShut {NoStop}%
\bibitem [{\citenamefont {{An}}\ \emph {et~al.}(2019)\citenamefont {{An}},
  \citenamefont {{Costa}}, \citenamefont {{Xiao}}, \citenamefont {{Zhang}},\
  and\ \citenamefont {{Wang}}}]{an2019mnras}%
  \BibitemOpen
  \bibfield  {author} {\bibinfo {author} {\bibfnamefont {R.}~\bibnamefont
  {{An}}}, \bibinfo {author} {\bibfnamefont {A.~A.}\ \bibnamefont {{Costa}}},
  \bibinfo {author} {\bibfnamefont {L.}~\bibnamefont {{Xiao}}}, \bibinfo
  {author} {\bibfnamefont {J.}~\bibnamefont {{Zhang}}}, \ and\ \bibinfo
  {author} {\bibfnamefont {B.}~\bibnamefont {{Wang}}},\ }\href {\doibase
  10.1093/mnras/stz2028} {\bibfield  {journal} {\bibinfo  {journal} {\mnras}\
  }\textbf {\bibinfo {volume} {489}},\ \bibinfo {pages} {297} (\bibinfo {year}
  {2019})},\ \Eprint {http://arxiv.org/abs/1809.03224} {arXiv:1809.03224
  [astro-ph.CO]} \BibitemShut {NoStop}%
\bibitem [{\citenamefont {{Abazajian}}\ \emph {et~al.}(2009)\citenamefont
  {{Abazajian}}, \citenamefont {{Adelman-McCarthy}}, \citenamefont
  {{Ag{\"u}eros}}, \citenamefont {{Allam}}, \citenamefont {{Allende Prieto}},
  \citenamefont {{An}}, \citenamefont {{Anderson}}, \citenamefont {{Anderson}},
  \citenamefont {{Annis}}, \citenamefont {{Bahcall}}, \citenamefont
  {{Bailer-Jones}}, \citenamefont {{Barentine}}, \citenamefont {{Bassett}},
  \citenamefont {{Becker}}, \citenamefont {{Beers}}, \citenamefont {{Bell}},
  \citenamefont {{Belokurov}}, \citenamefont {{Berlind}}, \citenamefont
  {{Berman}}, \citenamefont {{Bernardi}}, \citenamefont {{Bickerton}},
  \citenamefont {{Bizyaev}}, \citenamefont {{Blakeslee}}, \citenamefont
  {{Blanton}}, \citenamefont {{Bochanski}}, \citenamefont {{Boroski}},
  \citenamefont {{Brewington}}, \citenamefont {{Brinchmann}}, \citenamefont
  {{Brinkmann}}, \citenamefont {{Brunner}}, \citenamefont {{Budav{\'a}ri}},
  \citenamefont {{Carey}}, \citenamefont {{Carliles}}, \citenamefont {{Carr}},
  \citenamefont {{Castander}}, \citenamefont {{Cinabro}}, \citenamefont
  {{Connolly}}, \citenamefont {{Csabai}}, \citenamefont {{Cunha}},
  \citenamefont {{Czarapata}}, \citenamefont {{Davenport}}, \citenamefont {{de
  Haas}}, \citenamefont {{Dilday}}, \citenamefont {{Doi}}, \citenamefont
  {{Eisenstein}}, \citenamefont {{Evans}}, \citenamefont {{Evans}},
  \citenamefont {{Fan}}, \citenamefont {{Friedman}}, \citenamefont {{Frieman}},
  \citenamefont {{Fukugita}}, \citenamefont {{G{\"a}nsicke}}, \citenamefont
  {{Gates}}, \citenamefont {{Gillespie}}, \citenamefont {{Gilmore}},
  \citenamefont {{Gonzalez}}, \citenamefont {{Gonzalez}}, \citenamefont
  {{Grebel}}, \citenamefont {{Gunn}}, \citenamefont {{Gy{\"o}ry}},
  \citenamefont {{Hall}}, \citenamefont {{Harding}}, \citenamefont {{Harris}},
  \citenamefont {{Harvanek}}, \citenamefont {{Hawley}}, \citenamefont
  {{Hayes}}, \citenamefont {{Heckman}}, \citenamefont {{Hendry}}, \citenamefont
  {{Hennessy}}, \citenamefont {{Hindsley}}, \citenamefont {{Hoblitt}},
  \citenamefont {{Hogan}}, \citenamefont {{Hogg}}, \citenamefont {{Holtzman}},
  \citenamefont {{Hyde}}, \citenamefont {{Ichikawa}}, \citenamefont
  {{Ichikawa}}, \citenamefont {{Im}}, \citenamefont {{Ivezi{\'c}}},
  \citenamefont {{Jester}}, \citenamefont {{Jiang}}, \citenamefont {{Johnson}},
  \citenamefont {{Jorgensen}}, \citenamefont {{Juri{\'c}}}, \citenamefont
  {{Kent}}, \citenamefont {{Kessler}}, \citenamefont {{Kleinman}},
  \citenamefont {{Knapp}}, \citenamefont {{Konishi}}, \citenamefont {{Kron}},
  \citenamefont {{Krzesinski}}, \citenamefont {{Kuropatkin}}, \citenamefont
  {{Lampeitl}}, \citenamefont {{Lebedeva}}, \citenamefont {{Lee}},
  \citenamefont {{Lee}}, \citenamefont {{French Leger}}, \citenamefont
  {{L{\'e}pine}}, \citenamefont {{Li}}, \citenamefont {{Lima}}, \citenamefont
  {{Lin}}, \citenamefont {{Long}}, \citenamefont {{Loomis}}, \citenamefont
  {{Loveday}}, \citenamefont {{Lupton}}, \citenamefont {{Magnier}},
  \citenamefont {{Malanushenko}}, \citenamefont {{Malanushenko}}, \citenamefont
  {{Mand elbaum}}, \citenamefont {{Margon}}, \citenamefont {{Marriner}},
  \citenamefont {{Mart{\'\i}nez-Delgado}}, \citenamefont {{Matsubara}},
  \citenamefont {{McGehee}}, \citenamefont {{McKay}}, \citenamefont
  {{Meiksin}}, \citenamefont {{Morrison}}, \citenamefont {{Mullally}},
  \citenamefont {{Munn}}, \citenamefont {{Murphy}}, \citenamefont {{Nash}},
  \citenamefont {{Nebot}}, \citenamefont {{Neilsen}}, \citenamefont
  {{Newberg}}, \citenamefont {{Newman}}, \citenamefont {{Nichol}},
  \citenamefont {{Nicinski}}, \citenamefont {{Nieto-Santisteban}},
  \citenamefont {{Nitta}}, \citenamefont {{Okamura}}, \citenamefont
  {{Oravetz}}, \citenamefont {{Ostriker}}, \citenamefont {{Owen}},
  \citenamefont {{Padmanabhan}}, \citenamefont {{Pan}}, \citenamefont {{Park}},
  \citenamefont {{Pauls}}, \citenamefont {{Peoples}}, \citenamefont
  {{Percival}}, \citenamefont {{Pier}}, \citenamefont {{Pope}}, \citenamefont
  {{Pourbaix}}, \citenamefont {{Price}}, \citenamefont {{Purger}},
  \citenamefont {{Quinn}}, \citenamefont {{Raddick}}, \citenamefont {{Re
  Fiorentin}}, \citenamefont {{Richards}}, \citenamefont {{Richmond}},
  \citenamefont {{Riess}}, \citenamefont {{Rix}}, \citenamefont {{Rockosi}},
  \citenamefont {{Sako}}, \citenamefont {{Schlegel}}, \citenamefont
  {{Schneider}}, \citenamefont {{Scholz}}, \citenamefont {{Schreiber}},
  \citenamefont {{Schwope}}, \citenamefont {{Seljak}}, \citenamefont {{Sesar}},
  \citenamefont {{Sheldon}}, \citenamefont {{Shimasaku}}, \citenamefont
  {{Sibley}}, \citenamefont {{Simmons}}, \citenamefont {{Sivarani}},
  \citenamefont {{Allyn Smith}}, \citenamefont {{Smith}}, \citenamefont
  {{Smol{\v{c}}i{\'c}}}, \citenamefont {{Snedden}}, \citenamefont {{Stebbins}},
  \citenamefont {{Steinmetz}}, \citenamefont {{Stoughton}}, \citenamefont
  {{Strauss}}, \citenamefont {{SubbaRao}}, \citenamefont {{Suto}},
  \citenamefont {{Szalay}}, \citenamefont {{Szapudi}}, \citenamefont
  {{Szkody}}, \citenamefont {{Tanaka}}, \citenamefont {{Tegmark}},
  \citenamefont {{Teodoro}}, \citenamefont {{Thakar}}, \citenamefont
  {{Tremonti}}, \citenamefont {{Tucker}}, \citenamefont {{Uomoto}},
  \citenamefont {{Vanden Berk}}, \citenamefont {{Vandenberg}}, \citenamefont
  {{Vidrih}}, \citenamefont {{Vogeley}}, \citenamefont {{Voges}}, \citenamefont
  {{Vogt}}, \citenamefont {{Wadadekar}}, \citenamefont {{Watters}},
  \citenamefont {{Weinberg}}, \citenamefont {{West}}, \citenamefont {{White}},
  \citenamefont {{Wilhite}}, \citenamefont {{Wonders}}, \citenamefont
  {{Yanny}}, \citenamefont {{Yocum}}, \citenamefont {{York}}, \citenamefont
  {{Zehavi}}, \citenamefont {{Zibetti}},\ and\ \citenamefont
  {{Zucker}}}]{sdss7}%
  \BibitemOpen
  \bibfield  {author} {\bibinfo {author} {\bibfnamefont {K.~N.}\ \bibnamefont
  {{Abazajian}}}, \bibinfo {author} {\bibfnamefont {J.~K.}\ \bibnamefont
  {{Adelman-McCarthy}}}, \bibinfo {author} {\bibfnamefont {M.~A.}\ \bibnamefont
  {{Ag{\"u}eros}}}, \bibinfo {author} {\bibfnamefont {S.~S.}\ \bibnamefont
  {{Allam}}}, \bibinfo {author} {\bibfnamefont {C.}~\bibnamefont {{Allende
  Prieto}}}, \bibinfo {author} {\bibfnamefont {D.}~\bibnamefont {{An}}},
  \bibinfo {author} {\bibfnamefont {K.~S.~J.}\ \bibnamefont {{Anderson}}},
  \bibinfo {author} {\bibfnamefont {S.~F.}\ \bibnamefont {{Anderson}}},
  \bibinfo {author} {\bibfnamefont {J.}~\bibnamefont {{Annis}}}, \bibinfo
  {author} {\bibfnamefont {N.~A.}\ \bibnamefont {{Bahcall}}}, \bibinfo {author}
  {\bibfnamefont {C.~A.~L.}\ \bibnamefont {{Bailer-Jones}}}, \bibinfo {author}
  {\bibfnamefont {J.~C.}\ \bibnamefont {{Barentine}}}, \bibinfo {author}
  {\bibfnamefont {B.~A.}\ \bibnamefont {{Bassett}}}, \bibinfo {author}
  {\bibfnamefont {A.~C.}\ \bibnamefont {{Becker}}}, \bibinfo {author}
  {\bibfnamefont {T.~C.}\ \bibnamefont {{Beers}}}, \bibinfo {author}
  {\bibfnamefont {E.~F.}\ \bibnamefont {{Bell}}}, \bibinfo {author}
  {\bibfnamefont {V.}~\bibnamefont {{Belokurov}}}, \bibinfo {author}
  {\bibfnamefont {A.~A.}\ \bibnamefont {{Berlind}}}, \bibinfo {author}
  {\bibfnamefont {E.~F.}\ \bibnamefont {{Berman}}}, \bibinfo {author}
  {\bibfnamefont {M.}~\bibnamefont {{Bernardi}}}, \bibinfo {author}
  {\bibfnamefont {S.~J.}\ \bibnamefont {{Bickerton}}}, \bibinfo {author}
  {\bibfnamefont {D.}~\bibnamefont {{Bizyaev}}}, \bibinfo {author}
  {\bibfnamefont {J.~P.}\ \bibnamefont {{Blakeslee}}}, \bibinfo {author}
  {\bibfnamefont {M.~R.}\ \bibnamefont {{Blanton}}}, \bibinfo {author}
  {\bibfnamefont {J.~J.}\ \bibnamefont {{Bochanski}}}, \bibinfo {author}
  {\bibfnamefont {W.~N.}\ \bibnamefont {{Boroski}}}, \bibinfo {author}
  {\bibfnamefont {H.~J.}\ \bibnamefont {{Brewington}}}, \bibinfo {author}
  {\bibfnamefont {J.}~\bibnamefont {{Brinchmann}}}, \bibinfo {author}
  {\bibfnamefont {J.}~\bibnamefont {{Brinkmann}}}, \bibinfo {author}
  {\bibfnamefont {R.~J.}\ \bibnamefont {{Brunner}}}, \bibinfo {author}
  {\bibfnamefont {T.}~\bibnamefont {{Budav{\'a}ri}}}, \bibinfo {author}
  {\bibfnamefont {L.~N.}\ \bibnamefont {{Carey}}}, \bibinfo {author}
  {\bibfnamefont {S.}~\bibnamefont {{Carliles}}}, \bibinfo {author}
  {\bibfnamefont {M.~A.}\ \bibnamefont {{Carr}}}, \bibinfo {author}
  {\bibfnamefont {F.~J.}\ \bibnamefont {{Castander}}}, \bibinfo {author}
  {\bibfnamefont {D.}~\bibnamefont {{Cinabro}}}, \bibinfo {author}
  {\bibfnamefont {A.~J.}\ \bibnamefont {{Connolly}}}, \bibinfo {author}
  {\bibfnamefont {I.}~\bibnamefont {{Csabai}}}, \bibinfo {author}
  {\bibfnamefont {C.~E.}\ \bibnamefont {{Cunha}}}, \bibinfo {author}
  {\bibfnamefont {P.~C.}\ \bibnamefont {{Czarapata}}}, \bibinfo {author}
  {\bibfnamefont {J.~R.~A.}\ \bibnamefont {{Davenport}}}, \bibinfo {author}
  {\bibfnamefont {E.}~\bibnamefont {{de Haas}}}, \bibinfo {author}
  {\bibfnamefont {B.}~\bibnamefont {{Dilday}}}, \bibinfo {author}
  {\bibfnamefont {M.}~\bibnamefont {{Doi}}}, \bibinfo {author} {\bibfnamefont
  {D.~J.}\ \bibnamefont {{Eisenstein}}}, \bibinfo {author} {\bibfnamefont
  {M.~L.}\ \bibnamefont {{Evans}}}, \bibinfo {author} {\bibfnamefont {N.~W.}\
  \bibnamefont {{Evans}}}, \bibinfo {author} {\bibfnamefont {X.}~\bibnamefont
  {{Fan}}}, \bibinfo {author} {\bibfnamefont {S.~D.}\ \bibnamefont
  {{Friedman}}}, \bibinfo {author} {\bibfnamefont {J.~A.}\ \bibnamefont
  {{Frieman}}}, \bibinfo {author} {\bibfnamefont {M.}~\bibnamefont
  {{Fukugita}}}, \bibinfo {author} {\bibfnamefont {B.~T.}\ \bibnamefont
  {{G{\"a}nsicke}}}, \bibinfo {author} {\bibfnamefont {E.}~\bibnamefont
  {{Gates}}}, \bibinfo {author} {\bibfnamefont {B.}~\bibnamefont
  {{Gillespie}}}, \bibinfo {author} {\bibfnamefont {G.}~\bibnamefont
  {{Gilmore}}}, \bibinfo {author} {\bibfnamefont {B.}~\bibnamefont
  {{Gonzalez}}}, \bibinfo {author} {\bibfnamefont {C.~F.}\ \bibnamefont
  {{Gonzalez}}}, \bibinfo {author} {\bibfnamefont {E.~K.}\ \bibnamefont
  {{Grebel}}}, \bibinfo {author} {\bibfnamefont {J.~E.}\ \bibnamefont
  {{Gunn}}}, \bibinfo {author} {\bibfnamefont {Z.}~\bibnamefont {{Gy{\"o}ry}}},
  \bibinfo {author} {\bibfnamefont {P.~B.}\ \bibnamefont {{Hall}}}, \bibinfo
  {author} {\bibfnamefont {P.}~\bibnamefont {{Harding}}}, \bibinfo {author}
  {\bibfnamefont {F.~H.}\ \bibnamefont {{Harris}}}, \bibinfo {author}
  {\bibfnamefont {M.}~\bibnamefont {{Harvanek}}}, \bibinfo {author}
  {\bibfnamefont {S.~L.}\ \bibnamefont {{Hawley}}}, \bibinfo {author}
  {\bibfnamefont {J.~J.~E.}\ \bibnamefont {{Hayes}}}, \bibinfo {author}
  {\bibfnamefont {T.~M.}\ \bibnamefont {{Heckman}}}, \bibinfo {author}
  {\bibfnamefont {J.~S.}\ \bibnamefont {{Hendry}}}, \bibinfo {author}
  {\bibfnamefont {G.~S.}\ \bibnamefont {{Hennessy}}}, \bibinfo {author}
  {\bibfnamefont {R.~B.}\ \bibnamefont {{Hindsley}}}, \bibinfo {author}
  {\bibfnamefont {J.}~\bibnamefont {{Hoblitt}}}, \bibinfo {author}
  {\bibfnamefont {C.~J.}\ \bibnamefont {{Hogan}}}, \bibinfo {author}
  {\bibfnamefont {D.~W.}\ \bibnamefont {{Hogg}}}, \bibinfo {author}
  {\bibfnamefont {J.~A.}\ \bibnamefont {{Holtzman}}}, \bibinfo {author}
  {\bibfnamefont {J.~B.}\ \bibnamefont {{Hyde}}}, \bibinfo {author}
  {\bibfnamefont {S.-i.}\ \bibnamefont {{Ichikawa}}}, \bibinfo {author}
  {\bibfnamefont {T.}~\bibnamefont {{Ichikawa}}}, \bibinfo {author}
  {\bibfnamefont {M.}~\bibnamefont {{Im}}}, \bibinfo {author} {\bibfnamefont
  {{\v{Z}}.}~\bibnamefont {{Ivezi{\'c}}}}, \bibinfo {author} {\bibfnamefont
  {S.}~\bibnamefont {{Jester}}}, \bibinfo {author} {\bibfnamefont
  {L.}~\bibnamefont {{Jiang}}}, \bibinfo {author} {\bibfnamefont {J.~A.}\
  \bibnamefont {{Johnson}}}, \bibinfo {author} {\bibfnamefont {A.~M.}\
  \bibnamefont {{Jorgensen}}}, \bibinfo {author} {\bibfnamefont
  {M.}~\bibnamefont {{Juri{\'c}}}}, \bibinfo {author} {\bibfnamefont {S.~M.}\
  \bibnamefont {{Kent}}}, \bibinfo {author} {\bibfnamefont {R.}~\bibnamefont
  {{Kessler}}}, \bibinfo {author} {\bibfnamefont {S.~J.}\ \bibnamefont
  {{Kleinman}}}, \bibinfo {author} {\bibfnamefont {G.~R.}\ \bibnamefont
  {{Knapp}}}, \bibinfo {author} {\bibfnamefont {K.}~\bibnamefont {{Konishi}}},
  \bibinfo {author} {\bibfnamefont {R.~G.}\ \bibnamefont {{Kron}}}, \bibinfo
  {author} {\bibfnamefont {J.}~\bibnamefont {{Krzesinski}}}, \bibinfo {author}
  {\bibfnamefont {N.}~\bibnamefont {{Kuropatkin}}}, \bibinfo {author}
  {\bibfnamefont {H.}~\bibnamefont {{Lampeitl}}}, \bibinfo {author}
  {\bibfnamefont {S.}~\bibnamefont {{Lebedeva}}}, \bibinfo {author}
  {\bibfnamefont {M.~G.}\ \bibnamefont {{Lee}}}, \bibinfo {author}
  {\bibfnamefont {Y.~S.}\ \bibnamefont {{Lee}}}, \bibinfo {author}
  {\bibfnamefont {R.}~\bibnamefont {{French Leger}}}, \bibinfo {author}
  {\bibfnamefont {S.}~\bibnamefont {{L{\'e}pine}}}, \bibinfo {author}
  {\bibfnamefont {N.}~\bibnamefont {{Li}}}, \bibinfo {author} {\bibfnamefont
  {M.}~\bibnamefont {{Lima}}}, \bibinfo {author} {\bibfnamefont
  {H.}~\bibnamefont {{Lin}}}, \bibinfo {author} {\bibfnamefont {D.~C.}\
  \bibnamefont {{Long}}}, \bibinfo {author} {\bibfnamefont {C.~P.}\
  \bibnamefont {{Loomis}}}, \bibinfo {author} {\bibfnamefont {J.}~\bibnamefont
  {{Loveday}}}, \bibinfo {author} {\bibfnamefont {R.~H.}\ \bibnamefont
  {{Lupton}}}, \bibinfo {author} {\bibfnamefont {E.}~\bibnamefont {{Magnier}}},
  \bibinfo {author} {\bibfnamefont {O.}~\bibnamefont {{Malanushenko}}},
  \bibinfo {author} {\bibfnamefont {V.}~\bibnamefont {{Malanushenko}}},
  \bibinfo {author} {\bibfnamefont {R.}~\bibnamefont {{Mand elbaum}}}, \bibinfo
  {author} {\bibfnamefont {B.}~\bibnamefont {{Margon}}}, \bibinfo {author}
  {\bibfnamefont {J.~P.}\ \bibnamefont {{Marriner}}}, \bibinfo {author}
  {\bibfnamefont {D.}~\bibnamefont {{Mart{\'\i}nez-Delgado}}}, \bibinfo
  {author} {\bibfnamefont {T.}~\bibnamefont {{Matsubara}}}, \bibinfo {author}
  {\bibfnamefont {P.~M.}\ \bibnamefont {{McGehee}}}, \bibinfo {author}
  {\bibfnamefont {T.~A.}\ \bibnamefont {{McKay}}}, \bibinfo {author}
  {\bibfnamefont {A.}~\bibnamefont {{Meiksin}}}, \bibinfo {author}
  {\bibfnamefont {H.~L.}\ \bibnamefont {{Morrison}}}, \bibinfo {author}
  {\bibfnamefont {F.}~\bibnamefont {{Mullally}}}, \bibinfo {author}
  {\bibfnamefont {J.~A.}\ \bibnamefont {{Munn}}}, \bibinfo {author}
  {\bibfnamefont {T.}~\bibnamefont {{Murphy}}}, \bibinfo {author}
  {\bibfnamefont {T.}~\bibnamefont {{Nash}}}, \bibinfo {author} {\bibfnamefont
  {A.}~\bibnamefont {{Nebot}}}, \bibinfo {author} {\bibfnamefont
  {J.}~\bibnamefont {{Neilsen}}, \bibfnamefont {Eric~H.}}, \bibinfo {author}
  {\bibfnamefont {H.~J.}\ \bibnamefont {{Newberg}}}, \bibinfo {author}
  {\bibfnamefont {P.~R.}\ \bibnamefont {{Newman}}}, \bibinfo {author}
  {\bibfnamefont {R.~C.}\ \bibnamefont {{Nichol}}}, \bibinfo {author}
  {\bibfnamefont {T.}~\bibnamefont {{Nicinski}}}, \bibinfo {author}
  {\bibfnamefont {M.}~\bibnamefont {{Nieto-Santisteban}}}, \bibinfo {author}
  {\bibfnamefont {A.}~\bibnamefont {{Nitta}}}, \bibinfo {author} {\bibfnamefont
  {S.}~\bibnamefont {{Okamura}}}, \bibinfo {author} {\bibfnamefont {D.~J.}\
  \bibnamefont {{Oravetz}}}, \bibinfo {author} {\bibfnamefont {J.~P.}\
  \bibnamefont {{Ostriker}}}, \bibinfo {author} {\bibfnamefont
  {R.}~\bibnamefont {{Owen}}}, \bibinfo {author} {\bibfnamefont
  {N.}~\bibnamefont {{Padmanabhan}}}, \bibinfo {author} {\bibfnamefont
  {K.}~\bibnamefont {{Pan}}}, \bibinfo {author} {\bibfnamefont
  {C.}~\bibnamefont {{Park}}}, \bibinfo {author} {\bibfnamefont
  {G.}~\bibnamefont {{Pauls}}}, \bibinfo {author} {\bibfnamefont
  {J.}~\bibnamefont {{Peoples}}, \bibfnamefont {John}}, \bibinfo {author}
  {\bibfnamefont {W.~J.}\ \bibnamefont {{Percival}}}, \bibinfo {author}
  {\bibfnamefont {J.~R.}\ \bibnamefont {{Pier}}}, \bibinfo {author}
  {\bibfnamefont {A.~C.}\ \bibnamefont {{Pope}}}, \bibinfo {author}
  {\bibfnamefont {D.}~\bibnamefont {{Pourbaix}}}, \bibinfo {author}
  {\bibfnamefont {P.~A.}\ \bibnamefont {{Price}}}, \bibinfo {author}
  {\bibfnamefont {N.}~\bibnamefont {{Purger}}}, \bibinfo {author}
  {\bibfnamefont {T.}~\bibnamefont {{Quinn}}}, \bibinfo {author} {\bibfnamefont
  {M.~J.}\ \bibnamefont {{Raddick}}}, \bibinfo {author} {\bibfnamefont
  {P.}~\bibnamefont {{Re Fiorentin}}}, \bibinfo {author} {\bibfnamefont
  {G.~T.}\ \bibnamefont {{Richards}}}, \bibinfo {author} {\bibfnamefont
  {M.~W.}\ \bibnamefont {{Richmond}}}, \bibinfo {author} {\bibfnamefont
  {A.~G.}\ \bibnamefont {{Riess}}}, \bibinfo {author} {\bibfnamefont {H.-W.}\
  \bibnamefont {{Rix}}}, \bibinfo {author} {\bibfnamefont {C.~M.}\ \bibnamefont
  {{Rockosi}}}, \bibinfo {author} {\bibfnamefont {M.}~\bibnamefont {{Sako}}},
  \bibinfo {author} {\bibfnamefont {D.~J.}\ \bibnamefont {{Schlegel}}},
  \bibinfo {author} {\bibfnamefont {D.~P.}\ \bibnamefont {{Schneider}}},
  \bibinfo {author} {\bibfnamefont {R.-D.}\ \bibnamefont {{Scholz}}}, \bibinfo
  {author} {\bibfnamefont {M.~R.}\ \bibnamefont {{Schreiber}}}, \bibinfo
  {author} {\bibfnamefont {A.~D.}\ \bibnamefont {{Schwope}}}, \bibinfo {author}
  {\bibfnamefont {U.}~\bibnamefont {{Seljak}}}, \bibinfo {author}
  {\bibfnamefont {B.}~\bibnamefont {{Sesar}}}, \bibinfo {author} {\bibfnamefont
  {E.}~\bibnamefont {{Sheldon}}}, \bibinfo {author} {\bibfnamefont
  {K.}~\bibnamefont {{Shimasaku}}}, \bibinfo {author} {\bibfnamefont {V.~C.}\
  \bibnamefont {{Sibley}}}, \bibinfo {author} {\bibfnamefont {A.~E.}\
  \bibnamefont {{Simmons}}}, \bibinfo {author} {\bibfnamefont {T.}~\bibnamefont
  {{Sivarani}}}, \bibinfo {author} {\bibfnamefont {J.}~\bibnamefont {{Allyn
  Smith}}}, \bibinfo {author} {\bibfnamefont {M.~C.}\ \bibnamefont {{Smith}}},
  \bibinfo {author} {\bibfnamefont {V.}~\bibnamefont {{Smol{\v{c}}i{\'c}}}},
  \bibinfo {author} {\bibfnamefont {S.~A.}\ \bibnamefont {{Snedden}}}, \bibinfo
  {author} {\bibfnamefont {A.}~\bibnamefont {{Stebbins}}}, \bibinfo {author}
  {\bibfnamefont {M.}~\bibnamefont {{Steinmetz}}}, \bibinfo {author}
  {\bibfnamefont {C.}~\bibnamefont {{Stoughton}}}, \bibinfo {author}
  {\bibfnamefont {M.~A.}\ \bibnamefont {{Strauss}}}, \bibinfo {author}
  {\bibfnamefont {M.}~\bibnamefont {{SubbaRao}}}, \bibinfo {author}
  {\bibfnamefont {Y.}~\bibnamefont {{Suto}}}, \bibinfo {author} {\bibfnamefont
  {A.~S.}\ \bibnamefont {{Szalay}}}, \bibinfo {author} {\bibfnamefont
  {I.}~\bibnamefont {{Szapudi}}}, \bibinfo {author} {\bibfnamefont
  {P.}~\bibnamefont {{Szkody}}}, \bibinfo {author} {\bibfnamefont
  {M.}~\bibnamefont {{Tanaka}}}, \bibinfo {author} {\bibfnamefont
  {M.}~\bibnamefont {{Tegmark}}}, \bibinfo {author} {\bibfnamefont {L.~F.~A.}\
  \bibnamefont {{Teodoro}}}, \bibinfo {author} {\bibfnamefont {A.~R.}\
  \bibnamefont {{Thakar}}}, \bibinfo {author} {\bibfnamefont {C.~A.}\
  \bibnamefont {{Tremonti}}}, \bibinfo {author} {\bibfnamefont {D.~L.}\
  \bibnamefont {{Tucker}}}, \bibinfo {author} {\bibfnamefont {A.}~\bibnamefont
  {{Uomoto}}}, \bibinfo {author} {\bibfnamefont {D.~E.}\ \bibnamefont {{Vanden
  Berk}}}, \bibinfo {author} {\bibfnamefont {J.}~\bibnamefont {{Vandenberg}}},
  \bibinfo {author} {\bibfnamefont {S.}~\bibnamefont {{Vidrih}}}, \bibinfo
  {author} {\bibfnamefont {M.~S.}\ \bibnamefont {{Vogeley}}}, \bibinfo {author}
  {\bibfnamefont {W.}~\bibnamefont {{Voges}}}, \bibinfo {author} {\bibfnamefont
  {N.~P.}\ \bibnamefont {{Vogt}}}, \bibinfo {author} {\bibfnamefont
  {Y.}~\bibnamefont {{Wadadekar}}}, \bibinfo {author} {\bibfnamefont
  {S.}~\bibnamefont {{Watters}}}, \bibinfo {author} {\bibfnamefont {D.~H.}\
  \bibnamefont {{Weinberg}}}, \bibinfo {author} {\bibfnamefont {A.~A.}\
  \bibnamefont {{West}}}, \bibinfo {author} {\bibfnamefont {S.~D.~M.}\
  \bibnamefont {{White}}}, \bibinfo {author} {\bibfnamefont {B.~C.}\
  \bibnamefont {{Wilhite}}}, \bibinfo {author} {\bibfnamefont {A.~C.}\
  \bibnamefont {{Wonders}}}, \bibinfo {author} {\bibfnamefont {B.}~\bibnamefont
  {{Yanny}}}, \bibinfo {author} {\bibfnamefont {D.~R.}\ \bibnamefont
  {{Yocum}}}, \bibinfo {author} {\bibfnamefont {D.~G.}\ \bibnamefont {{York}}},
  \bibinfo {author} {\bibfnamefont {I.}~\bibnamefont {{Zehavi}}}, \bibinfo
  {author} {\bibfnamefont {S.}~\bibnamefont {{Zibetti}}}, \ and\ \bibinfo
  {author} {\bibfnamefont {D.~B.}\ \bibnamefont {{Zucker}}},\ }\href {\doibase
  10.1088/0067-0049/182/2/543} {\bibfield  {journal} {\bibinfo  {journal}
  {\apjs}\ }\textbf {\bibinfo {volume} {182}},\ \bibinfo {pages} {543}
  (\bibinfo {year} {2009})},\ \Eprint {http://arxiv.org/abs/0812.0649}
  {arXiv:0812.0649 [astro-ph]} \BibitemShut {NoStop}%
\bibitem [{\citenamefont {{Luo}}\ \emph {et~al.}(2017)\citenamefont {{Luo}},
  \citenamefont {{Yang}}, \citenamefont {{Zhang}}, \citenamefont {{Tweed}},
  \citenamefont {{Fu}}, \citenamefont {{Mo}}, \citenamefont {{van den Bosch}},
  \citenamefont {{Shu}}, \citenamefont {{Li}}, \citenamefont {{Li}},
  \citenamefont {{Liu}}, \citenamefont {{Pan}}, \citenamefont {{Wang}},\ and\
  \citenamefont {{Radovich}}}]{luo2017apj}%
  \BibitemOpen
  \bibfield  {author} {\bibinfo {author} {\bibfnamefont {W.}~\bibnamefont
  {{Luo}}}, \bibinfo {author} {\bibfnamefont {X.}~\bibnamefont {{Yang}}},
  \bibinfo {author} {\bibfnamefont {J.}~\bibnamefont {{Zhang}}}, \bibinfo
  {author} {\bibfnamefont {D.}~\bibnamefont {{Tweed}}}, \bibinfo {author}
  {\bibfnamefont {L.}~\bibnamefont {{Fu}}}, \bibinfo {author} {\bibfnamefont
  {H.~J.}\ \bibnamefont {{Mo}}}, \bibinfo {author} {\bibfnamefont {F.~C.}\
  \bibnamefont {{van den Bosch}}}, \bibinfo {author} {\bibfnamefont
  {C.}~\bibnamefont {{Shu}}}, \bibinfo {author} {\bibfnamefont
  {R.}~\bibnamefont {{Li}}}, \bibinfo {author} {\bibfnamefont {N.}~\bibnamefont
  {{Li}}}, \bibinfo {author} {\bibfnamefont {X.}~\bibnamefont {{Liu}}},
  \bibinfo {author} {\bibfnamefont {C.}~\bibnamefont {{Pan}}}, \bibinfo
  {author} {\bibfnamefont {Y.}~\bibnamefont {{Wang}}}, \ and\ \bibinfo {author}
  {\bibfnamefont {M.}~\bibnamefont {{Radovich}}},\ }\href {\doibase
  10.3847/1538-4357/836/1/38} {\bibfield  {journal} {\bibinfo  {journal}
  {\apj}\ }\textbf {\bibinfo {volume} {836}},\ \bibinfo {eid} {38} (\bibinfo
  {year} {2017})},\ \Eprint {http://arxiv.org/abs/1607.05406}
  {arXiv:1607.05406} \BibitemShut {NoStop}%
\bibitem [{\citenamefont {{Barreira}}\ \emph {et~al.}(2014)\citenamefont
  {{Barreira}}, \citenamefont {{Li}}, \citenamefont {{Hellwing}}, \citenamefont
  {{Lombriser}}, \citenamefont {{Baugh}},\ and\ \citenamefont
  {{Pascoli}}}]{barreira2014jcap}%
  \BibitemOpen
  \bibfield  {author} {\bibinfo {author} {\bibfnamefont {A.}~\bibnamefont
  {{Barreira}}}, \bibinfo {author} {\bibfnamefont {B.}~\bibnamefont {{Li}}},
  \bibinfo {author} {\bibfnamefont {W.~A.}\ \bibnamefont {{Hellwing}}},
  \bibinfo {author} {\bibfnamefont {L.}~\bibnamefont {{Lombriser}}}, \bibinfo
  {author} {\bibfnamefont {C.~M.}\ \bibnamefont {{Baugh}}}, \ and\ \bibinfo
  {author} {\bibfnamefont {S.}~\bibnamefont {{Pascoli}}},\ }\href {\doibase
  10.1088/1475-7516/2014/04/029} {\bibfield  {journal} {\bibinfo  {journal}
  {\jcap}\ }\textbf {\bibinfo {volume} {2014}},\ \bibinfo {eid} {029} (\bibinfo
  {year} {2014})},\ \Eprint {http://arxiv.org/abs/1401.1497} {arXiv:1401.1497
  [astro-ph.CO]} \BibitemShut {NoStop}%
\bibitem [{\citenamefont {{Zhang}}\ \emph {et~al.}(2018)\citenamefont
  {{Zhang}}, \citenamefont {{An}}, \citenamefont {{Liao}}, \citenamefont
  {{Luo}}, \citenamefont {{Li}},\ and\ \citenamefont {{Wang}}}]{megadget}%
  \BibitemOpen
  \bibfield  {author} {\bibinfo {author} {\bibfnamefont {J.}~\bibnamefont
  {{Zhang}}}, \bibinfo {author} {\bibfnamefont {R.}~\bibnamefont {{An}}},
  \bibinfo {author} {\bibfnamefont {S.}~\bibnamefont {{Liao}}}, \bibinfo
  {author} {\bibfnamefont {W.}~\bibnamefont {{Luo}}}, \bibinfo {author}
  {\bibfnamefont {Z.}~\bibnamefont {{Li}}}, \ and\ \bibinfo {author}
  {\bibfnamefont {B.}~\bibnamefont {{Wang}}},\ }\href {\doibase
  10.1103/PhysRevD.98.103530} {\bibfield  {journal} {\bibinfo  {journal}
  {\prd}\ }\textbf {\bibinfo {volume} {98}},\ \bibinfo {eid} {103530} (\bibinfo
  {year} {2018})},\ \Eprint {http://arxiv.org/abs/1811.01519} {arXiv:1811.01519
  [astro-ph.CO]} \BibitemShut {NoStop}%
\bibitem [{\citenamefont {{LSST Science Collaboration}}\ \emph
  {et~al.}(2009)\citenamefont {{LSST Science Collaboration}}, \citenamefont
  {{Abell}}, \citenamefont {{Allison}}, \citenamefont {{Anderson}},
  \citenamefont {{Andrew}}, \citenamefont {{Angel}}, \citenamefont {{Armus}},
  \citenamefont {{Arnett}}, \citenamefont {{Asztalos}}, \citenamefont
  {{Axelrod}}, \citenamefont {{Bailey}}, \citenamefont {{Ballantyne}},
  \citenamefont {{Bankert}}, \citenamefont {{Barkhouse}}, \citenamefont
  {{Barr}}, \citenamefont {{Barrientos}}, \citenamefont {{Barth}},
  \citenamefont {{Bartlett}}, \citenamefont {{Becker}}, \citenamefont
  {{Becla}}, \citenamefont {{Beers}}, \citenamefont {{Bernstein}},
  \citenamefont {{Biswas}}, \citenamefont {{Blanton}}, \citenamefont {{Bloom}},
  \citenamefont {{Bochanski}}, \citenamefont {{Boeshaar}}, \citenamefont
  {{Borne}}, \citenamefont {{Bradac}}, \citenamefont {{Brandt}}, \citenamefont
  {{Bridge}}, \citenamefont {{Brown}}, \citenamefont {{Brunner}}, \citenamefont
  {{Bullock}}, \citenamefont {{Burgasser}}, \citenamefont {{Burge}},
  \citenamefont {{Burke}}, \citenamefont {{Cargile}}, \citenamefont {{Chand
  rasekharan}}, \citenamefont {{Chartas}}, \citenamefont {{Chesley}},
  \citenamefont {{Chu}}, \citenamefont {{Cinabro}}, \citenamefont {{Claire}},
  \citenamefont {{Claver}}, \citenamefont {{Clowe}}, \citenamefont
  {{Connolly}}, \citenamefont {{Cook}}, \citenamefont {{Cooke}}, \citenamefont
  {{Cooray}}, \citenamefont {{Covey}}, \citenamefont {{Culliton}},
  \citenamefont {{de Jong}}, \citenamefont {{de Vries}}, \citenamefont
  {{Debattista}}, \citenamefont {{Delgado}}, \citenamefont {{Dell'Antonio}},
  \citenamefont {{Dhital}}, \citenamefont {{Di Stefano}}, \citenamefont
  {{Dickinson}}, \citenamefont {{Dilday}}, \citenamefont {{Djorgovski}},
  \citenamefont {{Dobler}}, \citenamefont {{Donalek}}, \citenamefont
  {{Dubois-Felsmann}}, \citenamefont {{Durech}}, \citenamefont {{Eliasdottir}},
  \citenamefont {{Eracleous}}, \citenamefont {{Eyer}}, \citenamefont {{Falco}},
  \citenamefont {{Fan}}, \citenamefont {{Fassnacht}}, \citenamefont
  {{Ferguson}}, \citenamefont {{Fernandez}}, \citenamefont {{Fields}},
  \citenamefont {{Finkbeiner}}, \citenamefont {{Figueroa}}, \citenamefont
  {{Fox}}, \citenamefont {{Francke}}, \citenamefont {{Frank}}, \citenamefont
  {{Frieman}}, \citenamefont {{Fromenteau}}, \citenamefont {{Furqan}},
  \citenamefont {{Galaz}}, \citenamefont {{Gal-Yam}}, \citenamefont
  {{Garnavich}}, \citenamefont {{Gawiser}}, \citenamefont {{Geary}},
  \citenamefont {{Gee}}, \citenamefont {{Gibson}}, \citenamefont {{Gilmore}},
  \citenamefont {{Grace}}, \citenamefont {{Green}}, \citenamefont {{Gressler}},
  \citenamefont {{Grillmair}}, \citenamefont {{Habib}}, \citenamefont
  {{Haggerty}}, \citenamefont {{Hamuy}}, \citenamefont {{Harris}},
  \citenamefont {{Hawley}}, \citenamefont {{Heavens}}, \citenamefont {{Hebb}},
  \citenamefont {{Henry}}, \citenamefont {{Hileman}}, \citenamefont {{Hilton}},
  \citenamefont {{Hoadley}}, \citenamefont {{Holberg}}, \citenamefont
  {{Holman}}, \citenamefont {{Howell}}, \citenamefont {{Infante}},
  \citenamefont {{Ivezic}}, \citenamefont {{Jacoby}}, \citenamefont {{Jain}},
  \citenamefont {{R}}, \citenamefont {{Jedicke}}, \citenamefont {{Jee}},
  \citenamefont {{Garrett Jernigan}}, \citenamefont {{Jha}}, \citenamefont
  {{Johnston}}, \citenamefont {{Jones}}, \citenamefont {{Juric}}, \citenamefont
  {{Kaasalainen}}, \citenamefont {{Styliani}}, \citenamefont {{Kafka}},
  \citenamefont {{Kahn}}, \citenamefont {{Kaib}}, \citenamefont {{Kalirai}},
  \citenamefont {{Kantor}}, \citenamefont {{Kasliwal}}, \citenamefont
  {{Keeton}}, \citenamefont {{Kessler}}, \citenamefont {{Knezevic}},
  \citenamefont {{Kowalski}}, \citenamefont {{Krabbendam}}, \citenamefont
  {{Krughoff}}, \citenamefont {{Kulkarni}}, \citenamefont {{Kuhlman}},
  \citenamefont {{Lacy}}, \citenamefont {{Lepine}}, \citenamefont {{Liang}},
  \citenamefont {{Lien}}, \citenamefont {{Lira}}, \citenamefont {{Long}},
  \citenamefont {{Lorenz}}, \citenamefont {{Lotz}}, \citenamefont {{Lupton}},
  \citenamefont {{Lutz}}, \citenamefont {{Macri}}, \citenamefont {{Mahabal}},
  \citenamefont {{Mandelbaum}}, \citenamefont {{Marshall}}, \citenamefont
  {{May}}, \citenamefont {{McGehee}}, \citenamefont {{Meadows}}, \citenamefont
  {{Meert}}, \citenamefont {{Milani}}, \citenamefont {{Miller}}, \citenamefont
  {{Miller}}, \citenamefont {{Mills}}, \citenamefont {{Minniti}}, \citenamefont
  {{Monet}}, \citenamefont {{Mukadam}}, \citenamefont {{Nakar}}, \citenamefont
  {{Neill}}, \citenamefont {{Newman}}, \citenamefont {{Nikolaev}},
  \citenamefont {{Nordby}}, \citenamefont {{O'Connor}}, \citenamefont
  {{Oguri}}, \citenamefont {{Oliver}}, \citenamefont {{Olivier}}, \citenamefont
  {{Olsen}}, \citenamefont {{Olsen}}, \citenamefont {{Olszewski}},
  \citenamefont {{Oluseyi}}, \citenamefont {{Padilla}}, \citenamefont
  {{Parker}}, \citenamefont {{Pepper}}, \citenamefont {{Peterson}},
  \citenamefont {{Petry}}, \citenamefont {{Pinto}}, \citenamefont {{Pizagno}},
  \citenamefont {{Popescu}}, \citenamefont {{Prsa}}, \citenamefont {{Radcka}},
  \citenamefont {{Raddick}}, \citenamefont {{Rasmussen}}, \citenamefont
  {{Rau}}, \citenamefont {{Rho}}, \citenamefont {{Rhoads}}, \citenamefont
  {{Richards}}, \citenamefont {{Ridgway}}, \citenamefont {{Robertson}},
  \citenamefont {{Roskar}}, \citenamefont {{Saha}}, \citenamefont
  {{Sarajedini}}, \citenamefont {{Scannapieco}}, \citenamefont {{Schalk}},
  \citenamefont {{Schindler}}, \citenamefont {{Schmidt}}, \citenamefont
  {{Schmidt}}, \citenamefont {{Schneider}}, \citenamefont {{Schumacher}},
  \citenamefont {{Scranton}}, \citenamefont {{Sebag}}, \citenamefont
  {{Seppala}}, \citenamefont {{Shemmer}}, \citenamefont {{Simon}},
  \citenamefont {{Sivertz}}, \citenamefont {{Smith}}, \citenamefont {{Allyn
  Smith}}, \citenamefont {{Smith}}, \citenamefont {{Spitz}}, \citenamefont
  {{Stanford}}, \citenamefont {{Stassun}}, \citenamefont {{Strader}},
  \citenamefont {{Strauss}}, \citenamefont {{Stubbs}}, \citenamefont
  {{Sweeney}}, \citenamefont {{Szalay}}, \citenamefont {{Szkody}},
  \citenamefont {{Takada}}, \citenamefont {{Thorman}}, \citenamefont
  {{Trilling}}, \citenamefont {{Trimble}}, \citenamefont {{Tyson}},
  \citenamefont {{Van Berg}}, \citenamefont {{Vand en Berk}}, \citenamefont
  {{VanderPlas}}, \citenamefont {{Verde}}, \citenamefont {{Vrsnak}},
  \citenamefont {{Walkowicz}}, \citenamefont {{Wand elt}}, \citenamefont
  {{Wang}}, \citenamefont {{Wang}}, \citenamefont {{Warner}}, \citenamefont
  {{Wechsler}}, \citenamefont {{West}}, \citenamefont {{Wiecha}}, \citenamefont
  {{Williams}}, \citenamefont {{Willman}}, \citenamefont {{Wittman}},
  \citenamefont {{Wolff}}, \citenamefont {{Wood-Vasey}}, \citenamefont
  {{Wozniak}}, \citenamefont {{Young}}, \citenamefont {{Zentner}},\ and\
  \citenamefont {{Zhan}}}]{lsst}%
  \BibitemOpen
  \bibfield  {author} {\bibinfo {author} {\bibnamefont {{LSST Science
  Collaboration}}}, \bibinfo {author} {\bibfnamefont {P.~A.}\ \bibnamefont
  {{Abell}}}, \bibinfo {author} {\bibfnamefont {J.}~\bibnamefont {{Allison}}},
  \bibinfo {author} {\bibfnamefont {S.~F.}\ \bibnamefont {{Anderson}}},
  \bibinfo {author} {\bibfnamefont {J.~R.}\ \bibnamefont {{Andrew}}}, \bibinfo
  {author} {\bibfnamefont {J.~R.~P.}\ \bibnamefont {{Angel}}}, \bibinfo
  {author} {\bibfnamefont {L.}~\bibnamefont {{Armus}}}, \bibinfo {author}
  {\bibfnamefont {D.}~\bibnamefont {{Arnett}}}, \bibinfo {author}
  {\bibfnamefont {S.~J.}\ \bibnamefont {{Asztalos}}}, \bibinfo {author}
  {\bibfnamefont {T.~S.}\ \bibnamefont {{Axelrod}}}, \bibinfo {author}
  {\bibfnamefont {S.}~\bibnamefont {{Bailey}}}, \bibinfo {author}
  {\bibfnamefont {D.~R.}\ \bibnamefont {{Ballantyne}}}, \bibinfo {author}
  {\bibfnamefont {J.~R.}\ \bibnamefont {{Bankert}}}, \bibinfo {author}
  {\bibfnamefont {W.~A.}\ \bibnamefont {{Barkhouse}}}, \bibinfo {author}
  {\bibfnamefont {J.~D.}\ \bibnamefont {{Barr}}}, \bibinfo {author}
  {\bibfnamefont {L.~F.}\ \bibnamefont {{Barrientos}}}, \bibinfo {author}
  {\bibfnamefont {A.~J.}\ \bibnamefont {{Barth}}}, \bibinfo {author}
  {\bibfnamefont {J.~G.}\ \bibnamefont {{Bartlett}}}, \bibinfo {author}
  {\bibfnamefont {A.~C.}\ \bibnamefont {{Becker}}}, \bibinfo {author}
  {\bibfnamefont {J.}~\bibnamefont {{Becla}}}, \bibinfo {author} {\bibfnamefont
  {T.~C.}\ \bibnamefont {{Beers}}}, \bibinfo {author} {\bibfnamefont {J.~P.}\
  \bibnamefont {{Bernstein}}}, \bibinfo {author} {\bibfnamefont
  {R.}~\bibnamefont {{Biswas}}}, \bibinfo {author} {\bibfnamefont {M.~R.}\
  \bibnamefont {{Blanton}}}, \bibinfo {author} {\bibfnamefont {J.~S.}\
  \bibnamefont {{Bloom}}}, \bibinfo {author} {\bibfnamefont {J.~J.}\
  \bibnamefont {{Bochanski}}}, \bibinfo {author} {\bibfnamefont
  {P.}~\bibnamefont {{Boeshaar}}}, \bibinfo {author} {\bibfnamefont {K.~D.}\
  \bibnamefont {{Borne}}}, \bibinfo {author} {\bibfnamefont {M.}~\bibnamefont
  {{Bradac}}}, \bibinfo {author} {\bibfnamefont {W.~N.}\ \bibnamefont
  {{Brandt}}}, \bibinfo {author} {\bibfnamefont {C.~R.}\ \bibnamefont
  {{Bridge}}}, \bibinfo {author} {\bibfnamefont {M.~E.}\ \bibnamefont
  {{Brown}}}, \bibinfo {author} {\bibfnamefont {R.~J.}\ \bibnamefont
  {{Brunner}}}, \bibinfo {author} {\bibfnamefont {J.~S.}\ \bibnamefont
  {{Bullock}}}, \bibinfo {author} {\bibfnamefont {A.~J.}\ \bibnamefont
  {{Burgasser}}}, \bibinfo {author} {\bibfnamefont {J.~H.}\ \bibnamefont
  {{Burge}}}, \bibinfo {author} {\bibfnamefont {D.~L.}\ \bibnamefont
  {{Burke}}}, \bibinfo {author} {\bibfnamefont {P.~A.}\ \bibnamefont
  {{Cargile}}}, \bibinfo {author} {\bibfnamefont {S.}~\bibnamefont {{Chand
  rasekharan}}}, \bibinfo {author} {\bibfnamefont {G.}~\bibnamefont
  {{Chartas}}}, \bibinfo {author} {\bibfnamefont {S.~R.}\ \bibnamefont
  {{Chesley}}}, \bibinfo {author} {\bibfnamefont {Y.-H.}\ \bibnamefont
  {{Chu}}}, \bibinfo {author} {\bibfnamefont {D.}~\bibnamefont {{Cinabro}}},
  \bibinfo {author} {\bibfnamefont {M.~W.}\ \bibnamefont {{Claire}}}, \bibinfo
  {author} {\bibfnamefont {C.~F.}\ \bibnamefont {{Claver}}}, \bibinfo {author}
  {\bibfnamefont {D.}~\bibnamefont {{Clowe}}}, \bibinfo {author} {\bibfnamefont
  {A.~J.}\ \bibnamefont {{Connolly}}}, \bibinfo {author} {\bibfnamefont
  {K.~H.}\ \bibnamefont {{Cook}}}, \bibinfo {author} {\bibfnamefont
  {J.}~\bibnamefont {{Cooke}}}, \bibinfo {author} {\bibfnamefont
  {A.}~\bibnamefont {{Cooray}}}, \bibinfo {author} {\bibfnamefont {K.~R.}\
  \bibnamefont {{Covey}}}, \bibinfo {author} {\bibfnamefont {C.~S.}\
  \bibnamefont {{Culliton}}}, \bibinfo {author} {\bibfnamefont
  {R.}~\bibnamefont {{de Jong}}}, \bibinfo {author} {\bibfnamefont {W.~H.}\
  \bibnamefont {{de Vries}}}, \bibinfo {author} {\bibfnamefont {V.~P.}\
  \bibnamefont {{Debattista}}}, \bibinfo {author} {\bibfnamefont
  {F.}~\bibnamefont {{Delgado}}}, \bibinfo {author} {\bibfnamefont {I.~P.}\
  \bibnamefont {{Dell'Antonio}}}, \bibinfo {author} {\bibfnamefont
  {S.}~\bibnamefont {{Dhital}}}, \bibinfo {author} {\bibfnamefont
  {R.}~\bibnamefont {{Di Stefano}}}, \bibinfo {author} {\bibfnamefont
  {M.}~\bibnamefont {{Dickinson}}}, \bibinfo {author} {\bibfnamefont
  {B.}~\bibnamefont {{Dilday}}}, \bibinfo {author} {\bibfnamefont {S.~G.}\
  \bibnamefont {{Djorgovski}}}, \bibinfo {author} {\bibfnamefont
  {G.}~\bibnamefont {{Dobler}}}, \bibinfo {author} {\bibfnamefont
  {C.}~\bibnamefont {{Donalek}}}, \bibinfo {author} {\bibfnamefont
  {G.}~\bibnamefont {{Dubois-Felsmann}}}, \bibinfo {author} {\bibfnamefont
  {J.}~\bibnamefont {{Durech}}}, \bibinfo {author} {\bibfnamefont
  {A.}~\bibnamefont {{Eliasdottir}}}, \bibinfo {author} {\bibfnamefont
  {M.}~\bibnamefont {{Eracleous}}}, \bibinfo {author} {\bibfnamefont
  {L.}~\bibnamefont {{Eyer}}}, \bibinfo {author} {\bibfnamefont {E.~E.}\
  \bibnamefont {{Falco}}}, \bibinfo {author} {\bibfnamefont {X.}~\bibnamefont
  {{Fan}}}, \bibinfo {author} {\bibfnamefont {C.~D.}\ \bibnamefont
  {{Fassnacht}}}, \bibinfo {author} {\bibfnamefont {H.~C.}\ \bibnamefont
  {{Ferguson}}}, \bibinfo {author} {\bibfnamefont {Y.~R.}\ \bibnamefont
  {{Fernandez}}}, \bibinfo {author} {\bibfnamefont {B.~D.}\ \bibnamefont
  {{Fields}}}, \bibinfo {author} {\bibfnamefont {D.}~\bibnamefont
  {{Finkbeiner}}}, \bibinfo {author} {\bibfnamefont {E.~E.}\ \bibnamefont
  {{Figueroa}}}, \bibinfo {author} {\bibfnamefont {D.~B.}\ \bibnamefont
  {{Fox}}}, \bibinfo {author} {\bibfnamefont {H.}~\bibnamefont {{Francke}}},
  \bibinfo {author} {\bibfnamefont {J.~S.}\ \bibnamefont {{Frank}}}, \bibinfo
  {author} {\bibfnamefont {J.}~\bibnamefont {{Frieman}}}, \bibinfo {author}
  {\bibfnamefont {S.}~\bibnamefont {{Fromenteau}}}, \bibinfo {author}
  {\bibfnamefont {M.}~\bibnamefont {{Furqan}}}, \bibinfo {author}
  {\bibfnamefont {G.}~\bibnamefont {{Galaz}}}, \bibinfo {author} {\bibfnamefont
  {A.}~\bibnamefont {{Gal-Yam}}}, \bibinfo {author} {\bibfnamefont
  {P.}~\bibnamefont {{Garnavich}}}, \bibinfo {author} {\bibfnamefont
  {E.}~\bibnamefont {{Gawiser}}}, \bibinfo {author} {\bibfnamefont
  {J.}~\bibnamefont {{Geary}}}, \bibinfo {author} {\bibfnamefont
  {P.}~\bibnamefont {{Gee}}}, \bibinfo {author} {\bibfnamefont {R.~R.}\
  \bibnamefont {{Gibson}}}, \bibinfo {author} {\bibfnamefont {K.}~\bibnamefont
  {{Gilmore}}}, \bibinfo {author} {\bibfnamefont {E.~A.}\ \bibnamefont
  {{Grace}}}, \bibinfo {author} {\bibfnamefont {R.~F.}\ \bibnamefont
  {{Green}}}, \bibinfo {author} {\bibfnamefont {W.~J.}\ \bibnamefont
  {{Gressler}}}, \bibinfo {author} {\bibfnamefont {C.~J.}\ \bibnamefont
  {{Grillmair}}}, \bibinfo {author} {\bibfnamefont {S.}~\bibnamefont
  {{Habib}}}, \bibinfo {author} {\bibfnamefont {J.~S.}\ \bibnamefont
  {{Haggerty}}}, \bibinfo {author} {\bibfnamefont {M.}~\bibnamefont {{Hamuy}}},
  \bibinfo {author} {\bibfnamefont {A.~W.}\ \bibnamefont {{Harris}}}, \bibinfo
  {author} {\bibfnamefont {S.~L.}\ \bibnamefont {{Hawley}}}, \bibinfo {author}
  {\bibfnamefont {A.~F.}\ \bibnamefont {{Heavens}}}, \bibinfo {author}
  {\bibfnamefont {L.}~\bibnamefont {{Hebb}}}, \bibinfo {author} {\bibfnamefont
  {T.~J.}\ \bibnamefont {{Henry}}}, \bibinfo {author} {\bibfnamefont
  {E.}~\bibnamefont {{Hileman}}}, \bibinfo {author} {\bibfnamefont {E.~J.}\
  \bibnamefont {{Hilton}}}, \bibinfo {author} {\bibfnamefont {K.}~\bibnamefont
  {{Hoadley}}}, \bibinfo {author} {\bibfnamefont {J.~B.}\ \bibnamefont
  {{Holberg}}}, \bibinfo {author} {\bibfnamefont {M.~J.}\ \bibnamefont
  {{Holman}}}, \bibinfo {author} {\bibfnamefont {S.~B.}\ \bibnamefont
  {{Howell}}}, \bibinfo {author} {\bibfnamefont {L.}~\bibnamefont {{Infante}}},
  \bibinfo {author} {\bibfnamefont {Z.}~\bibnamefont {{Ivezic}}}, \bibinfo
  {author} {\bibfnamefont {S.~H.}\ \bibnamefont {{Jacoby}}}, \bibinfo {author}
  {\bibfnamefont {B.}~\bibnamefont {{Jain}}}, \bibinfo {author} {\bibnamefont
  {{R}}}, \bibinfo {author} {\bibnamefont {{Jedicke}}}, \bibinfo {author}
  {\bibfnamefont {M.~J.}\ \bibnamefont {{Jee}}}, \bibinfo {author}
  {\bibfnamefont {J.}~\bibnamefont {{Garrett Jernigan}}}, \bibinfo {author}
  {\bibfnamefont {S.~W.}\ \bibnamefont {{Jha}}}, \bibinfo {author}
  {\bibfnamefont {K.~V.}\ \bibnamefont {{Johnston}}}, \bibinfo {author}
  {\bibfnamefont {R.~L.}\ \bibnamefont {{Jones}}}, \bibinfo {author}
  {\bibfnamefont {M.}~\bibnamefont {{Juric}}}, \bibinfo {author} {\bibfnamefont
  {M.}~\bibnamefont {{Kaasalainen}}}, \bibinfo {author} {\bibnamefont
  {{Styliani}}}, \bibinfo {author} {\bibnamefont {{Kafka}}}, \bibinfo {author}
  {\bibfnamefont {S.~M.}\ \bibnamefont {{Kahn}}}, \bibinfo {author}
  {\bibfnamefont {N.~A.}\ \bibnamefont {{Kaib}}}, \bibinfo {author}
  {\bibfnamefont {J.}~\bibnamefont {{Kalirai}}}, \bibinfo {author}
  {\bibfnamefont {J.}~\bibnamefont {{Kantor}}}, \bibinfo {author}
  {\bibfnamefont {M.~M.}\ \bibnamefont {{Kasliwal}}}, \bibinfo {author}
  {\bibfnamefont {C.~R.}\ \bibnamefont {{Keeton}}}, \bibinfo {author}
  {\bibfnamefont {R.}~\bibnamefont {{Kessler}}}, \bibinfo {author}
  {\bibfnamefont {Z.}~\bibnamefont {{Knezevic}}}, \bibinfo {author}
  {\bibfnamefont {A.}~\bibnamefont {{Kowalski}}}, \bibinfo {author}
  {\bibfnamefont {V.~L.}\ \bibnamefont {{Krabbendam}}}, \bibinfo {author}
  {\bibfnamefont {K.~S.}\ \bibnamefont {{Krughoff}}}, \bibinfo {author}
  {\bibfnamefont {S.}~\bibnamefont {{Kulkarni}}}, \bibinfo {author}
  {\bibfnamefont {S.}~\bibnamefont {{Kuhlman}}}, \bibinfo {author}
  {\bibfnamefont {M.}~\bibnamefont {{Lacy}}}, \bibinfo {author} {\bibfnamefont
  {S.}~\bibnamefont {{Lepine}}}, \bibinfo {author} {\bibfnamefont
  {M.}~\bibnamefont {{Liang}}}, \bibinfo {author} {\bibfnamefont
  {A.}~\bibnamefont {{Lien}}}, \bibinfo {author} {\bibfnamefont
  {P.}~\bibnamefont {{Lira}}}, \bibinfo {author} {\bibfnamefont {K.~S.}\
  \bibnamefont {{Long}}}, \bibinfo {author} {\bibfnamefont {S.}~\bibnamefont
  {{Lorenz}}}, \bibinfo {author} {\bibfnamefont {J.~M.}\ \bibnamefont
  {{Lotz}}}, \bibinfo {author} {\bibfnamefont {R.~H.}\ \bibnamefont
  {{Lupton}}}, \bibinfo {author} {\bibfnamefont {J.}~\bibnamefont {{Lutz}}},
  \bibinfo {author} {\bibfnamefont {L.~M.}\ \bibnamefont {{Macri}}}, \bibinfo
  {author} {\bibfnamefont {A.~A.}\ \bibnamefont {{Mahabal}}}, \bibinfo {author}
  {\bibfnamefont {R.}~\bibnamefont {{Mandelbaum}}}, \bibinfo {author}
  {\bibfnamefont {P.}~\bibnamefont {{Marshall}}}, \bibinfo {author}
  {\bibfnamefont {M.}~\bibnamefont {{May}}}, \bibinfo {author} {\bibfnamefont
  {P.~M.}\ \bibnamefont {{McGehee}}}, \bibinfo {author} {\bibfnamefont {B.~T.}\
  \bibnamefont {{Meadows}}}, \bibinfo {author} {\bibfnamefont {A.}~\bibnamefont
  {{Meert}}}, \bibinfo {author} {\bibfnamefont {A.}~\bibnamefont {{Milani}}},
  \bibinfo {author} {\bibfnamefont {C.~J.}\ \bibnamefont {{Miller}}}, \bibinfo
  {author} {\bibfnamefont {M.}~\bibnamefont {{Miller}}}, \bibinfo {author}
  {\bibfnamefont {D.}~\bibnamefont {{Mills}}}, \bibinfo {author} {\bibfnamefont
  {D.}~\bibnamefont {{Minniti}}}, \bibinfo {author} {\bibfnamefont
  {D.}~\bibnamefont {{Monet}}}, \bibinfo {author} {\bibfnamefont {A.~S.}\
  \bibnamefont {{Mukadam}}}, \bibinfo {author} {\bibfnamefont {E.}~\bibnamefont
  {{Nakar}}}, \bibinfo {author} {\bibfnamefont {D.~R.}\ \bibnamefont
  {{Neill}}}, \bibinfo {author} {\bibfnamefont {J.~A.}\ \bibnamefont
  {{Newman}}}, \bibinfo {author} {\bibfnamefont {S.}~\bibnamefont
  {{Nikolaev}}}, \bibinfo {author} {\bibfnamefont {M.}~\bibnamefont
  {{Nordby}}}, \bibinfo {author} {\bibfnamefont {P.}~\bibnamefont
  {{O'Connor}}}, \bibinfo {author} {\bibfnamefont {M.}~\bibnamefont {{Oguri}}},
  \bibinfo {author} {\bibfnamefont {J.}~\bibnamefont {{Oliver}}}, \bibinfo
  {author} {\bibfnamefont {S.~S.}\ \bibnamefont {{Olivier}}}, \bibinfo {author}
  {\bibfnamefont {J.~K.}\ \bibnamefont {{Olsen}}}, \bibinfo {author}
  {\bibfnamefont {K.}~\bibnamefont {{Olsen}}}, \bibinfo {author} {\bibfnamefont
  {E.~W.}\ \bibnamefont {{Olszewski}}}, \bibinfo {author} {\bibfnamefont
  {H.}~\bibnamefont {{Oluseyi}}}, \bibinfo {author} {\bibfnamefont {N.~D.}\
  \bibnamefont {{Padilla}}}, \bibinfo {author} {\bibfnamefont {A.}~\bibnamefont
  {{Parker}}}, \bibinfo {author} {\bibfnamefont {J.}~\bibnamefont {{Pepper}}},
  \bibinfo {author} {\bibfnamefont {J.~R.}\ \bibnamefont {{Peterson}}},
  \bibinfo {author} {\bibfnamefont {C.}~\bibnamefont {{Petry}}}, \bibinfo
  {author} {\bibfnamefont {P.~A.}\ \bibnamefont {{Pinto}}}, \bibinfo {author}
  {\bibfnamefont {J.~L.}\ \bibnamefont {{Pizagno}}}, \bibinfo {author}
  {\bibfnamefont {B.}~\bibnamefont {{Popescu}}}, \bibinfo {author}
  {\bibfnamefont {A.}~\bibnamefont {{Prsa}}}, \bibinfo {author} {\bibfnamefont
  {V.}~\bibnamefont {{Radcka}}}, \bibinfo {author} {\bibfnamefont {M.~J.}\
  \bibnamefont {{Raddick}}}, \bibinfo {author} {\bibfnamefont {A.}~\bibnamefont
  {{Rasmussen}}}, \bibinfo {author} {\bibfnamefont {A.}~\bibnamefont {{Rau}}},
  \bibinfo {author} {\bibfnamefont {J.}~\bibnamefont {{Rho}}}, \bibinfo
  {author} {\bibfnamefont {J.~E.}\ \bibnamefont {{Rhoads}}}, \bibinfo {author}
  {\bibfnamefont {G.~T.}\ \bibnamefont {{Richards}}}, \bibinfo {author}
  {\bibfnamefont {S.~T.}\ \bibnamefont {{Ridgway}}}, \bibinfo {author}
  {\bibfnamefont {B.~E.}\ \bibnamefont {{Robertson}}}, \bibinfo {author}
  {\bibfnamefont {R.}~\bibnamefont {{Roskar}}}, \bibinfo {author}
  {\bibfnamefont {A.}~\bibnamefont {{Saha}}}, \bibinfo {author} {\bibfnamefont
  {A.}~\bibnamefont {{Sarajedini}}}, \bibinfo {author} {\bibfnamefont
  {E.}~\bibnamefont {{Scannapieco}}}, \bibinfo {author} {\bibfnamefont
  {T.}~\bibnamefont {{Schalk}}}, \bibinfo {author} {\bibfnamefont
  {R.}~\bibnamefont {{Schindler}}}, \bibinfo {author} {\bibfnamefont
  {S.}~\bibnamefont {{Schmidt}}}, \bibinfo {author} {\bibfnamefont
  {S.}~\bibnamefont {{Schmidt}}}, \bibinfo {author} {\bibfnamefont {D.~P.}\
  \bibnamefont {{Schneider}}}, \bibinfo {author} {\bibfnamefont
  {G.}~\bibnamefont {{Schumacher}}}, \bibinfo {author} {\bibfnamefont
  {R.}~\bibnamefont {{Scranton}}}, \bibinfo {author} {\bibfnamefont
  {J.}~\bibnamefont {{Sebag}}}, \bibinfo {author} {\bibfnamefont {L.~G.}\
  \bibnamefont {{Seppala}}}, \bibinfo {author} {\bibfnamefont {O.}~\bibnamefont
  {{Shemmer}}}, \bibinfo {author} {\bibfnamefont {J.~D.}\ \bibnamefont
  {{Simon}}}, \bibinfo {author} {\bibfnamefont {M.}~\bibnamefont {{Sivertz}}},
  \bibinfo {author} {\bibfnamefont {H.~A.}\ \bibnamefont {{Smith}}}, \bibinfo
  {author} {\bibfnamefont {J.}~\bibnamefont {{Allyn Smith}}}, \bibinfo {author}
  {\bibfnamefont {N.}~\bibnamefont {{Smith}}}, \bibinfo {author} {\bibfnamefont
  {A.~H.}\ \bibnamefont {{Spitz}}}, \bibinfo {author} {\bibfnamefont
  {A.}~\bibnamefont {{Stanford}}}, \bibinfo {author} {\bibfnamefont {K.~G.}\
  \bibnamefont {{Stassun}}}, \bibinfo {author} {\bibfnamefont {J.}~\bibnamefont
  {{Strader}}}, \bibinfo {author} {\bibfnamefont {M.~A.}\ \bibnamefont
  {{Strauss}}}, \bibinfo {author} {\bibfnamefont {C.~W.}\ \bibnamefont
  {{Stubbs}}}, \bibinfo {author} {\bibfnamefont {D.~W.}\ \bibnamefont
  {{Sweeney}}}, \bibinfo {author} {\bibfnamefont {A.}~\bibnamefont {{Szalay}}},
  \bibinfo {author} {\bibfnamefont {P.}~\bibnamefont {{Szkody}}}, \bibinfo
  {author} {\bibfnamefont {M.}~\bibnamefont {{Takada}}}, \bibinfo {author}
  {\bibfnamefont {P.}~\bibnamefont {{Thorman}}}, \bibinfo {author}
  {\bibfnamefont {D.~E.}\ \bibnamefont {{Trilling}}}, \bibinfo {author}
  {\bibfnamefont {V.}~\bibnamefont {{Trimble}}}, \bibinfo {author}
  {\bibfnamefont {A.}~\bibnamefont {{Tyson}}}, \bibinfo {author} {\bibfnamefont
  {R.}~\bibnamefont {{Van Berg}}}, \bibinfo {author} {\bibfnamefont
  {D.}~\bibnamefont {{Vand en Berk}}}, \bibinfo {author} {\bibfnamefont
  {J.}~\bibnamefont {{VanderPlas}}}, \bibinfo {author} {\bibfnamefont
  {L.}~\bibnamefont {{Verde}}}, \bibinfo {author} {\bibfnamefont
  {B.}~\bibnamefont {{Vrsnak}}}, \bibinfo {author} {\bibfnamefont {L.~M.}\
  \bibnamefont {{Walkowicz}}}, \bibinfo {author} {\bibfnamefont {B.~D.}\
  \bibnamefont {{Wand elt}}}, \bibinfo {author} {\bibfnamefont
  {S.}~\bibnamefont {{Wang}}}, \bibinfo {author} {\bibfnamefont
  {Y.}~\bibnamefont {{Wang}}}, \bibinfo {author} {\bibfnamefont
  {M.}~\bibnamefont {{Warner}}}, \bibinfo {author} {\bibfnamefont {R.~H.}\
  \bibnamefont {{Wechsler}}}, \bibinfo {author} {\bibfnamefont {A.~A.}\
  \bibnamefont {{West}}}, \bibinfo {author} {\bibfnamefont {O.}~\bibnamefont
  {{Wiecha}}}, \bibinfo {author} {\bibfnamefont {B.~F.}\ \bibnamefont
  {{Williams}}}, \bibinfo {author} {\bibfnamefont {B.}~\bibnamefont
  {{Willman}}}, \bibinfo {author} {\bibfnamefont {D.}~\bibnamefont
  {{Wittman}}}, \bibinfo {author} {\bibfnamefont {S.~C.}\ \bibnamefont
  {{Wolff}}}, \bibinfo {author} {\bibfnamefont {W.~M.}\ \bibnamefont
  {{Wood-Vasey}}}, \bibinfo {author} {\bibfnamefont {P.}~\bibnamefont
  {{Wozniak}}}, \bibinfo {author} {\bibfnamefont {P.}~\bibnamefont {{Young}}},
  \bibinfo {author} {\bibfnamefont {A.}~\bibnamefont {{Zentner}}}, \ and\
  \bibinfo {author} {\bibfnamefont {H.}~\bibnamefont {{Zhan}}},\ }\href@noop {}
  {\bibfield  {journal} {\bibinfo  {journal} {arXiv e-prints}\ ,\ \bibinfo
  {eid} {arXiv:0912.0201}} (\bibinfo {year} {2009})},\ \Eprint
  {http://arxiv.org/abs/0912.0201} {arXiv:0912.0201 [astro-ph.IM]} \BibitemShut
  {NoStop}%
\bibitem [{\citenamefont {{Laureijs}}\ \emph {et~al.}(2011)\citenamefont
  {{Laureijs}}, \citenamefont {{Amiaux}}, \citenamefont {{Arduini}},
  \citenamefont {{Augu{\`e}res}}, \citenamefont {{Brinchmann}}, \citenamefont
  {{Cole}}, \citenamefont {{Cropper}}, \citenamefont {{Dabin}}, \citenamefont
  {{Duvet}}, \citenamefont {{Ealet}}, \citenamefont {{Garilli}}, \citenamefont
  {{Gondoin}}, \citenamefont {{Guzzo}}, \citenamefont {{Hoar}}, \citenamefont
  {{Hoekstra}}, \citenamefont {{Holmes}}, \citenamefont {{Kitching}},
  \citenamefont {{Maciaszek}}, \citenamefont {{Mellier}}, \citenamefont
  {{Pasian}}, \citenamefont {{Percival}}, \citenamefont {{Rhodes}},
  \citenamefont {{Saavedra Criado}}, \citenamefont {{Sauvage}}, \citenamefont
  {{Scaramella}}, \citenamefont {{Valenziano}}, \citenamefont {{Warren}},
  \citenamefont {{Bender}}, \citenamefont {{Castander}}, \citenamefont
  {{Cimatti}}, \citenamefont {{Le F{\`e}vre}}, \citenamefont {{Kurki-Suonio}},
  \citenamefont {{Levi}}, \citenamefont {{Lilje}}, \citenamefont {{Meylan}},
  \citenamefont {{Nichol}}, \citenamefont {{Pedersen}}, \citenamefont {{Popa}},
  \citenamefont {{Rebolo Lopez}}, \citenamefont {{Rix}}, \citenamefont
  {{Rottgering}}, \citenamefont {{Zeilinger}}, \citenamefont {{Grupp}},
  \citenamefont {{Hudelot}}, \citenamefont {{Massey}}, \citenamefont
  {{Meneghetti}}, \citenamefont {{Miller}}, \citenamefont {{Paltani}},
  \citenamefont {{Paulin-Henriksson}}, \citenamefont {{Pires}}, \citenamefont
  {{Saxton}}, \citenamefont {{Schrabback}}, \citenamefont {{Seidel}},
  \citenamefont {{Walsh}}, \citenamefont {{Aghanim}}, \citenamefont
  {{Amendola}}, \citenamefont {{Bartlett}}, \citenamefont {{Baccigalupi}},
  \citenamefont {{Beaulieu}}, \citenamefont {{Benabed}}, \citenamefont
  {{Cuby}}, \citenamefont {{Elbaz}}, \citenamefont {{Fosalba}}, \citenamefont
  {{Gavazzi}}, \citenamefont {{Helmi}}, \citenamefont {{Hook}}, \citenamefont
  {{Irwin}}, \citenamefont {{Kneib}}, \citenamefont {{Kunz}}, \citenamefont
  {{Mannucci}}, \citenamefont {{Moscardini}}, \citenamefont {{Tao}},
  \citenamefont {{Teyssier}}, \citenamefont {{Weller}}, \citenamefont
  {{Zamorani}}, \citenamefont {{Zapatero Osorio}}, \citenamefont {{Boulade}},
  \citenamefont {{Foumond}}, \citenamefont {{Di Giorgio}}, \citenamefont
  {{Guttridge}}, \citenamefont {{James}}, \citenamefont {{Kemp}}, \citenamefont
  {{Martignac}}, \citenamefont {{Spencer}}, \citenamefont {{Walton}},
  \citenamefont {{Bl{\"u}mchen}}, \citenamefont {{Bonoli}}, \citenamefont
  {{Bortoletto}}, \citenamefont {{Cerna}}, \citenamefont {{Corcione}},
  \citenamefont {{Fabron}}, \citenamefont {{Jahnke}}, \citenamefont {{Ligori}},
  \citenamefont {{Madrid}}, \citenamefont {{Martin}}, \citenamefont
  {{Morgante}}, \citenamefont {{Pamplona}}, \citenamefont {{Prieto}},
  \citenamefont {{Riva}}, \citenamefont {{Toledo}}, \citenamefont
  {{Trifoglio}}, \citenamefont {{Zerbi}}, \citenamefont {{Abdalla}},
  \citenamefont {{Douspis}}, \citenamefont {{Grenet}}, \citenamefont
  {{Borgani}}, \citenamefont {{Bouwens}}, \citenamefont {{Courbin}},
  \citenamefont {{Delouis}}, \citenamefont {{Dubath}}, \citenamefont
  {{Fontana}}, \citenamefont {{Frailis}}, \citenamefont {{Grazian}},
  \citenamefont {{Koppenh{\"o}fer}}, \citenamefont {{Mansutti}}, \citenamefont
  {{Melchior}}, \citenamefont {{Mignoli}}, \citenamefont {{Mohr}},
  \citenamefont {{Neissner}}, \citenamefont {{Noddle}}, \citenamefont
  {{Poncet}}, \citenamefont {{Scodeggio}}, \citenamefont {{Serrano}},
  \citenamefont {{Shane}}, \citenamefont {{Starck}}, \citenamefont {{Surace}},
  \citenamefont {{Taylor}}, \citenamefont {{Verdoes-Kleijn}}, \citenamefont
  {{Vuerli}}, \citenamefont {{Williams}}, \citenamefont {{Zacchei}},
  \citenamefont {{Altieri}}, \citenamefont {{Escudero Sanz}}, \citenamefont
  {{Kohley}}, \citenamefont {{Oosterbroek}}, \citenamefont {{Astier}},
  \citenamefont {{Bacon}}, \citenamefont {{Bardelli}}, \citenamefont {{Baugh}},
  \citenamefont {{Bellagamba}}, \citenamefont {{Benoist}}, \citenamefont
  {{Bianchi}}, \citenamefont {{Biviano}}, \citenamefont {{Branchini}},
  \citenamefont {{Carbone}}, \citenamefont {{Cardone}}, \citenamefont
  {{Clements}}, \citenamefont {{Colombi}}, \citenamefont {{Conselice}},
  \citenamefont {{Cresci}}, \citenamefont {{Deacon}}, \citenamefont {{Dunlop}},
  \citenamefont {{Fedeli}}, \citenamefont {{Fontanot}}, \citenamefont
  {{Franzetti}}, \citenamefont {{Giocoli}}, \citenamefont {{Garcia-Bellido}},
  \citenamefont {{Gow}}, \citenamefont {{Heavens}}, \citenamefont {{Hewett}},
  \citenamefont {{Heymans}}, \citenamefont {{Holland}}, \citenamefont
  {{Huang}}, \citenamefont {{Ilbert}}, \citenamefont {{Joachimi}},
  \citenamefont {{Jennins}}, \citenamefont {{Kerins}}, \citenamefont
  {{Kiessling}}, \citenamefont {{Kirk}}, \citenamefont {{Kotak}}, \citenamefont
  {{Krause}}, \citenamefont {{Lahav}}, \citenamefont {{van Leeuwen}},
  \citenamefont {{Lesgourgues}}, \citenamefont {{Lombardi}}, \citenamefont
  {{Magliocchetti}}, \citenamefont {{Maguire}}, \citenamefont {{Majerotto}},
  \citenamefont {{Maoli}}, \citenamefont {{Marulli}}, \citenamefont
  {{Maurogordato}}, \citenamefont {{McCracken}}, \citenamefont {{McLure}},
  \citenamefont {{Melchiorri}}, \citenamefont {{Merson}}, \citenamefont
  {{Moresco}}, \citenamefont {{Nonino}}, \citenamefont {{Norberg}},
  \citenamefont {{Peacock}}, \citenamefont {{Pello}}, \citenamefont {{Penny}},
  \citenamefont {{Pettorino}}, \citenamefont {{Di Porto}}, \citenamefont
  {{Pozzetti}}, \citenamefont {{Quercellini}}, \citenamefont {{Radovich}},
  \citenamefont {{Rassat}}, \citenamefont {{Roche}}, \citenamefont
  {{Ronayette}}, \citenamefont {{Rossetti}}, \citenamefont {{Sartoris}},
  \citenamefont {{Schneider}}, \citenamefont {{Semboloni}}, \citenamefont
  {{Serjeant}}, \citenamefont {{Simpson}}, \citenamefont {{Skordis}},
  \citenamefont {{Smadja}}, \citenamefont {{Smartt}}, \citenamefont {{Spano}},
  \citenamefont {{Spiro}}, \citenamefont {{Sullivan}}, \citenamefont
  {{Tilquin}}, \citenamefont {{Trotta}}, \citenamefont {{Verde}}, \citenamefont
  {{Wang}}, \citenamefont {{Williger}}, \citenamefont {{Zhao}}, \citenamefont
  {{Zoubian}},\ and\ \citenamefont {{Zucca}}}]{euclid1}%
  \BibitemOpen
  \bibfield  {author} {\bibinfo {author} {\bibfnamefont {R.}~\bibnamefont
  {{Laureijs}}}, \bibinfo {author} {\bibfnamefont {J.}~\bibnamefont
  {{Amiaux}}}, \bibinfo {author} {\bibfnamefont {S.}~\bibnamefont {{Arduini}}},
  \bibinfo {author} {\bibfnamefont {J.~L.}\ \bibnamefont {{Augu{\`e}res}}},
  \bibinfo {author} {\bibfnamefont {J.}~\bibnamefont {{Brinchmann}}}, \bibinfo
  {author} {\bibfnamefont {R.}~\bibnamefont {{Cole}}}, \bibinfo {author}
  {\bibfnamefont {M.}~\bibnamefont {{Cropper}}}, \bibinfo {author}
  {\bibfnamefont {C.}~\bibnamefont {{Dabin}}}, \bibinfo {author} {\bibfnamefont
  {L.}~\bibnamefont {{Duvet}}}, \bibinfo {author} {\bibfnamefont
  {A.}~\bibnamefont {{Ealet}}}, \bibinfo {author} {\bibfnamefont
  {B.}~\bibnamefont {{Garilli}}}, \bibinfo {author} {\bibfnamefont
  {P.}~\bibnamefont {{Gondoin}}}, \bibinfo {author} {\bibfnamefont
  {L.}~\bibnamefont {{Guzzo}}}, \bibinfo {author} {\bibfnamefont
  {J.}~\bibnamefont {{Hoar}}}, \bibinfo {author} {\bibfnamefont
  {H.}~\bibnamefont {{Hoekstra}}}, \bibinfo {author} {\bibfnamefont
  {R.}~\bibnamefont {{Holmes}}}, \bibinfo {author} {\bibfnamefont
  {T.}~\bibnamefont {{Kitching}}}, \bibinfo {author} {\bibfnamefont
  {T.}~\bibnamefont {{Maciaszek}}}, \bibinfo {author} {\bibfnamefont
  {Y.}~\bibnamefont {{Mellier}}}, \bibinfo {author} {\bibfnamefont
  {F.}~\bibnamefont {{Pasian}}}, \bibinfo {author} {\bibfnamefont
  {W.}~\bibnamefont {{Percival}}}, \bibinfo {author} {\bibfnamefont
  {J.}~\bibnamefont {{Rhodes}}}, \bibinfo {author} {\bibfnamefont
  {G.}~\bibnamefont {{Saavedra Criado}}}, \bibinfo {author} {\bibfnamefont
  {M.}~\bibnamefont {{Sauvage}}}, \bibinfo {author} {\bibfnamefont
  {R.}~\bibnamefont {{Scaramella}}}, \bibinfo {author} {\bibfnamefont
  {L.}~\bibnamefont {{Valenziano}}}, \bibinfo {author} {\bibfnamefont
  {S.}~\bibnamefont {{Warren}}}, \bibinfo {author} {\bibfnamefont
  {R.}~\bibnamefont {{Bender}}}, \bibinfo {author} {\bibfnamefont
  {F.}~\bibnamefont {{Castander}}}, \bibinfo {author} {\bibfnamefont
  {A.}~\bibnamefont {{Cimatti}}}, \bibinfo {author} {\bibfnamefont
  {O.}~\bibnamefont {{Le F{\`e}vre}}}, \bibinfo {author} {\bibfnamefont
  {H.}~\bibnamefont {{Kurki-Suonio}}}, \bibinfo {author} {\bibfnamefont
  {M.}~\bibnamefont {{Levi}}}, \bibinfo {author} {\bibfnamefont
  {P.}~\bibnamefont {{Lilje}}}, \bibinfo {author} {\bibfnamefont
  {G.}~\bibnamefont {{Meylan}}}, \bibinfo {author} {\bibfnamefont
  {R.}~\bibnamefont {{Nichol}}}, \bibinfo {author} {\bibfnamefont
  {K.}~\bibnamefont {{Pedersen}}}, \bibinfo {author} {\bibfnamefont
  {V.}~\bibnamefont {{Popa}}}, \bibinfo {author} {\bibfnamefont
  {R.}~\bibnamefont {{Rebolo Lopez}}}, \bibinfo {author} {\bibfnamefont
  {H.~W.}\ \bibnamefont {{Rix}}}, \bibinfo {author} {\bibfnamefont
  {H.}~\bibnamefont {{Rottgering}}}, \bibinfo {author} {\bibfnamefont
  {W.}~\bibnamefont {{Zeilinger}}}, \bibinfo {author} {\bibfnamefont
  {F.}~\bibnamefont {{Grupp}}}, \bibinfo {author} {\bibfnamefont
  {P.}~\bibnamefont {{Hudelot}}}, \bibinfo {author} {\bibfnamefont
  {R.}~\bibnamefont {{Massey}}}, \bibinfo {author} {\bibfnamefont
  {M.}~\bibnamefont {{Meneghetti}}}, \bibinfo {author} {\bibfnamefont
  {L.}~\bibnamefont {{Miller}}}, \bibinfo {author} {\bibfnamefont
  {S.}~\bibnamefont {{Paltani}}}, \bibinfo {author} {\bibfnamefont
  {S.}~\bibnamefont {{Paulin-Henriksson}}}, \bibinfo {author} {\bibfnamefont
  {S.}~\bibnamefont {{Pires}}}, \bibinfo {author} {\bibfnamefont
  {C.}~\bibnamefont {{Saxton}}}, \bibinfo {author} {\bibfnamefont
  {T.}~\bibnamefont {{Schrabback}}}, \bibinfo {author} {\bibfnamefont
  {G.}~\bibnamefont {{Seidel}}}, \bibinfo {author} {\bibfnamefont
  {J.}~\bibnamefont {{Walsh}}}, \bibinfo {author} {\bibfnamefont
  {N.}~\bibnamefont {{Aghanim}}}, \bibinfo {author} {\bibfnamefont
  {L.}~\bibnamefont {{Amendola}}}, \bibinfo {author} {\bibfnamefont
  {J.}~\bibnamefont {{Bartlett}}}, \bibinfo {author} {\bibfnamefont
  {C.}~\bibnamefont {{Baccigalupi}}}, \bibinfo {author} {\bibfnamefont {J.~P.}\
  \bibnamefont {{Beaulieu}}}, \bibinfo {author} {\bibfnamefont
  {K.}~\bibnamefont {{Benabed}}}, \bibinfo {author} {\bibfnamefont {J.~G.}\
  \bibnamefont {{Cuby}}}, \bibinfo {author} {\bibfnamefont {D.}~\bibnamefont
  {{Elbaz}}}, \bibinfo {author} {\bibfnamefont {P.}~\bibnamefont {{Fosalba}}},
  \bibinfo {author} {\bibfnamefont {G.}~\bibnamefont {{Gavazzi}}}, \bibinfo
  {author} {\bibfnamefont {A.}~\bibnamefont {{Helmi}}}, \bibinfo {author}
  {\bibfnamefont {I.}~\bibnamefont {{Hook}}}, \bibinfo {author} {\bibfnamefont
  {M.}~\bibnamefont {{Irwin}}}, \bibinfo {author} {\bibfnamefont {J.~P.}\
  \bibnamefont {{Kneib}}}, \bibinfo {author} {\bibfnamefont {M.}~\bibnamefont
  {{Kunz}}}, \bibinfo {author} {\bibfnamefont {F.}~\bibnamefont {{Mannucci}}},
  \bibinfo {author} {\bibfnamefont {L.}~\bibnamefont {{Moscardini}}}, \bibinfo
  {author} {\bibfnamefont {C.}~\bibnamefont {{Tao}}}, \bibinfo {author}
  {\bibfnamefont {R.}~\bibnamefont {{Teyssier}}}, \bibinfo {author}
  {\bibfnamefont {J.}~\bibnamefont {{Weller}}}, \bibinfo {author}
  {\bibfnamefont {G.}~\bibnamefont {{Zamorani}}}, \bibinfo {author}
  {\bibfnamefont {M.~R.}\ \bibnamefont {{Zapatero Osorio}}}, \bibinfo {author}
  {\bibfnamefont {O.}~\bibnamefont {{Boulade}}}, \bibinfo {author}
  {\bibfnamefont {J.~J.}\ \bibnamefont {{Foumond}}}, \bibinfo {author}
  {\bibfnamefont {A.}~\bibnamefont {{Di Giorgio}}}, \bibinfo {author}
  {\bibfnamefont {P.}~\bibnamefont {{Guttridge}}}, \bibinfo {author}
  {\bibfnamefont {A.}~\bibnamefont {{James}}}, \bibinfo {author} {\bibfnamefont
  {M.}~\bibnamefont {{Kemp}}}, \bibinfo {author} {\bibfnamefont
  {J.}~\bibnamefont {{Martignac}}}, \bibinfo {author} {\bibfnamefont
  {A.}~\bibnamefont {{Spencer}}}, \bibinfo {author} {\bibfnamefont
  {D.}~\bibnamefont {{Walton}}}, \bibinfo {author} {\bibfnamefont
  {T.}~\bibnamefont {{Bl{\"u}mchen}}}, \bibinfo {author} {\bibfnamefont
  {C.}~\bibnamefont {{Bonoli}}}, \bibinfo {author} {\bibfnamefont
  {F.}~\bibnamefont {{Bortoletto}}}, \bibinfo {author} {\bibfnamefont
  {C.}~\bibnamefont {{Cerna}}}, \bibinfo {author} {\bibfnamefont
  {L.}~\bibnamefont {{Corcione}}}, \bibinfo {author} {\bibfnamefont
  {C.}~\bibnamefont {{Fabron}}}, \bibinfo {author} {\bibfnamefont
  {K.}~\bibnamefont {{Jahnke}}}, \bibinfo {author} {\bibfnamefont
  {S.}~\bibnamefont {{Ligori}}}, \bibinfo {author} {\bibfnamefont
  {F.}~\bibnamefont {{Madrid}}}, \bibinfo {author} {\bibfnamefont
  {L.}~\bibnamefont {{Martin}}}, \bibinfo {author} {\bibfnamefont
  {G.}~\bibnamefont {{Morgante}}}, \bibinfo {author} {\bibfnamefont
  {T.}~\bibnamefont {{Pamplona}}}, \bibinfo {author} {\bibfnamefont
  {E.}~\bibnamefont {{Prieto}}}, \bibinfo {author} {\bibfnamefont
  {M.}~\bibnamefont {{Riva}}}, \bibinfo {author} {\bibfnamefont
  {R.}~\bibnamefont {{Toledo}}}, \bibinfo {author} {\bibfnamefont
  {M.}~\bibnamefont {{Trifoglio}}}, \bibinfo {author} {\bibfnamefont
  {F.}~\bibnamefont {{Zerbi}}}, \bibinfo {author} {\bibfnamefont
  {F.}~\bibnamefont {{Abdalla}}}, \bibinfo {author} {\bibfnamefont
  {M.}~\bibnamefont {{Douspis}}}, \bibinfo {author} {\bibfnamefont
  {C.}~\bibnamefont {{Grenet}}}, \bibinfo {author} {\bibfnamefont
  {S.}~\bibnamefont {{Borgani}}}, \bibinfo {author} {\bibfnamefont
  {R.}~\bibnamefont {{Bouwens}}}, \bibinfo {author} {\bibfnamefont
  {F.}~\bibnamefont {{Courbin}}}, \bibinfo {author} {\bibfnamefont {J.~M.}\
  \bibnamefont {{Delouis}}}, \bibinfo {author} {\bibfnamefont {P.}~\bibnamefont
  {{Dubath}}}, \bibinfo {author} {\bibfnamefont {A.}~\bibnamefont {{Fontana}}},
  \bibinfo {author} {\bibfnamefont {M.}~\bibnamefont {{Frailis}}}, \bibinfo
  {author} {\bibfnamefont {A.}~\bibnamefont {{Grazian}}}, \bibinfo {author}
  {\bibfnamefont {J.}~\bibnamefont {{Koppenh{\"o}fer}}}, \bibinfo {author}
  {\bibfnamefont {O.}~\bibnamefont {{Mansutti}}}, \bibinfo {author}
  {\bibfnamefont {M.}~\bibnamefont {{Melchior}}}, \bibinfo {author}
  {\bibfnamefont {M.}~\bibnamefont {{Mignoli}}}, \bibinfo {author}
  {\bibfnamefont {J.}~\bibnamefont {{Mohr}}}, \bibinfo {author} {\bibfnamefont
  {C.}~\bibnamefont {{Neissner}}}, \bibinfo {author} {\bibfnamefont
  {K.}~\bibnamefont {{Noddle}}}, \bibinfo {author} {\bibfnamefont
  {M.}~\bibnamefont {{Poncet}}}, \bibinfo {author} {\bibfnamefont
  {M.}~\bibnamefont {{Scodeggio}}}, \bibinfo {author} {\bibfnamefont
  {S.}~\bibnamefont {{Serrano}}}, \bibinfo {author} {\bibfnamefont
  {N.}~\bibnamefont {{Shane}}}, \bibinfo {author} {\bibfnamefont {J.~L.}\
  \bibnamefont {{Starck}}}, \bibinfo {author} {\bibfnamefont {C.}~\bibnamefont
  {{Surace}}}, \bibinfo {author} {\bibfnamefont {A.}~\bibnamefont {{Taylor}}},
  \bibinfo {author} {\bibfnamefont {G.}~\bibnamefont {{Verdoes-Kleijn}}},
  \bibinfo {author} {\bibfnamefont {C.}~\bibnamefont {{Vuerli}}}, \bibinfo
  {author} {\bibfnamefont {O.~R.}\ \bibnamefont {{Williams}}}, \bibinfo
  {author} {\bibfnamefont {A.}~\bibnamefont {{Zacchei}}}, \bibinfo {author}
  {\bibfnamefont {B.}~\bibnamefont {{Altieri}}}, \bibinfo {author}
  {\bibfnamefont {I.}~\bibnamefont {{Escudero Sanz}}}, \bibinfo {author}
  {\bibfnamefont {R.}~\bibnamefont {{Kohley}}}, \bibinfo {author}
  {\bibfnamefont {T.}~\bibnamefont {{Oosterbroek}}}, \bibinfo {author}
  {\bibfnamefont {P.}~\bibnamefont {{Astier}}}, \bibinfo {author}
  {\bibfnamefont {D.}~\bibnamefont {{Bacon}}}, \bibinfo {author} {\bibfnamefont
  {S.}~\bibnamefont {{Bardelli}}}, \bibinfo {author} {\bibfnamefont
  {C.}~\bibnamefont {{Baugh}}}, \bibinfo {author} {\bibfnamefont
  {F.}~\bibnamefont {{Bellagamba}}}, \bibinfo {author} {\bibfnamefont
  {C.}~\bibnamefont {{Benoist}}}, \bibinfo {author} {\bibfnamefont
  {D.}~\bibnamefont {{Bianchi}}}, \bibinfo {author} {\bibfnamefont
  {A.}~\bibnamefont {{Biviano}}}, \bibinfo {author} {\bibfnamefont
  {E.}~\bibnamefont {{Branchini}}}, \bibinfo {author} {\bibfnamefont
  {C.}~\bibnamefont {{Carbone}}}, \bibinfo {author} {\bibfnamefont
  {V.}~\bibnamefont {{Cardone}}}, \bibinfo {author} {\bibfnamefont
  {D.}~\bibnamefont {{Clements}}}, \bibinfo {author} {\bibfnamefont
  {S.}~\bibnamefont {{Colombi}}}, \bibinfo {author} {\bibfnamefont
  {C.}~\bibnamefont {{Conselice}}}, \bibinfo {author} {\bibfnamefont
  {G.}~\bibnamefont {{Cresci}}}, \bibinfo {author} {\bibfnamefont
  {N.}~\bibnamefont {{Deacon}}}, \bibinfo {author} {\bibfnamefont
  {J.}~\bibnamefont {{Dunlop}}}, \bibinfo {author} {\bibfnamefont
  {C.}~\bibnamefont {{Fedeli}}}, \bibinfo {author} {\bibfnamefont
  {F.}~\bibnamefont {{Fontanot}}}, \bibinfo {author} {\bibfnamefont
  {P.}~\bibnamefont {{Franzetti}}}, \bibinfo {author} {\bibfnamefont
  {C.}~\bibnamefont {{Giocoli}}}, \bibinfo {author} {\bibfnamefont
  {J.}~\bibnamefont {{Garcia-Bellido}}}, \bibinfo {author} {\bibfnamefont
  {J.}~\bibnamefont {{Gow}}}, \bibinfo {author} {\bibfnamefont
  {A.}~\bibnamefont {{Heavens}}}, \bibinfo {author} {\bibfnamefont
  {P.}~\bibnamefont {{Hewett}}}, \bibinfo {author} {\bibfnamefont
  {C.}~\bibnamefont {{Heymans}}}, \bibinfo {author} {\bibfnamefont
  {A.}~\bibnamefont {{Holland}}}, \bibinfo {author} {\bibfnamefont
  {Z.}~\bibnamefont {{Huang}}}, \bibinfo {author} {\bibfnamefont
  {O.}~\bibnamefont {{Ilbert}}}, \bibinfo {author} {\bibfnamefont
  {B.}~\bibnamefont {{Joachimi}}}, \bibinfo {author} {\bibfnamefont
  {E.}~\bibnamefont {{Jennins}}}, \bibinfo {author} {\bibfnamefont
  {E.}~\bibnamefont {{Kerins}}}, \bibinfo {author} {\bibfnamefont
  {A.}~\bibnamefont {{Kiessling}}}, \bibinfo {author} {\bibfnamefont
  {D.}~\bibnamefont {{Kirk}}}, \bibinfo {author} {\bibfnamefont
  {R.}~\bibnamefont {{Kotak}}}, \bibinfo {author} {\bibfnamefont
  {O.}~\bibnamefont {{Krause}}}, \bibinfo {author} {\bibfnamefont
  {O.}~\bibnamefont {{Lahav}}}, \bibinfo {author} {\bibfnamefont
  {F.}~\bibnamefont {{van Leeuwen}}}, \bibinfo {author} {\bibfnamefont
  {J.}~\bibnamefont {{Lesgourgues}}}, \bibinfo {author} {\bibfnamefont
  {M.}~\bibnamefont {{Lombardi}}}, \bibinfo {author} {\bibfnamefont
  {M.}~\bibnamefont {{Magliocchetti}}}, \bibinfo {author} {\bibfnamefont
  {K.}~\bibnamefont {{Maguire}}}, \bibinfo {author} {\bibfnamefont
  {E.}~\bibnamefont {{Majerotto}}}, \bibinfo {author} {\bibfnamefont
  {R.}~\bibnamefont {{Maoli}}}, \bibinfo {author} {\bibfnamefont
  {F.}~\bibnamefont {{Marulli}}}, \bibinfo {author} {\bibfnamefont
  {S.}~\bibnamefont {{Maurogordato}}}, \bibinfo {author} {\bibfnamefont
  {H.}~\bibnamefont {{McCracken}}}, \bibinfo {author} {\bibfnamefont
  {R.}~\bibnamefont {{McLure}}}, \bibinfo {author} {\bibfnamefont
  {A.}~\bibnamefont {{Melchiorri}}}, \bibinfo {author} {\bibfnamefont
  {A.}~\bibnamefont {{Merson}}}, \bibinfo {author} {\bibfnamefont
  {M.}~\bibnamefont {{Moresco}}}, \bibinfo {author} {\bibfnamefont
  {M.}~\bibnamefont {{Nonino}}}, \bibinfo {author} {\bibfnamefont
  {P.}~\bibnamefont {{Norberg}}}, \bibinfo {author} {\bibfnamefont
  {J.}~\bibnamefont {{Peacock}}}, \bibinfo {author} {\bibfnamefont
  {R.}~\bibnamefont {{Pello}}}, \bibinfo {author} {\bibfnamefont
  {M.}~\bibnamefont {{Penny}}}, \bibinfo {author} {\bibfnamefont
  {V.}~\bibnamefont {{Pettorino}}}, \bibinfo {author} {\bibfnamefont
  {C.}~\bibnamefont {{Di Porto}}}, \bibinfo {author} {\bibfnamefont
  {L.}~\bibnamefont {{Pozzetti}}}, \bibinfo {author} {\bibfnamefont
  {C.}~\bibnamefont {{Quercellini}}}, \bibinfo {author} {\bibfnamefont
  {M.}~\bibnamefont {{Radovich}}}, \bibinfo {author} {\bibfnamefont
  {A.}~\bibnamefont {{Rassat}}}, \bibinfo {author} {\bibfnamefont
  {N.}~\bibnamefont {{Roche}}}, \bibinfo {author} {\bibfnamefont
  {S.}~\bibnamefont {{Ronayette}}}, \bibinfo {author} {\bibfnamefont
  {E.}~\bibnamefont {{Rossetti}}}, \bibinfo {author} {\bibfnamefont
  {B.}~\bibnamefont {{Sartoris}}}, \bibinfo {author} {\bibfnamefont
  {P.}~\bibnamefont {{Schneider}}}, \bibinfo {author} {\bibfnamefont
  {E.}~\bibnamefont {{Semboloni}}}, \bibinfo {author} {\bibfnamefont
  {S.}~\bibnamefont {{Serjeant}}}, \bibinfo {author} {\bibfnamefont
  {F.}~\bibnamefont {{Simpson}}}, \bibinfo {author} {\bibfnamefont
  {C.}~\bibnamefont {{Skordis}}}, \bibinfo {author} {\bibfnamefont
  {G.}~\bibnamefont {{Smadja}}}, \bibinfo {author} {\bibfnamefont
  {S.}~\bibnamefont {{Smartt}}}, \bibinfo {author} {\bibfnamefont
  {P.}~\bibnamefont {{Spano}}}, \bibinfo {author} {\bibfnamefont
  {S.}~\bibnamefont {{Spiro}}}, \bibinfo {author} {\bibfnamefont
  {M.}~\bibnamefont {{Sullivan}}}, \bibinfo {author} {\bibfnamefont
  {A.}~\bibnamefont {{Tilquin}}}, \bibinfo {author} {\bibfnamefont
  {R.}~\bibnamefont {{Trotta}}}, \bibinfo {author} {\bibfnamefont
  {L.}~\bibnamefont {{Verde}}}, \bibinfo {author} {\bibfnamefont
  {Y.}~\bibnamefont {{Wang}}}, \bibinfo {author} {\bibfnamefont
  {G.}~\bibnamefont {{Williger}}}, \bibinfo {author} {\bibfnamefont
  {G.}~\bibnamefont {{Zhao}}}, \bibinfo {author} {\bibfnamefont
  {J.}~\bibnamefont {{Zoubian}}}, \ and\ \bibinfo {author} {\bibfnamefont
  {E.}~\bibnamefont {{Zucca}}},\ }\href@noop {} {\bibfield  {journal} {\bibinfo
   {journal} {arXiv e-prints}\ ,\ \bibinfo {eid} {arXiv:1110.3193}} (\bibinfo
  {year} {2011})},\ \Eprint {http://arxiv.org/abs/1110.3193} {arXiv:1110.3193
  [astro-ph.CO]} \BibitemShut {NoStop}%
\bibitem [{\citenamefont {{Amendola}}\ \emph {et~al.}(2018)\citenamefont
  {{Amendola}}, \citenamefont {{Appleby}}, \citenamefont {{Avgoustidis}},
  \citenamefont {{Bacon}}, \citenamefont {{Baker}}, \citenamefont {{Baldi}},
  \citenamefont {{Bartolo}}, \citenamefont {{Blanchard}}, \citenamefont
  {{Bonvin}}, \citenamefont {{Borgani}}, \citenamefont {{Branchini}},
  \citenamefont {{Burrage}}, \citenamefont {{Camera}}, \citenamefont
  {{Carbone}}, \citenamefont {{Casarini}}, \citenamefont {{Cropper}},
  \citenamefont {{de Rham}}, \citenamefont {{Dietrich}}, \citenamefont {{Di
  Porto}}, \citenamefont {{Durrer}}, \citenamefont {{Ealet}}, \citenamefont
  {{Ferreira}}, \citenamefont {{Finelli}}, \citenamefont
  {{Garc{\'\i}a-Bellido}}, \citenamefont {{Giannantonio}}, \citenamefont
  {{Guzzo}}, \citenamefont {{Heavens}}, \citenamefont {{Heisenberg}},
  \citenamefont {{Heymans}}, \citenamefont {{Hoekstra}}, \citenamefont
  {{Hollenstein}}, \citenamefont {{Holmes}}, \citenamefont {{Hwang}},
  \citenamefont {{Jahnke}}, \citenamefont {{Kitching}}, \citenamefont
  {{Koivisto}}, \citenamefont {{Kunz}}, \citenamefont {{La Vacca}},
  \citenamefont {{Linder}}, \citenamefont {{March}}, \citenamefont {{Marra}},
  \citenamefont {{Martins}}, \citenamefont {{Majerotto}}, \citenamefont
  {{Markovic}}, \citenamefont {{Marsh}}, \citenamefont {{Marulli}},
  \citenamefont {{Massey}}, \citenamefont {{Mellier}}, \citenamefont
  {{Montanari}}, \citenamefont {{Mota}}, \citenamefont {{Nunes}}, \citenamefont
  {{Percival}}, \citenamefont {{Pettorino}}, \citenamefont {{Porciani}},
  \citenamefont {{Quercellini}}, \citenamefont {{Read}}, \citenamefont
  {{Rinaldi}}, \citenamefont {{Sapone}}, \citenamefont {{Sawicki}},
  \citenamefont {{Scaramella}}, \citenamefont {{Skordis}}, \citenamefont
  {{Simpson}}, \citenamefont {{Taylor}}, \citenamefont {{Thomas}},
  \citenamefont {{Trotta}}, \citenamefont {{Verde}}, \citenamefont
  {{Vernizzi}}, \citenamefont {{Vollmer}}, \citenamefont {{Wang}},
  \citenamefont {{Weller}},\ and\ \citenamefont {{Zlosnik}}}]{euclid2}%
  \BibitemOpen
  \bibfield  {author} {\bibinfo {author} {\bibfnamefont {L.}~\bibnamefont
  {{Amendola}}}, \bibinfo {author} {\bibfnamefont {S.}~\bibnamefont
  {{Appleby}}}, \bibinfo {author} {\bibfnamefont {A.}~\bibnamefont
  {{Avgoustidis}}}, \bibinfo {author} {\bibfnamefont {D.}~\bibnamefont
  {{Bacon}}}, \bibinfo {author} {\bibfnamefont {T.}~\bibnamefont {{Baker}}},
  \bibinfo {author} {\bibfnamefont {M.}~\bibnamefont {{Baldi}}}, \bibinfo
  {author} {\bibfnamefont {N.}~\bibnamefont {{Bartolo}}}, \bibinfo {author}
  {\bibfnamefont {A.}~\bibnamefont {{Blanchard}}}, \bibinfo {author}
  {\bibfnamefont {C.}~\bibnamefont {{Bonvin}}}, \bibinfo {author}
  {\bibfnamefont {S.}~\bibnamefont {{Borgani}}}, \bibinfo {author}
  {\bibfnamefont {E.}~\bibnamefont {{Branchini}}}, \bibinfo {author}
  {\bibfnamefont {C.}~\bibnamefont {{Burrage}}}, \bibinfo {author}
  {\bibfnamefont {S.}~\bibnamefont {{Camera}}}, \bibinfo {author}
  {\bibfnamefont {C.}~\bibnamefont {{Carbone}}}, \bibinfo {author}
  {\bibfnamefont {L.}~\bibnamefont {{Casarini}}}, \bibinfo {author}
  {\bibfnamefont {M.}~\bibnamefont {{Cropper}}}, \bibinfo {author}
  {\bibfnamefont {C.}~\bibnamefont {{de Rham}}}, \bibinfo {author}
  {\bibfnamefont {J.~P.}\ \bibnamefont {{Dietrich}}}, \bibinfo {author}
  {\bibfnamefont {C.}~\bibnamefont {{Di Porto}}}, \bibinfo {author}
  {\bibfnamefont {R.}~\bibnamefont {{Durrer}}}, \bibinfo {author}
  {\bibfnamefont {A.}~\bibnamefont {{Ealet}}}, \bibinfo {author} {\bibfnamefont
  {P.~G.}\ \bibnamefont {{Ferreira}}}, \bibinfo {author} {\bibfnamefont
  {F.}~\bibnamefont {{Finelli}}}, \bibinfo {author} {\bibfnamefont
  {J.}~\bibnamefont {{Garc{\'\i}a-Bellido}}}, \bibinfo {author} {\bibfnamefont
  {T.}~\bibnamefont {{Giannantonio}}}, \bibinfo {author} {\bibfnamefont
  {L.}~\bibnamefont {{Guzzo}}}, \bibinfo {author} {\bibfnamefont
  {A.}~\bibnamefont {{Heavens}}}, \bibinfo {author} {\bibfnamefont
  {L.}~\bibnamefont {{Heisenberg}}}, \bibinfo {author} {\bibfnamefont
  {C.}~\bibnamefont {{Heymans}}}, \bibinfo {author} {\bibfnamefont
  {H.}~\bibnamefont {{Hoekstra}}}, \bibinfo {author} {\bibfnamefont
  {L.}~\bibnamefont {{Hollenstein}}}, \bibinfo {author} {\bibfnamefont
  {R.}~\bibnamefont {{Holmes}}}, \bibinfo {author} {\bibfnamefont
  {Z.}~\bibnamefont {{Hwang}}}, \bibinfo {author} {\bibfnamefont
  {K.}~\bibnamefont {{Jahnke}}}, \bibinfo {author} {\bibfnamefont {T.~D.}\
  \bibnamefont {{Kitching}}}, \bibinfo {author} {\bibfnamefont
  {T.}~\bibnamefont {{Koivisto}}}, \bibinfo {author} {\bibfnamefont
  {M.}~\bibnamefont {{Kunz}}}, \bibinfo {author} {\bibfnamefont
  {G.}~\bibnamefont {{La Vacca}}}, \bibinfo {author} {\bibfnamefont
  {E.}~\bibnamefont {{Linder}}}, \bibinfo {author} {\bibfnamefont
  {M.}~\bibnamefont {{March}}}, \bibinfo {author} {\bibfnamefont
  {V.}~\bibnamefont {{Marra}}}, \bibinfo {author} {\bibfnamefont
  {C.}~\bibnamefont {{Martins}}}, \bibinfo {author} {\bibfnamefont
  {E.}~\bibnamefont {{Majerotto}}}, \bibinfo {author} {\bibfnamefont
  {D.}~\bibnamefont {{Markovic}}}, \bibinfo {author} {\bibfnamefont
  {D.}~\bibnamefont {{Marsh}}}, \bibinfo {author} {\bibfnamefont
  {F.}~\bibnamefont {{Marulli}}}, \bibinfo {author} {\bibfnamefont
  {R.}~\bibnamefont {{Massey}}}, \bibinfo {author} {\bibfnamefont
  {Y.}~\bibnamefont {{Mellier}}}, \bibinfo {author} {\bibfnamefont
  {F.}~\bibnamefont {{Montanari}}}, \bibinfo {author} {\bibfnamefont {D.~F.}\
  \bibnamefont {{Mota}}}, \bibinfo {author} {\bibfnamefont {N.~J.}\
  \bibnamefont {{Nunes}}}, \bibinfo {author} {\bibfnamefont {W.}~\bibnamefont
  {{Percival}}}, \bibinfo {author} {\bibfnamefont {V.}~\bibnamefont
  {{Pettorino}}}, \bibinfo {author} {\bibfnamefont {C.}~\bibnamefont
  {{Porciani}}}, \bibinfo {author} {\bibfnamefont {C.}~\bibnamefont
  {{Quercellini}}}, \bibinfo {author} {\bibfnamefont {J.}~\bibnamefont
  {{Read}}}, \bibinfo {author} {\bibfnamefont {M.}~\bibnamefont {{Rinaldi}}},
  \bibinfo {author} {\bibfnamefont {D.}~\bibnamefont {{Sapone}}}, \bibinfo
  {author} {\bibfnamefont {I.}~\bibnamefont {{Sawicki}}}, \bibinfo {author}
  {\bibfnamefont {R.}~\bibnamefont {{Scaramella}}}, \bibinfo {author}
  {\bibfnamefont {C.}~\bibnamefont {{Skordis}}}, \bibinfo {author}
  {\bibfnamefont {F.}~\bibnamefont {{Simpson}}}, \bibinfo {author}
  {\bibfnamefont {A.}~\bibnamefont {{Taylor}}}, \bibinfo {author}
  {\bibfnamefont {S.}~\bibnamefont {{Thomas}}}, \bibinfo {author}
  {\bibfnamefont {R.}~\bibnamefont {{Trotta}}}, \bibinfo {author}
  {\bibfnamefont {L.}~\bibnamefont {{Verde}}}, \bibinfo {author} {\bibfnamefont
  {F.}~\bibnamefont {{Vernizzi}}}, \bibinfo {author} {\bibfnamefont
  {A.}~\bibnamefont {{Vollmer}}}, \bibinfo {author} {\bibfnamefont
  {Y.}~\bibnamefont {{Wang}}}, \bibinfo {author} {\bibfnamefont
  {J.}~\bibnamefont {{Weller}}}, \ and\ \bibinfo {author} {\bibfnamefont
  {T.}~\bibnamefont {{Zlosnik}}},\ }\href {\doibase 10.1007/s41114-017-0010-3}
  {\bibfield  {journal} {\bibinfo  {journal} {Living Reviews in Relativity}\
  }\textbf {\bibinfo {volume} {21}},\ \bibinfo {eid} {2} (\bibinfo {year}
  {2018})},\ \Eprint {http://arxiv.org/abs/1606.00180} {arXiv:1606.00180
  [astro-ph.CO]} \BibitemShut {NoStop}%
\bibitem [{\citenamefont {{Dey}}\ \emph {et~al.}(2019)\citenamefont {{Dey}},
  \citenamefont {{Schlegel}}, \citenamefont {{Lang}}, \citenamefont {{Blum}},
  \citenamefont {{Burleigh}}, \citenamefont {{Fan}}, \citenamefont {{Findlay}},
  \citenamefont {{Finkbeiner}}, \citenamefont {{Herrera}}, \citenamefont
  {{Juneau}}, \citenamefont {{Landriau}}, \citenamefont {{Levi}}, \citenamefont
  {{McGreer}}, \citenamefont {{Meisner}}, \citenamefont {{Myers}},
  \citenamefont {{Moustakas}}, \citenamefont {{Nugent}}, \citenamefont
  {{Patej}}, \citenamefont {{Schlafly}}, \citenamefont {{Walker}},
  \citenamefont {{Valdes}}, \citenamefont {{Weaver}}, \citenamefont
  {{Y{\`e}che}}, \citenamefont {{Zou}}, \citenamefont {{Zhou}}, \citenamefont
  {{Abareshi}}, \citenamefont {{Abbott}}, \citenamefont {{Abolfathi}},
  \citenamefont {{Aguilera}}, \citenamefont {{Alam}}, \citenamefont {{Allen}},
  \citenamefont {{Alvarez}}, \citenamefont {{Annis}}, \citenamefont
  {{Ansarinejad}}, \citenamefont {{Aubert}}, \citenamefont {{Beechert}},
  \citenamefont {{Bell}}, \citenamefont {{BenZvi}}, \citenamefont {{Beutler}},
  \citenamefont {{Bielby}}, \citenamefont {{Bolton}}, \citenamefont
  {{Brice{\~n}o}}, \citenamefont {{Buckley-Geer}}, \citenamefont {{Butler}},
  \citenamefont {{Calamida}}, \citenamefont {{Carlberg}}, \citenamefont
  {{Carter}}, \citenamefont {{Casas}}, \citenamefont {{Castander}},
  \citenamefont {{Choi}}, \citenamefont {{Comparat}}, \citenamefont
  {{Cukanovaite}}, \citenamefont {{Delubac}}, \citenamefont {{DeVries}},
  \citenamefont {{Dey}}, \citenamefont {{Dhungana}}, \citenamefont
  {{Dickinson}}, \citenamefont {{Ding}}, \citenamefont {{Donaldson}},
  \citenamefont {{Duan}}, \citenamefont {{Duckworth}}, \citenamefont
  {{Eftekharzadeh}}, \citenamefont {{Eisenstein}}, \citenamefont {{Etourneau}},
  \citenamefont {{Fagrelius}}, \citenamefont {{Farihi}}, \citenamefont
  {{Fitzpatrick}}, \citenamefont {{Font-Ribera}}, \citenamefont {{Fulmer}},
  \citenamefont {{G{\"a}nsicke}}, \citenamefont {{Gaztanaga}}, \citenamefont
  {{George}}, \citenamefont {{Gerdes}}, \citenamefont {{Gontcho}},
  \citenamefont {{Gorgoni}}, \citenamefont {{Green}}, \citenamefont {{Guy}},
  \citenamefont {{Harmer}}, \citenamefont {{Hernand ez}}, \citenamefont
  {{Honscheid}}, \citenamefont {{Huang}}, \citenamefont {{James}},
  \citenamefont {{Jannuzi}}, \citenamefont {{Jiang}}, \citenamefont {{Joyce}},
  \citenamefont {{Karcher}}, \citenamefont {{Karkar}}, \citenamefont {{Kehoe}},
  \citenamefont {{Kneib}}, \citenamefont {{Kueter-Young}}, \citenamefont
  {{Lan}}, \citenamefont {{Lauer}}, \citenamefont {{Le Guillou}}, \citenamefont
  {{Le Van Suu}}, \citenamefont {{Lee}}, \citenamefont {{Lesser}},
  \citenamefont {{Perreault Levasseur}}, \citenamefont {{Li}}, \citenamefont
  {{Mann}}, \citenamefont {{Marshall}}, \citenamefont
  {{Mart{\'\i}nez-V{\'a}zquez}}, \citenamefont {{Martini}}, \citenamefont {{du
  Mas des Bourboux}}, \citenamefont {{McManus}}, \citenamefont {{Meier}},
  \citenamefont {{M{\'e}nard}}, \citenamefont {{Metcalfe}}, \citenamefont
  {{Mu{\~n}oz-Guti{\'e}rrez}}, \citenamefont {{Najita}}, \citenamefont
  {{Napier}}, \citenamefont {{Narayan}}, \citenamefont {{Newman}},
  \citenamefont {{Nie}}, \citenamefont {{Nord}}, \citenamefont {{Norman}},
  \citenamefont {{Olsen}}, \citenamefont {{Paat}}, \citenamefont
  {{Palanque-Delabrouille}}, \citenamefont {{Peng}}, \citenamefont {{Poppett}},
  \citenamefont {{Poremba}}, \citenamefont {{Prakash}}, \citenamefont
  {{Rabinowitz}}, \citenamefont {{Raichoor}}, \citenamefont {{Rezaie}},
  \citenamefont {{Robertson}}, \citenamefont {{Roe}}, \citenamefont {{Ross}},
  \citenamefont {{Ross}}, \citenamefont {{Rudnick}}, \citenamefont
  {{Safonova}}, \citenamefont {{Saha}}, \citenamefont {{S{\'a}nchez}},
  \citenamefont {{Savary}}, \citenamefont {{Schweiker}}, \citenamefont
  {{Scott}}, \citenamefont {{Seo}}, \citenamefont {{Shan}}, \citenamefont
  {{Silva}}, \citenamefont {{Slepian}}, \citenamefont {{Soto}}, \citenamefont
  {{Sprayberry}}, \citenamefont {{Staten}}, \citenamefont {{Stillman}},
  \citenamefont {{Stupak}}, \citenamefont {{Summers}}, \citenamefont {{Sien
  Tie}}, \citenamefont {{Tirado}}, \citenamefont {{Vargas-Maga{\~n}a}},
  \citenamefont {{Vivas}}, \citenamefont {{Wechsler}}, \citenamefont
  {{Williams}}, \citenamefont {{Yang}}, \citenamefont {{Yang}}, \citenamefont
  {{Yapici}}, \citenamefont {{Zaritsky}}, \citenamefont {{Zenteno}},
  \citenamefont {{Zhang}}, \citenamefont {{Zhang}}, \citenamefont {{Zhou}},\
  and\ \citenamefont {{Zhou}}}]{desilegacy}%
  \BibitemOpen
  \bibfield  {author} {\bibinfo {author} {\bibfnamefont {A.}~\bibnamefont
  {{Dey}}}, \bibinfo {author} {\bibfnamefont {D.~J.}\ \bibnamefont
  {{Schlegel}}}, \bibinfo {author} {\bibfnamefont {D.}~\bibnamefont {{Lang}}},
  \bibinfo {author} {\bibfnamefont {R.}~\bibnamefont {{Blum}}}, \bibinfo
  {author} {\bibfnamefont {K.}~\bibnamefont {{Burleigh}}}, \bibinfo {author}
  {\bibfnamefont {X.}~\bibnamefont {{Fan}}}, \bibinfo {author} {\bibfnamefont
  {J.~R.}\ \bibnamefont {{Findlay}}}, \bibinfo {author} {\bibfnamefont
  {D.}~\bibnamefont {{Finkbeiner}}}, \bibinfo {author} {\bibfnamefont
  {D.}~\bibnamefont {{Herrera}}}, \bibinfo {author} {\bibfnamefont
  {S.}~\bibnamefont {{Juneau}}}, \bibinfo {author} {\bibfnamefont
  {M.}~\bibnamefont {{Landriau}}}, \bibinfo {author} {\bibfnamefont
  {M.}~\bibnamefont {{Levi}}}, \bibinfo {author} {\bibfnamefont
  {I.}~\bibnamefont {{McGreer}}}, \bibinfo {author} {\bibfnamefont
  {A.}~\bibnamefont {{Meisner}}}, \bibinfo {author} {\bibfnamefont {A.~D.}\
  \bibnamefont {{Myers}}}, \bibinfo {author} {\bibfnamefont {J.}~\bibnamefont
  {{Moustakas}}}, \bibinfo {author} {\bibfnamefont {P.}~\bibnamefont
  {{Nugent}}}, \bibinfo {author} {\bibfnamefont {A.}~\bibnamefont {{Patej}}},
  \bibinfo {author} {\bibfnamefont {E.~F.}\ \bibnamefont {{Schlafly}}},
  \bibinfo {author} {\bibfnamefont {A.~R.}\ \bibnamefont {{Walker}}}, \bibinfo
  {author} {\bibfnamefont {F.}~\bibnamefont {{Valdes}}}, \bibinfo {author}
  {\bibfnamefont {B.~A.}\ \bibnamefont {{Weaver}}}, \bibinfo {author}
  {\bibfnamefont {C.}~\bibnamefont {{Y{\`e}che}}}, \bibinfo {author}
  {\bibfnamefont {H.}~\bibnamefont {{Zou}}}, \bibinfo {author} {\bibfnamefont
  {X.}~\bibnamefont {{Zhou}}}, \bibinfo {author} {\bibfnamefont
  {B.}~\bibnamefont {{Abareshi}}}, \bibinfo {author} {\bibfnamefont {T.~M.~C.}\
  \bibnamefont {{Abbott}}}, \bibinfo {author} {\bibfnamefont {B.}~\bibnamefont
  {{Abolfathi}}}, \bibinfo {author} {\bibfnamefont {C.}~\bibnamefont
  {{Aguilera}}}, \bibinfo {author} {\bibfnamefont {S.}~\bibnamefont {{Alam}}},
  \bibinfo {author} {\bibfnamefont {L.}~\bibnamefont {{Allen}}}, \bibinfo
  {author} {\bibfnamefont {A.}~\bibnamefont {{Alvarez}}}, \bibinfo {author}
  {\bibfnamefont {J.}~\bibnamefont {{Annis}}}, \bibinfo {author} {\bibfnamefont
  {B.}~\bibnamefont {{Ansarinejad}}}, \bibinfo {author} {\bibfnamefont
  {M.}~\bibnamefont {{Aubert}}}, \bibinfo {author} {\bibfnamefont
  {J.}~\bibnamefont {{Beechert}}}, \bibinfo {author} {\bibfnamefont {E.~F.}\
  \bibnamefont {{Bell}}}, \bibinfo {author} {\bibfnamefont {S.~Y.}\
  \bibnamefont {{BenZvi}}}, \bibinfo {author} {\bibfnamefont {F.}~\bibnamefont
  {{Beutler}}}, \bibinfo {author} {\bibfnamefont {R.~M.}\ \bibnamefont
  {{Bielby}}}, \bibinfo {author} {\bibfnamefont {A.~S.}\ \bibnamefont
  {{Bolton}}}, \bibinfo {author} {\bibfnamefont {C.}~\bibnamefont
  {{Brice{\~n}o}}}, \bibinfo {author} {\bibfnamefont {E.~J.}\ \bibnamefont
  {{Buckley-Geer}}}, \bibinfo {author} {\bibfnamefont {K.}~\bibnamefont
  {{Butler}}}, \bibinfo {author} {\bibfnamefont {A.}~\bibnamefont
  {{Calamida}}}, \bibinfo {author} {\bibfnamefont {R.~G.}\ \bibnamefont
  {{Carlberg}}}, \bibinfo {author} {\bibfnamefont {P.}~\bibnamefont
  {{Carter}}}, \bibinfo {author} {\bibfnamefont {R.}~\bibnamefont {{Casas}}},
  \bibinfo {author} {\bibfnamefont {F.~J.}\ \bibnamefont {{Castander}}},
  \bibinfo {author} {\bibfnamefont {Y.}~\bibnamefont {{Choi}}}, \bibinfo
  {author} {\bibfnamefont {J.}~\bibnamefont {{Comparat}}}, \bibinfo {author}
  {\bibfnamefont {E.}~\bibnamefont {{Cukanovaite}}}, \bibinfo {author}
  {\bibfnamefont {T.}~\bibnamefont {{Delubac}}}, \bibinfo {author}
  {\bibfnamefont {K.}~\bibnamefont {{DeVries}}}, \bibinfo {author}
  {\bibfnamefont {S.}~\bibnamefont {{Dey}}}, \bibinfo {author} {\bibfnamefont
  {G.}~\bibnamefont {{Dhungana}}}, \bibinfo {author} {\bibfnamefont
  {M.}~\bibnamefont {{Dickinson}}}, \bibinfo {author} {\bibfnamefont
  {Z.}~\bibnamefont {{Ding}}}, \bibinfo {author} {\bibfnamefont {J.~B.}\
  \bibnamefont {{Donaldson}}}, \bibinfo {author} {\bibfnamefont
  {Y.}~\bibnamefont {{Duan}}}, \bibinfo {author} {\bibfnamefont {C.~J.}\
  \bibnamefont {{Duckworth}}}, \bibinfo {author} {\bibfnamefont
  {S.}~\bibnamefont {{Eftekharzadeh}}}, \bibinfo {author} {\bibfnamefont
  {D.~J.}\ \bibnamefont {{Eisenstein}}}, \bibinfo {author} {\bibfnamefont
  {T.}~\bibnamefont {{Etourneau}}}, \bibinfo {author} {\bibfnamefont {P.~A.}\
  \bibnamefont {{Fagrelius}}}, \bibinfo {author} {\bibfnamefont
  {J.}~\bibnamefont {{Farihi}}}, \bibinfo {author} {\bibfnamefont
  {M.}~\bibnamefont {{Fitzpatrick}}}, \bibinfo {author} {\bibfnamefont
  {A.}~\bibnamefont {{Font-Ribera}}}, \bibinfo {author} {\bibfnamefont
  {L.}~\bibnamefont {{Fulmer}}}, \bibinfo {author} {\bibfnamefont {B.~T.}\
  \bibnamefont {{G{\"a}nsicke}}}, \bibinfo {author} {\bibfnamefont
  {E.}~\bibnamefont {{Gaztanaga}}}, \bibinfo {author} {\bibfnamefont
  {K.}~\bibnamefont {{George}}}, \bibinfo {author} {\bibfnamefont {D.~W.}\
  \bibnamefont {{Gerdes}}}, \bibinfo {author} {\bibfnamefont {S.~G.~A.}\
  \bibnamefont {{Gontcho}}}, \bibinfo {author} {\bibfnamefont {C.}~\bibnamefont
  {{Gorgoni}}}, \bibinfo {author} {\bibfnamefont {G.}~\bibnamefont {{Green}}},
  \bibinfo {author} {\bibfnamefont {J.}~\bibnamefont {{Guy}}}, \bibinfo
  {author} {\bibfnamefont {D.}~\bibnamefont {{Harmer}}}, \bibinfo {author}
  {\bibfnamefont {M.}~\bibnamefont {{Hernand ez}}}, \bibinfo {author}
  {\bibfnamefont {K.}~\bibnamefont {{Honscheid}}}, \bibinfo {author}
  {\bibfnamefont {L.~W.}\ \bibnamefont {{Huang}}}, \bibinfo {author}
  {\bibfnamefont {D.~J.}\ \bibnamefont {{James}}}, \bibinfo {author}
  {\bibfnamefont {B.~T.}\ \bibnamefont {{Jannuzi}}}, \bibinfo {author}
  {\bibfnamefont {L.}~\bibnamefont {{Jiang}}}, \bibinfo {author} {\bibfnamefont
  {R.}~\bibnamefont {{Joyce}}}, \bibinfo {author} {\bibfnamefont
  {A.}~\bibnamefont {{Karcher}}}, \bibinfo {author} {\bibfnamefont
  {S.}~\bibnamefont {{Karkar}}}, \bibinfo {author} {\bibfnamefont
  {R.}~\bibnamefont {{Kehoe}}}, \bibinfo {author} {\bibfnamefont {J.-P.}\
  \bibnamefont {{Kneib}}}, \bibinfo {author} {\bibfnamefont {A.}~\bibnamefont
  {{Kueter-Young}}}, \bibinfo {author} {\bibfnamefont {T.-W.}\ \bibnamefont
  {{Lan}}}, \bibinfo {author} {\bibfnamefont {T.~R.}\ \bibnamefont {{Lauer}}},
  \bibinfo {author} {\bibfnamefont {L.}~\bibnamefont {{Le Guillou}}}, \bibinfo
  {author} {\bibfnamefont {A.}~\bibnamefont {{Le Van Suu}}}, \bibinfo {author}
  {\bibfnamefont {J.~H.}\ \bibnamefont {{Lee}}}, \bibinfo {author}
  {\bibfnamefont {M.}~\bibnamefont {{Lesser}}}, \bibinfo {author}
  {\bibfnamefont {L.}~\bibnamefont {{Perreault Levasseur}}}, \bibinfo {author}
  {\bibfnamefont {T.~S.}\ \bibnamefont {{Li}}}, \bibinfo {author}
  {\bibfnamefont {J.~L.}\ \bibnamefont {{Mann}}}, \bibinfo {author}
  {\bibfnamefont {R.}~\bibnamefont {{Marshall}}}, \bibinfo {author}
  {\bibfnamefont {C.~E.}\ \bibnamefont {{Mart{\'\i}nez-V{\'a}zquez}}}, \bibinfo
  {author} {\bibfnamefont {P.}~\bibnamefont {{Martini}}}, \bibinfo {author}
  {\bibfnamefont {H.}~\bibnamefont {{du Mas des Bourboux}}}, \bibinfo {author}
  {\bibfnamefont {S.}~\bibnamefont {{McManus}}}, \bibinfo {author}
  {\bibfnamefont {T.~G.}\ \bibnamefont {{Meier}}}, \bibinfo {author}
  {\bibfnamefont {B.}~\bibnamefont {{M{\'e}nard}}}, \bibinfo {author}
  {\bibfnamefont {N.}~\bibnamefont {{Metcalfe}}}, \bibinfo {author}
  {\bibfnamefont {A.}~\bibnamefont {{Mu{\~n}oz-Guti{\'e}rrez}}}, \bibinfo
  {author} {\bibfnamefont {J.}~\bibnamefont {{Najita}}}, \bibinfo {author}
  {\bibfnamefont {K.}~\bibnamefont {{Napier}}}, \bibinfo {author}
  {\bibfnamefont {G.}~\bibnamefont {{Narayan}}}, \bibinfo {author}
  {\bibfnamefont {J.~A.}\ \bibnamefont {{Newman}}}, \bibinfo {author}
  {\bibfnamefont {J.}~\bibnamefont {{Nie}}}, \bibinfo {author} {\bibfnamefont
  {B.}~\bibnamefont {{Nord}}}, \bibinfo {author} {\bibfnamefont {D.~J.}\
  \bibnamefont {{Norman}}}, \bibinfo {author} {\bibfnamefont {K.~A.~G.}\
  \bibnamefont {{Olsen}}}, \bibinfo {author} {\bibfnamefont {A.}~\bibnamefont
  {{Paat}}}, \bibinfo {author} {\bibfnamefont {N.}~\bibnamefont
  {{Palanque-Delabrouille}}}, \bibinfo {author} {\bibfnamefont
  {X.}~\bibnamefont {{Peng}}}, \bibinfo {author} {\bibfnamefont {C.~L.}\
  \bibnamefont {{Poppett}}}, \bibinfo {author} {\bibfnamefont {M.~R.}\
  \bibnamefont {{Poremba}}}, \bibinfo {author} {\bibfnamefont {A.}~\bibnamefont
  {{Prakash}}}, \bibinfo {author} {\bibfnamefont {D.}~\bibnamefont
  {{Rabinowitz}}}, \bibinfo {author} {\bibfnamefont {A.}~\bibnamefont
  {{Raichoor}}}, \bibinfo {author} {\bibfnamefont {M.}~\bibnamefont
  {{Rezaie}}}, \bibinfo {author} {\bibfnamefont {A.~N.}\ \bibnamefont
  {{Robertson}}}, \bibinfo {author} {\bibfnamefont {N.~A.}\ \bibnamefont
  {{Roe}}}, \bibinfo {author} {\bibfnamefont {A.~J.}\ \bibnamefont {{Ross}}},
  \bibinfo {author} {\bibfnamefont {N.~P.}\ \bibnamefont {{Ross}}}, \bibinfo
  {author} {\bibfnamefont {G.}~\bibnamefont {{Rudnick}}}, \bibinfo {author}
  {\bibfnamefont {S.}~\bibnamefont {{Safonova}}}, \bibinfo {author}
  {\bibfnamefont {A.}~\bibnamefont {{Saha}}}, \bibinfo {author} {\bibfnamefont
  {F.~J.}\ \bibnamefont {{S{\'a}nchez}}}, \bibinfo {author} {\bibfnamefont
  {E.}~\bibnamefont {{Savary}}}, \bibinfo {author} {\bibfnamefont
  {H.}~\bibnamefont {{Schweiker}}}, \bibinfo {author} {\bibfnamefont
  {A.}~\bibnamefont {{Scott}}}, \bibinfo {author} {\bibfnamefont {H.-J.}\
  \bibnamefont {{Seo}}}, \bibinfo {author} {\bibfnamefont {H.}~\bibnamefont
  {{Shan}}}, \bibinfo {author} {\bibfnamefont {D.~R.}\ \bibnamefont {{Silva}}},
  \bibinfo {author} {\bibfnamefont {Z.}~\bibnamefont {{Slepian}}}, \bibinfo
  {author} {\bibfnamefont {C.}~\bibnamefont {{Soto}}}, \bibinfo {author}
  {\bibfnamefont {D.}~\bibnamefont {{Sprayberry}}}, \bibinfo {author}
  {\bibfnamefont {R.}~\bibnamefont {{Staten}}}, \bibinfo {author}
  {\bibfnamefont {C.~M.}\ \bibnamefont {{Stillman}}}, \bibinfo {author}
  {\bibfnamefont {R.~J.}\ \bibnamefont {{Stupak}}}, \bibinfo {author}
  {\bibfnamefont {D.~L.}\ \bibnamefont {{Summers}}}, \bibinfo {author}
  {\bibfnamefont {S.}~\bibnamefont {{Sien Tie}}}, \bibinfo {author}
  {\bibfnamefont {H.}~\bibnamefont {{Tirado}}}, \bibinfo {author}
  {\bibfnamefont {M.}~\bibnamefont {{Vargas-Maga{\~n}a}}}, \bibinfo {author}
  {\bibfnamefont {A.~K.}\ \bibnamefont {{Vivas}}}, \bibinfo {author}
  {\bibfnamefont {R.~H.}\ \bibnamefont {{Wechsler}}}, \bibinfo {author}
  {\bibfnamefont {D.}~\bibnamefont {{Williams}}}, \bibinfo {author}
  {\bibfnamefont {J.}~\bibnamefont {{Yang}}}, \bibinfo {author} {\bibfnamefont
  {Q.}~\bibnamefont {{Yang}}}, \bibinfo {author} {\bibfnamefont
  {T.}~\bibnamefont {{Yapici}}}, \bibinfo {author} {\bibfnamefont
  {D.}~\bibnamefont {{Zaritsky}}}, \bibinfo {author} {\bibfnamefont
  {A.}~\bibnamefont {{Zenteno}}}, \bibinfo {author} {\bibfnamefont
  {K.}~\bibnamefont {{Zhang}}}, \bibinfo {author} {\bibfnamefont
  {T.}~\bibnamefont {{Zhang}}}, \bibinfo {author} {\bibfnamefont
  {R.}~\bibnamefont {{Zhou}}}, \ and\ \bibinfo {author} {\bibfnamefont
  {Z.}~\bibnamefont {{Zhou}}},\ }\href {\doibase 10.3847/1538-3881/ab089d}
  {\bibfield  {journal} {\bibinfo  {journal} {\aj}\ }\textbf {\bibinfo {volume}
  {157}},\ \bibinfo {eid} {168} (\bibinfo {year} {2019})},\ \Eprint
  {http://arxiv.org/abs/1804.08657} {arXiv:1804.08657 [astro-ph.IM]}
  \BibitemShut {NoStop}%
\bibitem [{\citenamefont {{Costa}}\ \emph {et~al.}(2019)\citenamefont
  {{Costa}}, \citenamefont {{Marcondes}}, \citenamefont {{Landim}},
  \citenamefont {{Abdalla}}, \citenamefont {{Abramo}}, \citenamefont
  {{Xavier}}, \citenamefont {{Orsi}}, \citenamefont {{Devi}}, \citenamefont
  {{Cenarro}}, \citenamefont {{Crist{\'o}bal-Hornillos}}, \citenamefont
  {{Dupke}}, \citenamefont {{Ederoclite}}, \citenamefont {{Mar{\'\i}n-Franch}},
  \citenamefont {{Oliveira}}, \citenamefont {{V{\'a}zquez Rami{\'o}}},
  \citenamefont {{Taylor}},\ and\ \citenamefont {{Varela}}}]{jpas1}%
  \BibitemOpen
  \bibfield  {author} {\bibinfo {author} {\bibfnamefont {A.~A.}\ \bibnamefont
  {{Costa}}}, \bibinfo {author} {\bibfnamefont {R.~J.~F.}\ \bibnamefont
  {{Marcondes}}}, \bibinfo {author} {\bibfnamefont {R.~G.}\ \bibnamefont
  {{Landim}}}, \bibinfo {author} {\bibfnamefont {E.}~\bibnamefont {{Abdalla}}},
  \bibinfo {author} {\bibfnamefont {L.~R.}\ \bibnamefont {{Abramo}}}, \bibinfo
  {author} {\bibfnamefont {H.~S.}\ \bibnamefont {{Xavier}}}, \bibinfo {author}
  {\bibfnamefont {A.~A.}\ \bibnamefont {{Orsi}}}, \bibinfo {author}
  {\bibfnamefont {N.~C.}\ \bibnamefont {{Devi}}}, \bibinfo {author}
  {\bibfnamefont {A.~J.}\ \bibnamefont {{Cenarro}}}, \bibinfo {author}
  {\bibfnamefont {D.}~\bibnamefont {{Crist{\'o}bal-Hornillos}}}, \bibinfo
  {author} {\bibfnamefont {R.~A.}\ \bibnamefont {{Dupke}}}, \bibinfo {author}
  {\bibfnamefont {A.}~\bibnamefont {{Ederoclite}}}, \bibinfo {author}
  {\bibfnamefont {A.}~\bibnamefont {{Mar{\'\i}n-Franch}}}, \bibinfo {author}
  {\bibfnamefont {C.~M.}\ \bibnamefont {{Oliveira}}}, \bibinfo {author}
  {\bibfnamefont {H.}~\bibnamefont {{V{\'a}zquez Rami{\'o}}}}, \bibinfo
  {author} {\bibfnamefont {K.}~\bibnamefont {{Taylor}}}, \ and\ \bibinfo
  {author} {\bibfnamefont {J.}~\bibnamefont {{Varela}}},\ }\href {\doibase
  10.1093/mnras/stz1675} {\bibfield  {journal} {\bibinfo  {journal} {\mnras}\
  }\textbf {\bibinfo {volume} {488}},\ \bibinfo {pages} {78} (\bibinfo {year}
  {2019})},\ \Eprint {http://arxiv.org/abs/1901.02540} {arXiv:1901.02540
  [astro-ph.CO]} \BibitemShut {NoStop}%
\bibitem [{\citenamefont {{Aparicio Resco}}\ \emph {et~al.}(2020)\citenamefont
  {{Aparicio Resco}}, \citenamefont {{Maroto}}, \citenamefont {{Alcaniz}},
  \citenamefont {{Abramo}}, \citenamefont {{Hern{\'a}ndez-Monteagudo}},
  \citenamefont {{Ben{\'\i}tez}}, \citenamefont {{Carneiro}}, \citenamefont
  {{Cenarro}}, \citenamefont {{Crist{\'o}bal-Hornillos}}, \citenamefont
  {{Dupke}}, \citenamefont {{Ederoclite}}, \citenamefont {{L{\'o}pez-Sanjuan}},
  \citenamefont {{Mar{\'\i}n-Franch}}, \citenamefont {{Moles}}, \citenamefont
  {{Oliveira}}, \citenamefont {{Sodr{\'e}}}, \citenamefont {{Taylor}},
  \citenamefont {{Varela}},\ and\ \citenamefont {{V{\'a}zquez
  Rami{\'o}}}}]{jpas2}%
  \BibitemOpen
  \bibfield  {author} {\bibinfo {author} {\bibfnamefont {M.}~\bibnamefont
  {{Aparicio Resco}}}, \bibinfo {author} {\bibfnamefont {A.~L.}\ \bibnamefont
  {{Maroto}}}, \bibinfo {author} {\bibfnamefont {J.~S.}\ \bibnamefont
  {{Alcaniz}}}, \bibinfo {author} {\bibfnamefont {L.~R.}\ \bibnamefont
  {{Abramo}}}, \bibinfo {author} {\bibfnamefont {C.}~\bibnamefont
  {{Hern{\'a}ndez-Monteagudo}}}, \bibinfo {author} {\bibfnamefont
  {N.}~\bibnamefont {{Ben{\'\i}tez}}}, \bibinfo {author} {\bibfnamefont
  {S.}~\bibnamefont {{Carneiro}}}, \bibinfo {author} {\bibfnamefont {A.~J.}\
  \bibnamefont {{Cenarro}}}, \bibinfo {author} {\bibfnamefont {D.}~\bibnamefont
  {{Crist{\'o}bal-Hornillos}}}, \bibinfo {author} {\bibfnamefont {R.~A.}\
  \bibnamefont {{Dupke}}}, \bibinfo {author} {\bibfnamefont {A.}~\bibnamefont
  {{Ederoclite}}}, \bibinfo {author} {\bibfnamefont {C.}~\bibnamefont
  {{L{\'o}pez-Sanjuan}}}, \bibinfo {author} {\bibfnamefont {A.}~\bibnamefont
  {{Mar{\'\i}n-Franch}}}, \bibinfo {author} {\bibfnamefont {M.}~\bibnamefont
  {{Moles}}}, \bibinfo {author} {\bibfnamefont {C.~M.}\ \bibnamefont
  {{Oliveira}}}, \bibinfo {author} {\bibfnamefont {J.}~\bibnamefont
  {{Sodr{\'e}}}, \bibfnamefont {L.}}, \bibinfo {author} {\bibfnamefont
  {K.}~\bibnamefont {{Taylor}}}, \bibinfo {author} {\bibfnamefont
  {J.}~\bibnamefont {{Varela}}}, \ and\ \bibinfo {author} {\bibfnamefont
  {H.}~\bibnamefont {{V{\'a}zquez Rami{\'o}}}},\ }\href {\doibase
  10.1093/mnras/staa367} {\bibfield  {journal} {\bibinfo  {journal} {\mnras}\
  }\textbf {\bibinfo {volume} {493}},\ \bibinfo {pages} {3616} (\bibinfo {year}
  {2020})},\ \Eprint {http://arxiv.org/abs/1910.02694} {arXiv:1910.02694
  [astro-ph.CO]} \BibitemShut {NoStop}%
\bibitem [{\citenamefont {Caldwell}\ and\ \citenamefont
  {Linder}(2005)}]{Caldwell:2005tm}%
  \BibitemOpen
  \bibfield  {author} {\bibinfo {author} {\bibfnamefont {R.~R.}\ \bibnamefont
  {Caldwell}}\ and\ \bibinfo {author} {\bibfnamefont {E.~V.}\ \bibnamefont
  {Linder}},\ }\href {\doibase 10.1103/PhysRevLett.95.141301} {\bibfield
  {journal} {\bibinfo  {journal} {Phys. Rev. Lett.}\ }\textbf {\bibinfo
  {volume} {95}},\ \bibinfo {pages} {141301} (\bibinfo {year} {2005})},\
  \Eprint {http://arxiv.org/abs/astro-ph/0505494} {arXiv:astro-ph/0505494
  [astro-ph]} \BibitemShut {NoStop}%
\bibitem [{\citenamefont {Linder}(2006)}]{Linder:2006sv}%
  \BibitemOpen
  \bibfield  {author} {\bibinfo {author} {\bibfnamefont {E.~V.}\ \bibnamefont
  {Linder}},\ }\href {\doibase 10.1103/PhysRevD.73.063010} {\bibfield
  {journal} {\bibinfo  {journal} {Phys. Rev.}\ }\textbf {\bibinfo {volume}
  {D73}},\ \bibinfo {pages} {063010} (\bibinfo {year} {2006})},\ \Eprint
  {http://arxiv.org/abs/astro-ph/0601052} {arXiv:astro-ph/0601052 [astro-ph]}
  \BibitemShut {NoStop}%
\bibitem [{\citenamefont {Tsujikawa}(2011)}]{Tsujikawa:2010sc}%
  \BibitemOpen
  \bibfield  {author} {\bibinfo {author} {\bibfnamefont {S.}~\bibnamefont
  {Tsujikawa}},\ }\href {\doibase 10.1007/978-90-481-8685-3_8} {\ \textbf
  {\bibinfo {volume} {370}},\ \bibinfo {pages} {331} (\bibinfo {year}
  {2011})},\ \Eprint {http://arxiv.org/abs/1004.1493} {arXiv:1004.1493
  [astro-ph.CO]} \BibitemShut {NoStop}%
\bibitem [{\citenamefont {Scherrer}\ and\ \citenamefont
  {Sen}(2008)}]{Scherrer:2007pu}%
  \BibitemOpen
  \bibfield  {author} {\bibinfo {author} {\bibfnamefont {R.~J.}\ \bibnamefont
  {Scherrer}}\ and\ \bibinfo {author} {\bibfnamefont {A.~A.}\ \bibnamefont
  {Sen}},\ }\href {\doibase 10.1103/PhysRevD.77.083515} {\bibfield  {journal}
  {\bibinfo  {journal} {Phys. Rev.}\ }\textbf {\bibinfo {volume} {D77}},\
  \bibinfo {pages} {083515} (\bibinfo {year} {2008})},\ \Eprint
  {http://arxiv.org/abs/0712.3450} {arXiv:0712.3450 [astro-ph]} \BibitemShut
  {NoStop}%
\bibitem [{\citenamefont {Dinda}\ and\ \citenamefont
  {Sen}(2018)}]{Dinda:2016ibo}%
  \BibitemOpen
  \bibfield  {author} {\bibinfo {author} {\bibfnamefont {B.~R.}\ \bibnamefont
  {Dinda}}\ and\ \bibinfo {author} {\bibfnamefont {A.~A.}\ \bibnamefont
  {Sen}},\ }\href {\doibase 10.1103/PhysRevD.97.083506} {\bibfield  {journal}
  {\bibinfo  {journal} {Phys. Rev.}\ }\textbf {\bibinfo {volume} {D97}},\
  \bibinfo {pages} {083506} (\bibinfo {year} {2018})},\ \Eprint
  {http://arxiv.org/abs/1607.05123} {arXiv:1607.05123 [astro-ph.CO]}
  \BibitemShut {NoStop}%
\bibitem [{\citenamefont {Dinda}(2018)}]{Dinda:2018eyt}%
  \BibitemOpen
  \bibfield  {author} {\bibinfo {author} {\bibfnamefont {B.~R.}\ \bibnamefont
  {Dinda}},\ }\href {\doibase 10.1088/1475-7516/2018/06/017} {\bibfield
  {journal} {\bibinfo  {journal} {JCAP}\ }\textbf {\bibinfo {volume} {06}},\
  \bibinfo {pages} {017} (\bibinfo {year} {2018})},\ \Eprint
  {http://arxiv.org/abs/1801.01741} {arXiv:1801.01741 [astro-ph.CO]}
  \BibitemShut {NoStop}%
\bibitem [{\citenamefont {Costa}\ \emph {et~al.}(2017)\citenamefont {Costa},
  \citenamefont {Xu}, \citenamefont {Wang},\ and\ \citenamefont
  {Abdalla}}]{Costa:2016tpb}%
  \BibitemOpen
  \bibfield  {author} {\bibinfo {author} {\bibfnamefont {A.~A.}\ \bibnamefont
  {Costa}}, \bibinfo {author} {\bibfnamefont {X.-D.}\ \bibnamefont {Xu}},
  \bibinfo {author} {\bibfnamefont {B.}~\bibnamefont {Wang}}, \ and\ \bibinfo
  {author} {\bibfnamefont {E.}~\bibnamefont {Abdalla}},\ }\href {\doibase
  10.1088/1475-7516/2017/01/028} {\bibfield  {journal} {\bibinfo  {journal}
  {JCAP}\ }\textbf {\bibinfo {volume} {01}},\ \bibinfo {pages} {028} (\bibinfo
  {year} {2017})},\ \Eprint {http://arxiv.org/abs/1605.04138} {arXiv:1605.04138
  [astro-ph.CO]} \BibitemShut {NoStop}%
\bibitem [{\citenamefont {{Crocce}}\ \emph {et~al.}(2006)\citenamefont
  {{Crocce}}, \citenamefont {{Pueblas}},\ and\ \citenamefont
  {{Scoccimarro}}}]{2lptic}%
  \BibitemOpen
  \bibfield  {author} {\bibinfo {author} {\bibfnamefont {M.}~\bibnamefont
  {{Crocce}}}, \bibinfo {author} {\bibfnamefont {S.}~\bibnamefont {{Pueblas}}},
  \ and\ \bibinfo {author} {\bibfnamefont {R.}~\bibnamefont {{Scoccimarro}}},\
  }\href {\doibase 10.1111/j.1365-2966.2006.11040.x} {\bibfield  {journal}
  {\bibinfo  {journal} {\mnras}\ }\textbf {\bibinfo {volume} {373}},\ \bibinfo
  {pages} {369} (\bibinfo {year} {2006})},\ \Eprint
  {http://arxiv.org/abs/astro-ph/0606505} {astro-ph/0606505} \BibitemShut
  {NoStop}%
\bibitem [{\citenamefont {{Liao}}(2018)}]{liao2018ccvt}%
  \BibitemOpen
  \bibfield  {author} {\bibinfo {author} {\bibfnamefont {S.}~\bibnamefont
  {{Liao}}},\ }\href {\doibase 10.1093/mnras/sty2523} {\bibfield  {journal}
  {\bibinfo  {journal} {\mnras}\ }\textbf {\bibinfo {volume} {481}},\ \bibinfo
  {pages} {3750} (\bibinfo {year} {2018})},\ \Eprint
  {http://arxiv.org/abs/1807.03574} {arXiv:1807.03574} \BibitemShut {NoStop}%
\bibitem [{\citenamefont {{Massara}}\ \emph {et~al.}(2020)\citenamefont
  {{Massara}}, \citenamefont {{Villaescusa-Navarro}}, \citenamefont {{Ho}},
  \citenamefont {{Dalal}},\ and\ \citenamefont {{Spergel}}}]{massara2020}%
  \BibitemOpen
  \bibfield  {author} {\bibinfo {author} {\bibfnamefont {E.}~\bibnamefont
  {{Massara}}}, \bibinfo {author} {\bibfnamefont {F.}~\bibnamefont
  {{Villaescusa-Navarro}}}, \bibinfo {author} {\bibfnamefont {S.}~\bibnamefont
  {{Ho}}}, \bibinfo {author} {\bibfnamefont {N.}~\bibnamefont {{Dalal}}}, \
  and\ \bibinfo {author} {\bibfnamefont {D.~N.}\ \bibnamefont {{Spergel}}},\
  }\href@noop {} {\bibfield  {journal} {\bibinfo  {journal} {arXiv e-prints}\
  ,\ \bibinfo {eid} {arXiv:2001.11024}} (\bibinfo {year} {2020})},\ \Eprint
  {http://arxiv.org/abs/2001.11024} {arXiv:2001.11024 [astro-ph.CO]}
  \BibitemShut {NoStop}%
\bibitem [{\citenamefont {{Villaescusa-Navarro}}\ \emph
  {et~al.}(2018)\citenamefont {{Villaescusa-Navarro}}, \citenamefont {{Genel}},
  \citenamefont {{Castorina}}, \citenamefont {{Obuljen}}, \citenamefont
  {{Spergel}}, \citenamefont {{Hernquist}}, \citenamefont {{Nelson}},
  \citenamefont {{Carucci}}, \citenamefont {{Pillepich}}, \citenamefont
  {{Marinacci}}, \citenamefont {{Diemer}}, \citenamefont {{Vogelsberger}},
  \citenamefont {{Weinberger}},\ and\ \citenamefont {{Pakmor}}}]{pylians}%
  \BibitemOpen
  \bibfield  {author} {\bibinfo {author} {\bibfnamefont {F.}~\bibnamefont
  {{Villaescusa-Navarro}}}, \bibinfo {author} {\bibfnamefont {S.}~\bibnamefont
  {{Genel}}}, \bibinfo {author} {\bibfnamefont {E.}~\bibnamefont
  {{Castorina}}}, \bibinfo {author} {\bibfnamefont {A.}~\bibnamefont
  {{Obuljen}}}, \bibinfo {author} {\bibfnamefont {D.~N.}\ \bibnamefont
  {{Spergel}}}, \bibinfo {author} {\bibfnamefont {L.}~\bibnamefont
  {{Hernquist}}}, \bibinfo {author} {\bibfnamefont {D.}~\bibnamefont
  {{Nelson}}}, \bibinfo {author} {\bibfnamefont {I.~P.}\ \bibnamefont
  {{Carucci}}}, \bibinfo {author} {\bibfnamefont {A.}~\bibnamefont
  {{Pillepich}}}, \bibinfo {author} {\bibfnamefont {F.}~\bibnamefont
  {{Marinacci}}}, \bibinfo {author} {\bibfnamefont {B.}~\bibnamefont
  {{Diemer}}}, \bibinfo {author} {\bibfnamefont {M.}~\bibnamefont
  {{Vogelsberger}}}, \bibinfo {author} {\bibfnamefont {R.}~\bibnamefont
  {{Weinberger}}}, \ and\ \bibinfo {author} {\bibfnamefont {R.}~\bibnamefont
  {{Pakmor}}},\ }\href {\doibase 10.3847/1538-4357/aadba0} {\bibfield
  {journal} {\bibinfo  {journal} {\apj}\ }\textbf {\bibinfo {volume} {866}},\
  \bibinfo {eid} {135} (\bibinfo {year} {2018})},\ \Eprint
  {http://arxiv.org/abs/1804.09180} {arXiv:1804.09180 [astro-ph.CO]}
  \BibitemShut {NoStop}%
\bibitem [{\citenamefont {{Angulo}}\ and\ \citenamefont
  {{Pontzen}}(2016)}]{Angulo2016MNRAS}%
  \BibitemOpen
  \bibfield  {author} {\bibinfo {author} {\bibfnamefont {R.~E.}\ \bibnamefont
  {{Angulo}}}\ and\ \bibinfo {author} {\bibfnamefont {A.}~\bibnamefont
  {{Pontzen}}},\ }\href {\doibase 10.1093/mnrasl/slw098} {\bibfield  {journal}
  {\bibinfo  {journal} {\mnras}\ }\textbf {\bibinfo {volume} {462}},\ \bibinfo
  {pages} {L1} (\bibinfo {year} {2016})},\ \Eprint
  {http://arxiv.org/abs/1603.05253} {arXiv:1603.05253 [astro-ph.CO]}
  \BibitemShut {NoStop}%
\bibitem [{\citenamefont {{Klypin}}\ \emph {et~al.}(2020)\citenamefont
  {{Klypin}}, \citenamefont {{Prada}},\ and\ \citenamefont
  {{Byun}}}]{Klypin2020MNRAS}%
  \BibitemOpen
  \bibfield  {author} {\bibinfo {author} {\bibfnamefont {A.}~\bibnamefont
  {{Klypin}}}, \bibinfo {author} {\bibfnamefont {F.}~\bibnamefont {{Prada}}}, \
  and\ \bibinfo {author} {\bibfnamefont {J.}~\bibnamefont {{Byun}}},\ }\href
  {\doibase 10.1093/mnras/staa734} {\bibfield  {journal} {\bibinfo  {journal}
  {\mnras}\ }\textbf {\bibinfo {volume} {496}},\ \bibinfo {pages} {3862}
  (\bibinfo {year} {2020})},\ \Eprint {http://arxiv.org/abs/1903.08518}
  {arXiv:1903.08518 [astro-ph.CO]} \BibitemShut {NoStop}%
\bibitem [{\citenamefont {{Krause}}\ \emph {et~al.}(2017)\citenamefont
  {{Krause}}, \citenamefont {{Eifler}}, \citenamefont {{Zuntz}}, \citenamefont
  {{Friedrich}}, \citenamefont {{Troxel}}, \citenamefont {{Dodelson}},
  \citenamefont {{Blazek}}, \citenamefont {{Secco}}, \citenamefont
  {{MacCrann}}, \citenamefont {{Baxter}}, \citenamefont {{Chang}},
  \citenamefont {{Chen}}, \citenamefont {{Crocce}}, \citenamefont {{DeRose}},
  \citenamefont {{Ferte}}, \citenamefont {{Kokron}}, \citenamefont {{Lacasa}},
  \citenamefont {{Miranda}}, \citenamefont {{Omori}}, \citenamefont
  {{Porredon}}, \citenamefont {{Rosenfeld}}, \citenamefont {{Samuroff}},
  \citenamefont {{Wang}}, \citenamefont {{Wechsler}}, \citenamefont {{Abbott}},
  \citenamefont {{Abdalla}}, \citenamefont {{Allam}}, \citenamefont {{Annis}},
  \citenamefont {{Bechtol}}, \citenamefont {{Benoit-Levy}}, \citenamefont
  {{Bernstein}}, \citenamefont {{Brooks}}, \citenamefont {{Burke}},
  \citenamefont {{Capozzi}}, \citenamefont {{Carrasco Kind}}, \citenamefont
  {{Carretero}}, \citenamefont {{D'Andrea}}, \citenamefont {{da Costa}},
  \citenamefont {{Davis}}, \citenamefont {{DePoy}}, \citenamefont {{Desai}},
  \citenamefont {{Diehl}}, \citenamefont {{Dietrich}}, \citenamefont
  {{Evrard}}, \citenamefont {{Flaugher}}, \citenamefont {{Fosalba}},
  \citenamefont {{Frieman}}, \citenamefont {{Garcia-Bellido}}, \citenamefont
  {{Gaztanaga}}, \citenamefont {{Giannantonio}}, \citenamefont {{Gruen}},
  \citenamefont {{Gruendl}}, \citenamefont {{Gschwend}}, \citenamefont
  {{Gutierrez}}, \citenamefont {{Honscheid}}, \citenamefont {{James}},
  \citenamefont {{Jeltema}}, \citenamefont {{Kuehn}}, \citenamefont
  {{Kuhlmann}}, \citenamefont {{Lahav}}, \citenamefont {{Lima}}, \citenamefont
  {{Maia}}, \citenamefont {{March}}, \citenamefont {{Marshall}}, \citenamefont
  {{Martini}}, \citenamefont {{Menanteau}}, \citenamefont {{Miquel}},
  \citenamefont {{Nichol}}, \citenamefont {{Plazas}}, \citenamefont {{Romer}},
  \citenamefont {{Rykoff}}, \citenamefont {{Sanchez}}, \citenamefont
  {{Scarpine}}, \citenamefont {{Schindler}}, \citenamefont {{Schubnell}},
  \citenamefont {{Sevilla-Noarbe}}, \citenamefont {{Smith}}, \citenamefont
  {{Soares-Santos}}, \citenamefont {{Sobreira}}, \citenamefont {{Suchyta}},
  \citenamefont {{Swanson}}, \citenamefont {{Tarle}}, \citenamefont {{Tucker}},
  \citenamefont {{Vikram}}, \citenamefont {{Walker}},\ and\ \citenamefont
  {{Weller}}}]{DES1}%
  \BibitemOpen
  \bibfield  {author} {\bibinfo {author} {\bibfnamefont {E.}~\bibnamefont
  {{Krause}}}, \bibinfo {author} {\bibfnamefont {T.~F.}\ \bibnamefont
  {{Eifler}}}, \bibinfo {author} {\bibfnamefont {J.}~\bibnamefont {{Zuntz}}},
  \bibinfo {author} {\bibfnamefont {O.}~\bibnamefont {{Friedrich}}}, \bibinfo
  {author} {\bibfnamefont {M.~A.}\ \bibnamefont {{Troxel}}}, \bibinfo {author}
  {\bibfnamefont {S.}~\bibnamefont {{Dodelson}}}, \bibinfo {author}
  {\bibfnamefont {J.}~\bibnamefont {{Blazek}}}, \bibinfo {author}
  {\bibfnamefont {L.~F.}\ \bibnamefont {{Secco}}}, \bibinfo {author}
  {\bibfnamefont {N.}~\bibnamefont {{MacCrann}}}, \bibinfo {author}
  {\bibfnamefont {E.}~\bibnamefont {{Baxter}}}, \bibinfo {author}
  {\bibfnamefont {C.}~\bibnamefont {{Chang}}}, \bibinfo {author} {\bibfnamefont
  {N.}~\bibnamefont {{Chen}}}, \bibinfo {author} {\bibfnamefont
  {M.}~\bibnamefont {{Crocce}}}, \bibinfo {author} {\bibfnamefont
  {J.}~\bibnamefont {{DeRose}}}, \bibinfo {author} {\bibfnamefont
  {A.}~\bibnamefont {{Ferte}}}, \bibinfo {author} {\bibfnamefont
  {N.}~\bibnamefont {{Kokron}}}, \bibinfo {author} {\bibfnamefont
  {F.}~\bibnamefont {{Lacasa}}}, \bibinfo {author} {\bibfnamefont
  {V.}~\bibnamefont {{Miranda}}}, \bibinfo {author} {\bibfnamefont
  {Y.}~\bibnamefont {{Omori}}}, \bibinfo {author} {\bibfnamefont
  {A.}~\bibnamefont {{Porredon}}}, \bibinfo {author} {\bibfnamefont
  {R.}~\bibnamefont {{Rosenfeld}}}, \bibinfo {author} {\bibfnamefont
  {S.}~\bibnamefont {{Samuroff}}}, \bibinfo {author} {\bibfnamefont
  {M.}~\bibnamefont {{Wang}}}, \bibinfo {author} {\bibfnamefont {R.~H.}\
  \bibnamefont {{Wechsler}}}, \bibinfo {author} {\bibfnamefont {T.~M.~C.}\
  \bibnamefont {{Abbott}}}, \bibinfo {author} {\bibfnamefont {F.~B.}\
  \bibnamefont {{Abdalla}}}, \bibinfo {author} {\bibfnamefont {S.}~\bibnamefont
  {{Allam}}}, \bibinfo {author} {\bibfnamefont {J.}~\bibnamefont {{Annis}}},
  \bibinfo {author} {\bibfnamefont {K.}~\bibnamefont {{Bechtol}}}, \bibinfo
  {author} {\bibfnamefont {A.}~\bibnamefont {{Benoit-Levy}}}, \bibinfo {author}
  {\bibfnamefont {G.~M.}\ \bibnamefont {{Bernstein}}}, \bibinfo {author}
  {\bibfnamefont {D.}~\bibnamefont {{Brooks}}}, \bibinfo {author}
  {\bibfnamefont {D.~L.}\ \bibnamefont {{Burke}}}, \bibinfo {author}
  {\bibfnamefont {D.}~\bibnamefont {{Capozzi}}}, \bibinfo {author}
  {\bibfnamefont {M.}~\bibnamefont {{Carrasco Kind}}}, \bibinfo {author}
  {\bibfnamefont {J.}~\bibnamefont {{Carretero}}}, \bibinfo {author}
  {\bibfnamefont {C.~B.}\ \bibnamefont {{D'Andrea}}}, \bibinfo {author}
  {\bibfnamefont {L.~N.}\ \bibnamefont {{da Costa}}}, \bibinfo {author}
  {\bibfnamefont {C.}~\bibnamefont {{Davis}}}, \bibinfo {author} {\bibfnamefont
  {D.~L.}\ \bibnamefont {{DePoy}}}, \bibinfo {author} {\bibfnamefont
  {S.}~\bibnamefont {{Desai}}}, \bibinfo {author} {\bibfnamefont {H.~T.}\
  \bibnamefont {{Diehl}}}, \bibinfo {author} {\bibfnamefont {J.~P.}\
  \bibnamefont {{Dietrich}}}, \bibinfo {author} {\bibfnamefont {A.~E.}\
  \bibnamefont {{Evrard}}}, \bibinfo {author} {\bibfnamefont {B.}~\bibnamefont
  {{Flaugher}}}, \bibinfo {author} {\bibfnamefont {P.}~\bibnamefont
  {{Fosalba}}}, \bibinfo {author} {\bibfnamefont {J.}~\bibnamefont
  {{Frieman}}}, \bibinfo {author} {\bibfnamefont {J.}~\bibnamefont
  {{Garcia-Bellido}}}, \bibinfo {author} {\bibfnamefont {E.}~\bibnamefont
  {{Gaztanaga}}}, \bibinfo {author} {\bibfnamefont {T.}~\bibnamefont
  {{Giannantonio}}}, \bibinfo {author} {\bibfnamefont {D.}~\bibnamefont
  {{Gruen}}}, \bibinfo {author} {\bibfnamefont {R.~A.}\ \bibnamefont
  {{Gruendl}}}, \bibinfo {author} {\bibfnamefont {J.}~\bibnamefont
  {{Gschwend}}}, \bibinfo {author} {\bibfnamefont {G.}~\bibnamefont
  {{Gutierrez}}}, \bibinfo {author} {\bibfnamefont {K.}~\bibnamefont
  {{Honscheid}}}, \bibinfo {author} {\bibfnamefont {D.~J.}\ \bibnamefont
  {{James}}}, \bibinfo {author} {\bibfnamefont {T.}~\bibnamefont {{Jeltema}}},
  \bibinfo {author} {\bibfnamefont {K.}~\bibnamefont {{Kuehn}}}, \bibinfo
  {author} {\bibfnamefont {S.}~\bibnamefont {{Kuhlmann}}}, \bibinfo {author}
  {\bibfnamefont {O.}~\bibnamefont {{Lahav}}}, \bibinfo {author} {\bibfnamefont
  {M.}~\bibnamefont {{Lima}}}, \bibinfo {author} {\bibfnamefont {M.~A.~G.}\
  \bibnamefont {{Maia}}}, \bibinfo {author} {\bibfnamefont {M.}~\bibnamefont
  {{March}}}, \bibinfo {author} {\bibfnamefont {J.~L.}\ \bibnamefont
  {{Marshall}}}, \bibinfo {author} {\bibfnamefont {P.}~\bibnamefont
  {{Martini}}}, \bibinfo {author} {\bibfnamefont {F.}~\bibnamefont
  {{Menanteau}}}, \bibinfo {author} {\bibfnamefont {R.}~\bibnamefont
  {{Miquel}}}, \bibinfo {author} {\bibfnamefont {R.~C.}\ \bibnamefont
  {{Nichol}}}, \bibinfo {author} {\bibfnamefont {A.~A.}\ \bibnamefont
  {{Plazas}}}, \bibinfo {author} {\bibfnamefont {A.~K.}\ \bibnamefont
  {{Romer}}}, \bibinfo {author} {\bibfnamefont {E.~S.}\ \bibnamefont
  {{Rykoff}}}, \bibinfo {author} {\bibfnamefont {E.}~\bibnamefont {{Sanchez}}},
  \bibinfo {author} {\bibfnamefont {V.}~\bibnamefont {{Scarpine}}}, \bibinfo
  {author} {\bibfnamefont {R.}~\bibnamefont {{Schindler}}}, \bibinfo {author}
  {\bibfnamefont {M.}~\bibnamefont {{Schubnell}}}, \bibinfo {author}
  {\bibfnamefont {I.}~\bibnamefont {{Sevilla-Noarbe}}}, \bibinfo {author}
  {\bibfnamefont {M.}~\bibnamefont {{Smith}}}, \bibinfo {author} {\bibfnamefont
  {M.}~\bibnamefont {{Soares-Santos}}}, \bibinfo {author} {\bibfnamefont
  {F.}~\bibnamefont {{Sobreira}}}, \bibinfo {author} {\bibfnamefont
  {E.}~\bibnamefont {{Suchyta}}}, \bibinfo {author} {\bibfnamefont {M.~E.~C.}\
  \bibnamefont {{Swanson}}}, \bibinfo {author} {\bibfnamefont {G.}~\bibnamefont
  {{Tarle}}}, \bibinfo {author} {\bibfnamefont {D.~L.}\ \bibnamefont
  {{Tucker}}}, \bibinfo {author} {\bibfnamefont {V.}~\bibnamefont {{Vikram}}},
  \bibinfo {author} {\bibfnamefont {A.~R.}\ \bibnamefont {{Walker}}}, \ and\
  \bibinfo {author} {\bibfnamefont {J.}~\bibnamefont {{Weller}}},\ }\href@noop
  {} {\bibfield  {journal} {\bibinfo  {journal} {arXiv e-prints}\ ,\ \bibinfo
  {eid} {arXiv:1706.09359}} (\bibinfo {year} {2017})},\ \Eprint
  {http://arxiv.org/abs/1706.09359} {arXiv:1706.09359 [astro-ph.CO]}
  \BibitemShut {NoStop}%
\bibitem [{\citenamefont {{Troxel}}\ \emph {et~al.}(2018)\citenamefont
  {{Troxel}}, \citenamefont {{MacCrann}}, \citenamefont {{Zuntz}},
  \citenamefont {{Eifler}}, \citenamefont {{Krause}}, \citenamefont
  {{Dodelson}}, \citenamefont {{Gruen}}, \citenamefont {{Blazek}},
  \citenamefont {{Friedrich}}, \citenamefont {{Samuroff}}, \citenamefont
  {{Prat}}, \citenamefont {{Secco}}, \citenamefont {{Davis}}, \citenamefont
  {{Fert{\'e}}}, \citenamefont {{DeRose}}, \citenamefont {{Alarcon}},
  \citenamefont {{Amara}}, \citenamefont {{Baxter}}, \citenamefont {{Becker}},
  \citenamefont {{Bernstein}}, \citenamefont {{Bridle}}, \citenamefont
  {{Cawthon}}, \citenamefont {{Chang}}, \citenamefont {{Choi}}, \citenamefont
  {{De Vicente}}, \citenamefont {{Drlica-Wagner}}, \citenamefont
  {{Elvin-Poole}}, \citenamefont {{Frieman}}, \citenamefont {{Gatti}},
  \citenamefont {{Hartley}}, \citenamefont {{Honscheid}}, \citenamefont
  {{Hoyle}}, \citenamefont {{Huff}}, \citenamefont {{Huterer}}, \citenamefont
  {{Jain}}, \citenamefont {{Jarvis}}, \citenamefont {{Kacprzak}}, \citenamefont
  {{Kirk}}, \citenamefont {{Kokron}}, \citenamefont {{Krawiec}}, \citenamefont
  {{Lahav}}, \citenamefont {{Liddle}}, \citenamefont {{Peacock}}, \citenamefont
  {{Rau}}, \citenamefont {{Refregier}}, \citenamefont {{Rollins}},
  \citenamefont {{Rozo}}, \citenamefont {{Rykoff}}, \citenamefont
  {{S{\'a}nchez}}, \citenamefont {{Sevilla-Noarbe}}, \citenamefont {{Sheldon}},
  \citenamefont {{Stebbins}}, \citenamefont {{Varga}}, \citenamefont
  {{Vielzeuf}}, \citenamefont {{Wang}}, \citenamefont {{Wechsler}},
  \citenamefont {{Yanny}}, \citenamefont {{Abbott}}, \citenamefont {{Abdalla}},
  \citenamefont {{Allam}}, \citenamefont {{Annis}}, \citenamefont {{Bechtol}},
  \citenamefont {{Benoit-L{\'e}vy}}, \citenamefont {{Bertin}}, \citenamefont
  {{Brooks}}, \citenamefont {{Buckley-Geer}}, \citenamefont {{Burke}},
  \citenamefont {{Carnero Rosell}}, \citenamefont {{Carrasco Kind}},
  \citenamefont {{Carretero}}, \citenamefont {{Castander}}, \citenamefont
  {{Crocce}}, \citenamefont {{Cunha}}, \citenamefont {{D'Andrea}},
  \citenamefont {{da Costa}}, \citenamefont {{DePoy}}, \citenamefont {{Desai}},
  \citenamefont {{Diehl}}, \citenamefont {{Dietrich}}, \citenamefont {{Doel}},
  \citenamefont {{Fernandez}}, \citenamefont {{Flaugher}}, \citenamefont
  {{Fosalba}}, \citenamefont {{Garc{\'\i}a-Bellido}}, \citenamefont
  {{Gaztanaga}}, \citenamefont {{Gerdes}}, \citenamefont {{Giannantonio}},
  \citenamefont {{Goldstein}}, \citenamefont {{Gruendl}}, \citenamefont
  {{Gschwend}}, \citenamefont {{Gutierrez}}, \citenamefont {{James}},
  \citenamefont {{Jeltema}}, \citenamefont {{Johnson}}, \citenamefont
  {{Johnson}}, \citenamefont {{Kent}}, \citenamefont {{Kuehn}}, \citenamefont
  {{Kuhlmann}}, \citenamefont {{Kuropatkin}}, \citenamefont {{Li}},
  \citenamefont {{Lima}}, \citenamefont {{Lin}}, \citenamefont {{Maia}},
  \citenamefont {{March}}, \citenamefont {{Marshall}}, \citenamefont
  {{Martini}}, \citenamefont {{Melchior}}, \citenamefont {{Menanteau}},
  \citenamefont {{Miquel}}, \citenamefont {{Mohr}}, \citenamefont {{Neilsen}},
  \citenamefont {{Nichol}}, \citenamefont {{Nord}}, \citenamefont
  {{Petravick}}, \citenamefont {{Plazas}}, \citenamefont {{Romer}},
  \citenamefont {{Roodman}}, \citenamefont {{Sako}}, \citenamefont {{Sanchez}},
  \citenamefont {{Scarpine}}, \citenamefont {{Schindler}}, \citenamefont
  {{Schubnell}}, \citenamefont {{Smith}}, \citenamefont {{Smith}},
  \citenamefont {{Soares-Santos}}, \citenamefont {{Sobreira}}, \citenamefont
  {{Suchyta}}, \citenamefont {{Swanson}}, \citenamefont {{Tarle}},
  \citenamefont {{Thomas}}, \citenamefont {{Tucker}}, \citenamefont {{Vikram}},
  \citenamefont {{Walker}}, \citenamefont {{Weller}}, \citenamefont {{Zhang}},\
  and\ \citenamefont {{DES Collaboration}}}]{DES2}%
  \BibitemOpen
  \bibfield  {author} {\bibinfo {author} {\bibfnamefont {M.~A.}\ \bibnamefont
  {{Troxel}}}, \bibinfo {author} {\bibfnamefont {N.}~\bibnamefont
  {{MacCrann}}}, \bibinfo {author} {\bibfnamefont {J.}~\bibnamefont {{Zuntz}}},
  \bibinfo {author} {\bibfnamefont {T.~F.}\ \bibnamefont {{Eifler}}}, \bibinfo
  {author} {\bibfnamefont {E.}~\bibnamefont {{Krause}}}, \bibinfo {author}
  {\bibfnamefont {S.}~\bibnamefont {{Dodelson}}}, \bibinfo {author}
  {\bibfnamefont {D.}~\bibnamefont {{Gruen}}}, \bibinfo {author} {\bibfnamefont
  {J.}~\bibnamefont {{Blazek}}}, \bibinfo {author} {\bibfnamefont
  {O.}~\bibnamefont {{Friedrich}}}, \bibinfo {author} {\bibfnamefont
  {S.}~\bibnamefont {{Samuroff}}}, \bibinfo {author} {\bibfnamefont
  {J.}~\bibnamefont {{Prat}}}, \bibinfo {author} {\bibfnamefont {L.~F.}\
  \bibnamefont {{Secco}}}, \bibinfo {author} {\bibfnamefont {C.}~\bibnamefont
  {{Davis}}}, \bibinfo {author} {\bibfnamefont {A.}~\bibnamefont
  {{Fert{\'e}}}}, \bibinfo {author} {\bibfnamefont {J.}~\bibnamefont
  {{DeRose}}}, \bibinfo {author} {\bibfnamefont {A.}~\bibnamefont {{Alarcon}}},
  \bibinfo {author} {\bibfnamefont {A.}~\bibnamefont {{Amara}}}, \bibinfo
  {author} {\bibfnamefont {E.}~\bibnamefont {{Baxter}}}, \bibinfo {author}
  {\bibfnamefont {M.~R.}\ \bibnamefont {{Becker}}}, \bibinfo {author}
  {\bibfnamefont {G.~M.}\ \bibnamefont {{Bernstein}}}, \bibinfo {author}
  {\bibfnamefont {S.~L.}\ \bibnamefont {{Bridle}}}, \bibinfo {author}
  {\bibfnamefont {R.}~\bibnamefont {{Cawthon}}}, \bibinfo {author}
  {\bibfnamefont {C.}~\bibnamefont {{Chang}}}, \bibinfo {author} {\bibfnamefont
  {A.}~\bibnamefont {{Choi}}}, \bibinfo {author} {\bibfnamefont
  {J.}~\bibnamefont {{De Vicente}}}, \bibinfo {author} {\bibfnamefont
  {A.}~\bibnamefont {{Drlica-Wagner}}}, \bibinfo {author} {\bibfnamefont
  {J.}~\bibnamefont {{Elvin-Poole}}}, \bibinfo {author} {\bibfnamefont
  {J.}~\bibnamefont {{Frieman}}}, \bibinfo {author} {\bibfnamefont
  {M.}~\bibnamefont {{Gatti}}}, \bibinfo {author} {\bibfnamefont {W.~G.}\
  \bibnamefont {{Hartley}}}, \bibinfo {author} {\bibfnamefont {K.}~\bibnamefont
  {{Honscheid}}}, \bibinfo {author} {\bibfnamefont {B.}~\bibnamefont
  {{Hoyle}}}, \bibinfo {author} {\bibfnamefont {E.~M.}\ \bibnamefont {{Huff}}},
  \bibinfo {author} {\bibfnamefont {D.}~\bibnamefont {{Huterer}}}, \bibinfo
  {author} {\bibfnamefont {B.}~\bibnamefont {{Jain}}}, \bibinfo {author}
  {\bibfnamefont {M.}~\bibnamefont {{Jarvis}}}, \bibinfo {author}
  {\bibfnamefont {T.}~\bibnamefont {{Kacprzak}}}, \bibinfo {author}
  {\bibfnamefont {D.}~\bibnamefont {{Kirk}}}, \bibinfo {author} {\bibfnamefont
  {N.}~\bibnamefont {{Kokron}}}, \bibinfo {author} {\bibfnamefont
  {C.}~\bibnamefont {{Krawiec}}}, \bibinfo {author} {\bibfnamefont
  {O.}~\bibnamefont {{Lahav}}}, \bibinfo {author} {\bibfnamefont {A.~R.}\
  \bibnamefont {{Liddle}}}, \bibinfo {author} {\bibfnamefont {J.}~\bibnamefont
  {{Peacock}}}, \bibinfo {author} {\bibfnamefont {M.~M.}\ \bibnamefont
  {{Rau}}}, \bibinfo {author} {\bibfnamefont {A.}~\bibnamefont {{Refregier}}},
  \bibinfo {author} {\bibfnamefont {R.~P.}\ \bibnamefont {{Rollins}}}, \bibinfo
  {author} {\bibfnamefont {E.}~\bibnamefont {{Rozo}}}, \bibinfo {author}
  {\bibfnamefont {E.~S.}\ \bibnamefont {{Rykoff}}}, \bibinfo {author}
  {\bibfnamefont {C.}~\bibnamefont {{S{\'a}nchez}}}, \bibinfo {author}
  {\bibfnamefont {I.}~\bibnamefont {{Sevilla-Noarbe}}}, \bibinfo {author}
  {\bibfnamefont {E.}~\bibnamefont {{Sheldon}}}, \bibinfo {author}
  {\bibfnamefont {A.}~\bibnamefont {{Stebbins}}}, \bibinfo {author}
  {\bibfnamefont {T.~N.}\ \bibnamefont {{Varga}}}, \bibinfo {author}
  {\bibfnamefont {P.}~\bibnamefont {{Vielzeuf}}}, \bibinfo {author}
  {\bibfnamefont {M.}~\bibnamefont {{Wang}}}, \bibinfo {author} {\bibfnamefont
  {R.~H.}\ \bibnamefont {{Wechsler}}}, \bibinfo {author} {\bibfnamefont
  {B.}~\bibnamefont {{Yanny}}}, \bibinfo {author} {\bibfnamefont {T.~M.~C.}\
  \bibnamefont {{Abbott}}}, \bibinfo {author} {\bibfnamefont {F.~B.}\
  \bibnamefont {{Abdalla}}}, \bibinfo {author} {\bibfnamefont {S.}~\bibnamefont
  {{Allam}}}, \bibinfo {author} {\bibfnamefont {J.}~\bibnamefont {{Annis}}},
  \bibinfo {author} {\bibfnamefont {K.}~\bibnamefont {{Bechtol}}}, \bibinfo
  {author} {\bibfnamefont {A.}~\bibnamefont {{Benoit-L{\'e}vy}}}, \bibinfo
  {author} {\bibfnamefont {E.}~\bibnamefont {{Bertin}}}, \bibinfo {author}
  {\bibfnamefont {D.}~\bibnamefont {{Brooks}}}, \bibinfo {author}
  {\bibfnamefont {E.}~\bibnamefont {{Buckley-Geer}}}, \bibinfo {author}
  {\bibfnamefont {D.~L.}\ \bibnamefont {{Burke}}}, \bibinfo {author}
  {\bibfnamefont {A.}~\bibnamefont {{Carnero Rosell}}}, \bibinfo {author}
  {\bibfnamefont {M.}~\bibnamefont {{Carrasco Kind}}}, \bibinfo {author}
  {\bibfnamefont {J.}~\bibnamefont {{Carretero}}}, \bibinfo {author}
  {\bibfnamefont {F.~J.}\ \bibnamefont {{Castander}}}, \bibinfo {author}
  {\bibfnamefont {M.}~\bibnamefont {{Crocce}}}, \bibinfo {author}
  {\bibfnamefont {C.~E.}\ \bibnamefont {{Cunha}}}, \bibinfo {author}
  {\bibfnamefont {C.~B.}\ \bibnamefont {{D'Andrea}}}, \bibinfo {author}
  {\bibfnamefont {L.~N.}\ \bibnamefont {{da Costa}}}, \bibinfo {author}
  {\bibfnamefont {D.~L.}\ \bibnamefont {{DePoy}}}, \bibinfo {author}
  {\bibfnamefont {S.}~\bibnamefont {{Desai}}}, \bibinfo {author} {\bibfnamefont
  {H.~T.}\ \bibnamefont {{Diehl}}}, \bibinfo {author} {\bibfnamefont {J.~P.}\
  \bibnamefont {{Dietrich}}}, \bibinfo {author} {\bibfnamefont
  {P.}~\bibnamefont {{Doel}}}, \bibinfo {author} {\bibfnamefont
  {E.}~\bibnamefont {{Fernandez}}}, \bibinfo {author} {\bibfnamefont
  {B.}~\bibnamefont {{Flaugher}}}, \bibinfo {author} {\bibfnamefont
  {P.}~\bibnamefont {{Fosalba}}}, \bibinfo {author} {\bibfnamefont
  {J.}~\bibnamefont {{Garc{\'\i}a-Bellido}}}, \bibinfo {author} {\bibfnamefont
  {E.}~\bibnamefont {{Gaztanaga}}}, \bibinfo {author} {\bibfnamefont {D.~W.}\
  \bibnamefont {{Gerdes}}}, \bibinfo {author} {\bibfnamefont {T.}~\bibnamefont
  {{Giannantonio}}}, \bibinfo {author} {\bibfnamefont {D.~A.}\ \bibnamefont
  {{Goldstein}}}, \bibinfo {author} {\bibfnamefont {R.~A.}\ \bibnamefont
  {{Gruendl}}}, \bibinfo {author} {\bibfnamefont {J.}~\bibnamefont
  {{Gschwend}}}, \bibinfo {author} {\bibfnamefont {G.}~\bibnamefont
  {{Gutierrez}}}, \bibinfo {author} {\bibfnamefont {D.~J.}\ \bibnamefont
  {{James}}}, \bibinfo {author} {\bibfnamefont {T.}~\bibnamefont {{Jeltema}}},
  \bibinfo {author} {\bibfnamefont {M.~W.~G.}\ \bibnamefont {{Johnson}}},
  \bibinfo {author} {\bibfnamefont {M.~D.}\ \bibnamefont {{Johnson}}}, \bibinfo
  {author} {\bibfnamefont {S.}~\bibnamefont {{Kent}}}, \bibinfo {author}
  {\bibfnamefont {K.}~\bibnamefont {{Kuehn}}}, \bibinfo {author} {\bibfnamefont
  {S.}~\bibnamefont {{Kuhlmann}}}, \bibinfo {author} {\bibfnamefont
  {N.}~\bibnamefont {{Kuropatkin}}}, \bibinfo {author} {\bibfnamefont {T.~S.}\
  \bibnamefont {{Li}}}, \bibinfo {author} {\bibfnamefont {M.}~\bibnamefont
  {{Lima}}}, \bibinfo {author} {\bibfnamefont {H.}~\bibnamefont {{Lin}}},
  \bibinfo {author} {\bibfnamefont {M.~A.~G.}\ \bibnamefont {{Maia}}}, \bibinfo
  {author} {\bibfnamefont {M.}~\bibnamefont {{March}}}, \bibinfo {author}
  {\bibfnamefont {J.~L.}\ \bibnamefont {{Marshall}}}, \bibinfo {author}
  {\bibfnamefont {P.}~\bibnamefont {{Martini}}}, \bibinfo {author}
  {\bibfnamefont {P.}~\bibnamefont {{Melchior}}}, \bibinfo {author}
  {\bibfnamefont {F.}~\bibnamefont {{Menanteau}}}, \bibinfo {author}
  {\bibfnamefont {R.}~\bibnamefont {{Miquel}}}, \bibinfo {author}
  {\bibfnamefont {J.~J.}\ \bibnamefont {{Mohr}}}, \bibinfo {author}
  {\bibfnamefont {E.}~\bibnamefont {{Neilsen}}}, \bibinfo {author}
  {\bibfnamefont {R.~C.}\ \bibnamefont {{Nichol}}}, \bibinfo {author}
  {\bibfnamefont {B.}~\bibnamefont {{Nord}}}, \bibinfo {author} {\bibfnamefont
  {D.}~\bibnamefont {{Petravick}}}, \bibinfo {author} {\bibfnamefont {A.~A.}\
  \bibnamefont {{Plazas}}}, \bibinfo {author} {\bibfnamefont {A.~K.}\
  \bibnamefont {{Romer}}}, \bibinfo {author} {\bibfnamefont {A.}~\bibnamefont
  {{Roodman}}}, \bibinfo {author} {\bibfnamefont {M.}~\bibnamefont {{Sako}}},
  \bibinfo {author} {\bibfnamefont {E.}~\bibnamefont {{Sanchez}}}, \bibinfo
  {author} {\bibfnamefont {V.}~\bibnamefont {{Scarpine}}}, \bibinfo {author}
  {\bibfnamefont {R.}~\bibnamefont {{Schindler}}}, \bibinfo {author}
  {\bibfnamefont {M.}~\bibnamefont {{Schubnell}}}, \bibinfo {author}
  {\bibfnamefont {M.}~\bibnamefont {{Smith}}}, \bibinfo {author} {\bibfnamefont
  {R.~C.}\ \bibnamefont {{Smith}}}, \bibinfo {author} {\bibfnamefont
  {M.}~\bibnamefont {{Soares-Santos}}}, \bibinfo {author} {\bibfnamefont
  {F.}~\bibnamefont {{Sobreira}}}, \bibinfo {author} {\bibfnamefont
  {E.}~\bibnamefont {{Suchyta}}}, \bibinfo {author} {\bibfnamefont {M.~E.~C.}\
  \bibnamefont {{Swanson}}}, \bibinfo {author} {\bibfnamefont {G.}~\bibnamefont
  {{Tarle}}}, \bibinfo {author} {\bibfnamefont {D.}~\bibnamefont {{Thomas}}},
  \bibinfo {author} {\bibfnamefont {D.~L.}\ \bibnamefont {{Tucker}}}, \bibinfo
  {author} {\bibfnamefont {V.}~\bibnamefont {{Vikram}}}, \bibinfo {author}
  {\bibfnamefont {A.~R.}\ \bibnamefont {{Walker}}}, \bibinfo {author}
  {\bibfnamefont {J.}~\bibnamefont {{Weller}}}, \bibinfo {author}
  {\bibfnamefont {Y.}~\bibnamefont {{Zhang}}}, \ and\ \bibinfo {author}
  {\bibnamefont {{DES Collaboration}}},\ }\href {\doibase
  10.1103/PhysRevD.98.043528} {\bibfield  {journal} {\bibinfo  {journal}
  {\prd}\ }\textbf {\bibinfo {volume} {98}},\ \bibinfo {eid} {043528} (\bibinfo
  {year} {2018})},\ \Eprint {http://arxiv.org/abs/1708.01538} {arXiv:1708.01538
  [astro-ph.CO]} \BibitemShut {NoStop}%
\bibitem [{\citenamefont {{Sheldon}}\ and\ \citenamefont
  {{Huff}}(2017)}]{Sheldon2017ApJ}%
  \BibitemOpen
  \bibfield  {author} {\bibinfo {author} {\bibfnamefont {E.~S.}\ \bibnamefont
  {{Sheldon}}}\ and\ \bibinfo {author} {\bibfnamefont {E.~M.}\ \bibnamefont
  {{Huff}}},\ }\href {\doibase 10.3847/1538-4357/aa704b} {\bibfield  {journal}
  {\bibinfo  {journal} {\apj}\ }\textbf {\bibinfo {volume} {841}},\ \bibinfo
  {eid} {24} (\bibinfo {year} {2017})},\ \Eprint
  {http://arxiv.org/abs/1702.02601} {arXiv:1702.02601 [astro-ph.CO]}
  \BibitemShut {NoStop}%
\bibitem [{\citenamefont {{Baker}}\ \emph {et~al.}(2018)\citenamefont
  {{Baker}}, \citenamefont {{Clampitt}}, \citenamefont {{Jain}},\ and\
  \citenamefont {{Trodden}}}]{baker2018prd}%
  \BibitemOpen
  \bibfield  {author} {\bibinfo {author} {\bibfnamefont {T.}~\bibnamefont
  {{Baker}}}, \bibinfo {author} {\bibfnamefont {J.}~\bibnamefont {{Clampitt}}},
  \bibinfo {author} {\bibfnamefont {B.}~\bibnamefont {{Jain}}}, \ and\ \bibinfo
  {author} {\bibfnamefont {M.}~\bibnamefont {{Trodden}}},\ }\href {\doibase
  10.1103/PhysRevD.98.023511} {\bibfield  {journal} {\bibinfo  {journal}
  {\prd}\ }\textbf {\bibinfo {volume} {98}},\ \bibinfo {eid} {023511} (\bibinfo
  {year} {2018})},\ \Eprint {http://arxiv.org/abs/1803.07533} {arXiv:1803.07533
  [astro-ph.CO]} \BibitemShut {NoStop}%
\bibitem [{\citenamefont {{Dong}}\ \emph {et~al.}(2019)\citenamefont {{Dong}},
  \citenamefont {{Zhang}}, \citenamefont {{Yu}}, \citenamefont {{Yang}},
  \citenamefont {{Li}}, \citenamefont {{Han}}, \citenamefont {{Luo}},
  \citenamefont {{Zhang}},\ and\ \citenamefont {{Fu}}}]{dong2019apj}%
  \BibitemOpen
  \bibfield  {author} {\bibinfo {author} {\bibfnamefont {F.}~\bibnamefont
  {{Dong}}}, \bibinfo {author} {\bibfnamefont {J.}~\bibnamefont {{Zhang}}},
  \bibinfo {author} {\bibfnamefont {Y.}~\bibnamefont {{Yu}}}, \bibinfo {author}
  {\bibfnamefont {X.}~\bibnamefont {{Yang}}}, \bibinfo {author} {\bibfnamefont
  {H.}~\bibnamefont {{Li}}}, \bibinfo {author} {\bibfnamefont {J.}~\bibnamefont
  {{Han}}}, \bibinfo {author} {\bibfnamefont {W.}~\bibnamefont {{Luo}}},
  \bibinfo {author} {\bibfnamefont {J.}~\bibnamefont {{Zhang}}}, \ and\
  \bibinfo {author} {\bibfnamefont {L.}~\bibnamefont {{Fu}}},\ }\href {\doibase
  10.3847/1538-4357/ab0648} {\bibfield  {journal} {\bibinfo  {journal} {\apj}\
  }\textbf {\bibinfo {volume} {874}},\ \bibinfo {eid} {7} (\bibinfo {year}
  {2019})},\ \Eprint {http://arxiv.org/abs/1809.00282} {arXiv:1809.00282
  [astro-ph.CO]} \BibitemShut {NoStop}%
\bibitem [{\citenamefont {{Knollmann}}\ and\ \citenamefont
  {{Knebe}}(2009)}]{ahf}%
  \BibitemOpen
  \bibfield  {author} {\bibinfo {author} {\bibfnamefont {S.~R.}\ \bibnamefont
  {{Knollmann}}}\ and\ \bibinfo {author} {\bibfnamefont {A.}~\bibnamefont
  {{Knebe}}},\ }\href {\doibase 10.1088/0067-0049/182/2/608} {\bibfield
  {journal} {\bibinfo  {journal} {\apjs}\ }\textbf {\bibinfo {volume} {182}},\
  \bibinfo {pages} {608} (\bibinfo {year} {2009})},\ \Eprint
  {http://arxiv.org/abs/0904.3662} {arXiv:0904.3662} \BibitemShut {NoStop}%
\bibitem [{\citenamefont {{Brouwer}}\ \emph {et~al.}(2017)\citenamefont
  {{Brouwer}}, \citenamefont {{Visser}}, \citenamefont {{Dvornik}},
  \citenamefont {{Hoekstra}}, \citenamefont {{Kuijken}}, \citenamefont
  {{Valentijn}}, \citenamefont {{Bilicki}}, \citenamefont {{Blake}},
  \citenamefont {{Brough}}, \citenamefont {{Buddelmeijer}}, \citenamefont
  {{Erben}}, \citenamefont {{Heymans}}, \citenamefont {{Hildebrandt}},
  \citenamefont {{Holwerda}}, \citenamefont {{Hopkins}}, \citenamefont
  {{Klaes}}, \citenamefont {{Liske}}, \citenamefont {{Loveday}}, \citenamefont
  {{McFarland}}, \citenamefont {{Nakajima}}, \citenamefont {{Sif{\'o}n}},\ and\
  \citenamefont {{Taylor}}}]{brouwer2017MNR}%
  \BibitemOpen
  \bibfield  {author} {\bibinfo {author} {\bibfnamefont {M.~M.}\ \bibnamefont
  {{Brouwer}}}, \bibinfo {author} {\bibfnamefont {M.~R.}\ \bibnamefont
  {{Visser}}}, \bibinfo {author} {\bibfnamefont {A.}~\bibnamefont {{Dvornik}}},
  \bibinfo {author} {\bibfnamefont {H.}~\bibnamefont {{Hoekstra}}}, \bibinfo
  {author} {\bibfnamefont {K.}~\bibnamefont {{Kuijken}}}, \bibinfo {author}
  {\bibfnamefont {E.~A.}\ \bibnamefont {{Valentijn}}}, \bibinfo {author}
  {\bibfnamefont {M.}~\bibnamefont {{Bilicki}}}, \bibinfo {author}
  {\bibfnamefont {C.}~\bibnamefont {{Blake}}}, \bibinfo {author} {\bibfnamefont
  {S.}~\bibnamefont {{Brough}}}, \bibinfo {author} {\bibfnamefont
  {H.}~\bibnamefont {{Buddelmeijer}}}, \bibinfo {author} {\bibfnamefont
  {T.}~\bibnamefont {{Erben}}}, \bibinfo {author} {\bibfnamefont
  {C.}~\bibnamefont {{Heymans}}}, \bibinfo {author} {\bibfnamefont
  {H.}~\bibnamefont {{Hildebrandt}}}, \bibinfo {author} {\bibfnamefont {B.~W.}\
  \bibnamefont {{Holwerda}}}, \bibinfo {author} {\bibfnamefont {A.~M.}\
  \bibnamefont {{Hopkins}}}, \bibinfo {author} {\bibfnamefont {D.}~\bibnamefont
  {{Klaes}}}, \bibinfo {author} {\bibfnamefont {J.}~\bibnamefont {{Liske}}},
  \bibinfo {author} {\bibfnamefont {J.}~\bibnamefont {{Loveday}}}, \bibinfo
  {author} {\bibfnamefont {J.}~\bibnamefont {{McFarland}}}, \bibinfo {author}
  {\bibfnamefont {R.}~\bibnamefont {{Nakajima}}}, \bibinfo {author}
  {\bibfnamefont {C.}~\bibnamefont {{Sif{\'o}n}}}, \ and\ \bibinfo {author}
  {\bibfnamefont {E.~N.}\ \bibnamefont {{Taylor}}},\ }\href {\doibase
  10.1093/mnras/stw3192} {\bibfield  {journal} {\bibinfo  {journal} {\mnras}\
  }\textbf {\bibinfo {volume} {466}},\ \bibinfo {pages} {2547} (\bibinfo {year}
  {2017})},\ \Eprint {http://arxiv.org/abs/1612.03034} {arXiv:1612.03034
  [astro-ph.CO]} \BibitemShut {NoStop}%
\bibitem [{\citenamefont {{Luo}}\ \emph {et~al.}(2020)\citenamefont {{Luo}},
  \citenamefont {{Zhang}}, \citenamefont {{Halenka}}, \citenamefont {{Yang}},
  \citenamefont {{More}}, \citenamefont {{Miller}}, \citenamefont {{Sunayama}},
  \citenamefont {{Liu}},\ and\ \citenamefont {{Shi}}}]{luo2020}%
  \BibitemOpen
  \bibfield  {author} {\bibinfo {author} {\bibfnamefont {W.}~\bibnamefont
  {{Luo}}}, \bibinfo {author} {\bibfnamefont {J.}~\bibnamefont {{Zhang}}},
  \bibinfo {author} {\bibfnamefont {V.}~\bibnamefont {{Halenka}}}, \bibinfo
  {author} {\bibfnamefont {X.}~\bibnamefont {{Yang}}}, \bibinfo {author}
  {\bibfnamefont {S.}~\bibnamefont {{More}}}, \bibinfo {author} {\bibfnamefont
  {C.}~\bibnamefont {{Miller}}}, \bibinfo {author} {\bibfnamefont
  {T.}~\bibnamefont {{Sunayama}}}, \bibinfo {author} {\bibfnamefont
  {L.}~\bibnamefont {{Liu}}}, \ and\ \bibinfo {author} {\bibfnamefont
  {F.}~\bibnamefont {{Shi}}},\ }\href@noop {} {\bibfield  {journal} {\bibinfo
  {journal} {arXiv e-prints}\ ,\ \bibinfo {eid} {arXiv:2003.09818}} (\bibinfo
  {year} {2020})},\ \Eprint {http://arxiv.org/abs/2003.09818} {arXiv:2003.09818
  [astro-ph.GA]} \BibitemShut {NoStop}%
\bibitem [{\citenamefont {{Chen}}\ \emph {et~al.}(2019)\citenamefont {{Chen}},
  \citenamefont {{Luo}}, \citenamefont {{Cai}},\ and\ \citenamefont
  {{Saridakis}}}]{chen2019arx}%
  \BibitemOpen
  \bibfield  {author} {\bibinfo {author} {\bibfnamefont {Z.}~\bibnamefont
  {{Chen}}}, \bibinfo {author} {\bibfnamefont {W.}~\bibnamefont {{Luo}}},
  \bibinfo {author} {\bibfnamefont {Y.-F.}\ \bibnamefont {{Cai}}}, \ and\
  \bibinfo {author} {\bibfnamefont {E.~N.}\ \bibnamefont {{Saridakis}}},\
  }\href@noop {} {\bibfield  {journal} {\bibinfo  {journal} {arXiv e-prints}\
  ,\ \bibinfo {eid} {arXiv:1907.12225}} (\bibinfo {year} {2019})},\ \Eprint
  {http://arxiv.org/abs/1907.12225} {arXiv:1907.12225 [astro-ph.CO]}
  \BibitemShut {NoStop}%
\bibitem [{\citenamefont {{Yang}}\ \emph {et~al.}(2007)\citenamefont {{Yang}},
  \citenamefont {{Mo}}, \citenamefont {{van den Bosch}}, \citenamefont
  {{Pasquali}}, \citenamefont {{Li}},\ and\ \citenamefont
  {{Barden}}}]{yang2007apj}%
  \BibitemOpen
  \bibfield  {author} {\bibinfo {author} {\bibfnamefont {X.}~\bibnamefont
  {{Yang}}}, \bibinfo {author} {\bibfnamefont {H.~J.}\ \bibnamefont {{Mo}}},
  \bibinfo {author} {\bibfnamefont {F.~C.}\ \bibnamefont {{van den Bosch}}},
  \bibinfo {author} {\bibfnamefont {A.}~\bibnamefont {{Pasquali}}}, \bibinfo
  {author} {\bibfnamefont {C.}~\bibnamefont {{Li}}}, \ and\ \bibinfo {author}
  {\bibfnamefont {M.}~\bibnamefont {{Barden}}},\ }\href {\doibase
  10.1086/522027} {\bibfield  {journal} {\bibinfo  {journal} {\apj}\ }\textbf
  {\bibinfo {volume} {671}},\ \bibinfo {pages} {153} (\bibinfo {year}
  {2007})},\ \Eprint {http://arxiv.org/abs/0707.4640} {arXiv:0707.4640}
  \BibitemShut {NoStop}%
\bibitem [{\citenamefont {{Luo}}\ \emph {et~al.}(2018)\citenamefont {{Luo}},
  \citenamefont {{Yang}}, \citenamefont {{Lu}}, \citenamefont {{Shi}},
  \citenamefont {{Zhang}}, \citenamefont {{Mo}}, \citenamefont {{Shu}},
  \citenamefont {{Fu}}, \citenamefont {{Radovich}}, \citenamefont {{Zhang}},
  \citenamefont {{Li}}, \citenamefont {{Sunayama}},\ and\ \citenamefont
  {{Wang}}}]{luo2018apj}%
  \BibitemOpen
  \bibfield  {author} {\bibinfo {author} {\bibfnamefont {W.}~\bibnamefont
  {{Luo}}}, \bibinfo {author} {\bibfnamefont {X.}~\bibnamefont {{Yang}}},
  \bibinfo {author} {\bibfnamefont {T.}~\bibnamefont {{Lu}}}, \bibinfo {author}
  {\bibfnamefont {F.}~\bibnamefont {{Shi}}}, \bibinfo {author} {\bibfnamefont
  {J.}~\bibnamefont {{Zhang}}}, \bibinfo {author} {\bibfnamefont {H.~J.}\
  \bibnamefont {{Mo}}}, \bibinfo {author} {\bibfnamefont {C.}~\bibnamefont
  {{Shu}}}, \bibinfo {author} {\bibfnamefont {L.}~\bibnamefont {{Fu}}},
  \bibinfo {author} {\bibfnamefont {M.}~\bibnamefont {{Radovich}}}, \bibinfo
  {author} {\bibfnamefont {J.}~\bibnamefont {{Zhang}}}, \bibinfo {author}
  {\bibfnamefont {N.}~\bibnamefont {{Li}}}, \bibinfo {author} {\bibfnamefont
  {T.}~\bibnamefont {{Sunayama}}}, \ and\ \bibinfo {author} {\bibfnamefont
  {L.}~\bibnamefont {{Wang}}},\ }\href {\doibase 10.3847/1538-4357/aacaf1}
  {\bibfield  {journal} {\bibinfo  {journal} {\apj}\ }\textbf {\bibinfo
  {volume} {862}},\ \bibinfo {eid} {4} (\bibinfo {year} {2018})},\ \Eprint
  {http://arxiv.org/abs/1712.09030} {arXiv:1712.09030} \BibitemShut {NoStop}%
\bibitem [{\citenamefont {{Cai}}\ \emph {et~al.}(2014)\citenamefont {{Cai}},
  \citenamefont {{Li}}, \citenamefont {{Cole}}, \citenamefont {{Frenk}},\ and\
  \citenamefont {{Neyrinck}}}]{cai2014mnras}%
  \BibitemOpen
  \bibfield  {author} {\bibinfo {author} {\bibfnamefont {Y.-C.}\ \bibnamefont
  {{Cai}}}, \bibinfo {author} {\bibfnamefont {B.}~\bibnamefont {{Li}}},
  \bibinfo {author} {\bibfnamefont {S.}~\bibnamefont {{Cole}}}, \bibinfo
  {author} {\bibfnamefont {C.~S.}\ \bibnamefont {{Frenk}}}, \ and\ \bibinfo
  {author} {\bibfnamefont {M.}~\bibnamefont {{Neyrinck}}},\ }\href {\doibase
  10.1093/mnras/stu154} {\bibfield  {journal} {\bibinfo  {journal} {\mnras}\
  }\textbf {\bibinfo {volume} {439}},\ \bibinfo {pages} {2978} (\bibinfo {year}
  {2014})},\ \Eprint {http://arxiv.org/abs/1310.6986} {arXiv:1310.6986
  [astro-ph.CO]} \BibitemShut {NoStop}%
\bibitem [{\citenamefont {{Lam}}\ \emph {et~al.}(2015)\citenamefont {{Lam}},
  \citenamefont {{Clampitt}}, \citenamefont {{Cai}},\ and\ \citenamefont
  {{Li}}}]{lam2015mnras}%
  \BibitemOpen
  \bibfield  {author} {\bibinfo {author} {\bibfnamefont {T.~Y.}\ \bibnamefont
  {{Lam}}}, \bibinfo {author} {\bibfnamefont {J.}~\bibnamefont {{Clampitt}}},
  \bibinfo {author} {\bibfnamefont {Y.-C.}\ \bibnamefont {{Cai}}}, \ and\
  \bibinfo {author} {\bibfnamefont {B.}~\bibnamefont {{Li}}},\ }\href {\doibase
  10.1093/mnras/stv797} {\bibfield  {journal} {\bibinfo  {journal} {\mnras}\
  }\textbf {\bibinfo {volume} {450}},\ \bibinfo {pages} {3319} (\bibinfo {year}
  {2015})},\ \Eprint {http://arxiv.org/abs/1408.5338} {arXiv:1408.5338
  [astro-ph.CO]} \BibitemShut {NoStop}%
\bibitem [{\citenamefont {{Voivodic}}\ \emph {et~al.}(2017)\citenamefont
  {{Voivodic}}, \citenamefont {{Lima}}, \citenamefont {{Llinares}},\ and\
  \citenamefont {{Mota}}}]{Voivodic2017PhRvD}%
  \BibitemOpen
  \bibfield  {author} {\bibinfo {author} {\bibfnamefont {R.}~\bibnamefont
  {{Voivodic}}}, \bibinfo {author} {\bibfnamefont {M.}~\bibnamefont {{Lima}}},
  \bibinfo {author} {\bibfnamefont {C.}~\bibnamefont {{Llinares}}}, \ and\
  \bibinfo {author} {\bibfnamefont {D.~F.}\ \bibnamefont {{Mota}}},\ }\href
  {\doibase 10.1103/PhysRevD.95.024018} {\bibfield  {journal} {\bibinfo
  {journal} {\prd}\ }\textbf {\bibinfo {volume} {95}},\ \bibinfo {eid} {024018}
  (\bibinfo {year} {2017})},\ \Eprint {http://arxiv.org/abs/1609.02544}
  {arXiv:1609.02544 [astro-ph.CO]} \BibitemShut {NoStop}%
\bibitem [{\citenamefont {{Sahl{\'e}n}}\ and\ \citenamefont
  {{Silk}}(2018)}]{Sahl2018PhRvD}%
  \BibitemOpen
  \bibfield  {author} {\bibinfo {author} {\bibfnamefont {M.}~\bibnamefont
  {{Sahl{\'e}n}}}\ and\ \bibinfo {author} {\bibfnamefont {J.}~\bibnamefont
  {{Silk}}},\ }\href {\doibase 10.1103/PhysRevD.97.103504} {\bibfield
  {journal} {\bibinfo  {journal} {\prd}\ }\textbf {\bibinfo {volume} {97}},\
  \bibinfo {eid} {103504} (\bibinfo {year} {2018})},\ \Eprint
  {http://arxiv.org/abs/1612.06595} {arXiv:1612.06595 [astro-ph.CO]}
  \BibitemShut {NoStop}%
\bibitem [{\citenamefont {{Falck}}\ \emph {et~al.}(2018)\citenamefont
  {{Falck}}, \citenamefont {{Koyama}}, \citenamefont {{Zhao}},\ and\
  \citenamefont {{Cautun}}}]{falck2018mnras}%
  \BibitemOpen
  \bibfield  {author} {\bibinfo {author} {\bibfnamefont {B.}~\bibnamefont
  {{Falck}}}, \bibinfo {author} {\bibfnamefont {K.}~\bibnamefont {{Koyama}}},
  \bibinfo {author} {\bibfnamefont {G.-B.}\ \bibnamefont {{Zhao}}}, \ and\
  \bibinfo {author} {\bibfnamefont {M.}~\bibnamefont {{Cautun}}},\ }\href
  {\doibase 10.1093/mnras/stx3288} {\bibfield  {journal} {\bibinfo  {journal}
  {\mnras}\ }\textbf {\bibinfo {volume} {475}},\ \bibinfo {pages} {3262}
  (\bibinfo {year} {2018})},\ \Eprint {http://arxiv.org/abs/1704.08942}
  {arXiv:1704.08942 [astro-ph.CO]} \BibitemShut {NoStop}%
\bibitem [{\citenamefont {{Davies}}\ \emph {et~al.}(2019)\citenamefont
  {{Davies}}, \citenamefont {{Cautun}},\ and\ \citenamefont
  {{Li}}}]{davies2019mnras}%
  \BibitemOpen
  \bibfield  {author} {\bibinfo {author} {\bibfnamefont {C.~T.}\ \bibnamefont
  {{Davies}}}, \bibinfo {author} {\bibfnamefont {M.}~\bibnamefont {{Cautun}}},
  \ and\ \bibinfo {author} {\bibfnamefont {B.}~\bibnamefont {{Li}}},\ }\href
  {\doibase 10.1093/mnras/stz2933} {\bibfield  {journal} {\bibinfo  {journal}
  {\mnras}\ }\textbf {\bibinfo {volume} {490}},\ \bibinfo {pages} {4907}
  (\bibinfo {year} {2019})},\ \Eprint {http://arxiv.org/abs/1907.06657}
  {arXiv:1907.06657 [astro-ph.CO]} \BibitemShut {NoStop}%
\bibitem [{\citenamefont {Unnikrishnan}\ \emph {et~al.}(2008)\citenamefont
  {Unnikrishnan}, \citenamefont {Jassal},\ and\ \citenamefont
  {Seshadri}}]{Unnikrishnan:2008qe}%
  \BibitemOpen
  \bibfield  {author} {\bibinfo {author} {\bibfnamefont {S.}~\bibnamefont
  {Unnikrishnan}}, \bibinfo {author} {\bibfnamefont {H.~K.}\ \bibnamefont
  {Jassal}}, \ and\ \bibinfo {author} {\bibfnamefont {T.~R.}\ \bibnamefont
  {Seshadri}},\ }\href {\doibase 10.1103/PhysRevD.78.123504} {\bibfield
  {journal} {\bibinfo  {journal} {Phys. Rev.}\ }\textbf {\bibinfo {volume}
  {D78}},\ \bibinfo {pages} {123504} (\bibinfo {year} {2008})},\ \Eprint
  {http://arxiv.org/abs/0801.2017} {arXiv:0801.2017 [astro-ph]} \BibitemShut
  {NoStop}%
\end{thebibliography}%
\end{document}